\documentclass[12pt,a4paper,twoside]{book}

\usepackage{etex} 

\usepackage[nottoc,numbib]{tocbibind}

\usepackage{amsmath,amssymb,calc,ifthen,capt-of,mathrsfs,eufrak}
\usepackage{multirow} 
\usepackage{makecell} 

\usepackage[raggedright]{titlesec} 

\usepackage[nomessages]{fp}

\usepackage[ampersand]{easylist}

\usepackage[table,usenames,dvipsnames]{xcolor} 

\usepackage{epsf}

\definecolor{blue3}{HTML}{86B7FC} 
\definecolor{blue1}{HTML}{B5F1FF} 
\definecolor{blue2}{HTML}{E0F9FF} 
\definecolor{blue4}{HTML}{0047bc} 

\usepackage[bottom]{footmisc} 

\usepackage[pdftex,
            pdfauthor={Razvan V. Marinescu},
            pdftitle={Modelling the Neuroanatomical Progression of Alzheimer's Disease and Posterior Cortical Atrophy},
            pdfsubject={Modelling the Neuroanatomical Progression of Alzheimer's Disease and Posterior Cortical Atrophy},
            pdfkeywords={Alzheimer's disease, Bayesian Modelling, Disease Progression Model}]{hyperref}
            
\hypersetup{
    colorlinks,
    citecolor=blue4,
    filecolor=black,
    linkcolor=blue4,
    urlcolor=blue4
}

\usepackage{pgfplots}
\usetikzlibrary{plotmarks}
\usepackage{amsmath,graphicx}
\usepackage{epstopdf}
\usepackage{caption}
\usepackage{subcaption}
\usepackage{float}
\usepackage[table]{xcolor}

\usepackage{mathtools}

\usepackage{arydshln} 
\usepackage{pdfpages} 

\usepackage{multicol} 

\usepackage{listings} 

\usepackage[toc,page]{appendix}

\usepackage{scalefnt} 

\pgfplotsset{compat = newest}

\usepackage{color}
\newcommand{\hilight}[1]{}

\newcommand{\HRule}{\rule{\linewidth}{0.5mm}}

\usepackage{array,booktabs}

\usepackage[export]{adjustbox}

\usepackage{pifont}

\usepackage[margin=1in,headheight=14pt]{geometry}

\usepackage[linesnumbered]{algorithm2e}

\SetCommentSty{mycommfont}

\usepackage{enumitem} 
\usepackage{pgfgantt} 
\usepackage{rotating}
\usepackage[graphicx]{realboxes}

\usepackage{pdflscape}

\newganttchartelement*{mymilestone}{
mymilestone/.style={
shape=isosceles triangle,
inner sep=0pt,
draw=cyan,
top color=white,
bottom color=cyan!50
},
mymilestone incomplete/.style={
/pgfgantt/mymilestone,
draw=yellow,
bottom color=yellow!50
},
mymilestone label font=\slshape,
mymilestone left shift=0pt,
mymilestone right shift=0pt
}

\newgantttimeslotformat{stardate}{%
\def\decomposestardate##1.##2\relax{%
\def\stardateyear{##1}\def\stardateday{##2}%
}%
\decomposestardate#1\relax%
\pgfcalendardatetojulian{\stardateyear-01-01}{#2}%
\advance#2 by-1\relax%
\advance#2 by\stardateday\relax%
}

\usepackage{tikz}
\usetikzlibrary{arrows,positioning, shapes.symbols,shapes.callouts,patterns,shapes,chains,calc,backgrounds,fadings}
\tikzstyle{state}=[circle,thick,draw=black, align=center, minimum size=2.1cm,
inner sep=0]
\tikzstyle{vertex}=[circle,thick,draw=black]
\tikzstyle{vertex2}=[circle,thick,draw=black,minimum size=0.3cm]
\tikzstyle{vertex3}=[circle,thick,draw=white,minimum size=0.3cm,fill=white]
\tikzstyle{terminal}=[rectangle,thick,draw=black]
\tikzstyle{edge} = [draw,thick]
\tikzstyle{lo} = [edge,dotted]
\tikzstyle{hi} = [edge]
\tikzstyle{trans} = [edge,->]
\tikzstyle{image}=[circle,thick,draw=black]

\usepackage[section]{placeins} 

\usepackage{color}

\usepackage{silence} 

\definecolor{mygreen}{rgb}{0,0.6,0}
\definecolor{mygray}{rgb}{0.5,0.5,0.5}
\definecolor{mymauve}{rgb}{0.58,0,0.82}

\usepackage{framed}
\usepackage{amsthm}
\usepackage{thmtools}

\definecolor{lightgray}{gray}{0.00}
\renewenvironment{leftbar}[1][\hsize]{%
  \MakeFramed{\hsize#1\advance\hsize-\width\FrameRestore}%
}
{
  \endMakeFramed%
}
\usepackage{tabularx, booktabs} 
\newcolumntype{Y}{>{\centering\arraybackslash}X}

\DeclareMathOperator*{\argmin}{arg\,min}
\DeclareMathOperator*{\argmax}{arg\,max}

\usepackage{fancyhdr}

\newcommand{\FrontPageStyle}{\pagestyle{empty}}
\newcommand{\MainPageStyle}{\pagestyle{main}}

\fancypagestyle{plain}{%
  \fancyhf{}%
}

\fancypagestyle{main}{%
  \fancyhf{}%
  \fancyhead[RE]{\nouppercase{\leftmark}}%
  \fancyhead[LO]{\nouppercase{\rightmark}}%
  \fancyhead[LE,RO]{\thepage}
}

\newcommand*{\pcaPaperFigs}{./pcaPaperFigs}

\newcommand{\voxDpmFolder}{./voxelwiseDPM/}

\newcommand\ci{\perp\!\!\!\perp} 
\newcommand{\Mu}{M}
\newcommand\independent{\protect\mathpalette{\protect\independenT}{\perp}}
\def\independenT#1#2{\mathrel{\rlap{$#1#2$}\mkern2mu{#1#2}}}

\newcommand{\beq}{\begin{equation}}
\newcommand{\eeq}{\end{equation}}

\DeclareSymbolFont{matha}{OML}{txmi}{m}{it}
\DeclareMathSymbol{\varv}{\mathord}{matha}{118}

\usepackage{filecontents}

\begin{document}
\belowdisplayskip=12pt plus 3pt minus 9pt
\belowdisplayshortskip=7pt plus 3pt minus 4pt

\sloppy

\FrontPageStyle{}

\begin{titlepage}
\begin{center}

%
%

{\Large 

\HRule \\[0.4cm]
{ \LARGE Modelling the Neuroanatomical Progression of Alzheimer's Disease and Posterior Cortical Atrophy\\[0.4cm] }

\HRule \\[1.5cm]

\begin{minipage}{0.4\textwidth}
\begin{flushleft} \Large
\emph{Author:}\\
R\u{a}zvan V. \textsc{Marinescu}
\end{flushleft}
\end{minipage}
\begin{minipage}{0.5\textwidth}
\begin{flushright} \Large
\emph{Supervisors:} \\
Prof. Daniel C. \textsc{Alexander}\\
Dr. Sebastian \textsc{Crutch}\\
Dr. Neil P. \textsc{Oxtoby} 
\end{flushright}
\end{minipage}

\vfill

\includegraphics[height=7cm]{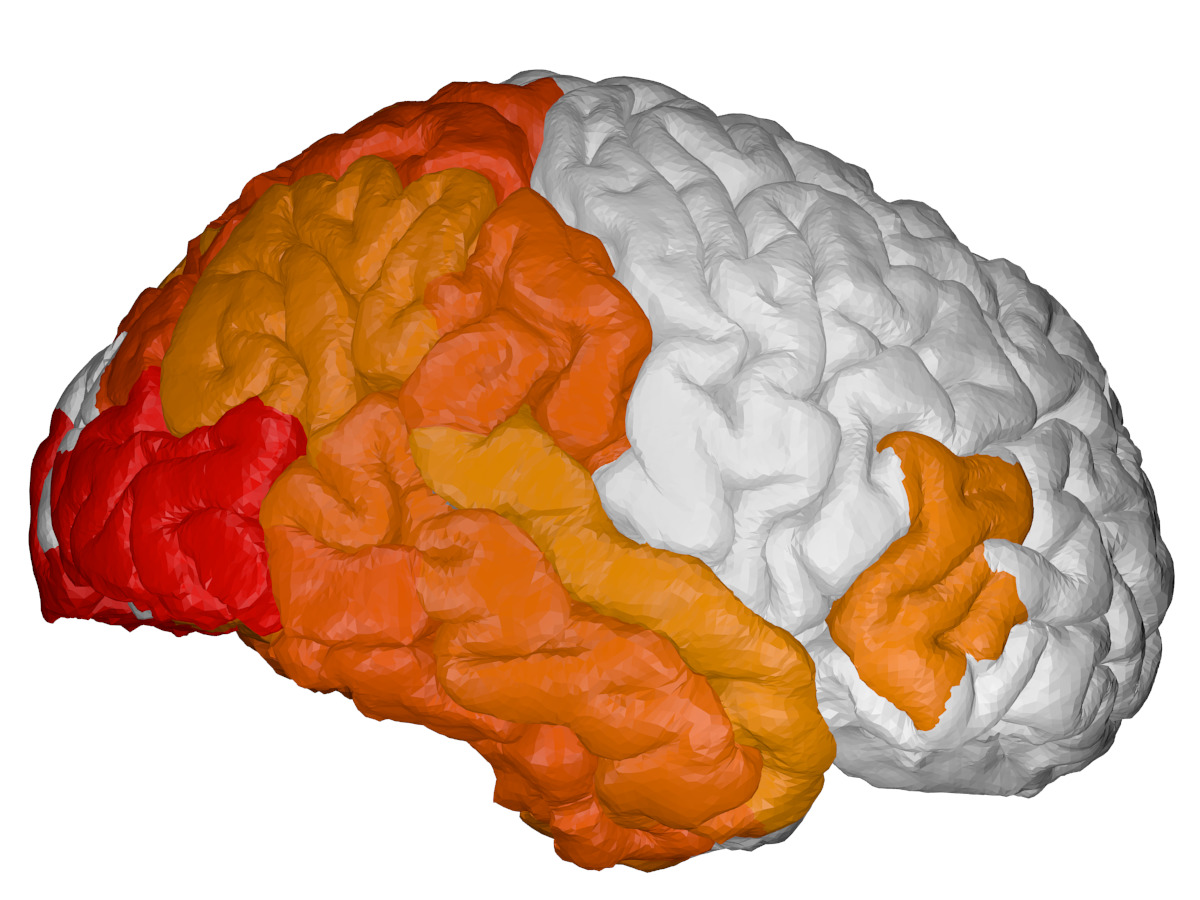}

\vfill
 
A dissertation submitted in partial fulfillment \\
of the requirements for the degree of\\[0.3cm]
\textbf{Doctor of Philosophy}\\[0.3cm]
of\\[0.3cm]
\textbf{University College London}\\[1cm]

Centre for Medical Image Computing, University College London
 
 \vfill
%
%

{\Large Defense date: January 23, 2019} 

}
\end{center}
\end{titlepage}

\clearpage

I, R\u{a}zvan Valentin Marinescu, confirm that the work presented in this thesis is my own. Where information has been derived from other sources, I confirm that this has been indicated in the thesis.

\clearpage

\chapter*{Abstract}

In order to find effective treatments for Alzheimer's disease (AD), a devastating neurodegenerative disease affecting millions of people worldwide, we need to identify subjects at risk of AD as early as possible. To this end, recently developed disease progression models can be used to perform early diagnosis, as well as predict the subjects' disease stages and future evolution. However, these models have not yet been applied to rare neurodegenerative diseases, are not suitable to understand the complex dynamics of biomarkers, work only on large multimodal datasets, and their predictive performance has not been objectively validated.

In this work I developed novel models of disease progression and applied them to estimate the progression of Alzheimer's disease and Posterior Cortical atrophy, a rare neurodegenerative syndrome causing visual deficits. My first contribution is a study on the progression of Posterior Cortical Atrophy, using models already developed: the Event-based Model (EBM) and the Differential Equation Model (DEM). My second contribution is the development of DIVE, a novel spatio-temporal model of disease progression that estimates fine-grained spatial patterns of pathology, potentially enabling us to understand complex disease mechanisms relating to pathology propagation along brain networks. My third contribution is the development of Disease Knowledge Transfer (DKT), a novel disease progression model that estimates the multimodal progression of rare neurodegenerative diseases from limited, unimodal datasets, by transferring information from larger, multimodal datasets of typical neurodegenerative diseases. My fourth contribution is the development of novel extensions for the EBM and the DEM, and the development of novel measures for performance evaluation of such models. My last contribution is the organization of the TADPOLE challenge, a competition which aims to identify algorithms and features that best predict the evolution of AD. 

\chapter*{Impact Statement}
\thispagestyle{empty}

The work presented in this thesis furthers our understanding of the temporal evolution of Posterior Cortical Atrophy and Alzheimer's disease. The disease progression models and evaluation techniques that we developed can help towards understanding underlying disease mechanisms, aid patient stratification and drug evaluation in clinical trials for Alzheimer's disease and Posterior Cortical Atrophy, and can be used in clinical practice for predicting the future evolution of subjects that are at risk of developing Alzheimer's disease. 

I published the work in this PhD thesis in two first-author papers (DIVE and TADPOLE chapters), and will soon submit another two papers (DKT and PCA chapters). I have also communicated my results in international conferences. I have also engaged with the broader scientific community by organising the TADPOLE Challenge, as well as a couple of hackathons at the PyConUK conference and the CMIC Summer School.


\chapter*{Acknowledgements}
\thispagestyle{empty}


There are many great people who have helped my PhD project become reality. First of all, I'd like to thank my supervisor Daniel Alexander, for his great advice, ideas and research directions. He has always encouraged me to pursue interesting ideas and supported me in developing them. Secondly, I'd also like to thank Alexandra Young and Neil Oxtoby for teaching me disease progression modelling, especially in the early years of my PhD. I'd also like to thank Sebastian Crutch, Tim Shakespeare, Keir Yong, and other DRC collaborators, for their help and advice on Posterior Cortical Atrophy and other clinical aspects of my work. I'd like to thank Marco Lorenzi, for trying to explain mathematics to a wanna-be mathematician like myself. Marco and Neil are also great guitar players, which I had the opportunity to hear a few times. I'd also like to thank Sara Garbarino, for her great spirit, for taking the time to repeatedly listen to my presentations when rehearsing them, and for reminding me that I was probably the biggest nerd in CMIC. I'd further like to thank the POND group, for the help they offered me throughout my PhD, for the great coffees we had after our meetings, and for reminding me that I can't deal with non-working technology in hotels during our trips in the Netherlands. I'd like to thank Gary Zhang for coaching me on how to present my work without putting half of the audience to sleep, as well as others in MIG and CMIC, for teaching me about diffusion MRI, machine learning and other imaging techniques. I remember coming to those meetings in early days of my PhD and not understanding what was being discussed.

In terms of the social aspect, I had a wonderful time at UCL. I'll miss the trips organised by Pawel Markiewicz around Wales and Cornwall, where we had a lot of fun surfing, playing frisbee and BBQ-ing on the beach. I'll also miss the great camping trips with the CMIC folks in Peak District and Lake district, when I attempted driving -- successfully! -- for the first time in the UK! I'll also miss the great time I had with Thore Bucking, Emma Hill and Kin Quan during the MRes year. I'll miss the dinners and lunches such as the EuroPOND celebratory lunch, when we got so excited that we each ordered 4 glasses of champagne, which got me tipsy. When we came back to UCL after lunch I realised I was actually breaking the code instead of doing anything useful.

Finally, I'd like to thank my parents, Aurora and Dan Marinescu, for their love and support, without which I wouldn't have been able to start the PhD in the first place. My brother Robert Marinescu, for his funny jokes and good spirit. My grandmother Anghelut\u{a} Constantina, for her funny and charismatic character. And my friends and housemates, in particular Vibhav Mishra, Carlos Gavidia and Mikael Brudfors, for the wonderful time spent in the Ifor residence, as well as Georgiana Ghetie, Alexandru Barbu and Oana Lang, for their light-hearted spirit, conversations and for the fun we had in the last few months of my PhD. 

\clearpage

\cleardoublepage{}
\MainPageStyle{}

\setcounter{tocdepth}{2}
\tableofcontents

\setcounter{tocdepth}{1}
\listoffigures

\listoftables

\newcommand{\pxgs}{\begin{equation}
  p(X|S) = \prod_{j=1}^J \left[ \sum_{k=0}^N p(k) \left( \prod_{i=1}^k p\left(x_{s(i),j} | E_{s(i)} \right) \prod_{i=k+1}^N p\left(x_{s(i),j} | \neg E_{s(i)}\right) \right) \right]
  \end{equation}}

\newcolumntype{C}[1]{>{\centering\let\newline\\\arraybackslash\hspace{0pt}}m{#1}}

\setlength{\tabcolsep}{0.2em}
  
\chapter{Introduction}
\label{chapter:intro}

\section{Alzheimer's Disease}

Alzheimer's disease (AD) is a chronic progressive neurodegenerative disorder that accounts for 60\% to 70\% of all cases of dementia worldwide \cite{Burns2009,world2013dementia}. In 2010 it was estimated that up to 35 million people worldwide suffered from AD \cite{world2013dementia}. Its symptoms include cognitive dysfunction such as memory loss, language difficulties and psychiatric symptoms such as depression, hallucinations, delusions and agitation. Diagnosis is usually based on the person's medical history, information from relatives and behavioural observations.

In terms of neuroimaging, Magnetic Resonance Imaging (MRI) shows early atrophy in the medial temporal lobes and fusifom gyrus, which then spreads to the posterior temporal lobe, parietal lobe, and finally to the frontal lobe \cite{whitwell2010progression}, with relative sparing of the sensorimotor cortex, visual cortex and the cerebellum. Imaging with Positron Emission Tomography (PET) shows reduced metabolism and increased uptake of amyloid proteins \cite{marcus2014brain}. The underlying disease mechanisms are currently not well understood -- it is currently believed that initial abnormalities in the folding of amyloid-$\beta$ and/or tau proteins leads to a cascade of events which results in neurodegeneration and cognitive decline \cite{mudher2002alzheimer}. These are known as the amyloid and tau hypotheses \cite{mudher2002alzheimer}.

There are no treatments that can stop or at least slow down cognitive decline, because all clinical trials so far have failed to prove any disease modifying effect \cite{mudher2002alzheimer}. One of the reasons why clinical trials have failed might be due to a lack of understanding of the underlying mechanisms, which results in wrong drug targets \cite{mehta2017trials}. For example, within the amyloid and tau hypotheses, it is not precisely understood what is the exact process underlying the formation of the misfolded amyloid and tau and what might be the cause of their misfolding \cite{mudher2002alzheimer}. Another reason why clinical trials in AD are believed to have failed is the late administration of the treatment to patients who were already in the symptomatic stage \cite{mehta2017trials}. It is currently believed that for clinical trials to be successful in AD, we need to fully understand the underlying disease mechanisms, in order to identify the right drug targets, and to administer the treatments early in the pre-symptomatic stages, and to the right subjects who will otherwise develop dementia in the future \cite{mehta2017trials}.

\section{Posterior Cortical Atrophy}

Alzheimer's disease is a very heterogeneous disease, which has been observed both clinically, with amnestic, visual, executive and aphasic types \cite{galton2000atypical} as well as pathologically, with hippocampal sparing and limbic predominant cases reported in the literature \cite{murray2011neuropathologically}. This heterogeneity can help us understand disease causes and underlying mechanisms, and identify risk- and protective-factors. For example, it has been observed that different speeds of progression can be due to differences in amyloid-$\beta$ fibrils among subjects \cite{qiang2017structural}. Another example is that different ages of onset in familial AD are associated with different underlying mutations in the PSEN1 gene \cite{larner2006clinical}.  

A notable example of phenotypic heterogeneity in Alzheimer's disease is given by Posterior Cortical Atrophy (PCA). PCA, also called Benson's syndrome \cite{benson1988posterior}, is a neurodegenerative disease similar to AD that results in disruptions of the visual and motor systems. Early symptoms include blurred vision, inability to read, difficulty with depth perception and problems navigating through space \cite{crutch2012posterior, borruat2013posterior}, while late-stage symptoms can include inability to recognise familiar faces and objects as well as visual hallucinations. Neuroanatomically, PCA is characterised by atrophy in the superior parietal, occipital and posterior temporal regions \cite{lehmann2011cortical, whitwell2007imaging}. However, due to the rarity of the disease, only a limited number of small studies have been done in PCA \cite{crutch2012posterior}. 

\section{Disease Progression Models}

For both PCA and typical AD (tAD), in order to understand the underlying disease mechanisms and to select the right subjects for clinical trials, we need to quantitatively map their longitudinal evolution. To this end, many biomarkers can be used, which are based on Magnetic Resonance Imaging (e.g. brain volumes, cortical thickness), Positron Emission Tomography (e.g. measures of hypometabolism, concentrations of amyloid and tau proteins), samples from cerebrospinal fluid (CSF) (e.g. concentrations of various molecular markers) or neuropsychiatric tests.  However, no single biomarker is sufficient for accurate staging and subject prediction, as they are not specific to one disease and can result in misdiagnosis, can be influenced by variability not related to the disease (e.g. the cognitive reserve theory \cite{stern2012cognitive}), show changes only in limited time windows, and have inherent noise. Therefore, holistic, quantitative models called \emph{disease progression models} are needed, which integrate a variety of biomarker data to estimate the subjects' disease stage and future evolution.

A hypothetical model of disease progression has been proposed by \cite{jack2010hypothetical}, describing the trajectory of key biomarkers along the progression of Alzheimer's disease. The model suggests that amyloid-beta and tau biomarkers become abnormal long before symptoms appear, followed by neurodegeneration and cognitive decline. Motivated by this idea, several data-driven disease progression models have been proposed, that reconstruct biomarker trajectories and can be used to stage subjects. One such model is the Event-Based Model \cite{fonteijn2012event, young2014data}, which estimates the progression of the disease as a sequence of discrete events, representing underlying biomarkers switching from a normal to abnormal state. Another model, the Differential Equation Model (DEM) \cite{villemagne2013amyloid}, reconstructs a continuous trajectory of biomarker measurements from changes in short-term follow-up data, which represent samples of the slope at different points along the trajectory. Other models such as the Disease Progression Score (DPS) \cite{jedynak2012},  Self-Modelling Regression \cite{donohue2014estimating} or Riemannian manifold techniques \cite{schiratti2015mixed} have been developed, that build continuous trajectories by "stitching" together short-term follow-up data.

While these models have shown great promise at identifying the earliest events in the Alzheimer's disease cascade \cite{young2014data, iturria2016early}, mapping the heterogeneity within Alzheimer's disease \cite{young2018uncovering} and showed increased performance in predictions compared to standard approaches \cite{oxtoby2018}, they have some limitations that need to be addressed. First of all, they have not been applied to some rare neurodegenerative diseases such as Posterior Cortical Atrophy. Secondly, they are not suitable for modelling the complex dynamics of biomarkers. This is because they work on extracted features, which generally lack important information present in the brain's morphology; also, they cannot exploit biomarker relationships shared across related diseases. Third, it is not yet clear how to measure the performance of such models, and no previous literature study has been done to establish the comparative performance of such models at different prediction tasks.

\section{Problem Statement}

In the field of Alzheimer's disease progression, there are several issues that need to be addressed:
\begin{itemize}
\item The longitudinal neuroanatomical progression of Posterior Cortical Atrophy has not been quantified in a comprehensive study.
\item Current disease progression models are not appropriate for modelling the complex dynamics of biomarker measurements.
\item The comparative performance of different models of disease prediction is yet to be established.
\end{itemize}

The work I present in this thesis tries to address these three aspects.

\section{Justification}

\subsection{Longitudinal Modelling of Posterior Cortical Atrophy}

The longitudinal neuroanatomical progression of Posterior Cortical Atrophy has not been quantified in a comprehensive study so far. Several case studies have been published, which described the brain pathological progression of PCA \cite{ross1996progressive, goethals2001posterior, giovagnoli2009neuropsychological, chang2015substance, kennedy2012visualizing, crutch2017consensus}. The only longitudinal study of PCA \cite{lehmann2012global} showed widespread gray matter loss in both PCA and tAD. However, the numbers were small (17 PCA and 16 tAD) and the time interval was short (1 year). Larger longitudinal studies are therefore required to robustly estimate the progression of brain pathology in PCA, which is important for understanding underlying disease mechanisms and for stratification of subjects clinical trials.

\subsection{Current Disease Progression Models Cannot Model Complex Dynamics}

Current disease progression models are not appropriate for modelling the complex dynamics of biomarker measurements. For example, many models such as the event-based model or the differential equation model cannot be applied to voxelwise biomarker data such as amyloid load or hypometabolism from PET, or cortical thickness/compression maps from MRI. While this can be mitigated by averaging these measures over pre-defined regions of interest, it has been shown that patterns of pathology in different types of dementia are dispersed and disconnected, as they follow underlying brain networks \cite{seeley2009neurodegenerative}. In order to study the link between neuroanatomical pathology and brain networks, we need to develop spatio-temporal models of disease progression that account for changes over the brain structure, as well as over the disease timeline. Such spatio-temporal models can help us understand more complex disease mechanisms and enable more accurate predictions of disease risk, which can aid stratification in clinical trials. 

Another limitation of current disease progression models is that it is challenging to apply them to study rare types of dementia such as PCA. These models generally require large multimodal datasets which are often not available for rare dementias. Therefore, there is a need to develop models that can transfer information from larger multimodal datasets. In particular for PCA, these transfer-learning approaches can enable us to estimate robust, multimodal biomarker trajectories, and to make more accurate predictions for each subject.

\subsection{Comparative Performance of Different Disease Progression Models}

The comparative performance of different models of disease prediction is yet to be established. More precisely, there has not been any study comparing the performance of algorithms and features at longitudinal prediction of subjects at risk of AD. While these questions are generally answered in the medical image community through grand challenges, most challenges so far have focused on classification of clinical diagnosis. For example, the recent CADDementia challenge \cite{bron2015standardized} aimed to predict clinical diagnosis from MRI scans, while a similar challenge, the "\emph{International challenge for automated prediction of MCI from MRI data}" \cite{sarica2018machine}, asked participants to predict diagnosis and conversion status from extracted MRI features. While these challenges are helpful in establishing which algorithms are best at predicting biomarkers at the current timepoint, they cannot identify algorithms that are best at predicting the continuous progression of subjects at risk of AD.

\section{Thesis Contributions}

In this thesis I contributed to the three key aspects mentioned above. My key contributions for each chapter are described in the following sections.

\subsection{Longitudinal Neuroanatomical Progression of Posterior Cortical Atrophy}

\begin{itemize}
\item I performed the first comprehensive study of longitudinal atrophy progression in Posterior Cortical Atrophy, and compared it with the atrophy progression in typical Alzheimer's disease, using data from the Dementia Research Centre (DRC), UK. Previous studies were limited to case series, or used small numbers of patients over short time-frames (1-year interval).
\item I estimated the ordering in which brain regions show volume reductions using the event-based model, and also estimated the rate and extent of volume loss using the differential equation model. I contrasted these between PCA and tAD, and showed differences both qualitatively and quantitatively, which were further supported by statistical tests.
\item I showed that three cognitively-defined PCA subgroups show different phenotype-specific patterns of early atrophy. This was the first study to show quantitative evidence of heterogeneity within PCA.
\end{itemize}

\subsection{DIVE: A Spatiotemporal Progression Model of Brain Pathology in Neurodegenerative Disorders}

\begin{itemize}
\item I developed DIVE, a novel spatiotemporal model that estimates fine-grained patterns of pathology at every point on the cortical surface, while also accounting for subject-specific time-shifts
\item I validated DIVE on simulations, in presence of ground truth. More precisely, I showed that DIVE can accurately estimate the true cluster assignments of each simulated vertex, biomarker trajectories and subject-specific time-shifts.
\item On patient data, I showed that DIVE estimates similar spatial patterns of pathology in two independent typical AD datasets: the Alzheimer's Disease Neuroimaging Initiative (ADNI) and the Dementia Research Centre (DRC), UK.
\item On patient data, I showed that DIVE estimates different spatial patterns of pathology for distinct diseases (typical Alzheimer's Disease vs Posterior Cortical Atrophy) and distinct imaging modalities (MRI vs PET).
\item I further validated DIVE on patient data, showing that it is robust under cross-validation and that the subjects' latent time-shifts, derived only from imaging data, are clinically meaningful as they correlate with four different cognitive tests.
\item I showed that DIVE has better or similar performance compared to standard approaches. 
\end{itemize}
 
\subsection{Disease Knowledge Transfer across Neurodegenerative Diseases}
\begin{itemize}
\item I developed DKT, a novel disease progression model that estimates multimodal biomarker progressions in rare neurodegenerative diseases even when only limited, unimodal data is available, by transferring information from larger multimodal datasets from common neurodegenerative diseases.
\item I validated DKT in a simulation in the presence of ground truth, where I showed that it can accurately estimate biomarker trajectories in one disease, where there is a complete lack of such data, by exploiting correlations with other known biomarkers.
\item I demonstrated DKT on Alzheimer's variants, where I showed it is able to infer plausible non-MRI biomarker trajectories in a rare dementia, i.e. Posterior Cortical Atrophy, by transferring such knowledge from a larger dataset of typical Alzheimer's disease.
\item I showed that DKT has favourable performance compared to standard models.
\end{itemize}
 
\subsection{Novel Extensions to the Event-based Model and Differential Equation Model}
\begin{itemize}
 \item I made novel extensions to the methodology of two disease progression models, the event-based model (EBM) and the differential equation model (DEM), which enable better estimation of their parameters. 
 \item I developed four novel performance metrics that were used to assess the performance of all the models evaluated. 
 \item I showed that the extended models had better or similar performance compared to the standard models.
 \item My results also indicate that the novel performance metrics are more sensitive than standard approaches based on the prediction accuracy of clinical diagnosis. 
\end{itemize} 
 
\subsection{TADPOLE Challenge: Prediction of Longitudinal Evolution in Alzheimer's Disease}
\begin{itemize}
\item I helped organise the TADPOLE Challenge, which aims to find algorithms and features that best predict the evolution of subjects at risk of Alzheimer's disease.
\item I helped build the website and I created the main training dataset.
\item I built a leaderboard system that enabled live evaluation of participants' submissions based on existing data.
\item I promoted the competition at medical imaging conferences, and I organised two TADPOLE mini-challenges, during the PyConUK 2017 conference and during the CMIC Medical Imaging Summer School, 2018.
\end{itemize}
 
\section{Thesis Structure}

The thesis has the following structure:
\begin{itemize}
 \item Chapter \ref{chapter:bck} contains background information on Alzheimer's disease and Posterior Cortical Atrophy.
 \item Chapter \ref{chapter:bckDpm} contains background information on disease progression models.
 \item Chapter \ref{chapter:pca} contains the clinical analysis regarding the progression of Posterior Cortical Atrophy as compared to typical Alzheimer's disease. 
  \item Chapter \ref{chapter:perf} presents novel extensions in the event-based model and differential equation model, which are evaluated against
  \item Chapter \ref{chapter:dive} presents the DIVE model formulation and results on four different datasets, along with model validation.
 \item Chapter \ref{chapter:dkt} presents the DKT model formulation, along with results on simulated data and patient data, and model validation.
 standard implementations based on performance metrics that I proposed.
 \item Chapter \ref{chapter:tadpole} presents the design of the TADPOLE Challenge.
 \item Chapter \ref{chapter:conclusions} presents a summary of the work in this thesis, and proposes directions for further research.
 \end{itemize}

\chapter{Background -- Alzheimer's Disease}
\label{chapter:bck}

\section{Alzheimer's Disease}
\label{sec:bckAd}

\begin{figure}
\centering
\includegraphics[width=0.7 \textwidth]{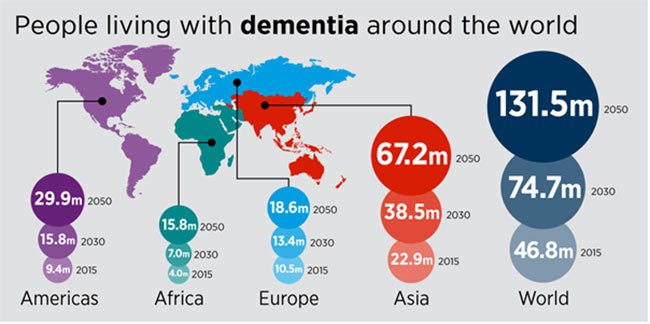}
\caption[Prevalence of dementia around the world]{Prevalence of dementia around the world, along with forecasts for 2030 and 2050. Source: \url{http://www.worldalzreport2015.org/}}
\label{fig:adPrevalence}
\end{figure}

Alzheimer's disease (AD) is a chronic progressive neurodegenerative disease that affects more than 35 million people worldwide \cite{querfurth2010mechanisms}, and this number is expected to triple by 2050 (Fig \ref{fig:adPrevalence}). Alzheimer's disease is the most common cause of dementia, accounting for 60\% to 70\% of the total cases of dementia \cite{Burns2009,world2013dementia}. It usually affects people over 65 years of age \cite{Burns2009,world2013dementia}, although early onset forms of the disease also exist. The disease was first described by German psychiatrist Alois Alzheimer's in 1906. The worldwide cost of dementia in 2018 was \$818 billion worldwide, which is more than 1\% of the aggregate global gross domestic product (GDP) \cite{princeglobal}.

\subsection{Symptoms}
\label{sec:bckSym}

Symptoms of AD vary depending on the stage of the disease. Some authors \cite{forstl1999clinical} split the symptoms into several categories: pre-dementia stage, mild dementia, moderate dementia and severe dementia. 

\subsubsection{Pre-dementia Phase}

In the pre-dementia stage, the first symptoms are usually attributed to stress and ageing. Careful neuropsychological investigations may reveal very mild cognitive impairment five years before the establishment of clinical diagnosis \cite{forstl1999clinical}. The performance of complex tasks might be reduced, and alterations of behaviour including social withdrawal and depressive dysphoria might also be already present \cite{forstl1999clinical}. 

\subsubsection{Mild Dementia Stage}

In the mild dementia stage, significant impairment of learning and memory are present \cite{forstl1999clinical}. However, short-term and implicit memory are less affected compared to declarative memory. Neuropsychological tests can reveal problems with object naming \cite{chobor1990semantic,locascio1995cognitive}, semantic difficulties with word generation \cite{chobor1990semantic,locascio1995cognitive} and inability to draw figures (i.e. constructional apraxia) \cite{moore1984drawing}. Non-cognitive disturbances are also present at this stage \cite{haupt1992psychopathologische}, where depression has been observed in these mild stages \cite{burns1990psychiatric}.

\subsubsection{Moderate Dementia Stage}

At the moderate dementia stage, the predominant features are severe short-term memory impairment \cite{beatty1988retrograde}, along with difficulties in logical reasoning, planning, language \cite{romero1995pragmatische}, reading \cite{cummings1986pattern} and writing \cite{neils1989descriptive}. More complex actions and activities such as using household appliances, dressing and eating are gradually lost. Vision-related symptoms triggered by cognitive deficits also develop, such as spatial disorientation, inability to recognise familiar faces or illusionary misidentification \cite{reisberg1996behavioral}. Around 20\% of patients also experience visual hallucinations, which may be associated with cholinergic deficits \cite{perry1990visual}.

Patients at this stage cannot survive in their community without help from caregivers. However, hospital or nursing home admission can be delayed if there is a good support system in place at the patient's home. 

\subsubsection{Severe Dementia Stage}

Specific cognitive dysfunctions cannot be disentangled at this stage, due to widespread cognitive deficits. Language is reduced to simple phrases. However, emotional signals can still be received and returned \cite{forstl1999clinical}. Patients need support for performing basic functions such as eating. 

The average life expectancy after clinical diagnosis is between three to nine years, although the speed of progression can vary \cite{querfurth2010mechanisms}. Pneumonia, myocardial infarction and septicaemia are the most frequent causes of death at this stage.

\subsection{Disease Causes and Mechanisms}
\label{sec:bckCau}

The causes for AD are poorly understood and around 70\% of them are thought to be genetic, with many genes involved which include APOE, GSK3$\beta$ and DYRK1A\cite{Ballard2011alzheimers}. Over the last few decades, several hypotheses have been proposed to explain the mechanisms of AD: amyloid hypothesis, tau hypothesis, cholinergic hypothesis and neurovascular hypothesis.

\subsubsection{Amyloid Hypothesis}
\label{sec:bckAmyHyp}

\begin{figure}
\centering
\includegraphics[width=0.6\textwidth]{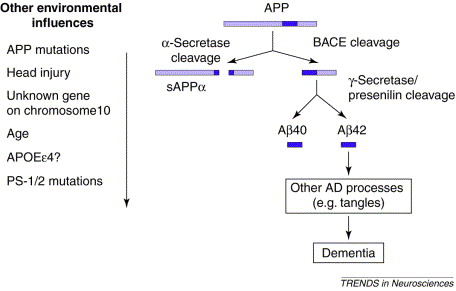}
\caption[Diagram showing the amyloid hypothesis]{Diagram showing the amyloid hypothesis. Amyloid precursor protein is split by $\alpha$-secretase resulting in sAPP$\alpha$, which might have a neuroprotective role. On the other hand, splitting by $\beta$-amyloid cleaving enzyme (BACE) results in amyloid-$\beta$, of which amyloid-$\beta$42 is more prone to self-aggregate and lead to pathogenesis. On the left, many other factors are shown that are believed to influence this pathway and lead to more pathology. Reproduced with permission from \cite{mudher2002alzheimer}.}
\label{fig:bckAmyloidHypothesis}
\end{figure}

In 1991, Hardy et al. \cite{hardy1991amyloid} postulated that amyloid-$\beta$ deposits are a central cause in the development of AD. The amyloid-$\beta$ protein, derived from the amyloid precursor protein (APP), is processed via two distinct pathways: the amyloidogenic pathway which produces amyloid-$\beta$ proteins (Fig \ref{fig:bckAmyloidHypothesis}) and the non-amyloidogenic pathway which prevents the formation of amyloid-$\beta$ and instead produces a secreted form of APP called sAPP$\alpha$ \cite{mudher2002alzheimer}. The amyloid hypothesis states that dysregulation in APP processing occurs early in the disease process, causing increased production of the more toxic amyloid-$\beta$42 protein, which aggregates into plaques \cite{mudher2002alzheimer}. The misfolded amyloid-$\beta$ then causes a chain of events leading to cognitive impairment, including tau aggregation, phosphorylation, neuronal damage and brain atrophy. However, the underlying mechanisms through which amyloid-$\beta$ induces neurodegeneration are not clear \cite{mudher2002alzheimer}. 

Different sources of evidence exist to support the amyloid hypothesis. First of all, mutations in the APP gene cause a rare, early-onset form of familial Alzheimer's disease which corresponds clinically and pathologically to AD. This suggests that changes in APP are an upstream \footnote{happening early in the chain of events leading to AD} event in the pathological cascade leading to AD \cite{mudher2002alzheimer}. Moreover, a locus on chromosome 10 which is linked to late onset AD is also associated with increased amyloid-beta production \cite{ertekin2000linkage}. 

Several studies have also established a clear link between amyloid and tau toxicity, another early event in the AD cascade \cite{gotz2001formation, lewis2001enhanced, roberson2007reducing, bloom2014amyloid}. Amyloid has been shown to enhance tau tangle formation in several mice studies \cite{gotz2001formation,lewis2001enhanced}. Evidence also exists that tau pathology is required for amyloid-$beta$ toxicity \cite{roberson2007reducing}, suggesting that there could be a feedback loop between amyloid and tau, or that tau pathology is also required for development of amyloid deficits \cite{bloom2014amyloid}. 

There are several aspects of the amyloid hypothesis that indicate it is not complete. For example, transgenic mouse models carrying the familial AD mutations have showed increases in amyloid toxicity, but no clear evidence of neuronal loss \cite{hsiao1995age,irizarry1997appsw} and tau aggregation as predicted by the amyloid hypothesis \cite{bloom2014amyloid}.

\subsubsection{Tau Hypothesis}
\label{sec:bckTauHyp}

Another key hypothesis about the cause of AD is the tau hypothesis, which  suggests that abnormalities related to the tau proteins initiate the disease cascade \cite{mudher2002alzheimer}. In this case, tau binding to microtubules is disrupted by phosphorylation, which results in free tau that aggregates into neurofibrillary tangles. This ends up destroying the cell's cytoskeleton which collapses the neuron's transport system, ultimately resulting in neuronal death. 

Support for the tau hypothesis was given by the fact that tau proteins aggregate and accumulate within neuronal cells and ultimately cause their death. Moreover, the number of tau tangles has been shown to correlate with cognitive decline \cite{nagy1995relative}, especially in memory-related areas \cite{braak1998evolution,braak1994sequence}. Furthermore, discoveries of tau aggregation in fronto-temporal degeneration (FTD) suggest that tau alone can cause degeneration \cite{heutink2000untangling}.  

There is also evidence that the tau hypothesis is incomplete. For example, the fact that mutations in tau-related genes give rise to tau tangles but no plaques, yet mutations in the APP gene result in both plaques and tangles suggest that amyloid toxicity might occur upstream, before tau toxicity \cite{mudher2002alzheimer}. 

\subsubsection{Cholinergic Hypothesis}
\label{sec:bckChoHyp}

An older hypothesis, on which current AD therapies rely, is the cholinergic hypothesis, which suggests that degeneration of cholinergic neurons and associated disruption of cholinergic neurotransmission are the main causes of pathology in AD \cite{francis1999cholinergic}. Support for the theory came in mid-1970s, where studies provided evidence of deficits in synthesis of neurotransmitter acetylcholine (ACh) and choline acetyltransferase (ChAT) \cite{davies1976selective}. However, the cholinergic hypothesis lost support due to unsatisfactory results of the cholinergic drugs \cite{martorana2010beyond}. Despite not being disease-modifying, cholinergic drugs have been shown to provide symptomatic benefits through improved memory and clinical function \cite{martorana2010beyond}.

\subsubsection{Vascular Hypothesis}
\label{sec:bckVasHyp}

A vascular hypothesis has also been proposed for AD \cite{de2004alzheimer}, suggesting that one of the incipient causes is related to vascular abnormalities, which leads to brain hypoperfusion, neurodegeneration and cognitive impairment. One study even indicates that this is an earlier event than amyloid and tau accumulation \cite{iturria2016early}. Evidence supporting this theory has been given by the close associations between dementia and stroke \cite{kalaria2003vascular,meyer2000risk}, cardiac diseases \cite{breteler2000vascular,aronson1990women,polidori2001heart} and atherosclerosis \cite{hofman1997atherosclerosis}.

\subsubsection{Genetic Causes}
\label{sec:bckGen}

\begin{figure}
\centering
\includegraphics[width=0.9\textwidth]{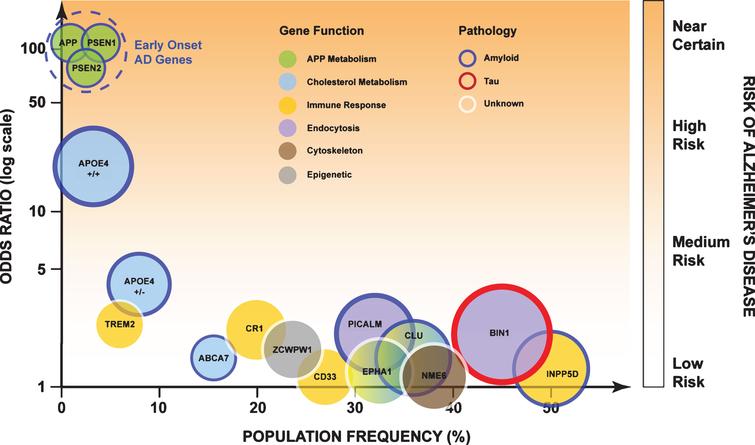}
\caption[Different genes and associated risk for AD]{Diagram showing different genes which increase the risk for AD (y-axis), as well as their frequency within the population (x-axis). EOAD genes APP, PSEN1 and PSEN2 (top-left) give a near-certain risk of developing AD, but are found in a very small minority of the AD population. APOE4 has a moderate risk, while the other genes have a lower risk, yet are found in a much larger population. Reproduced from \cite{robinson2017recent}, CC BY-NC.}
\label{fig:bckAdGenesMap}
\end{figure}

Alzheimer's disease has a strong genetic component, with many genes involved that alter the risk of developing the disease and the pace of progression. Twin studies show that disease heritability ranges between 60\% to 80\% \cite{bergem1997role,gatz1997heritability}. Currently there are two main forms of AD: familial AD and sporadic AD. Familial AD is a rare early-onset AD (EOAD) characterised by autosomal dominant disease transmission, and caused by mutations in three genes, APP, PSEN1 and PSEN2, which code for amyloid peptide precursor, presenilin 1 and 2 respectively \cite{chouraki2014genetics}. Sporadic AD is the most common, late-onset form of AD (LOAD), characterised by more complex, non-Mendelian transmission. 

In familial AD, several genetic risk factors have been identified so far. In 1980s, the discovery of amyloid-$\beta$ peptides in AD senile plaques and the identification of these peptides in the brains of people with Down's Syndrome, caused by abnormalities in chromosome 21 and where dementia was also observed, led the the hypothesis that mutations of a gene located on chromosome 21 might cause AD in people without Down's syndrome \cite{glenner1984alzheimer}. A few years later, a linkage peak was indeed found on chromosome 21 \cite{st1987genetic}, and the APP gene was identified \cite{goldgaber1987characterization} and confirmed in EOAD families \cite{chartier1991early}. However, the amount of heterogeneity observed in EOAD suggested additional genes were involved, and further genetic linkage analyses led to the discovery of PSEN1 \cite{sherrington1995cloning} and PSEN2 genes \cite{levy1995candidate}. As of March 2014, 40, 197 and 25 mutations were reported in APP, PSEN1 and PSEN2 genes respectively, all with autosomal dominant transmission with complete penetrance, with the exception of one mutation in the APP gene \cite{chouraki2014genetics}.

In sporadic, late-onset AD, the genetic landscape is much more complex. The most important risk factor is given by mutations in genes coding for Alipoprotein E (APOE) \cite{chouraki2014genetics}.  APOE is a protein whose key function is to transport lipids and cholesterols throughout the body, and has three major isoforms called APOE2, APOE3 and APOE4, corresponding to alleles $\epsilon2$, $\epsilon3$ and $\epsilon4$. Increased risk of AD due to APOE4 has been established in 1993 in three key studies \cite{strittmatter1993apolipoprotein, saunders1993association, corder1993gene}. Until 2005, more than 500 candidate genes other than APOE have been identified using association studies, with various pathways involved including tau phosphorylation, vacuolar sorting, glucose and insulin metabolism, nitrous oxide synthesis, oxidative stress, growth factors, inflammation and lipid-related pathways \cite{chouraki2014genetics}.  However, after the advent of genome-wide association studies (GWAS) in 2005, the first genes outside the APOE locus were identified in two independent studies \cite{harold2009genome,lambert2009genome}. Several genes including CLU \cite{harold2009genome,lambert2009genome}, CR1 \cite{lambert2009genome}, BIN1 \cite{seshadri2010genome}, PICALM \cite{harold2009genome}, ABCA7 \cite{hollingworth2011common} and CD2AP \cite{naj2011common} have been since identified. Moreover, associations were also found with quantitative endophenotypes, which provide more statistical power than yes/no disease status, such as early age of onset \cite{thambisetty2013alzheimerA, thambisetty2013alzheimerB}, greater burden of amyloid pathology \cite{biffi2012genetic,chibnik2011cr1}, abnormal levels of cerebro-spinal fluid (CSF) \cite{elias2013genetic, kauwe2011fine}, decrease in total brain volume \cite{bralten2011cr1, furney2011genome} and decreased cognitive scores \cite{barral2012genotype,chibnik2011cr1,mengel2013clu}.

\subsection{Other Risk Factors}
\label{sec:bckRisFac}

\begin{figure}
\centering
\includegraphics[width=0.6\textwidth]{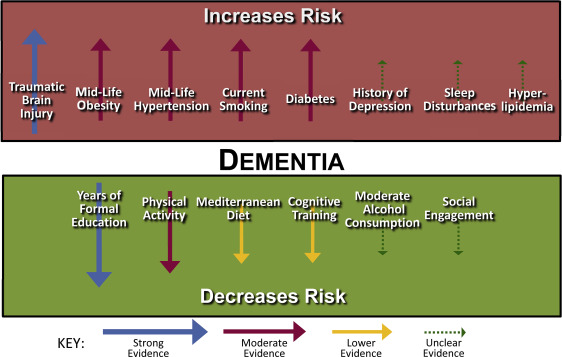}
\caption[Different risk factors for AD related to lifestyle and the associated level of evidence]{Diagram showing different risk factors for AD related to lifestyle and the associated level of evidence. Reproduced from \cite{baumgart2015summary}, CC-BY-NC-ND.}
\label{fig:bckAdRiskFactors}
\end{figure}

There are several known risk factors that are associated with AD. The principal risk factor is age, with incidence rates doubling every 5 years after 65 years of age \cite{querfurth2010mechanisms,todd2013survival}. Other risk factors include head injuries, depression and hypertension \cite{Burns2009}. Lifestyle factors such as smoking \cite{cataldo2010cigarette} also increase the risk for developing AD. There is  also potential evidence that living in polluted areas increases the risk for AD \cite{moulton2012air}. Physical exercise is also associated with lower risk of developing dementia \cite{ahlskog2011physical}. Other factors influencing the risk of dementia are shown in Fig. \ref{fig:bckAdRiskFactors}, and include traumatic brain injury, obesity, hypertension, diabetes (increases risk), with protective factors including higher levels of education \cite{baumgart2015summary}.

\subsection{Biomarkers}
\label{sec:bckBio}

The information in this section has been initially written by me for the TADPOLE Challenge website\footnote{\url{https://tadpole.grand-challenge.org/Data/}}, with feedback from Esther E. Bron and Daniel C. Alexander. The material has been subsequently adapted for this thesis. 

Over the last decades, various biomarkers have been developed to quantify the severity of Alzheimer's disease and track its progression:
\begin{itemize}
\item Cognitive tests such as the Mini-Mental State Examination (MMSE) \cite{mckhann1984clinical} are used to assess memory and cognitive performance (section \ref{sec:bckCog}). 
\item Magnetic Resonance Imaging (MRI) measures such as cortical volumes, thickness and atrophy rates detect shrinkage of individual brain areas that is caused by neurodegeneration (section \ref{sec:bckMri}). 
\item Positron Emission Tomography: can be used to measure neuronal metabolism through Fluorodeoxyglucose (FDG) PET \cite{herholz2012use}, amyloid uptake through the Pittsburgh compound B (PiB). \cite{klunk2004imaging} and, more recently, tau uptake through AV1451 PET (section \ref{sec:bckPet}).
\item Cerebro-spinal fluid markers: can be used to measure amyloid plaque deposits \cite{blennow2003csf} and neurofibrillary tangles through CSF total tau and phosphorylated tau \cite{blennow2003csf} (section \ref{sec:bckCsf}).
\end{itemize}

\subsubsection{Cognitive Tests}
\label{sec:bckCog}

\begin{figure}
\centering
\includegraphics[width=0.3\textwidth]{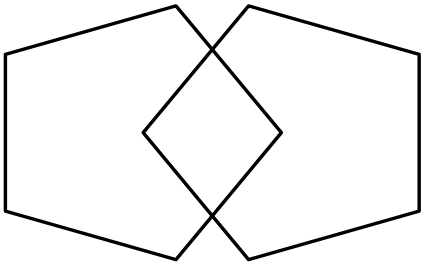}
\caption[Intercalated pentagons used in the Mini-Mental State Examination (MMSE)]{Intercalated pentagons used in the Mini-Mental State Examination (MMSE). Patients with dementia have difficulty drawing them. Image source: Wikipedia\footnotemark \ CC-SA.}
\label{fig:bckMmsePentagons}
\end{figure}
\footnotetext{\url{https://commons.wikimedia.org/wiki/File:InterlockingPentagons.svg}} 

Cognitive tests are neuropsychological tests performed by a clinical expect and can assess different cognitive domains such as general cognition, memory, language, vision. These give an overall sense of whether subjects are aware of their symptoms, surrounding environment and whether they can remember a short list of words, follow instructions and do simple calculations. For instance, in the Mini-Mental State Examination (MMSE), patients are asked to draw intercalated pentagons (Fig \ref{fig:bckMmsePentagons}). As there are no population standards, performance in many of these tests is measured relative to a control group, after adjusting for age, sex and education \cite{mckhann1984clinical}.

Cognitive tests are important in Alzheimer's disease because they measure cognitive decline in a direct and quantifiable manner. As a result, they are required for establishing a clinical diagnosis of probable AD in criteria such as NINCSD-ADRDA \cite{dubois2007research,dubois2010revising}. Apart from establishing the AD diagnosis, these tests are valuable also for establishing patterns of cognitive impairment, assessing changes over time, comparing drug efficacy and for establishing correspondences with other imaging, histopathology or molecular biomarkers \cite{mckhann1984clinical}. 

Cognitive tests have several limitations. First of all, they suffer from practice effects, i.e. patients who undertake the same test several times can learn/remember how to do it, and thus score higher at a follow-up visit. This limits the usefulness of the test in assessing dementia. Another limitations is that they have floor or ceiling effects, which means that many subjects might score the highest/lowest score possible. Finally, they can also be biased, as each subject is evaluated by a human expert who might be influenced by prior knowledge of the subject's cognitive abilities.

\subsubsection{Magnetic Resonance Imaging}
\label{sec:bckMri}

\begin{figure}
\centering
\includegraphics[width=1.0\textwidth]{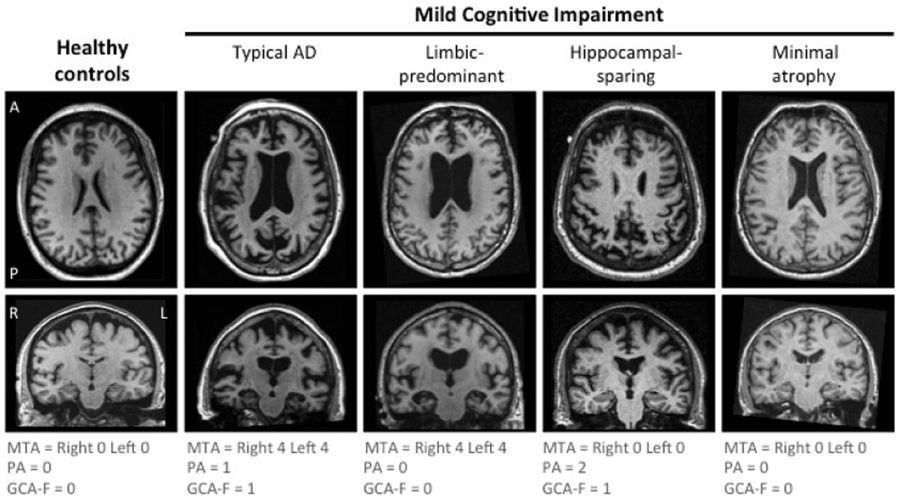}
\caption[Comparison between the MRI brain scans of healthy subjects and subjects with mild cognitive impairment]{Comparison between the MRI brain scan of a healthy subject (left) and subjects with different types of mild cognitive impairment (MCI) (middle-right), showing different patterns of atrophy for each group. MRI is a widely used technology for measuring the spatial distribution and extent of atrophy and for tracking the progression of Alzheimer's disease (AD). Reproduced from \cite{ekman2018t}, CC-BY license.}
\label{fig:bckAdAtrophy}
\end{figure}

Magnetic resonance imaging (MRI) is a technique used to image the anatomy and the physiological processes of the brain and other body parts. With MRI, brain structures can be quantified due to different contrast between gray matter (GM), white matter (WM), cerebrospinal fluid (CSF) and hard tissue such as the skull. The GM is the brain tissue that consists of the bodies of neurons, while the WM consists of fibres connecting the neurons. The cerebrospinal fluid is a clear, colourless fluid providing mechanical and immunological protection to the brain. Within MRI, different types of contrast between tissues can be obtained through T1, T2, T1-weighted and T2-weighted images. 

Brain MRI has been successfully applied to quantify neurodegeneration in Alzheimer's disease. Brain atrophy, which is caused by the death of neurons, can be visually assessed in MRI scans due to shrinkage of the brain (see Fig. \ref{fig:bckAdAtrophy}) and can be quantified using markers of volume, cortical thickness, surface areas, along with changes in these values between a baseline and a follow-up scan. These quantitative markers can be obtained with specialised software such as Freesurfer \cite{reuter2012within}.

MRI-derived biomarkers have both advantages and limitations. They are robust and have less noise compared to cognitive tests, and are non-invasive. Moreover, they are also a good indicator of progression from MCI to dementia in an individual subject because they become abnormal slightly earlier than the onset of dementia-specific symptoms \cite{jack2010hypothetical, jack2013update}. Limitations of these markers are that MRI scans are expensive,  require specialised equipment to be acquired, and can also suffer from motion artefacts.

\subsubsection{Positron Emission Tomography}
\label{sec:bckPet}

\begin{figure}
\centering
\includegraphics[width=0.4\textwidth]{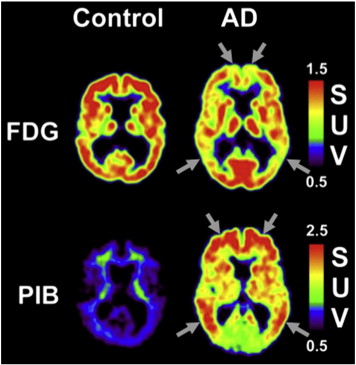}
\caption[FDG PET images of healthy and AD subjects]{(top) Fluorodeoxyglucose (FDG) PET images for a cognitively normal subject (left) and a subject with Alzheimer's disease (right). FDG PET measures cellular metabolism, which is known to decrease during the development of AD. There is decreased metabolism in parietal and frontal regions (gray arrows) in the AD subject compared to the cognitively normal subject. (bottom) Pittsburgh B (PiB) PET image measuring amyloid uptake in the brain of a healthy control (left) and AD subject (right). There is widespread amyloid presence in the brain of the AD subject. Image reproduced with permission from \cite{cohen2014early}.}
\label{fig:bckFdgAd}
\end{figure}

Positron Emission Tomography (PET) detects pairs of gamma rays emitted by a radioactive tracer, which is introduced into the body of a biologically active molecule. Three-dimensional images of tracer concentration within the body are then constructed by computer analysis. Before a PET scan, the patient is injected with a contrast agent (containing the tracer) which spreads throughout the brain and binds to abnormal proteins (amyloid and tau). This enables researchers to track the concentration of these proteins. PET scans can be of several types, depending on the cellular and molecular processes that are being measured:
\begin{itemize}
\item cell metabolism using Fluorodeoxyglucose (FDG) PET: Neuronal cell metabolism refers to the the activity going on inside neuronal cells such as the processing of food and elimination of waste. Metabolic processes use glucose, hence FDG PET quantifies metabolism by measuring the amount of glucose within each voxel. In Alzheimer's disease, neurons that are about to die will show reduced metabolism, hence FDG PET is also an early indicator of neurodegeneration.
\item levels of abnormal proteins such as amyloid-beta through AV45 PET: Amyloid-beta misfolding (i.e. errors in the construction of its 3D structure) is thought to be one of the causes of Alzheimer's disease (see section \ref{sec:bckAmyHyp}). AV45 PET can be used to measure the levels of amyloid in the brain, and is hence one of the earliest AD markers.
\item levels of abnormal tau proteins through AV1451 PET: Abnormal phosphorylated tau (i.e. tau protein and a phosphorus group) that gather together in an insoluble form eventually cause damage to the neuron's cytoskeleton, leading to the collapse of the neuron's transport system and eventually the neuron's death (see section \ref{sec:bckTauHyp}). AV45 PET can be used to measure the level of misfolded tau proteins and is also one of the earliest markers in AD.
\end{itemize}

PET-derived biomarkers are important because they give information about molecular processes that happen in the brain. These are usually the first to become abnormal in the cascade of events that lead to Alzheimer's disease, and are therefore important early markers of the disease that is about to unfold \cite{jack2010hypothetical,jack2013update}.  

PET scans have some limitations that need to be acknowledged. One main limitations is that the patient is exposed to ionising radiation, which limits the number of scans they can take in a specific time interval. PET scans also have a much lower spatial resolution compared to MRI scans. One other caveat with AV1451 PET (tau imaging) is that it is a very new tracer that is still under research, with some studies indicating evidence of some off-target binding in some tau conformations found in non-AD tauopathies \cite{lowe2016autoradiographic,marquie2015validating}. 

\subsubsection{Diffusion Tensor Imaging}
\label{sec:bckDti}

\begin{figure}
\centering
\includegraphics[width=1.0\textwidth]{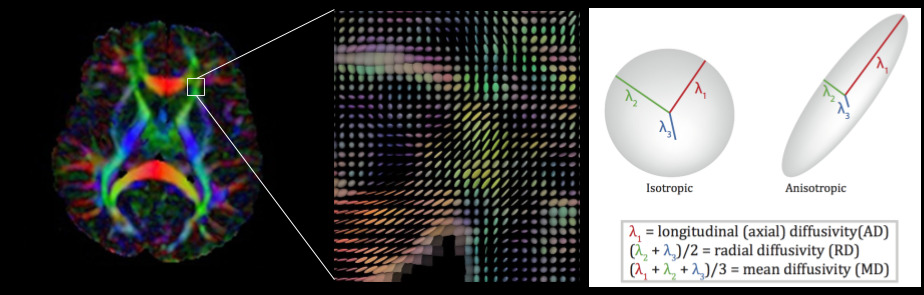}
\caption[Diffusion tensor image diagram]{(Left) Diffusion tensor image of a brain showing white matter fibre connections. The colours represent the direction of the connection (red for left-right, blue for superior-inferior, and green for anterior-posterior). (Middle) Zoomed image into the small region of interest (ROI), showing the diffusion tensor ellipses. Each ellipse indicates the direction where water molecules diffused (i.e. moved). (Right) Diagram showing the difference between isotropic diffusion (i.e. equal in all directions) versus anisotropic diffusion, along with the diffusivity measures that can be computed. Diagram assembled by me using images from several sources\footnotemark.}
\label{fig:bckDtiDiagram}
\end{figure}
\footnotetext{
\begin{tabular}{l}
Image sources:\\ 
\url{http://fmri.uib.no/index.php?option=com_content&view=article&id=68&Itemid=86}\\ 
\url{https://commons.wikimedia.org/wiki/File:DTI-axial-ellipsoids.jpg}\\
\url{http://www.diffusion-imaging.com/2012/10/voxel-based-versus-track-based.html}\\
\end{tabular}
}

Diffusion tensor imaging (DTI) is an MRI technique that can be used to measure the degeneration of white matter connections in the brain. This is done by analysing the diffusion of water molecules along the neuron fibre connections. Molecular diffusion in tissues is not free, but reflects interactions with many obstacles, such as macromolecules, fibers, and membranes. When a fiber connection degrades, the diffusion becomes more isotropic (i.e. equal in every direction), which can be quantified using a measure called fractional anisotropy (FA). Fig. \ref{fig:bckDtiDiagram} shows a diagram of a DTI image (left) which is made of diffusion tensors estimated at each voxel (middle). Diffusivities parallel and perpendicular to the fiber direction can then be measured (right).   

DTI is important for analysing the progression of Alzheimer's disease. It has been shown that AD affects white matter bundles \cite{sachdev2013alzheimer}. DTI has also shown great potential for aiding the diagnosis of dementia \cite{bozzali2002white,zhang2009white}. DTI tractography is also important for building brain structural connectomes which have been shown to be disrupted by different types of dementias including Alzheimer's disease \cite{seeley2009neurodegenerative, zhou2012predicting}.

DTI measures have some limitations. As with other MRI modalities, it is susceptible to motion artefacts and suffers from partial volume effects, i.e. measures at each voxel are biased due to averaging across many different cells and types of tissue that are contained in that voxel. Another limitation is that changes in DTI-derived measures such as FA are not specific, and can be attributed to many changes in the underlying cytoarchitecture, such as neurite density or dispersion \cite{zhang2012noddi}. 

\subsubsection{Cerebrospinal Fluid Markers}
\label{sec:bckCsf}

\begin{figure}
\centering
\includegraphics[width=0.5\textwidth]{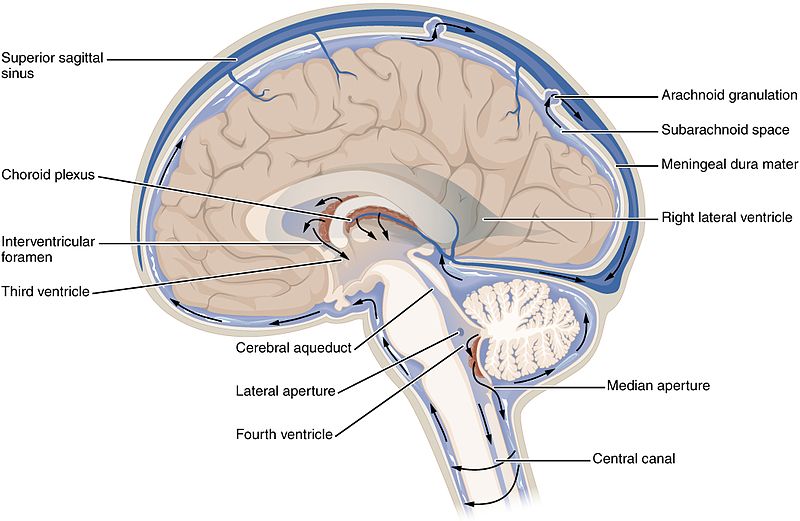}
\caption[Diagram showing the cerebro-spinal fluid (CSF).]{Diagram showing the cerebro-spinal fluid (CSF) coloured in blue, which is found in the subarachnoid space around the brain and spinal cord. Source: Wikipedia\footnotemark, CC license.}
\label{fig:bckCsfDiagram}
\end{figure}
\footnotetext{\url{https://en.wikipedia.org/wiki/File:1317_CFS_Circulation.jpg}}

The cerebrospinal fluid (CSF) is a clear, colourless body fluid found in the brain and spinal cord. It acts as a cushion or buffer for the brain, providing basic mechanical and immunological protection to the brain inside the skull. A sample of the CSF can be taken from patients invasively through lumbar puncture, which involves inserting a needle in the spinal cord.

Measures of CSF are very important for dementia research. In the CSF, the concentration of abnormal proteins such as amyloid-beta and tau is a strong indicator of AD. Abnormal levels of concentrations in these proteins are some of the earliest signs of Alzheimer's disease and can indicate abnormalities many years before symptom onset \cite{jack2010hypothetical}.

The CSF measures have some limitations. One key limitation is that the lumbar puncture is highly invasive and thus not performed in many studies. The CSF measures are also not specific to any particular part of the brain.

\subsection{Diagnosis}
\label{sec:bckDia}

A diagnosis of Alzheimer's disease is usually given based on the person's medical history, behaviour and information provided by the relatives. Medical imaging from Magnetic Resonance Imaging (MRI), Computer Tomography (CT) or Positron Emission Tomography (PET) can help exclude other types of brain pathologies or types of dementia. Memory tests from neuropsychological batteries can help characterise the stage of the disease \cite{waldemar2007recommendations}.

The most commonly used diagnostic criteria are from the National Institute of Neurological and Communicative Disorders and Stroke (NINCDS) and the Alzheimer's Disease and Related Disorders Association (ADRDA) \cite{dubois2007research,dubois2010revising}. This criteria, commonly called NINCSD-ADRDA, require evidence of cognitive impairment through neuropsychological testing for establishing a clinical diagnosis of \emph{probable} AD, while histopathologic confirmation is required for definite confirmation \cite{dubois2007research,dubois2010revising}.

\section{Progression of Alzheimer's Disease}
\label{sec:bckProgAd}

Several studies have been done so far on Alzheimer's disease progression \cite{jack2010hypothetical, ridha2006tracking,fox1999correlation, scahill2002mapping, braak1991neuropathological,schott2003assessing}. It is currently believed that abnormal changes in amyloid-beta and tau aggregation happen very early, long before symptoms occur, followed by hypometabolism and structural atrophy, and then cognitive decline such as memory loss and executive dysfunction \cite{jack2010hypothetical,jack2013tracking}.  Neuropathological staging of AD brains showed that the earliest change in brain structure are in the medial temporal lobe, particularly in the entorhinal cortex and hippocampus \cite{braak1991neuropathological}. This has also been confirmed with in-vivo MRI studies, which showed that even in mild AD the entorhinal area and hippocampus shrink by 20-25\% compared to controls \cite{jack1997medial, lehericy1994amygdalohippocampal, juottonen1999comparative, bobinski1999histological}. Results by Schott et al. \cite{schott2003assessing} and Ridha et al. \cite{ridha2006tracking} show that atrophy of the medial temporal lobe precedes the clinical onset of AD by approximately 3.5 years \cite{schott2003assessing}. 

\subsection{Braak Staging}
\label{sec:bckBraak}

In 1991, Braak and Braak \cite{braak1991neuropathological} proposed a staging system based on the spatial spread of amyloid plaques, neurofibrillary tangles (NFT) and neutropil threads (NT). It was the first attempt to build a staging scale for Alzheimer's disease from neuropathology, using a cross-sectional set of brains and without using clinical information.

The amyloid patterns of the Braak staging system proved to be of limited significance for the differentiation of neuropathological stages \cite{braak1991neuropathological}, but still enabled separation into three stages. In the initial stage, amyloid deposits are found in the basal portions of the isocortex, followed by fast spreading in virtually all isocortical association areas by the middle stage. In the late stage, amyloid plaques are found in all areas of the isocortex, including sensory and motor fields \cite{braak1991neuropathological}.

The spreading of neurofibrillary tangles allowed separation into six key stages. Stages I-II were characterised by mild or moderate accumulation in the transentorhinal layer Pre-$\alpha$\footnote{Pre-$\alpha$ is one of the layers from the principal stratum (Pre) of the entorhinal cortex. It is characterised by cellular islands of large projection cells, and its connections project to the hippocampus.} \cite{braak1991neuropathological}. Afterwards, stages III-IV were marked by the spread of NFTs into the transentorhinal region and proper entorhinal cortex, along with mild involvement of the first Ammon's horn sector \cite{braak1991neuropathological}. Finally, stages V-VI were marked by the spread of NFTs and NTs to almost all isocortical association areas. \cite{braak1991neuropathological}

\subsection{Neuroimaging}
\label{sec:bckNeu}

In AD, Magnetic Resonance Imaging (MRI) shows gray matter atrophy throughout the brain, in particular in the hippocampus and entorhinal cortex \cite{whitwell2010progression}. In terms of atrophy progression, it starts in the medial temporal lobe and fusifom gyrus at least 3 years before an AD diagnosis, and then spreads to the posterior temporal lobe, parietal lobe, and finally to the frontal lobe. However, the sensorimotor cortex, visual cortex and the cerebellum are relatively spared \cite{whitwell2010progression}.  

Imaging with Positron Emission Tomography (PET) shows reduced metabolism (FDG) and increased uptake of amyloid (e.g. AV45) proteins \cite{marcus2014brain}. In early stages of AD, hypometabolism affects the parietotemporal association areas, the posterior cingulate gyrus and the precuneus. In later stages, frontal cortices also become affected, while the striatum, thalamus, primary sensorymotor cortices, visual cortices and the cerebellum seem to be spared \cite{marcus2014brain}. In terms of amyloid deposition through amyloid PET, early deposits are found in the precuneus, orbitofrontal, inferior temporal and posterior cingulate, later followed by the entire prefrontal cortex, lateral temporal and parietal lobes \cite{marcus2014brain}. These patterns have been validated using autopsy studies\cite{clark2011use, clark2012cerebral}. 

\section{Posterior Cortical Atrophy}
\label{sec:bckPca}

\begin{figure}
\centering
\includegraphics[width=0.8\textwidth]{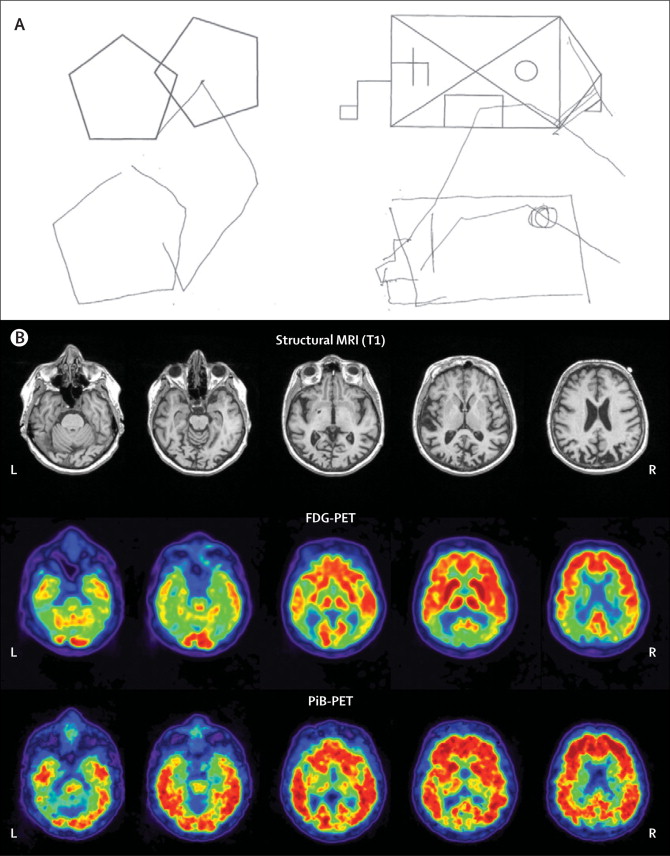}
\caption[Visual deficits and neuroimaging pathology in Posterior Cortical Atrophy]{(A) Visual deficits as shown when a 62-year old PCA patient was asked to copy the intersecting pentagons figure \cite{crutch2012posterior}. (B) Structural MRI, FDG PET and PiB PET scans of the same subject. Structural MRI shows atrophy predominant in the bilateral parietal, posterior temporal and lateral occipital regions (B, top), FDG PET shows reduced metabolism in the same regions (B, middle), while PiB-PET shows diffuse amyloid uptake throughout the entire brain (B, bottom) \cite{crutch2012posterior}.  }
\label{fig:bckPcaImg}
\end{figure}

Posterior cortical atrophy (PCA) is an early-onset neurodegenerative syndrome that affects the posterior part of the brain, resulting in the disruption of the visual cortex. The syndrome, also called Benson's syndrome, was first reported by Benson et al. \cite{benson1988posterior} in 1988 to describe five patients with fairly homogeneous but otherwise unclassified symptoms.  

\subsection{Symptoms}
\label{sec:bckPcaSym}

The most common symptoms include general visuospatial and visuoperceptual impairments such as inability to read, blurred vision, light sensitivity, trouble navigating through space and issues with depth perception \cite{crutch2012posterior,borruat2013posterior}. Additional symptoms also include apraxia (disorder of movement planning), visual agnosia (object recognition deficit) and agraphia (loss of writing ability) \cite{benson1988posterior, goethals2001posterior}. These symptoms get worse as the disease progresses, with patients becoming unable to recognise familiar people, objects, difficulty navigating familiar places and drawing (see Fig. \ref{fig:bckPcaImg}). Some studies  \cite{andrade2010visual, andrade2012visuospatial, andrade2013visuospatial} reported visual hemineglect (difficulty seeing one half of the visual field) to be frequent in PCA patients, especially if asymmetrical atrophy takes place in the occipital areas. 

PCA patients report higher-order visual problems related to object and space perception, compared to more basic visual impairments e.g. in colour and motion, although impairments in higher order visual functions might be due to lower-level disruption. One study \cite{lehmann2011basic} reported that all PCA subjects showed impairment in at least one low-level visual process, and that this correlates with higher-order visuospatial and visuoperceptual functions, but not with non-visual functions of the parietal lobe, including calculations and spelling.

\subsection{Causes}
\label{sec:bckPcaCau}

The causes of PCA are still unknown, due to the rarity of the disease, gradual onset of symptoms and no fully accepted diagnostic criteria \cite{borruat2013posterior,crutch2012posterior}. The progressive neurodegeneration that characterises PCA is often attributed to Alzheimer's disease pathology (i.e. aggregation of amyloid plaques and tau tangles), but alternative causes including dementia with Lewy bodies, corticobasal degeneration and prion disease have also been identified \cite{crutch2012posterior}. One study reported the PCA syndrome in a 4-sibling family with prion disease \cite{depaz2012long}, suggesting that prion propagation mechanisms might be involved in PCA.

Genetic factors that underlie PCA are also not well understood \cite{crutch2012posterior,borruat2013posterior}. Empirical findings suggest that there are no significant differences in the number of patients with a positive family history of PCA and typical AD \cite{crutch2012posterior}. Some studies also report no differences in Alipoprotein E (APOE) genotypes between PCA and typical AD \cite{mendez2002posterior, tang2004clinical, rosenbloom2011distinct, migliaccio2009clinical}, although other studies reported differences in APOE $\epsilon$4 allele status, with fewer PCA patients being $\epsilon$4-positive \cite{schott2006apolipoprotein,snowden2007cognitive}. These differences have been attributed to differences in inclusion criteria of PCA with respect to typical AD \cite{crutch2012posterior}.

\subsection{Diagnosis}
\label{sec:bckPcaDia}

PCA patients face difficulties in diagnosis due to the young age at onset and the fact that there are no fully accepted diagnostic criteria. Patients are sometimes misdiagnosed with depression, anxiety or even malingering in early stages of the disease \cite{crutch2012posterior}. They are often initially referred to opticians and ophthalmologists in the belief that ocular abnormalities are causing their visual deficits, often leading to unnecessary medical procedures such as cataract surgery. Neuroimaging modalities such as magnetic resonance imaging (MRI), positron emission tomography (PET) or single photon emission computed tomography (SPECT) can aid diagnosis of PCA \cite{goldstein2011posterior}. 

There are no widely accepted diagnostic criteria, although two criteria have been proposed so far by Mendez et al. \cite{mendez2002posterior} and Tang-Wai et al. \cite{tang2004clinical}. These criteria suggested presence of visual deficits in absence of other eye diseases, gradual progression, relative preservation of anterograde memory, absence of stroke or tumour, and other neuropsychological or imaging abnormalities that are related to parietal or occipital functions. 

However, these criteria have some limitations. They are yet to be thoroughly validated outside their centres, and need to be linked to underlying pathology, otherwise inconsistencies between studies and centres will occur. Moreover, the current criteria provide no guidance to the level of specificity required for a diagnosis of PCA \cite{crutch2012posterior}. It has been suggested that PCA, when caused by underlying AD pathology, lies on a continuum of phenotypical variation between AD and purely-visual PCA, with no clearly defined diagnosis boundary \cite{snowden2007cognitive, migliaccio2009clinical, lehmann2011basic}.

\subsection{Management}
\label{sec:bckPcaMan}

There is no known efficient treatment of PCA that will reverse or stop neurodegeneration \cite{borruat2013posterior}. Patients with PCA are usually treated with the same medication as for AD, namely cholinesterase inhibitors: tacrine, rivastigmine, galantamine and donepezil \cite{borruat2013posterior}. Crutch et al. \cite{crutch2012posterior} suggest that antidepressant drugs might also be appropriate in patients with low mood, and levodopa or carbidopa could aid individuals with Parkinsonism. However, there are no studies analysing the efficiency of these drugs in PCA patients \cite{borruat2013posterior}. 

A few non-pharmacological therapies have also been attempted recently in some patients that included psycho-educative programs \cite{videaud2012impact} or a combination of speech therapy, occupational therapy and physiotherapy \cite{weill2012physical}.

\subsection{Neuroimaging}
\label{sec:bckPcaNeu}

Several MRI studies in PCA have shown damage to posterior brain regions. Studies by Hof et al. \cite{hof1997atypical} and Tang-Wei et al. \cite{tang2004clinical} show a greater concentration of senile plaques and neurofibrillary tangles in the occipital and parietal lobes and at the occipito-temporal junction. Cross-sectional studies using voxel-based morphometry have also shown significant abnormalities in occipital and parietal lobes, followed by the temporal lobe \cite{lehmann2011basic,whitwell2007imaging}. When compared directly to typical AD subjects, PCA have shown greater atrophy in the right parietal lobe and less in the left temporal and hippocampal regions \cite{lehmann2011cortical,crutch2012posterior}. Some DTI studies also seem to suggest white matter damage in posterior regions \cite{duning2009pattern,yoshida2004white,migliaccio2012brain}. See Fig. \ref{fig:bckPcaImg} for MRI scans of a PCA patient, showing the typical posterior pattern of atrophy.

Non-MRI imaging studies in PCA have also shown similarly posterior abnormality patterns. Functional imaging studies using single photon emission computer tomography (SPECT) and FDG PET also show reduced function in occipital and parietal regions \cite{kas2011neural, gardini2011visuo, aharon1999posterior, pietrini1996preferential}. Amyloid pathology, as measured with PiB-PET, has been found in occipital and parietal areas, as compared to typical AD subjects \cite{ng2007evaluating, kambe2010posterior, tenovuo2008posterior, formaglio2011vivo}, although this finding was not confirmed in two other studies, which found more diffuse amyloid uptake \cite{rosenbloom2011distinct,de2011similar}.

\subsection{Heterogeneity}
\label{sec:bckPcaHet}

Some studies \cite{ross1996progressive, galton2000atypical} have shown that there is considerable heterogeneity within PCA itself, where three main PCA subgroups have been reported: primary visual (the striate cortex, caudal), parietal (dorsal) and occipitotemporal (ventral) \cite{ross1996progressive, galton2000atypical}. 

Patients with primary visual subtype showed poor vision deficits, with later problems with memory, attention and presence of visual hallucinations \cite{galton2000atypical,levine1993visual}. Imaging showed reduction in occipital lobe perfusion. In one of the studies, AD diagnosis was confirmed post-mortem, upon pathological examination \cite{galton2000atypical}. However, evidence for the existence of this subgroup is very limited, with only two patients identified so far in different case studies \cite{levine1993visual,galton2000atypical}, with another study having reported no "pure" visual deficits within a cohort of n=21 PCA subjects\cite{lehmann2011basic}.

Patients with the parietal (dorsal) PCA subtype generally show initial visuospatial symptoms, agraphia (inability to draw) and dyspraxia, but have preserved visual fields, basic perceptual abilities, object recognition and reading and show biparietal and occipital deficits, disrupting the dorsal or "where" stream \cite{ross1996progressive,galton2000atypical}. 

Patients with the occipitotemporal (ventral) PCA subtype generally show symptoms related to visual distortion, inability to recognise objects, general topography and written words and show occipitotemporal pathology, disrupting the ventral or "what" stream \cite{ross1996progressive,galton2000atypical}.

While all this evidence suggests that there is considerable heterogeneity within the PCA syndrome, evidence is very limited to a few case studies, with some patients also having no pathological confirmation of underlying AD pathology. Moreover, some \cite{crutch2012posterior,lehmann2011basic} have suggested that these subtypes should not be interpreted as distinct groups, but rather as points on a continuum of phenotypical variation.

\chapter{Background -- Disease Progression Models}
\label{chapter:bckDpm}

\begin{figure}
\centering
\includegraphics[width=1\textwidth]{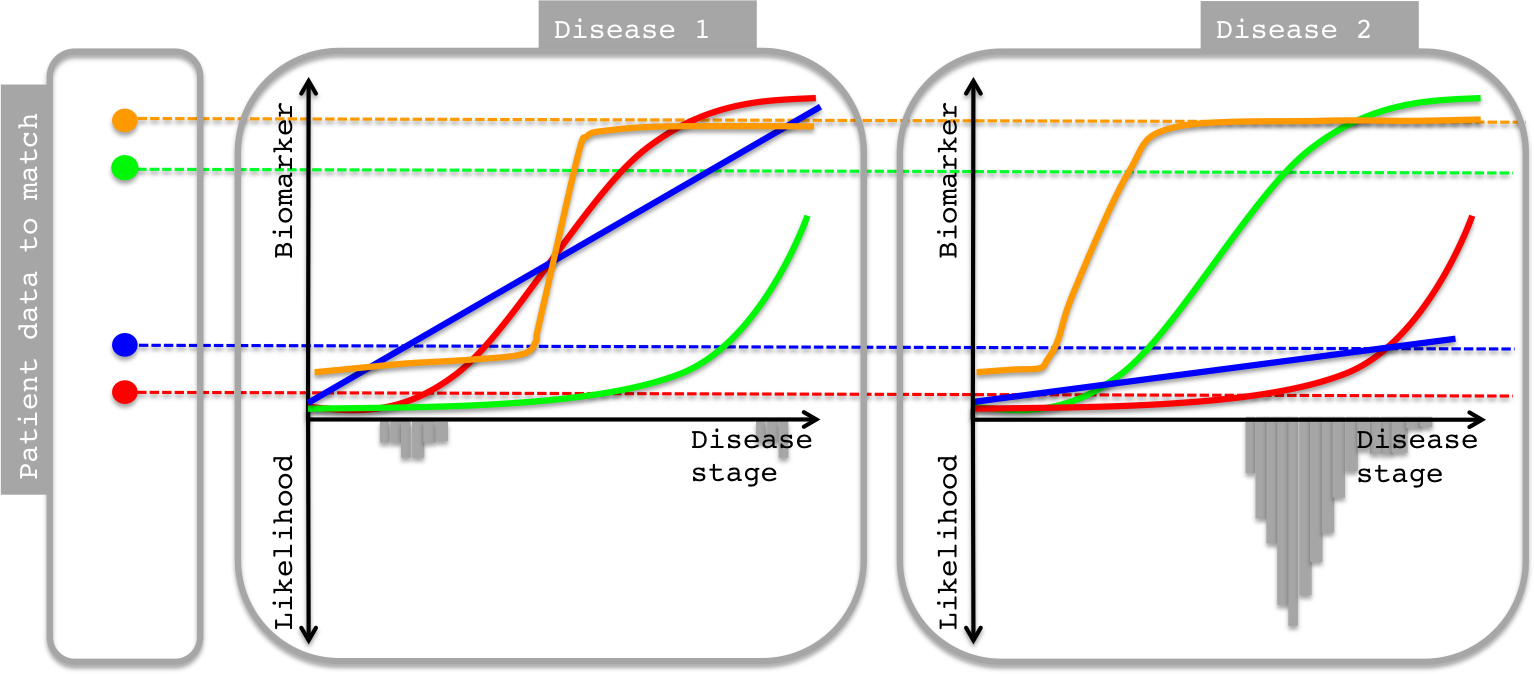}
\caption[Hypothetical biomarker signatures in two diseases]{Cartoon showing hypothetical biomarker signatures from two diseases, along with a cross-sectional snapshot of data from a patient (left). For one patient, disease staging implies finding the optimal time-shift along the horizontal axis that would match its data. On the negative y-axis, the histogram of possible stages is shown. Differential diagnosis can performed by evaluating the integral of the distribution of stages on the negative y-axis, and selecting the disease that has the largest integral. Deriving quantitative biomarker signatures using disease progression modelling can help with disease understanding, staging and differential diagnosis. Image courtesy of Neil Oxtoby and Daniel Alexander.}
\label{fig:bckDpmImg}
\end{figure}

A disease progression model is a mathematical model that describes the evolution of biomarkers in a neurodegenerative disease. Such quantitative models promise to enable early and precise diagnosis of dementia before symptoms appear, and will enable stratification of subjects in AD clinical trials. This is important, because it is currently believed that one of the reason why AD clinical trials failed is because treatments were not administered early enough, and to the right patients \cite{mehta2017trials}. Moreover, the advent of large multimodal biomarker datasets containing neuropsychological, imaging, genetic and molecular data can enable the development of specialised progression models that accurately predict the evolution of subjects. 

Quantitative biomarker signatures estimated through disease progression models have several other key benefits, which are illustrated in Fig. \ref{fig:bckDpmImg}. First of all, they enable disease understanding as well as testing and validation of hypotheses regarding underlying disease mechanisms. Secondly, they enable staging of patients along the progression axis (x-axis), along with prognosis estimates, which can be useful in clinical settings. Third, they also enable differential diagnosis by comparing the fit of the patient's data to signatures of different diseases. 

In this chapter we review the disease progression models that have been developed in the literature. We present the hypothetical model by Jack et al. \cite{jack2010hypothetical} (section \ref{sec:bckDpmHyp}), followed by early models of progression based on symptomatic groups (section \ref{sec:bckDpmSym}) and regression against a clinical marker (section \ref{sec:bckDpmReg}). We then review data-driven models of disease progression such as the event-based model \ref{sec:bckEbm}, the differential equation model (section \ref{sec:bckDem}), the disease progression score (section \ref{sec:bckDps}), self-modelling regression (section \ref{sec:bckSem}), the manifold-based model (section \ref{sec:bckMan}), the voxelwise mixed effects model (section \ref{sec:bckVox}) and the network diffusion model (section \ref{sec:bckNet}), as well as discriminative models used normally in machine learning (section \ref{sec:bckMac}).

\section{Hypothetical Models}
\label{sec:bckDpmHyp}

\begin{figure}
 \centering
 \includegraphics[width=0.85\textwidth,trim=5 5 5 5,clip]{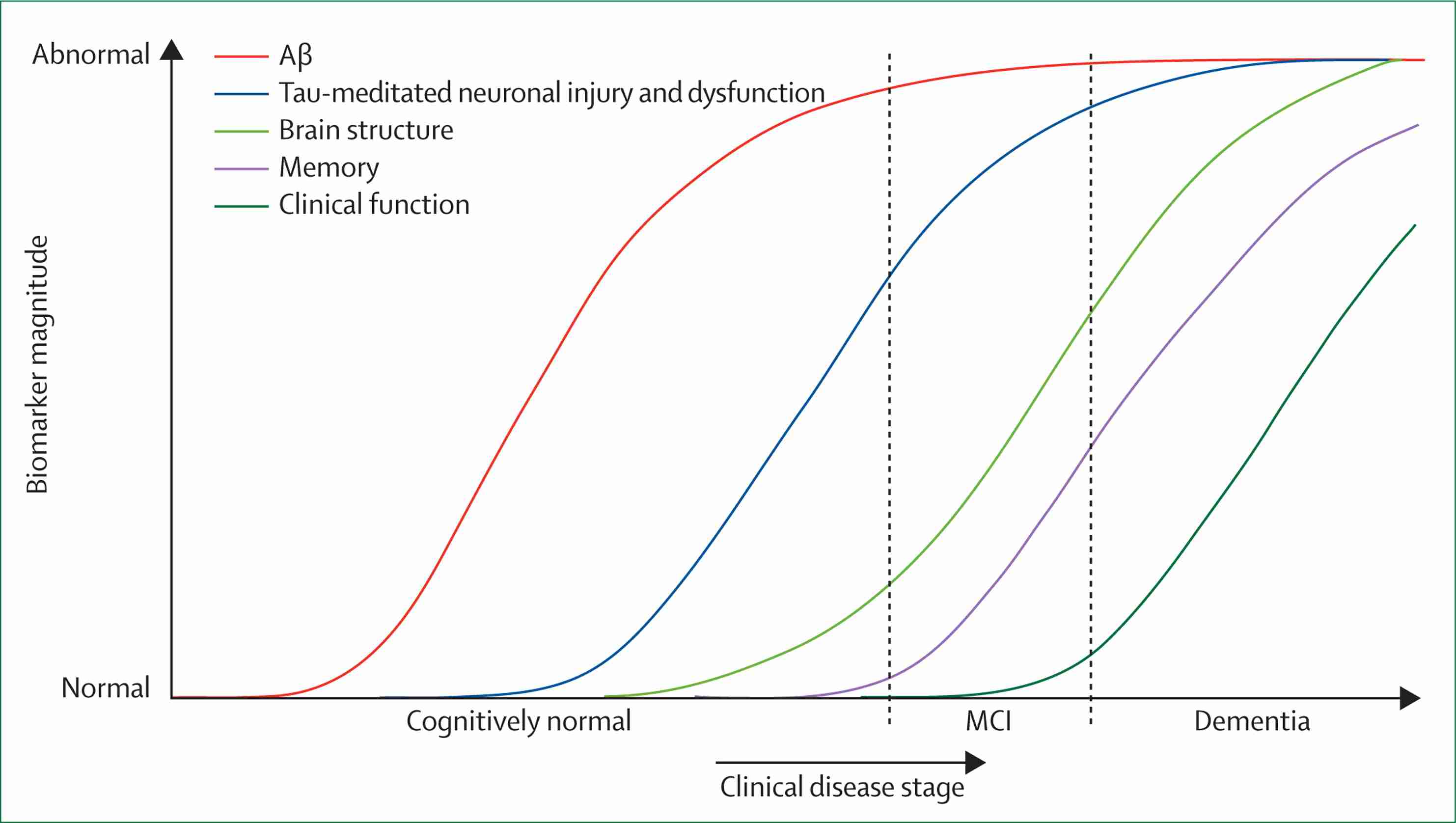}
 \caption[Biomarker cascade by Jack et al. \cite{jack2010hypothetical}]{Dynamic biomarkers of the AD cascade as hypothesised by Jack et al. \cite{jack2010hypothetical}. A$\beta$ and tau are thought to become abnormal before the onset of any dementia symptoms, while brain structure, memory and clinical function are thought to become abnormal later, during MCI and dementia stages. Reproduced with permission from \cite{jack2010hypothetical}.}
 \label{fig:biomk_cascade}
\end{figure}

A hypothetical model of disease progression has been proposed by Jack et al. \cite{jack2010hypothetical,jack2013tracking}, which describes the trajectory of several key biomarkers during the progression of Alzheimer's disease (fig. \ref{fig:biomk_cascade}). Using aggregated evidence from past literature, the model suggests that amyloid-$\beta$ and tau protein biomarkers become abnormal long before the onset of any dementia symptoms. Afterwards, during the mild cognitive impairment (MCI) phase, cognitive functions such as memory become abnormal along with brain structure measured using MRI. These biomarkers continue to be affected in the dementia stage, while A$\beta$ and tau seem to reach a plateau at this point. This hypothesised model of disease progression is shaping the current field of AD research \cite{donohue2014estimating}.

Apart from placing the biomarkers on a single time frame and suggesting the order in which they become abnormal, the model also made some key observations. First of all, the sequence of abnormality was not assumed to change then stop, rather some biomarkers gradually became abnormal simultaneously, although at different speeds and in an ordered manner. Secondly, amyloid plaques are necessary but not sufficient to develop AD pathology \cite{jack2013update}, with cognitive decline correlating less with amyloid deposition \cite{jack200811c} compared to tau and neurodegenerative markers \cite{hyman2011amyloid}. Third, the authors suggested biomarker trajectories follow non-linear curves, hypothesised to be similar in shape to sigmoids \cite{jack2013update, ridha2006tracking, jack2008atrophy}. Fourth, a time-lag exists between evidence of amyloid pathology and the appearance of cognitive deficits, probably mediated by brain resilience and cognitive reserve \cite{jack2013update}.

The hypothetical model nevertheless has some limitations. First of all, it is a hypothetical, theoretical model that is meant to be a guide for future researchers modelling disease progression in Alzheimer's disease. Hence, the model is not quantitative and cannot be used to e.g. stage patients. Another limitation is that the x-axis (disease progression) and y-axis (biomarker abnormality) are not well-defined. Various implementations that will be discussed next have made various assumptions about how to define this, such as computing Z-scores with respect to controls \cite{jedynak2012computational}, or used percentiles over the observed biomarker values \cite{donohue2014estimating}. In the next sections, we will present the development of quantitative models that address these limitations.

\section{Models of Progression using Symptomatic Groups}
\label{sec:bckDpmSym}

Some of the simplest disease progression model are based on symptomatic staging of patients into a small number of groups, e.g. "pre-symptomatic", "mild", "moderate" and "severe" \cite{fonteijn2012event}. They then describe the differences in biomarker measurements among these groups. Scahill et al. \cite{scahill2002mapping} devised such a method that finds changes in brain structure using voxel-based analysis of serial nonlinear-registered MRI images. Other models based on symptomatic staging are those of Dickerson et al., 2009 \cite{dickerson2009cortical} and Thomson et al., 2001 \cite{thompson2001cortical} and 2003 \cite{thompson2003dynamics}. 

These models have several key limitations. First of all, they rely on clinical assessment, which is usually subjective and biased. Secondly, they offer very limited temporal resolution and cannot model changes in pre-symptomatic phases of the disease.   

\section{Regression Against One Biomarker}
\label{sec:bckDpmReg}

In order to estimate longitudinal biomarker trajectories, some authors have proposed regressing against a clinical or age-related marker.  Sabuncu et al. \cite{sabuncu2011dynamics} regressed the cortical thinning rate against MMSE scores. Jack et al. \cite{jack2012shapes} also used regression against MMSE to estimate the shape of biomarker trajectories. Doody et al. \cite{doody2010predicting} regressed biomarkers against time since baseline visit. Driscoll et al. \cite{driscoll2009longitudinal} estimated brain volume trajectories using a mixed effects model against age, using other demographic variables such as gender and intracranial volume (ICV) as covariates. 

These methods have some limitations. Regression methods against clinical markers are limited by the fact that they cannot estimate biomarker dynamics in pre-clinical stages. On the other hand, regression against age or time since baseline visit assume that all subjects have the same age of disease onset or that disease onset is at baseline visit.

Another method for estimating biomarker trajectories, which is popular in familial AD, performs non-linear regression of mutation carriers' data against estimated years from parent's onset \cite{bateman2012clinical, benzinger2013regional}. However, this method can only be applied to dominantly inherited AD, which represents only a small percentage of the entire AD population.

\section{Survival Analysis Models}
\label{sec:bckDpmSur}

Survival analysis models are a class of models used to predict time until an event, in this case conversion to mild cognitive impairment (MCI) or AD. One popular type of survival model in AD is the Cox proportional hazards model, which assumes a multiplicative increase in the hazard rate with respect to a unit-increase in the covariate. Cox proportional hazards models have been used to estimate the probability of progression to AD in a variety of studies \cite{young2014data, dickerson2011alzheimer, bouwman2007csf, csernansky2005preclinical, hansson2006association, kawas2003visual}. A related model is the proportional odds model, which is more suitable for discrete data, and has also been applied to evaluate risk of developing in AD \cite{stoub2005mri, jack200811c, vemuri2009mri}.

Non-parametric survival models such as the Kaplan-Meyer estimator have also been used \cite{morris2001mild, modrego2004depression, carr2000value, khachaturian2004apolipoprotein}. However, the Kaplan-Meyer method can only use a single, binary predictor variable as opposed to the Cox regression method.

The main limitation of survival models is that they require accurate and reliable diagnostic classes, which are not always available and can sometimes be inaccurate due to human errors.

\section{Scalar Biomarker Models}
\label{sec:bckSca}

Over the last few years, a range of latent-time models of disease progression have also been proposed, which estimate scalar biomarker trajectories without relying on a-priori defined cognitive groups, diagnosis or clinical markers. Here, by latent-time models we mean models that estimate a latent temporal dimension of disease progression in an unsupervised manner. In this section we will present several such models: the event-based model (section \ref{sec:bckEbm}), the differential equation model (section \ref{sec:bckDem}), the disease progression score model (section \ref{sec:bckDps}), the self-modelling regression model (section \ref{sec:bckSem}) and the manifold-based mixed effects model (section \ref{sec:bckMan}). These models use scalar biomarker values which are assumed to be uncorrelated, as opposed to more complex spatial data such as brain images or cortical shapes.

The main premise behind there models is that they assume measurements are taken from subjects who are at various unknown points along the progression of the same disease. The models attempt to estimate simultaneously differences in the dynamics of the disease progression while also estimating the time shift and progression speed along the disease timeline, also called temporal heterogeneity. Some models go a step further and also estimate differences that are due to spatial heterogeneity of the subjects, using random effects estimating deviations from the population trajectory. Such combined modelling is challenging, as it introduces identifiability issues.

\subsection{The Event-Based Model}
\label{sec:bckEbm}

\begin{figure}
\begin{tikzpicture}[scale=1.4, every node/.style={scale=1}]

\node (tab1) at (-3,2.5) {
\begin{tabular}{c | c | c | c}
& Patient 1 & Patient 2 & Patient 3\\
\hline
  Region 1 & 1.1 & 0.9 & 0.1\\
  Region 2 & 0.95 & 0.0 & 0.05\\ 
\end{tabular}
};
\node (fig1) at (-2.5,0) {\includegraphics[height=5.0cm]{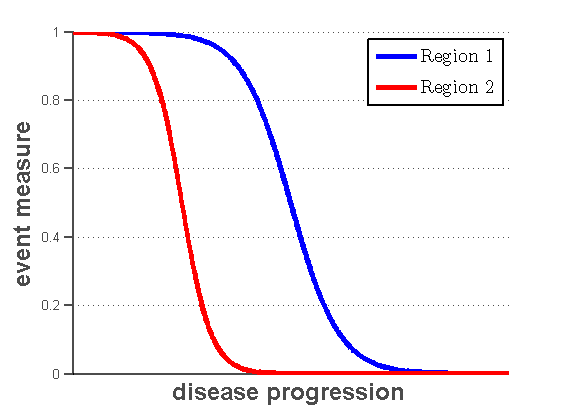}}; 
\draw[-,line width=0.4mm,] (-3.75,2) -- (-3.75,-1.5);
\draw[-,line width=0.4mm,] (-2.4,2) -- (-2.8,1.5) -- (-2.8,-1.5);
\draw[-,line width=0.4mm,] (-1.1,2) -- (-1.9,1.5) -- (-1.9,-1.5);

\node (mixModel) at (1.8,2.5) {
\includegraphics[height=3cm]{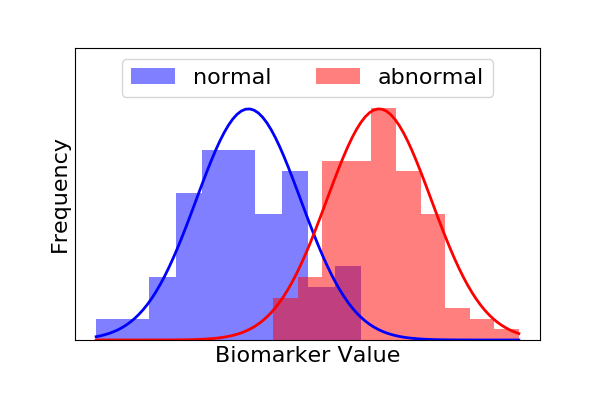}};
\node (mixModelLabel) at (mixModel.north) {Region 1};
\node (mixModel2) at (4.6,2.5) {
\includegraphics[height=3cm,trim=50 0 0 0, clip]{images/diagramPlots/abnormal1}};
\node (mixModelLabel2) at (mixModel2.north) {Region 2};
\draw[->,line width=0.3mm] (tab1) -- (mixModel);

\node (tab2) at (3,0.25) {
\begin{tabular}{c | c | c | c}
& Patient 1 & Patient 2 & Patient 3\\
\hline
  Region 1 & \textcolor{green!60!black}{normal} & \textcolor{green!60!black}{normal} & \textcolor{red}{abnormal}\\
  Region 2 & \textcolor{green!60!black}{normal} & \textcolor{red}{abnormal} & \textcolor{red}{abnormal}\\ 
\end{tabular}
};
\draw[->,line width=0.3mm] (3,1.5) -- (tab2);

\node (seq) at (3,-1.5) {Estimated Sequence: Region 2 $\rightarrow$ Region 1};
\node (seqDummyUp) at (3,-1.3) {};
\draw[->,line width=0.3mm] (tab2) -- (seqDummyUp);

\end{tikzpicture}
\caption[Event-based model diagram]{Diagram showing the key concepts behind the event-based model. We assume a toy dataset (top-left) of two region-of-interest biomarkers from three patients, which are at different stages along a hypothetical disease progression timeline (bottom-left). The aim is to estimate which region became abnormal earlier in the disease process. The event-based model solves this by fitting a mixture model to the data (top-right), where the two distributions are assumed to represent normal and abnormal biomarker values respectively. The measurements from each patient are then assessed according to each distribution (middle-right). Finally, the sequence of abnormality is estimated from these values, by placing earlier in the sequence the regions/biomarkers for which there are more abnormal values in the dataset. Diagram made by me.}
\label{fig:bckEbmDiagram}
\end{figure}

The event-based model was introduced by Fonteijn et al. \cite{fonteijn2012event} in 2012 and describes the disease as a sequence of discrete events. A key diagram describing the EBM is given in Fig. \ref{fig:bckEbmDiagram}. Given a small dataset of biomarker measurements from subjects who are assumed to lie at unknown shifts along the disease progression timeline (X-axis), the EBM aims to estimate the order in which brain regions, or more generally any biomarker measurements, become abnormal as the disease progresses. The disease is modelled as a sequence of events, where each event represents a change in the patient state, such as the onset of a new symptom (e.g. 'patient shows a drop in cognitive performance') or measurement of tissue pathology (e.g. 'lumbar puncture shows reduced amyloid beta'). 

In section \ref{sec:ebm_theory} we present the theory behind the EBM, in section \ref{sec:model_est} we present the methods that are used to estimate the abnormality sequence, in section \ref{sec:mix_models} we show how to fit the mixture model parameters and in section \ref{sec:staging} we show how to stage the subjects using the EBM. 

\subsubsection{Theory}
\label{sec:ebm_theory}

The event-based model consists of a series of events $E_1, E_2, \dots , E_N$ and an ordering $S = [s(1), \dots, s(N)]$ which is a permutation of the integers $1,\dots, N$ creating the event ordering $E_{s(1)}, E_{s(2)},\dots, E_{s(N)}$. The set of events is specified a-priori.  Moreover, the model uses a dataset $X$ which contains a set of $X_i$ measurements for each subject $i$. These measurements $X_i$ are defined as $X_i = \{x_{i1}, x_{i2}, \dots, x_{iN}\}$, where $x_{ij}$ represents the value of biomarker $j$ in subject $i$ and is informative of event $E_j$ in subject $i$. 

The event-based model makes two key assumptions: first, measurements are monotonic as the disease progresses and secondly, the event ordering is the same across all patients. The first assumption fits with the hypothetical model presented by Jack et al. \cite{jack2010hypothetical} in fig. \ref{fig:biomk_cascade}. Therefore, a patient for whom event $E_j$ has occurred cannot revert to a state where event $E_j$ did not occur. This assumption is essential because it ensures snapshots are informative about the event ordering \cite{fonteijn2012event}. The second assumption is necessary to be able to aggregate information about the event ordering from the entire set of subjects. 

The aim of the event-based model is to find the probability density function $p(S|X)$ of an event ordering given the biomarker data. One starts by fitting a model for the likelihood function $p(x_{ij}|E_j)$ the likelihood of measuring $x_{ij}$ given event $E_i$ occurred. A similar fit is obtained for $p(x_{ij}|\neg E_j)$, the likelihood of measuring $x_{ij}$ given event $E_j$ has not occurred. More information about mixture model fitting can be found in section \ref{sec:mix_models}. If a subject $i$ is at stage $k$ in the disease progression, events $E_{s(1)},\dots, E_{s(k)}$ have occurred while events $E_{s(k+1)},\dots, E_{s(N)}$ have not occurred. We can therefore define the likelihood of the data from subject $i$ given ordering $S$ as:

\begin{equation}
\label{eq:ebm1}
 p(X_i | S, k) = \prod_{j=1}^k p\left(x_{i,s(j)} | E_{s(j)} \right) \prod_{j=k+1}^N p\left(x_{i,s(j)} | \neg E_{s(j)}\right)
\end{equation}
 
where measurements $x_{ij}$ are assumed to be independent. Since the subject could potentially be at any stage $k$ in the progression, we integrate over $k$:

\begin{equation}
\label{eq:ebm2}
  p(X_i | S) = \sum_{k=0}^N p(k)p(X_i|S,k) 
\end{equation}

where $p(k)$ is the prior probability of the subject being at position $k$ in the sequence. A uniform prior is usually assumed here. Further assuming independence of measurements across patients we get:

\begin{equation}
\label{eq:ebm3}
 p(X|S) = \prod_{i=1}^P p(X_i | S)
\end{equation}

Combining equations \ref{eq:ebm1},\ref{eq:ebm2}, \ref{eq:ebm3} we get the total likelihood:

\begin{equation}
\label{eq:ebm4}
 p(X|S) = \prod_{i=1}^P \left[ \sum_{k=0}^N p(k) \left( \prod_{i=1}^k p\left(x_{i,s(j)} | E_{s(j)} \right) \prod_{i=k+1}^N p\left(x_{i,s(j)} | \neg E_{s(j)}\right) \right) \right]
\end{equation}

\subsubsection{Event Sequence Estimation}
\label{sec:model_est}

Applying Bayes' theorem  we can get the posterior on the sequence:

\begin{equation}
 p(S|X) = \frac{p(S)p(X|S)}{p(X)}
\end{equation}

As the marginal distribution $p(X)$ is analytically intractable, one uses a Markov-chain Monte Carlo (MCMC) algorithm to sample from the posterior distribution $p(S|X)$. One assumes flat priors on the sequence $S$ as any sequence could be equally likely. In the MCMC phase, at each iteration the sequence $S$ can be perturbed by swapping two randomly chosen events. This perturbation rule has been used by Fonteijn et al. \cite{fonteijn2012event}. However, another perturbation method used by Young et al. \cite{young2014data} randomly selects a source and target event and places the source event after the target event, sliding the other biomarkers accordingly (see Fig. \ref{fig:backgr_mcmc_perturb}). The resulting sequence $S^{new}$ is accepted with probability $p = min(1, a)$ where $ a = p(X|S^{new})/p(X|S)$. Otherwise the old sequence is stored and the process is repeated. As MCMC depends on accurate initialisation, one also runs a greedy ascent algorithm in order to find the sequence with the highest likelihood. The greedy ascent is very similar to the MCMC phase, the only difference being that $a$ is set to 1 if $p(X|S^{new}) > p(X|S)$ and to zero otherwise. Depending on the number of biomarkers, the greedy ascent is run for a few thousand iterations and repeated 10 times, with different random permutations of integers $1,\dots,N$ as the starting position. The maximum likelihood sequence obtained from greedy ascent is then used to initialise MCMC sampling, which usually runs for at least 100,000 iterations, again depending on problem size.

The resulting MCMC-sampled sequences are usually plotted in a positional variance matrix $M$ (Fig \ref{fig:bckEbmMcmc}), which is a compact way to represent uncertainty in the event ordering. Each element $M(i,j)$ represents the proportion of times event $E_{s(j)}$ appeared on position $i$ in the sampled sequences, given some \emph{master sequence} $S$. $S$ is usually set to be the maximum likelihood sequence or the \emph{characteristic ordering}, which is given by the average position of the events in the MCMC samples \cite{fonteijn2012event}.

\newcommand{\nodeRad}{0.4cm}

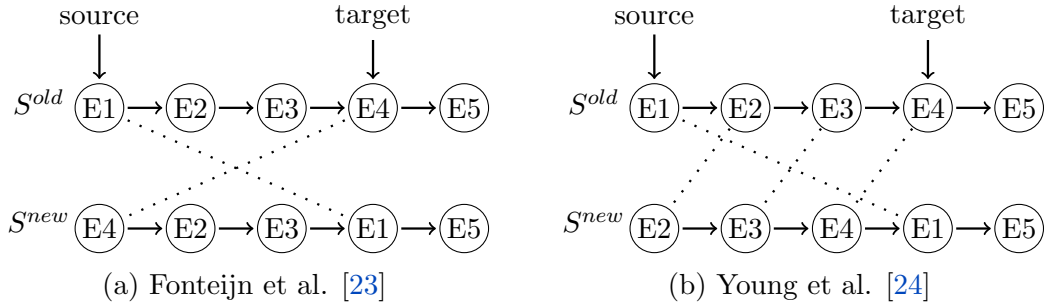
\begin{figure}
\centering
\begin{subfigure}[b]{0.45\textwidth}
 \centering
 \begin{tikzpicture}[scale=0.8]
  \tikzstyle{every node}=[font=\small]

  \draw (-5,3) circle [radius=\nodeRad] node (E1) {E1};
  \draw (-3.5,3) circle [radius=\nodeRad] node (E2) {E2};  
  \draw (-2,3) circle [radius=\nodeRad] node (E3) {E3};
  \draw (-0.5,3) circle [radius=\nodeRad] node (E4) {E4};
  \draw (1,3) circle [radius=\nodeRad] node (E5) {E5};  
  \node (descold) at (-6,3.1) {$S^{old}$};
  \node (src) at (-5,4.5) {source};
  \node (trg) at (-0.5,4.5) {target};
  
  \draw[->,line width=0.3mm] (E1) -- (E2);
  \draw[->,line width=0.3mm] (E2) -- (E3);
  \draw[->,line width=0.3mm] (E3) -- (E4);
  \draw[->,line width=0.3mm] (E4) -- (E5);
  \draw[->,line width=0.3mm,shorten >=3pt] (src) -- (E1);
  \draw[->,line width=0.3mm,shorten >=3pt] (trg) -- (E4);  
  
  \draw (-5,1) circle [radius=\nodeRad] node (E4b) {E4};
  \draw (-3.5,1) circle [radius=\nodeRad] node (E2b) {E2};  
  \draw (-2,1) circle [radius=\nodeRad] node (E3b) {E3};
  \draw (-0.5,1) circle [radius=\nodeRad] node (E1b) {E1};
  \draw (1,1) circle [radius=\nodeRad] node (E5b) {E5};
  \node (descnew) at (-6,1.1) {$S^{new}$};
  
  \draw[->,line width=0.3mm] (E4b) -- (E2b);
  \draw[->,line width=0.3mm] (E2b) -- (E3b);
  \draw[->,line width=0.3mm] (E3b) -- (E1b);
  \draw[->,line width=0.3mm] (E1b) -- (E5b);
  
  \draw[loosely dotted, line width=0.3mm] (E1) -- (E1b);
  \draw[loosely dotted, line width=0.3mm] (E4) -- (E4b);

\end{tikzpicture}
\caption{Fonteijn et al. \cite{fonteijn2012event}}
\end{subfigure}
\begin{subfigure}[b]{0.45\textwidth}
 \centering
 \begin{tikzpicture}[scale=0.8]
  \tikzstyle{every node}=[font=\small]

  \draw (-5,3) circle [radius=\nodeRad] node (E1) {E1};
  \draw (-3.5,3) circle [radius=\nodeRad] node (E2) {E2};  
  \draw (-2,3) circle [radius=\nodeRad] node (E3) {E3};
  \draw (-0.5,3) circle [radius=\nodeRad] node (E4) {E4};
  \draw (1,3) circle [radius=\nodeRad] node (E5) {E5};  
  \node (descold) at (-6,3.1) {$S^{old}$};
  \node (src) at (-5,4.5) {source};
  \node (trg) at (-0.5,4.5) {target};
  
  \draw[->,line width=0.3mm] (E1) -- (E2);
  \draw[->,line width=0.3mm] (E2) -- (E3);
  \draw[->,line width=0.3mm] (E3) -- (E4);
  \draw[->,line width=0.3mm] (E4) -- (E5);
  \draw[->,line width=0.3mm,shorten >=3pt] (src) -- (E1);
  \draw[->,line width=0.3mm,shorten >=3pt] (trg) -- (E4);  
  
  \draw (-5,1) circle [radius=\nodeRad] node (E2b) {E2};
  \draw (-3.5,1) circle [radius=\nodeRad] node (E3b) {E3};  
  \draw (-2,1) circle [radius=\nodeRad] node (E4b) {E4};
  \draw (-0.5,1) circle [radius=\nodeRad] node (E1b) {E1};
  \draw (1,1) circle [radius=\nodeRad] node (E5b) {E5};
  \node (descnew) at (-6,1.1) {$S^{new}$};
  
  \draw[->,line width=0.3mm] (E2b) -- (E3b);
  \draw[->,line width=0.3mm] (E3b) -- (E4b);
  \draw[->,line width=0.3mm] (E4b) -- (E1b);
  \draw[->,line width=0.3mm] (E1b) -- (E5b);
  
  \draw[loosely dotted, line width=0.3mm] (E1) -- (E1b);
  \draw[loosely dotted, line width=0.3mm] (E2) -- (E2b);
  \draw[loosely dotted, line width=0.3mm] (E3) -- (E3b);
  \draw[loosely dotted, line width=0.3mm] (E4) -- (E4b);
  
\end{tikzpicture}
\caption{Young et al. \cite{young2014data}}
\end{subfigure}

\caption[MCMC perturbation rules in the event-based model]{MCMC perturbation rules used by (a) Fonteijn et al. \cite{fonteijn2012event} and (b) Young et al. \cite{young2014data}. Both methods assume randomly selected source and target events. The method by Fonteijn et al. only swaps the source event (E1) with the target event (E4). On the other hand, the perturbation used by Young et al. moves a source event after a target event and slides the other biomarkers accordingly.  Diagram made by me.}
\label{fig:backgr_mcmc_perturb}
\end{figure}

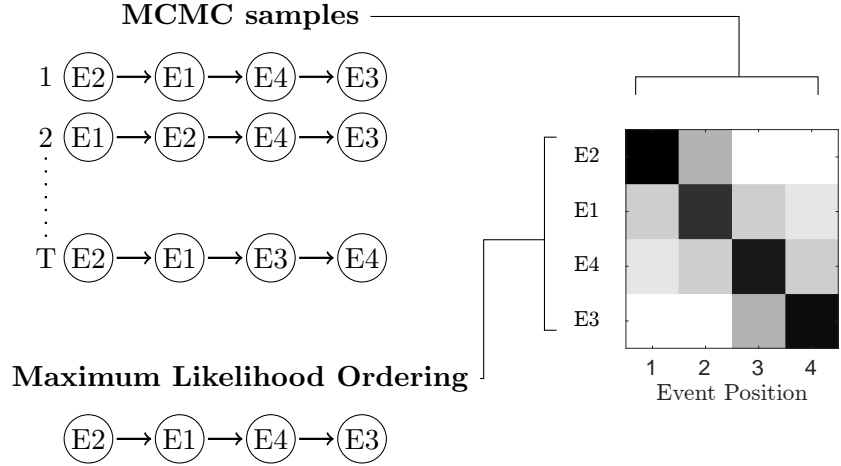
\begin{figure}
\centering
\begin{tikzpicture}[scale = 0.8]
  \tikzstyle{every node}=[font=\small]

  \node (samples) at (-2.5,4) {\textbf{MCMC samples}};
  \draw (-5,3) circle [radius=\nodeRad] node (E2) {E2};
  \draw (-3.5,3) circle [radius=\nodeRad] node (E1) {E1};  
  \draw (-2,3) circle [radius=\nodeRad] node (E4) {E4};
  \draw (-0.5,3) circle [radius=\nodeRad] node (E3) {E3};
  
  \draw (-5,2) circle [radius=\nodeRad] node (E1m) {E1};
  \draw (-3.5,2) circle [radius=\nodeRad] node (E2m) {E2};  
  \draw (-2,2) circle [radius=\nodeRad] node (E4m) {E4};
  \draw (-0.5,2) circle [radius=\nodeRad] node (E3m) {E3};
  
  \draw (-5,0) circle [radius=\nodeRad] node (E2s) {E2};
  \draw (-3.5,0) circle [radius=\nodeRad] node (E1s) {E1};
  \draw (-2,0) circle [radius=\nodeRad] node (E3s) {E3};
  \draw (-0.5,0) circle [radius=\nodeRad] node (E4s) {E4};
  
  \node (1) at (-5.7,3) {1};
  \node (2) at (-5.7,2) {2};
  \node (T) at (-5.7,0) {T};
  
  \draw[loosely dotted, line width=0.3mm] (2) -- (T);

  \draw[->,line width=0.3mm] (E2) -- (E1);
  \draw[->,line width=0.3mm] (E1) -- (E4);
  \draw[->,line width=0.3mm] (E4) -- (E3);
  
  \draw[->,line width=0.3mm] (E1m) -- (E2m);
  \draw[->,line width=0.3mm] (E2m) -- (E4m);
  \draw[->,line width=0.3mm] (E4m) -- (E3m);
  
  \draw[->,line width=0.3mm] (E2s) -- (E1s);
  \draw[->,line width=0.3mm] (E1s) -- (E3s);
  \draw[->,line width=0.3mm] (E3s) -- (E4s);
  
  
  \node (ordering) at (-2.5,-2) {\textbf{Maximum Likelihood Ordering}};
  \draw (-5,-3) circle [radius=\nodeRad] node (E2c) {E2};
  \draw (-3.5,-3) circle [radius=\nodeRad] node (E1c) {E1};  
  \draw (-2,-3) circle [radius=\nodeRad] node (E4c) {E4};
  \draw (-0.5,-3) circle [radius=\nodeRad] node (E3c) {E3};

  \draw[->,line width=0.3mm] (E2c) -- (E1c);
  \draw[->,line width=0.3mm] (E1c) -- (E4c);
  \draw[->,line width=0.3mm] (E4c) -- (E3c);
  
  \node (pic) at (5,0) {\includegraphics[scale=0.4]{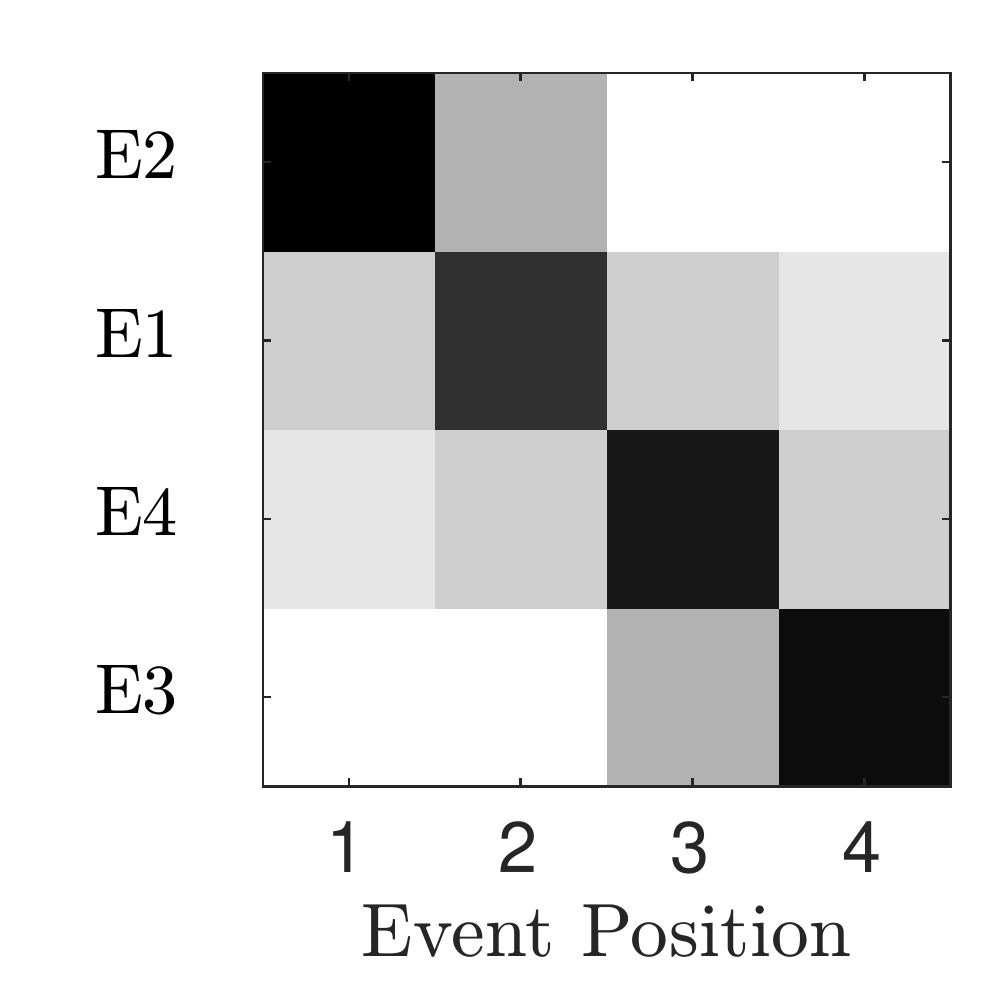}};
  
  \draw (ordering.east) -| (1.5,0.3) -- (2.5,0.3);
  \draw (samples.east) -| (5.7,3);
  
  \draw (4,2.7) |- (7,3) -- (7,2.7);
  \draw (2.7,2) -| (2.5,-1.2) -- (2.7,-1.2);
  
\end{tikzpicture}
\caption[Event-based model - MCMC sampling diagram]{MCMC sampling and positional variance computation. MCMC sampling finds a series of $T$ samples, which are then used to derive the \emph{characteristic ordering}, where events are ordered according to their average position in the MCMC samples. Entries $M(i,j)$ in the positional variance matrix stores the relative number of times each event appeared in each position in the sequence. The events in the positional variance matrix are ordered according to the characteristic ordering.  Diagram made by me.}
\label{fig:bckEbmMcmc}
\end{figure}

\subsubsection{Mixture Models for Data Likelihood}
\label{sec:mix_models}

In equation \ref{eq:ebm1} we need to model the distributions $p\left(x_{i,j} | E_{j} \right)$ and $p\left(x_{i,j} | \neg E_{j}\right)$ of abnormal and normal biomarker values, using the measurements in $X$. Fonteijn et al. \cite{fonteijn2012event} used a Gaussian distribution for $p\left(x_{i,j} | \neg E_{j}\right)$ and a uniform distribution for $p\left(x_{i,j} | E_{j} \right)$. The parameters for the Gaussian distribution were set as the mean and standard deviation of biomarker values corresponding to controls, while the limits of the uniform distribution were set to be the minimum and maximum observed biomarker values. While this works in familial AD and Huntington's disease \cite{fonteijn2012event} due to well-defined control populations, this does not work well in sporadic AD due to the control population being not well-defined -- e.g. some controls can already have abnormal amyloid levels, which could result in the distribution for normal values encompassing all observed values. Therefore, the approach of Young et al. \cite{young2014data} for sporadic AD involved optimising the mixture model parameters based on the subjects' data in a data-driven manner. In this case, prior constraints were used on the mixture model parameters, i.e. the mean and standard deviation of the Gaussian distributions, for biomarkers that did not change from healthy to diseased subjects.

\subsubsection{Patient Staging and Diagnosis Prediction}
\label{sec:staging}

After the maximum likelihood sequence has been found using the greedy ascent method described in section \ref{sec:model_est}, each subject can be assigned a disease stage $k$ as follows:

\begin{equation}
\label{eq:staging}
 k = \argmax_k p(k) p(X_i | S, k) = p(k) \prod_{i=1}^k p\left(x_{i,s(j)} | E_{s(j)} \right) \prod_{i=k+1}^N p\left(x_{i,s(j)} | \neg E_{s(j)}\right)
\end{equation}
 
As before, the prior $p(k)$ is assumed to be uniform. It should be noted that stages range from zero to $N$, the number of events. If a subject is at stage $k$ it means that all events up to and including $k$ have occurred while the events after $k$ have not occurred. 

The event-based model can also be used to classify subjects into controls and AD, or any other symptomatic subgroups \cite{young2014data}. Given a threshold stage $t$, one can predict all subjects having a stage less than or equal to $t$ to be controls and all subjects with stages greater than $t$ to be patients. The optimal threshold is the one which maximises the balanced accuracy, defined as follows:

\begin{equation}
 Accuracy = \frac{TP + TN}{TP + FP + FN + TN}
\end{equation}
where $TP$, $FP$, $FN$, $TN$ represent the number of true positive, false positive, false negative and true negative subjects respectively.

\subsubsection{Discussion}

The EBM is a useful tool for modelling the progression of diseases when only limited, cross-sectional data is available. The model can also be used to stage subjects, in discrete units, along the disease progression timeline. The model parameters are estimated using Markov Chain Monte Carlo sampling, based on optimising a conditional likelihood. 

\subsubsection{Advantages and Limitations}

The event-based model by Fonteijn et al. \cite{fonteijn2012event} has several advantages. It is a data-driven progression model which does not use a-priori defined clinical stages, which can often be unreliable and can limit the temporal resolution of the model. Moreover, it does not require longitudinal data, which makes it very useful for analysing rare types of dementia for which comprehensive longitudinal datasets do not exist. The Bayesian framework in which it is formulated also allows it to estimate uncertainty in the abnormality sequence. The model can also easily combine data from different modalities.  

The current model has several limitations. The trajectory parameters are modelled as step functions, which is a strong assumption given the continuous nature many biomarkers used in AD. Secondly, in the fitting process the conditional probability of the sequence given a-priori estimated distribution parameters is optimised, instead of the joint distribution over the sequence and the distribution parameters, which can result in a suboptimal solution. Third, the model cannot use longitudinal biomarker  measurements in order to enable more precise staging and prognosis estimates.  Finally, the model also assumes that all subjects follow the same progression sequence, which is not the case in heterogeneous datasets due to differences in the underlying pathology, genetics and environmental factors. Identifiability can also be an issue, mostly when, for a certain biomarker, the distributions of normal and abnormal values overlap -- this can result in the biomarker's event being placed either towards the beginning or the end of the sequence, even if the true position of that event is in the middle of the sequence.

\subsection{Differential Equation Model}
\label{sec:bckDem}

\begin{figure}[h]
 \centering
 
 \par{\huge $\xRightarrow{\text{Forward Model}}$}
 
 \begin{subfigure}{0.3\textwidth}
    \centering
    What we want\\
    \vspace{1em}
    \includegraphics[width=0.90\textwidth]{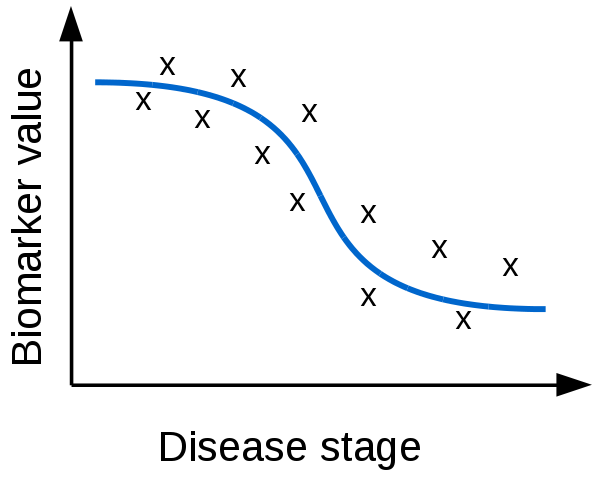}
    \vspace{1em}
 \end{subfigure}
 \begin{subfigure}{0.3\textwidth}
     \centering
     \vspace{1em}
     \includegraphics[width=\textwidth]{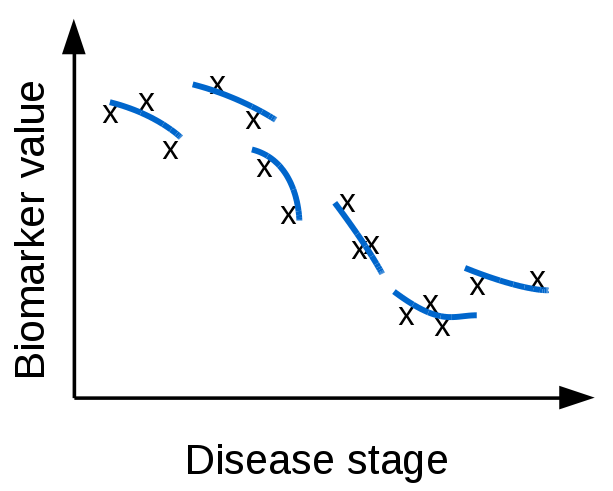}
     \vspace{1em}
 \end{subfigure}
 \begin{subfigure}{0.3\textwidth}
     \centering
     What we have\\
     \vspace{1em}
     \includegraphics[width=0.90\textwidth,trim= 0 0 0 30]{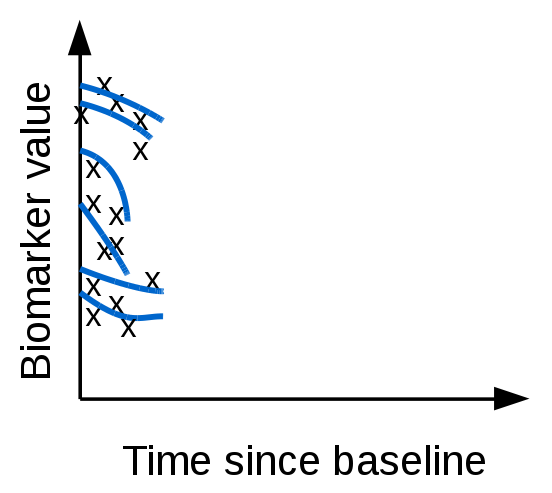}
     \vspace{1em}
 \end{subfigure}
 
 \begin{subfigure}{0.34\textwidth}
    \centering
    \includegraphics[width=\textwidth]{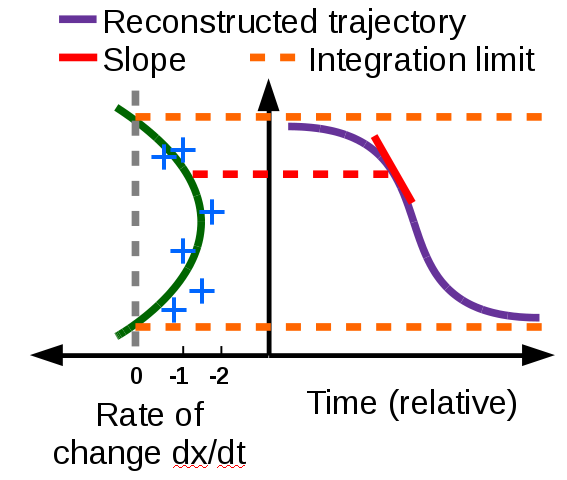}
    \vspace{1em}
 \end{subfigure}
  \begin{subfigure}{0.3\textwidth}
  \centering
  \vspace{-3.5em}
  $$
  lim_{\Delta t \xrightarrow{}  0} \frac{\Delta x}{\Delta t} = \frac{\delta x}{\delta t} = f(x)
  $$
  
  Solve for $x$ using the Euler method:
  \begin{align*}
  t_1 &= t_0 + \delta t \\
  x_1 &= x_0 + f(x_0) \delta t \label{eq:dem3}
  \end{align*}
 \end{subfigure}
 \begin{subfigure}{0.3\textwidth}
     \centering
     \includegraphics[width=\textwidth]{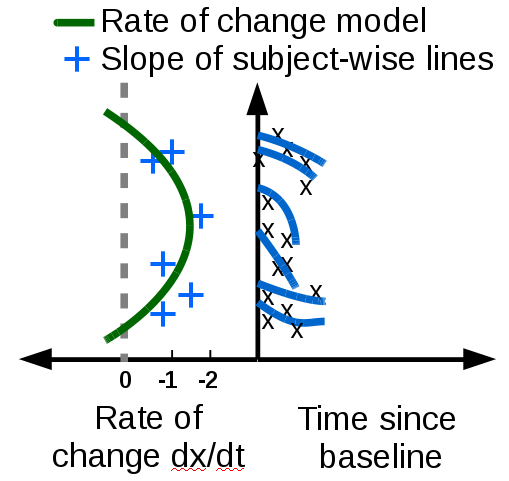}
     \vspace{1em}
 \end{subfigure}
 \vspace{-2em}
 \par{\huge $\xLeftarrow[\text{Inverse Problem}]{}$}
 
 \caption[Diagram of the Differential Equation Model (DEM)]{Diagram of the Differential Equation Model (DEM). (top-left) Hypothetical biomarker signature that needs to be reconstructed, along with subject measurements. (top-middle) To make the model more realistic, each subject is made to follow a slightly different trajectory due to heterogeneity. (top-right) In practice, we don't know the disease stage, so we align the measurements at time since baseline visit. (bottom-right) The DEM model estimates a rate of change model from the slopes of lines fitted to each subject's biomarker data. At least two measurements per subject are required in order to estimate this slope. (top-middle) The DEM then performs a line integral using the Euler method to recover the biomarker trajectory (top-right).  Diagram made by me.}
\label{fig:bckDEM}
\end{figure}

The differential equation model (DEM) \cite{ashford2001modeling, yang2011quantifying, sabuncu2011dynamics, villemagne2013amyloid, oxtoby2018} constructs the biomarker trajectories from the change in biomarker values between different visits (Fig \ref{fig:bckDEM}). In many medical settings we only have short-term longitudinal data, hence the biomarker scores $s$ are observed for each subject over a few visits. By determining how these scores change ($\Delta s$) over time ($t$) during a specified time interval $\Delta t$, the temporal rate of progression ($\Delta s/\Delta t$) can be modelled as a function of the mean biomarker value $f(s)$ \cite{ashford2001modeling}:

\begin{equation}
\label{eq:dem_1}
 \frac{\Delta s}{\Delta t} \approx f(s)
\end{equation}

The model given by $f(s)$ can be parametric (e.g. linear, polynomial) or non-parametric such as Gaussian Processes (GP). We then perform a line integral along $f(s)$ to recover $s(t)$. More explicitly, if we take the limit as $\Delta t \xrightarrow{} 0$ from Eq. \ref{eq:dem_1}, we get that:

\begin{equation}
\label{eq:dem2}
lim_{\Delta t \xrightarrow{}  0} \frac{\Delta s}{\Delta t} = \frac{\delta s}{\delta t} = f(s)
\end{equation}

Solving this numerically is done using the Euler method. We set an initial $(t_0,s_0)$ and small increment step $\delta t$ and find the next pair $(t_1, s_1)$ as follows:

\newcommand\numberthis{\addtocounter{equation}{1}\tag{\theequation}}

\begin{align*}
 t_1 &= t_0 + \delta t \\
 s_1 &= s_0 + f(s_0) \delta t \numberthis \label{eq:dem3}
\end{align*}

This is repeated until the full curve defined by $(t_0, s_0), (t_1, s_1), \dots, (t_n, s_n)$ is reconstructed. Since the DEM model is univariate, the process is repeated independently for the other biomarkers. 

\subsubsection{Advantages and Limitations}

The differential equation model has several advantages. It is a fully data-driven method that does not require a-priori defined clinical categories. In contrast to the event-based model, it can estimate non-parametric biomarker trajectories which make minimal assumptions on the shape of the biomarker trajectories. Moreover, the DEM can use any model to estimate the change in biomarker values. While \cite{oxtoby2018} used Gaussian Processes to estimate the change in values, others \cite{villemagne2013amyloid} used polynomial functions.

The model has several limitations. First of all, the DEM is univariate, so the biomarker trajectories are fit independently. This requires alignment on the temporal axis after they are recovered, and makes it susceptible to noise within that biomarker\footnote{A multivariate model would've been able to use information from other biomarkers to help estimate such a noisy trajectory, hence are more robust in theory.}. Secondly, in this formulation the model does not allow one to directly estimate uncertainty in the trajectory values along the y-axis. One option for estimating uncertainty is to integrate posterior samples of the differential model and then align them all on the temporal axis. However, this does not result in true confidence intervals, since at the anchor point there will be zero uncertainty.

\subsection{The Disease Progression Score Model}
\label{sec:bckDps}

\begin{figure}
\centering
\includegraphics[width=\textwidth]{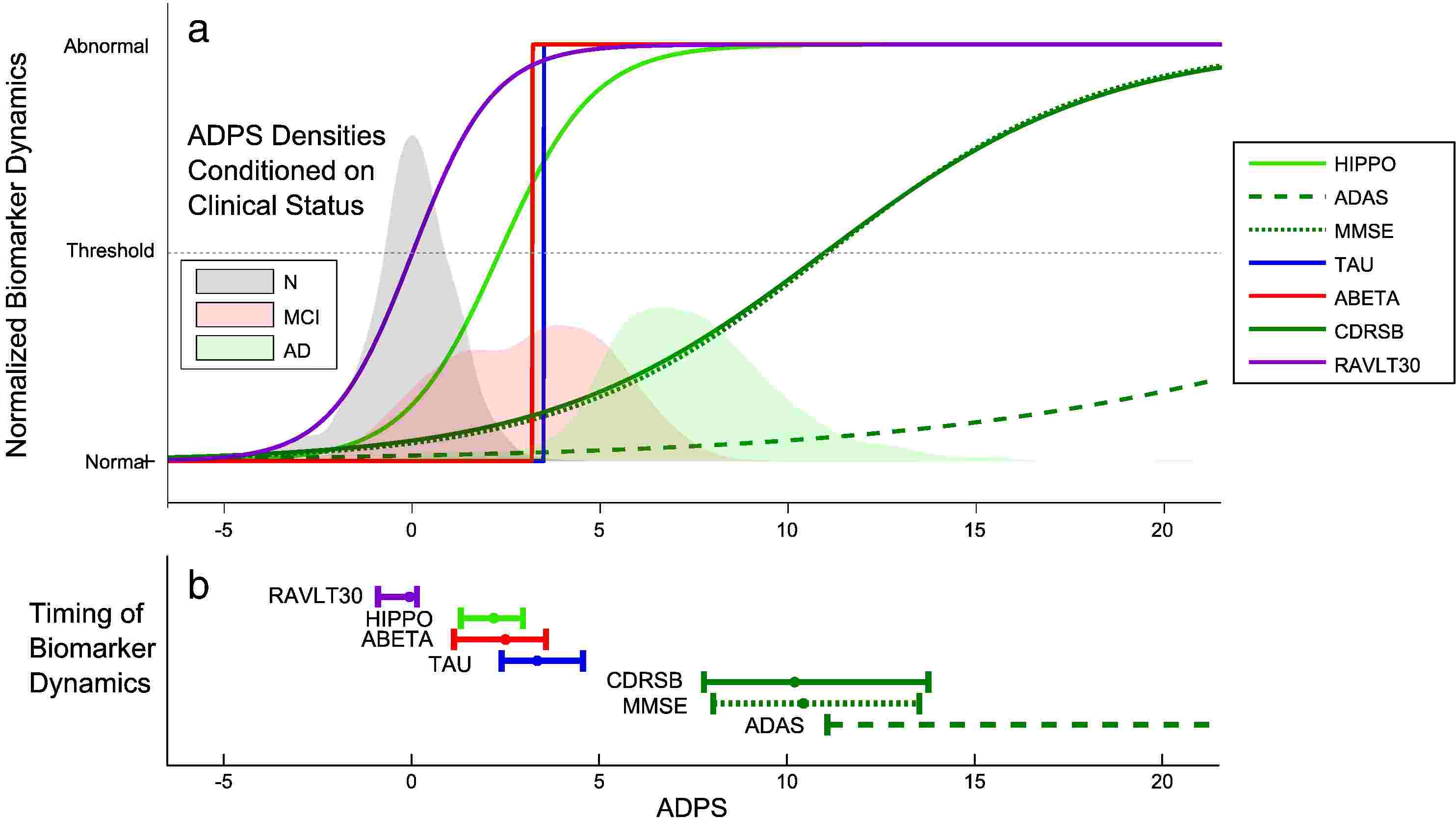}
\caption[ADNI biomarker trajectories estimated by Jedynak et al. \cite{jedynak2012computational}]{Biomarker trajectories estimated by the disease progression model by Jedynak et al. \cite{jedynak2012computational}. Reproduced with permission from \cite{jedynak2012computational}.}
\label{fig:bckDps}
\end{figure}

The disease progression score (DPS) model was proposed by Jedynak et al. \cite{jedynak2012computational}. It is based on three main assumptions:
\begin{itemize}
 \item Subjects follow a common disease progression but they have a different age at onset and progression speed.
 \item Each biomarker trajectory is a monotonic curve that follows a sigmoidal shape
 \item The speed of progression of each subject is the same across the entire disease time-course.
\end{itemize}

Biomarker trajectories estimated by the model for typical AD progression are shown in Fig. \ref{fig:bckDps}. The model estimates the optimal shape\footnote{within a parametric family, in this case sigmoidal family} of the biomarker trajectories, while estimating a disease progression score for each subject, which is the stage along the disease time course. The disease progression score $s_{ij}$ for subject $i$ at visit $j$ is defined as a linear transformation of age $t_{ij}$:

\begin{equation}
\label{eq:dps}
 s_{ij} = \alpha_i t_{ij} + \beta_i
\end{equation}
where $\alpha_i$ and $\beta_i$ represent the speed of progression and time shift (i.e. disease onset) of subject $i$. 

The DPS model assumes that biomarker measurements are independent and follow a sigmoidal trajectory $f(s)$ given the disease progression score $s$. The sigmoidal function for biomarker $k$ with parameters $\theta_k = [a_k,b_k,c_k,d_k]$ is defined as:

\begin{equation}
 f(s;\theta_k) = \frac{a_k}{1+exp(-b_k(s-c_k))} + d_k
\end{equation}
where $d_k$ is the minimum value, $d_k+a_k$ is the maximum value, $a_kb_k/4$ is the maximum slope and $c_k$ is the inflexion point. Authors choose to model the biomarker trajectory as parametric sigmoidal curves because they provide a better fit than linear models \cite{sabuncu2011dynamics, caroli2010dynamics}, and can account for floor and ceiling effects. The value $y_{ijk}$ of biomarker $k$ from subject $i$ at visit $j$ is a normally distributed random variable:
\begin{equation}
 p(y_{ijk} | \alpha_i, \beta_i, \theta_k, \sigma_k) = N(y_{ijk} | f(\alpha_i t_{ij} + \beta_i; \theta_k), \sigma_k )
\end{equation}
We further define $I$ to be the set of all triplets $(i,j,k)$ for which measurements are available. Assuming independence across all measurements, we get the following model conditional likelihood:
\begin{equation}
 p(y | \alpha, \beta, \theta, \sigma) = \prod_{(i,j,k) \in I} p(y_{ijk} | \alpha_i, \beta_i, \theta_k, \sigma_k)
\end{equation}
where $y = [y_{ijk}]$ for $(i,j,k) \in I$. Vectors $\alpha = [\alpha_1, \dots, \alpha_S]$ and $\beta = [\beta_1, \dots, \beta_S]$, where $S$ is the number of subjects, denote the stacked parameters for the subject shifts. Vectors $\theta = [\theta_1, \dots, \theta_K]$ and $\sigma = [\sigma_1, \dots, \sigma_K]$, with $K$ being the number of biomarkers, represent the stacked parameters for the sigmoidal trajectories and measurement noise specific to each biomarker.

The parameters of the model are therefore $\Theta = [\alpha, \beta, \theta, \sigma]$ and the log-likelihood function associated with it is:
\begin{equation}
 l(\alpha, \beta, \theta, \sigma) = \sum_{(i,j,k) \in I} log \sigma_k + \frac{1}{2 \sigma_k ^2} \left( y_{ijk} - f(\alpha_i t_{ij})\right)
\end{equation}

\subsubsection{Model Fitting}

Model fitting is done by loopy belief propagation, which alternates between optimising the sigmoidal parameters $\sigma$ and the subject specific parameters $\alpha, \beta$. Alg. \ref{fig:algo_dps} shows the fitting procedure. In line \ref{alg:init}, we initialise $\alpha_i = 1$, $\beta_i = 0$ for every $i$. On lines \ref{alg:theta} and \ref{alg:sigma} the optimal parameters for every biomarker trajectory are optimised. On line \ref{alg:alpha}, the subject specific shifts and progression speeds are also optimised. On line \ref{alg:identif}, a transformation is performed in order to make the model identifiable. For a similar reason, parameters $\alpha_i$ and $\beta_i$ are rescaled for every subject (line \ref{alg:rescale}), so that the disease progression scores of healthy controls have a mean $\mu_N$ of 0 and a standard deviation $\sigma_N$ of 1.

\begin{algorithm}
 Initialise $\alpha^{(0)}$, $\beta^{(0)}$\;\label{alg:init}
 \For{$l=1$ to $L$}{
    \For{$k=1$ to $K$}{
      $\theta_k^{(1)} = \argmin_{\theta_k} \sum_{(i,j) \in I_k} (y_{ijk} - f(\alpha_i^{(0)} t_{ij} + \beta_i^{(0)} | \theta_k))^2$\label{alg:theta}\\
      $\sigma_k^{(1)^2} = \frac{1}{|I_k-2I-4|} \sum_{(i,j) \in I_k} (y_{ijk} - f(\alpha_i^{(0)} t_{ij} + \beta_i^{(0)} | \theta_k))^2$\label{alg:sigma}
    }
    \For{$i=1$ to $I$}{
      $\alpha_i^{(1)}, \beta_i^{(1)} = \argmin_{\alpha_i, \beta_i}  \sum_{(j,k) \in I_i} \frac{1}{\sigma_k^{(1)^2}}(y_{ijk} - f(\alpha_i t_{ij} + \beta_i | \theta_k^{(1)}))^2$\label{alg:alpha}
    }
    $\alpha^{(0)}=\alpha^{(1)}, \beta^{(0)}=\beta^{(1)}$\label{alg:update}
 }

  \For{$k=1$ to $K$}{
    \If{$b_k < 0$}{
      $a_k^{(1)} = -a_k^{(1)}, b_k^{(1)} = -b_k^{(1)}, d_k^{(1)} = d_k^{(1)}+a_k^{(1)}$\label{alg:identif}\\
    }
  }
  
  \For{$i=1$ to $I$}{
    $\alpha_i = \frac{\alpha_i}{\sigma_N}, \beta_i = \frac{\beta_i - \mu_N}{\sigma_N}$\label{alg:rescale}
  }
 \caption{The optimisation procedure for the disease progression score by \cite{jedynak2012computational}.}
 \label{fig:algo_dps}
\end{algorithm}

\subsubsection{Advantages and Limitations}

The model by Jedynak et al. \cite{jedynak2012computational} has several advantages. As opposed to the differential equation model, the model is multivariate and automatically aligns biomarker trajectories on the same temporal axis. Furthermore, compared to the event-based model by \cite{fonteijn2012event}, the biomarker trajectories are modelled as continuous sigmoidal trajectories instead of step functions. Moreover, each subject has an associated time shift and progression speed. Parameter estimation is performed with loopy belief propagation which is very similar to the Expectation-Maximisation framework \cite{bishop2007pattern}, but using hard assignments of the latent variables at each iteration.

The model has several limitations. The main limitation of this model is that the trajectories are assumed to be sigmoidal, which is not necessarily the case for many biomarkers such as cognitive tests, which show continuous decline even in late stages of the disease. Furthermore, each subject is assumed to follow the same progression pattern, which is not true in many heterogeneous datasets such as ADNI. The DPS model can also suffer identifiability issues when it attempts to stage very early-stage or late-stage subjects, as in these time-windows the biomarker trajectories are mostly flat. This issue can generally be addressed by setting priors on the time-shift and progression-speed of the subjects.

\subsection{The Self-Modelling Regression Model}
\label{sec:bckSem}

\begin{figure}
\centering
\includegraphics[width=0.9\textwidth]{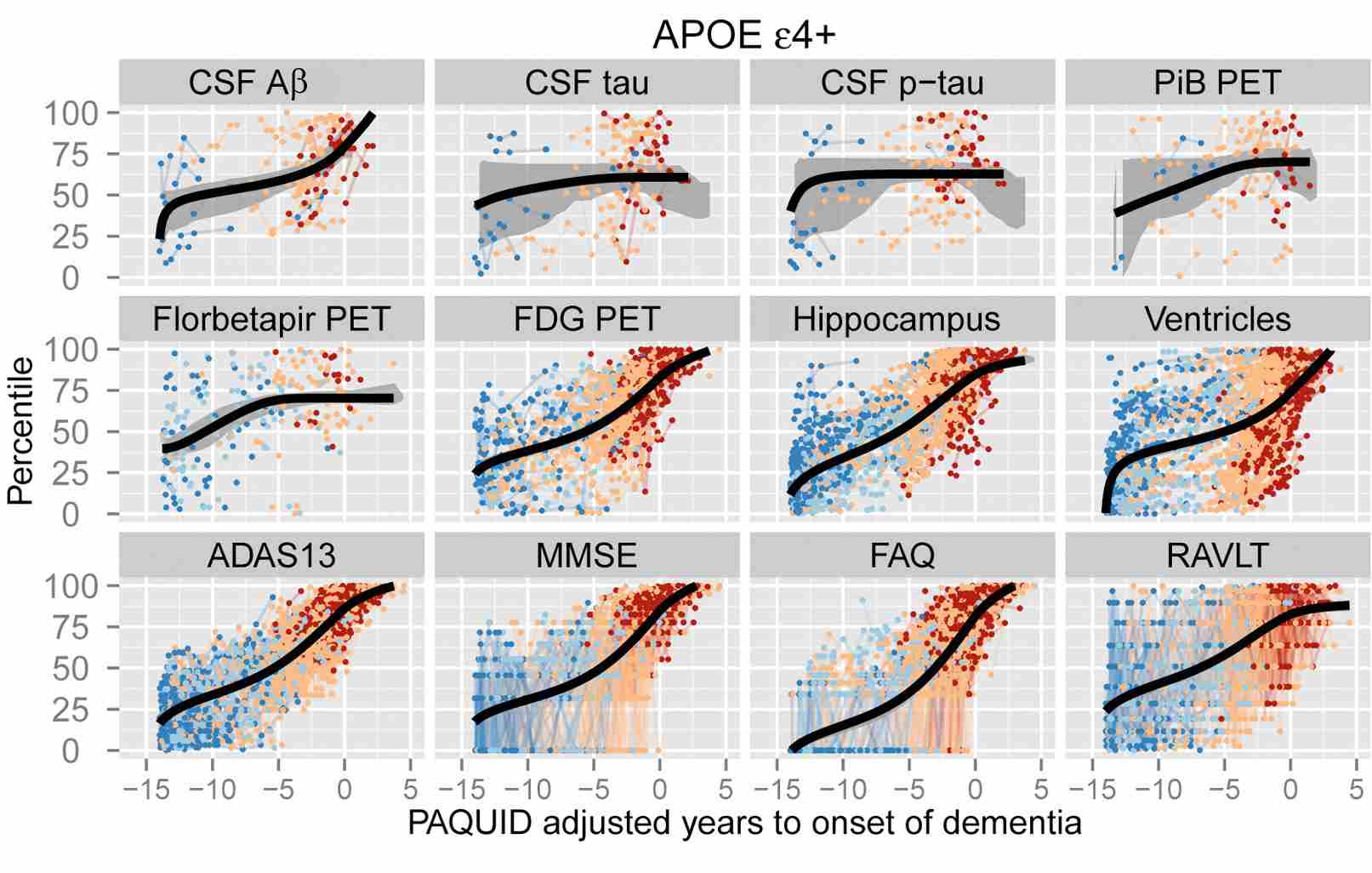}
\caption[ADNI Biomarker trajectories estimated by Donohue et al. \cite{donohue2014estimating}]{Biomarker trajectories estimated using the self-modelling regression approach by \cite{donohue2014estimating}. Reproduced with permission from \cite{donohue2014estimating}.}
\label{fig:bckSem}
\end{figure}

Self-modelling regression (SEMOR) is a method that fits several curves under the assumption of a common shape \cite{donohue2014estimating}. This approach has been used by Donohue et al. \cite{donohue2014estimating} to estimate non-parametric biomarker trajectories with linear subject-specific effects (Fig \ref{fig:bckSem}). Compared to the model by Jedynak et al. \cite{jedynak2012computational}, this model estimates non-parametric biomarker trajectories as fixed effects and includes subject-specific random effects. No subject-specific progression speed is modelled in the original formulation \cite{donohue2014estimating}. 

We assume that $Y_{ij}$ is the measurement of biomarker $j$ in subject $i$, $g_j$ is a continuously differentiable monotone function, $\gamma_i \sim N(0, \sigma_{\gamma}^2)$ is the time shift for subject $i$. The model is defined as follows:
\begin{equation}
 \label{eq:semor1}
 Y_{ij}(t) = g_j(t +\gamma_i)+\alpha_{0ij} + \alpha_{1ij}t+\epsilon_{ij}(t)
\end{equation}
where parameters $\alpha_{0ij}, \alpha_{1ij} \sim N(0, \Sigma_j)$ model a linear perturbation of the non-parametric trajectory $g_j$ for subject $i$ and biomarker $j$, $t$ is the time and $\epsilon_{ij}(t) \sim N(0, \sigma_j)$ is the measurement noise.  

\subsubsection{Parameter Fitting}

Fitting the model is also done by loopy belief propagation -- one iteratively estimates each set of parameters $(g_j, \gamma_i, \alpha)$ until convergence of the Residual Sum of Squares (RSS). The algorithm makes use of the following residuals:
\begin{equation}
\begin{cases}
  R_{ij}^g(t) = Y_{ij}(t) - \alpha_{0ij} - \alpha_{1ij}t & \qquad E\left[ R_{ij}^g(t) | g_j, t, \gamma_i \right] = g_j(t+\gamma_i)\\
  R_{ij}^{\alpha}(t) = Y_{ij}(t) - g_j(t+\gamma_i) & \qquad E\left[ R_{ij}^{\alpha}(t) | \alpha_{0ij}, \alpha_{1ij}, t \right] = \alpha_{0ij} + \alpha_{1ij}t\\
  R_{ij}^{\gamma}(t) = t-g_j^{-1}(Y_{ij}(t)) & \qquad E\left[ R_{ij}^{\gamma}(t) | \gamma_i \right] \approx g_j^{-1}(g_j(t+\gamma_i))-t = \gamma_i\\  
\end{cases}  
\end{equation}

Using the above residuals, the model is fit by initialising $\gamma_i$ and iterating the following steps\cite{donohue2014estimating}:
\begin{enumerate}
 \item Given $\gamma_i$, estimate $g_j$ by setting $\alpha_{0ij} = \alpha_{1ij} = 0$ and iterating the following subroutine:
 \begin{enumerate}
  \item Estimate $g_j$ by a monotone curve fit on $R_{ij}^g(t)$
  \item Estimate $\alpha_{0ij}, \alpha_{1ij}$ using the linear mixed model of $R_{ij}^{\alpha}(t)$. Repeat steps a and b until convergence of each $RSS_j = \sum_{it} \left[ Y_{ij}(t)-g_j(t+\gamma_i)-\alpha_{0ij}-\alpha_{1ij}t \right]^2 $
 \end{enumerate}
 
 \item Given the estimated $g_j$, set $\alpha_{0ij} = \alpha_{1ij} = \epsilon_{ij}(t) = 0$ and estimate each $\gamma_i$ with the average of $R_{ij}$ over all $j$ and $t$. Steps 1 and 2 are repeated until convergence of the total $RSS = \sum_{ijt} \left[ Y_{ij}(t)-g_j(t+\gamma_i)-\alpha_{0ij}-\alpha_{1ij}t \right]^2 $

\end{enumerate}

\subsubsection{Advantages and limitations}

The SEMOR model by \cite{donohue2014estimating} has many advantages. It robustly estimates biomarker trajectories using non-parametric curves, in contrast with previous approaches \cite{jedynak2012computational, fonteijn2012event}. Moreover, it also models unique trajectories for each subject as linear deviations from the average trajectories using a mixed effects model. Each subject also has its own temporal shift along the disease progression timeline. Parameter estimation is done with loopy belief propagation, carefully alternating the optimisation of certain groups of parameters in order to minimise the residual of the cost function.

The model has some limitations. In its basic formulation it does not model different progression speeds for different individuals. Moreover, estimating all the subject-specific parameters $(\alpha_{0ij}, \alpha_{1ij}, \gamma_i)$ requires suitable priors and at least two longitudinal measurements per subject, which might not be available in some datasets. The high number of parameters that need to be estimated, especially the subject-specific parameters, can result in overfitting of the data, although this is mitigated to some extent by the priors placed on these parameters. Moreover, it also assumes that biomarker measurements are independent, which makes it unsuitable for modelling large sets of correlated biomarkers such as voxelwise measurements. Identifiability can also be an issue with the SEMOR model when the population trajectory becomes almost linear, such as in the case of biomarkers that don't show differences between healthy subjects and patients.

\subsection{The Manifold-based Mixed Effects Model}
\label{sec:bckMan}

The manifold-based mixed effects model was introduced by Schiratti et al. in 2015 \cite{schiratti2015mixed}. The model generalises a previous linear mixed effects model by \cite{datar2012mixed} to account for time shifts and describes it in a Riemannian manifold setting. Each subject $i$ is assumed to have a trajectory $\gamma_{i}$ which is a deviation from the average population trajectory $\gamma$. The deviation is modelled as a time shift $\tau_i$ and progression speed $\alpha_i$ of subject $i$ along the disease time-course. 

Let us assume we observe $p$ individuals, each having $n_i$ observations obtained at times $t_{i,1} < \dots < t_{i,n_i}$, each having $y_{i,1}, \dots, y_{i,n_i}$ biomarker measurements. For a geodesic $M$ and a point $p_0 \in M$ at time $t_0$ with velocity $v_0 \in T_{p_0}M$, we define $\gamma_{p_0,t_0,v_0} = Exp_{t_0,p_0}(\alpha_iv_0)(.)$ as the geodesic which passes through point $p_0$ at time $t_0$ with velocity $v_0$\footnote{In Riemannian geometry, $Exp$ refers to the exponential map, which is a function from the tangent space $T_{p}M$ into $M$ itself.}. We also consider $t_{i,j}$ and $y_{i,j}$ as the age and biomarker measurement for subject $i$ at visit $j$. This gives us the following model:

\begin{equation}
\label{eq:manifold1}
 y_{i,j} = Exp_{t_0+\tau_i,p_0}(\alpha_iv_0)(t_i,j) + \epsilon_{i,j}
\end{equation}
where
\begin{equation}
\label{eq:manifold2}
\begin{cases}
  \alpha_i = exp(\eta_i),\ \eta_i \sim \bigotimes_{i=1}^p N(0, \sigma_{\eta}^2)\\
    
  \tau_i \sim \bigotimes_{i=1}^p N(0, \sigma_{\tau}^2)\\  
  \epsilon_{i,j} \sim \bigotimes_{i,j} N(0, \sigma^2)\\  
\end{cases}
\end{equation}
The model assumes that each $\eta_i$ and $\tau_i$ are independent. The parameters of the model are $\theta = [p_0, t_0, v_0, \sigma_{\eta}, \sigma_{\tau}, \sigma]$. The model above can be re-written as:
\begin{align*}
  y_{i,j} &= \gamma_{p_0,t_0,v_0}(\alpha_iv_0(t_{i,j}-t_0-\tau_i))+\epsilon_{i,j}\\
          &= \gamma_i(t_{i,j})+\epsilon_{i,j} \numberthis \label{eq:manifold3}
\end{align*}
where $\gamma_i$ is the subject specific trajectory that is modelled as an affine reparametrisation of the average trajectory $\gamma_{p_0,t_0,v_0}$. The model described here is univariate. However, an extension of the model has been published by Schiratti et al. \cite{schiratti2015learning} which extends this framework to a multivariate analysis. 

Parameter estimation in the model by Schiratti et al. \cite{schiratti2015mixed} is performed using maximum likelihood estimation (MLE) using the Gauss-Hermite quadrature approximation, which is equivalent to the Laplace approximation of the observed likelihood. Authors used the Nelder-Mead method for estimating the numerical optimisation.

\subsubsection{Advantages and Limitations}

The model has several strengths and weaknesses. The main strength lies in the flexible Riemannian manifold framework, that allows one to create different models depending on how the inner product is defined. Moreover, the model estimates subject specific trajectories $\gamma_i$, time shifts $\tau_i$ and progression speeds $\alpha_i$. However, one of the limitations of the model is that it assumes a parametric form of the biomarker trajectories (i.e. sigmoidal).

\section{Spatiotemporal Disease Progression Models}
\label{sec:bckSpa}

Spatiotemporal models of disease progression have been proposed over the last few years. In the following section, we will present two key spatiotemporal models of progression that estimate voxelwise patterns of pathology, while also estimating latent subject-specific time shifts.

\subsection{The Voxelwise Disease Progression Model}
\label{sec:bckVox}

\begin{figure}
\centering
\includegraphics[width=0.8\textwidth]{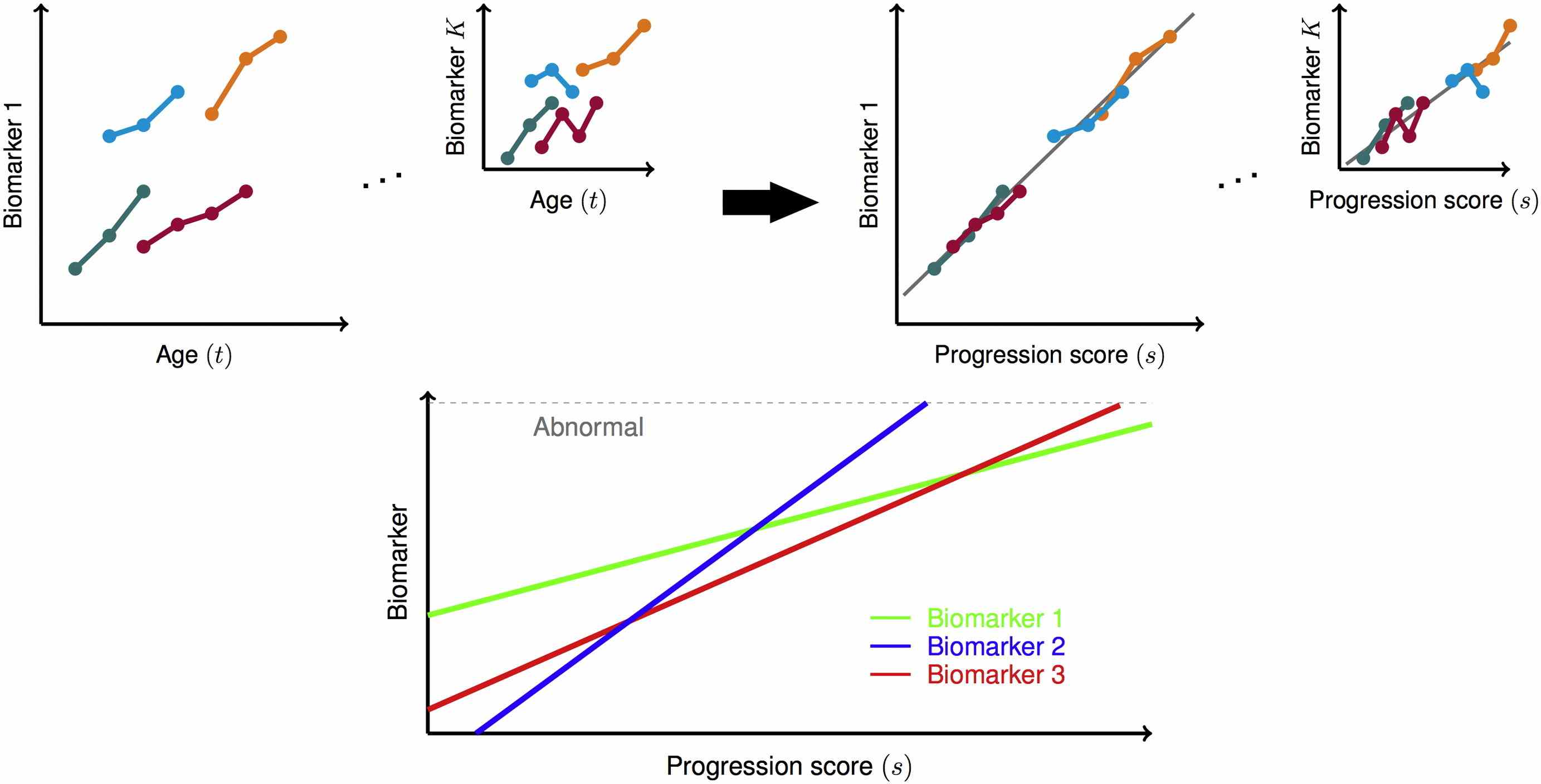}
\caption[Voxelwise disease progression model by Bilgel et al. \cite{bilgel2016multivariate}]{Diagram of the voxelwise disease progression model by Bilgel et al. \cite{bilgel2016multivariate}. The model places biomarker measurements along a latent "progression score" axis, and then models the dynamics of these measurements using linear functions. Reproduced with permission from \cite{bilgel2016multivariate}.}
\label{fig:bckBil}
\end{figure}

A voxelwise disease progression model has been introduced in 2016 by Bilgel et al. \cite{bilgel2016multivariate}. This model allows the discovery of patterns of atrophy that are not confined to a given region of interest (ROI). Since the input data is represented by voxel-wise measurements such as amyloid burden, a spatial correlation function is used to model correlation between voxel values. The model is built on the framework of the disease progression score by Jedynak et al. \cite{jedynak2012computational}. 

A diagram of the model is given in Fig \ref{fig:bckBil}. The model aligns biomarker measurements along a latent "progression score" axis, and then models the dynamics of these measurements using linear functions. Let us assume that $t_{ij}$ represents the age for subject $i$ at visit $j$. The progression score $s_{ij}$ for subject $i$ at visit $j$ is an affine transformation of age $t_{ij}$:

\begin{equation}
 \begin{cases}
 s_{ij} = \alpha_i t_{ij} + \beta_i = \textbf{q}_{ij}^T\textbf{u}_i\\
 \textbf{u}_i \sim N_2(\textbf{m}, V(\boldsymbol{\nu}))
\end{cases}
\end{equation}
where $\alpha_i$, $\beta_i$ are the progression speed and time shift of subject $i$, $\textbf{q}_{ij} = [t_{ij}, 1]^T$ and $\textbf{u}_{i} = [\alpha_i, \beta_i]^T$. The prior covariance matrix $V$ is modelled as a 2 $\times$ 2 positive definite matrix that has a Log-Cholesky parametrisation given by $\boldsymbol{\nu}$. More precisely, if we consider a matrix $U$ such that $V = UU^T$ and $U = \begin{bmatrix}
    U_{11} & U_{12} \\
    0 & U_{22}\\
\end{bmatrix}$, then $\boldsymbol{\nu} = [logU_{11}, logU_{12}, logU_{22}]$

Furthermore, let us denote by $\textbf{y}_{ij}$ the $K \times 1$ vector of biomarker measurements for subject $i$ at visit $j$. The longitudinal trajectories corresponding to these measurements are modelled as follows:

\begin{equation}
 \begin{cases}
  \textbf{y}_{ij} = \textbf{a}s_{ij}+\textbf{b}+\epsilon_{ij}\\
  \epsilon_{ij} \sim N_K(0,R(\boldsymbol{\lambda},\boldsymbol{\rho}))
 \end{cases}
\end{equation}
where $\textbf{a} = [a_1, \dots, a_K]^T$, $\textbf{b} = [b_1, \dots, b_K]^T$ are the coefficients of the linear model and $\epsilon_{ij}$ is the measurement noise that is independent and identically distributed across different subjects and visits. The matrix $R(\boldsymbol{\lambda},\boldsymbol{\rho})$ is the spatial covariance that is assumed to have the form $R = \Lambda C \Lambda$, where $\Lambda$ is a diagonal matrix with diagonal elements $\boldsymbol{\lambda}$ and $C$ is a correlation matrix that is parameterised by $\boldsymbol{\rho}$ \cite{bilgel2016multivariate}. This ensures that the matrix $R(\boldsymbol{\lambda},\boldsymbol{\rho})$ is positive definite. In order to model correlation among voxel measurements, the elements $C_{kk'}$ of matrix $C$ must be a function of the distance $d \equiv d(k,k')$ between voxels $k$ and $k'$. Several such options exist:
\begin{itemize}
 \item Exponential: $C_{kk'} = \text{exp}(-d/\rho)$
 \item Gaussian: $C_{kk'} = \text{exp}(-(d/\rho)^2)$
 \item Exponential: $C_{kk'} = \left(1+(d/\rho)^2\right)^{-1}$
 \item Spherical: $C_{kk'} = \left( 1 - \frac{3}{2}\frac{d}{\rho} + \frac{1}{2}\left(\frac{d}{\rho}\right)^{3} \right)$ if $d<\rho$
\end{itemize}
The model parameters are therefore $\boldsymbol{\theta} = [\textbf{m},\boldsymbol{\nu},\textbf{a},\textbf{b},\boldsymbol{\lambda},\boldsymbol{\rho}]$. The model is a mixed effects model where $\textbf{a}$, $\textbf{b}$ are the fixed effects and $\textbf{u}_i$ are the random effects. 

\subsubsection{Model Fitting}

The model is fit using the Expectation-Maximisation (EM), described below. In line with the standard EM framework \cite{bishop2007pattern}, the algorithm optimises the expected value of the full log-likelihood $\mathbb{E}_{p(\textbf{u}|\theta')}\left[l(\textbf{y}, \textbf{u}; \boldsymbol{\theta})\right]$ given the current estimate of the latent variables $\textbf{u}$. The complete data log-likelihood is:
\begin{align*}
 & l(\textbf{y}, \textbf{u}; \boldsymbol{\theta}) = \sum_i l(\textbf{y}_i, \textbf{u}_i; \boldsymbol{\theta}) =\\
 & -\frac{1}{2}\sum_{i,j}log|2\pi R|-\frac{1}{2}\sum_{i,j}\left(\textbf{y}_{ij}-Z_{ij}\textbf{u}_i-\textbf{b}\right)^T \\
 & -\frac{1}{2}\sum_{i} log|2\pi V|-\frac{1}{2}\sum_{i}(\textbf{u}_i-\textbf{m})^T V^{-1}(\textbf{u}_i-\textbf{m})
 \numberthis \label{eq:bilgel3}
\end{align*}

\textbf{E-step}

Let $(\textbf{y}, \textbf{u})$ be the complete data and $\boldsymbol{\theta}' = [\textbf{m}',\boldsymbol{\nu}',\textbf{a}',\textbf{b}',\boldsymbol{\lambda}',\boldsymbol{\rho}']$ be the parameters estimated at the previous EM iteration. Bilgel et al. \cite{bilgel2016multivariate} show that the E-step integral $Q(\boldsymbol{\theta}, \boldsymbol{\theta}')$ is proportional to $\sum_i \int \Phi(\tilde{\textbf{u}_i};\hat{\textbf{u}}'_i, \Sigma_i') l(\textbf{y}_i, \tilde{\textbf{u}}_i; \boldsymbol{\theta})d\tilde{\textbf{u}}_i$, where $\Phi$ is a multivariate normal probability density function with mean:

\begin{equation}
 \hat{\textbf{u}}'_i = \left( \sum_{j} Z'^T_{ij} R'^{-1} Z'_{ij} + V'^{-1}\right)^{-1}\left( \sum_{j} Z'^T_{ij} R'^{-1} (\textbf{y}_{ij}-\textbf{b}') + V'^{-1}\textbf{m}' \right)
\end{equation}
and covariance matrix $\Sigma'_{i}=\left( \sum_{j} Z'^T_{ij} R'^{-1} Z'_{ij} + V'^{-1}\right)^{-1}$. Evaluating the integral gives the following final form:

\begin{align*}
 & Q(\boldsymbol{\theta}, \boldsymbol{\theta}') = -\frac{1}{2} \sum_{ij}log|R| - \frac{1}{2}\sum_{ij}\left( \textbf{y}_{ij} - Z_{ij}\hat{\textbf{u}}'_i -\textbf{b}\right) - \frac{1}{2}\sum_{ij}Tr\left( Z^T_{ij} R^{-1} Z_{ij}\Sigma_i' \right) - \\
 & \frac{1}{2}\sum_{i}log|V| - \frac{1}{2}\sum_{i}\left( \hat{\textbf{u}}'_i - \textbf{m} \right)^T V^{-1} \left( \hat{\textbf{u}}'_i - \textbf{m} \right) - \frac{1}{2}\sum_i Tr\left( V^{-1} \Sigma_i'\right) \numberthis \label{eq:bilgel_e-step}
\end{align*}

\textbf{M-step}

At the M-step we need to find $\boldsymbol{\theta} = \argmax_{\boldsymbol{\theta}}Q(\boldsymbol{\theta}, \boldsymbol{\theta}')$. The full derivations are given in \cite{bilgel2016multivariate}, yielding the following updates:

\begin{equation}
\label{eq:bilgel_mstep1}
 \textbf{a} = \frac{\left( \sum_i \nu_i \right) \left( \sum_{ij} \textbf{y}_{ij}s'_{ij} \right) - \left( \sum_{ij} \textbf{y}_{ij} \right) \left( \sum_{ij}s'_{ij} \right)}{\left( \sum_i \nu_i \right) \left( \sum_{ij} \textbf{q}_{ij}^T \Sigma_i' \textbf{q}_{ij} + s_{ij}'^2 \right) - \left( \sum_{ij}s'_{ij} \right)^2}
\end{equation}

\begin{equation}
\label{eq:bilgel_mstep2}
 \textbf{b} = \frac{\left( \sum_{ij} \textbf{y}_{ij} \right) \left( \sum_{ij} \textbf{q}_{ij}^T \Sigma_i' \textbf{q}_{ij} + s_{ij}'^2 \right) - \left( \sum_{ij} \textbf{y}_{ij}s'_{ij} \right) \left( \sum_{ij} s'_{ij} \right)}{\left( \sum_i \nu_i \right) \left( \sum_{ij} \textbf{q}_{ij}^T \Sigma_i' \textbf{q}_{ij} + s_{ij}'^2 \right) - \left( \sum_{ij}s'_{ij} \right)^2}
\end{equation}

\begin{equation}
 \textbf{m} = \frac{1}{n}\sum_i \hat{\textbf{u}}'_i
\end{equation}

\begin{equation}
 \boldsymbol{\nu} = \argmax_{\boldsymbol{\nu}} Q(\boldsymbol{\theta}, \boldsymbol{\theta}')
\end{equation}

\begin{equation}
 \boldsymbol{\lambda}, \boldsymbol{\rho} = \argmax_{\boldsymbol{\lambda}, \boldsymbol{\rho}} Q(\boldsymbol{\theta}, \boldsymbol{\theta}')
\end{equation}

\subsubsection{Advantages and Limitations}

The model by Bilgel et al. \cite{bilgel2016multivariate} has several advantages. First of all, it is specifically tailored for dealing with voxelwise measurements such as amyloid load by modelling the spatial correlations. Secondly, like the disease progression score by Jedynak et al. \cite{jedynak2012computational}, it estimates subject specific temporal shifts and progression speeds. 

The model has several limitations. First of all, the biomarker trajectories are assumed to be linear, which is a strong assumption especially for early biomarkers such as amyloid, which start to plateau when subjects are in the MCI stages. The linearity was required in order to make the model inference with EM computationally tractable. Moreover, the biomarker correlation structure based on a spatial distance function does not allow one to recover fine-grained, disconnected patterns of pathology, as has been found for various types of dementias due to disruption in underlying brain networks \cite{seeley2009neurodegenerative}. The model also doesn't account for inter-subject differences by estimating deviations from the common population-wide trajectory.

\subsection{Cortical Atrophy Progression Model}
\label{sec:bckCor}

\begin{figure}
\centering
\includegraphics[width=0.8\textwidth]{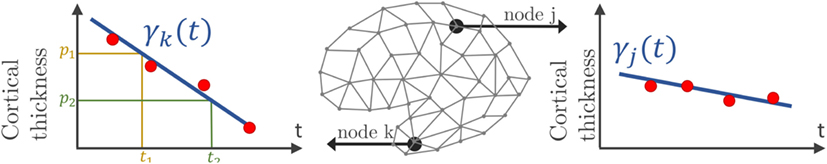}

\vspace{2em}
\includegraphics[width=0.8\textwidth]{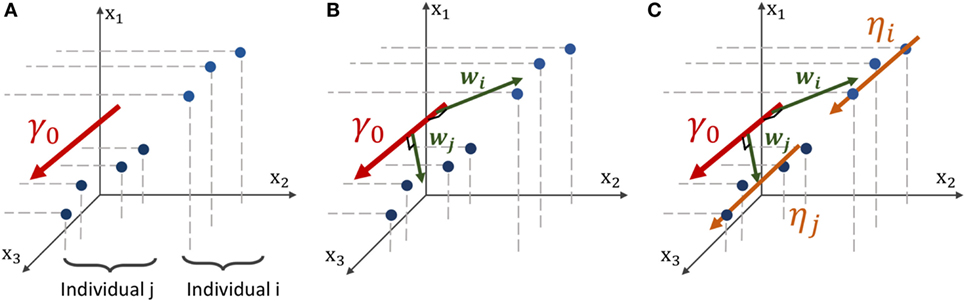}
\caption[Diagram of the cortical atrophy progression model by Koval et al. \cite{koval2017statistical}]{Diagram of the cortical atrophy progression model by Koval et al. \cite{koval2017statistical}. (top) The model estimates a unique, linear trajectory for the dynamics of cortical thickness measurements at each point on the brain cortical surface. (bottom) Subject-specific trajectories $\eta_i$ and $\eta_j$ are modelled by a shift of the population trajectory $\gamma_0$ through vectors $w_i$ and $w_j$. Reproduced with permission from \cite{koval2017statistical}.}
\label{fig:bckKov}
\end{figure}

The cortical atrophy progression model was introduced by Koval et al.  \cite{koval2018spatiotemporal}. A diagram of the model is given in Fig. \ref{fig:bckKov}. The model estimates vertexwise linear trajectories of cortical thickness over the entire population, accounting for latent subject specific time-shifts. The equation modelling a biomarker measurement $y_{ijk}$ for subject $i$ at visit $j$ for location $k$ on the brain surface is given as:

\begin{equation}
y_{ijk} = p_k + w_{ik} + \nu_k\alpha_i(t_{ij} - \tau_i - t_0) + \epsilon_{ijk}
\end{equation} 
where $p_k$ and $\nu_k$ are parameters of the linear trajectory over the latent space specific to location $k$, $w_{ik}$ is a subject and location specific intercept, $\tau_i$ is the time-shift for subject $i$, $t_0$ is a time-shift reference and $\epsilon_{ijk}$ is the Gaussian noise for the $y_{ijk}$ measurement. In order to account for spatial correlation, a set of control nodes $\mathbb{V_c}$ is defined, which is a subset of all nodes $V$. Only for these control nodes will parameters $p_k$ and $\nu_k$ be estimated. For the other nodes, the parameters will be an interpolation of the parameters of the control nodes, weighted by the distance of that node to the control nodes. 

\subsection{Parameter Estimation}

Parameter estimation is done using the Monte-Carlos Markov-Chain Stochastic Approximation Expectation Maximisation (MCMC-SAEM) algorithm. This method essentially approximates the intractable E-step using the MCMC sampler. The optimisation method is proven to converge if the model belongs to the exponential family \cite{koval2017statistical}.

\subsection{Advantages and Limitations}

The model by Koval et al. has several advantages. It can estimate spatiotemporal patterns of atrophy, and can be extended to other types of voxelwise biomarkers such as amyloid load or DTI measures. The model also estimates subject-specific latent time-shifts, accounting for different but unknown ages of disease onset in distinct subjects. 

The model also has some limitations that need to be addressed in future work. One limitation is that the authors need to define a-priori the number of control points, which can affect the final smoothness level of the estimated patterns of pathology. While the authors only applied it to a brain surface made of 2,000 nodes, it is unclear whether the model can scale to higher resolutions.

\section{Mechanistic Models}
\label{sec:bckMec}

\subsection{The Network Diffusion Model}
\label{sec:bckNet}

\begin{figure}[h]
\centering
\includegraphics[width=0.8\textwidth]{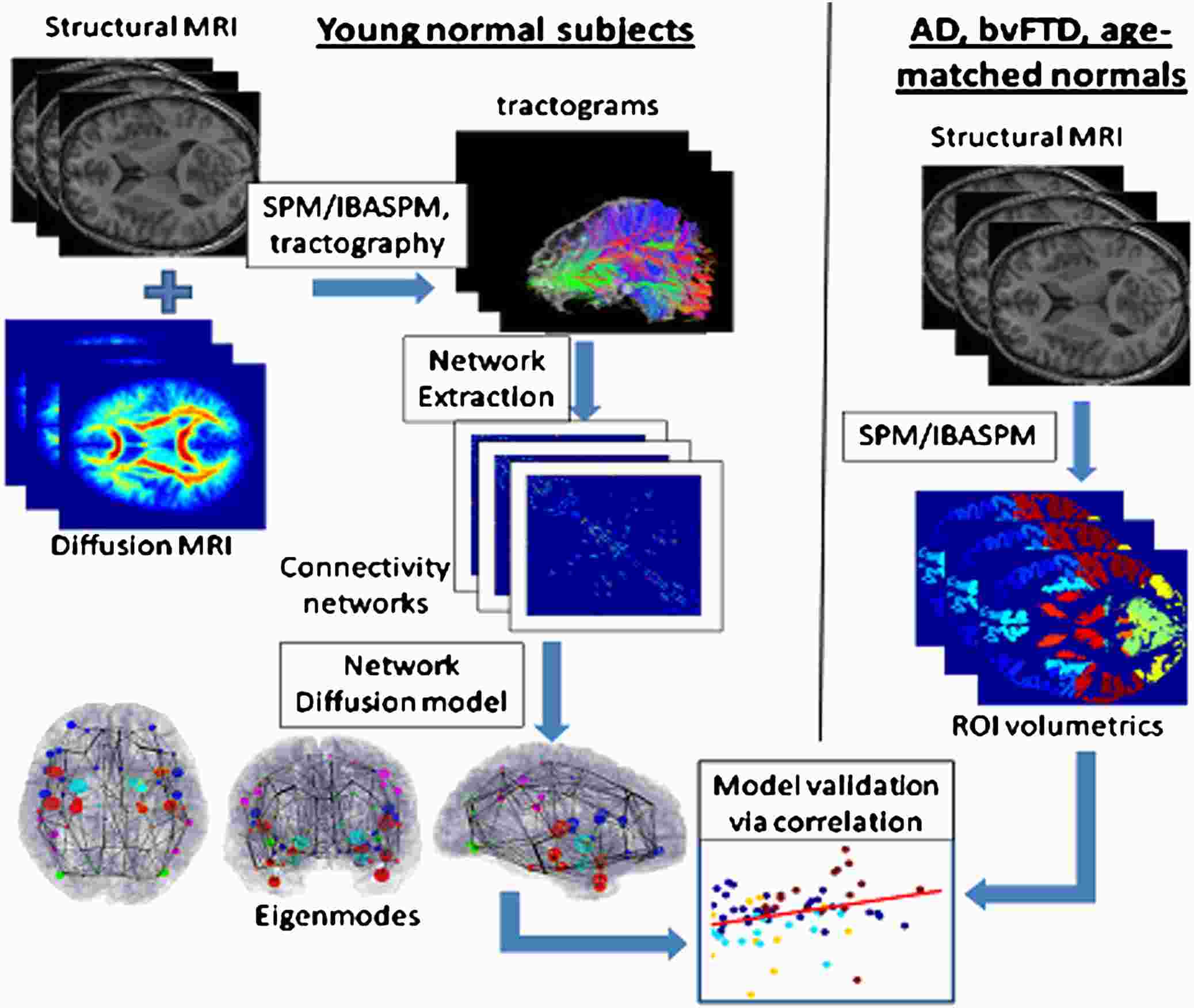}
\caption[Diagram of the network diffusion model by Raj et al. \cite{raj2012network}.]{Diagram of the network diffusion model by Raj et al. \cite{raj2012network}. The model uses MRI and DTI data to extract a structural connectome from healthy subjects through tractography, then computes a connectivity network. Each network is represented as a graph where nodes represent brain ROIs where there is a certain concentration of toxic pathogens and edges represent the connectivity strength. Using this matrix, the authors estimate the eigenvectors of the graph, also called eigenmodes, which are then shown to correlate with atrophy patterns in normal ageing, AD and bvFTD. More precisely, for each disease they compute the amount of atrophy within each ROI corresponding to the graph nodes, and then correlate with the eigenmodes. Reproduced with permission from \cite{raj2012network}.}
\label{fig:bckRaj}
\end{figure}

The network diffusion model was introduced by Raj et al. in 2012 \cite{raj2012network} and later extended in 2015 \cite{raj2015network}. The model is inspired by evidence that Alzheimer's disease pathology spreads along vulnerable pathways in a prion-like manner rather than by spatial proximity \cite{villain2010sequential, englund1988white, kuczynski2010white}. The model works by simulating the diffusion process of a pathogenic protein along a structural connectivity graph from healthy controls. Atrophy and other higher-level pathogenic processes are assumed to be a product of the lower-level diffusion process. See Fig \ref{fig:bckRaj} for a diagram of the model.

The diffusion process is modelled on a hypothetical brain network $G = \{V, E\}$ whose nodes $v_i \in V$ represent regions-of-interest and edges $(i,j)$ having weight $c_{ij}$ represent fibre connections between regions $i$ and $j$. Structures $v_i$ are parcellated regions-of-interest obtained from an atlas, while the connection strength $c_{ij}$ is measured using tractography \cite{behrens2007probabilistic}. If we denote the disease factor in region $i$ as $x_i$, then the flow of the disease from a region $i$ to another region $j$ in a short time interval $\delta t$ is $\beta(x_i - x_j)c_{ij}\delta t$, where $\beta$ is a diffusivity constant controlling the diffusion speed. As $\delta t \to 0$, this results in the following first-order differential equation:
\begin{equation}
\label{eq:raj1}
 \frac{dx_j}{dt} = \beta c_{ij}(x_i - x_j)
\end{equation}
Now let us denote by $\textbf{x}(t) = \{x(v,t), v \in V \}$ the disease factor at time $t$ in every node of the network. Eq. \ref{eq:raj1} will then translate into the "network heat equation" \cite{kondor2002diffusion}:

\begin{equation}
\label{eq:raj2}
 \frac{d\textbf{x}(t)}{dt} = \beta H \textbf{x}(t)
\end{equation}
where $H$ is the Laplacian matrix of $G$ defined as:

\begin{equation}
\label{eq:raj3}
 H(i,j) = \begin{cases} 
 \sum_{j' \neq i} c_{ij'} & \mbox{for } i = j \\ 
 -c_{ij} & \mbox{otherwise} \\ 
\end{cases} 
\end{equation}
We model the cortical atrophy in region $k$ as the accumulation of the disease process:
\begin{equation}
\label{eq:raj4}
 \phi_k(t) = \int_0^t x_k(\tau)d\tau
\end{equation}
Extending this to the whole brain gives:
\begin{equation}
 \boldsymbol{\Phi(t)} = \int_0^t \textbf{x}(\tau)d\tau
\end{equation}
The solution to the above equation is given by:
\begin{equation}
 \textbf{x}(t) = \text{exp}(-\beta H t) \textbf{x}_0
\end{equation}
where $\textbf{x}_0$ is the initial disease concentration, where the term $\text{exp}(-\beta H t)$ acts as a smoothing operator. Performing eigenvalue decomposition on $H = U\Lambda U^{\dagger}$, where $U = [\textbf{u}_1,\dots,\textbf{u}_N]$ is the matrix of eigenmodes, allows to express $\textbf{x}(t)$ as:
\begin{equation}
 \textbf{x}(t) = U \text{exp}(-\Lambda \beta t) U^{\dagger} \textbf{x}_0 = \sum_{i=1}^N \left( \text{exp}(-\beta \lambda_i t) \textbf{u}_i^{\dagger} \textbf{x}_0 \right) \textbf{u}_i
\end{equation}
The time evolution of atrophy can then be described as:
\begin{equation} 
 \boldsymbol{\Phi}(t) = \int_0^t \sum_{i=1}^N \left( \text{exp}(-\beta \lambda_i t) \textbf{u}_i^{\dagger} \textbf{x}_0 \right) \textbf{u}_i\ dt = \sum_{i=1}^N \frac{1}{\beta \lambda_i} \left( 1-\text{exp}(-\beta \lambda_i t)\right) \textbf{u}_i^{\dagger} \textbf{x}_0 \textbf{u}_i 
\end{equation}
Raj et al. \cite{raj2012network} present evidence to suggest that the eigenmodes $\textbf{u}_i$ with the highest corresponding eigenvalues $\lambda_i$ represent the areas that are normally affected by key neurodegenerative processes or diseases, such as normal ageing, AD and behavioural variant frontotemporal dementia (bvFTD) respectively. They suggest that these areas are selectively vulnerable to these types of dementia, in line with previous theories in the field \cite{seeley2009neurodegenerative, zhou2010divergent, zhou2012predicting}. 

\subsubsection{Advantages and Limitations}

The diffusion model by Raj et al. \cite{raj2012network} has several advantages. In contrast with the models presented above, it is able to model the propagation of atrophy along brain connectomes, which can be used to test the prion hypothesis or other related mechanisms. Secondly, this approach allows one to test for other hypotheses of network-based pathology spread such as nodal stress, transneuronal spread, trophic failure, and shared vulnerability \cite{zhou2012predicting}.

The model has several limitations. The model assumes static networks, even though the network dynamically evolves during the time course of the disease. The model also assumes a parametric form of the biomarker trajectories, either exponential or sigmoidal. 

\section{Machine Learning Methods}
\label{sec:bckMac}

Popular machine learning methods that are normally used for discriminative tasks can also be extended to model disease progression by estimating continuous variables. One such method is the Support Vector Machine, which is a non-probabilistic binary linear classifier that was originally developed by Vladimir N. Vapnik and Alexey Ya. Chervonenkis \cite{vapnik2006estimation}. They can perform non-linear classification by mapping the input data into higher-dimensional feature spaces using the \emph{kernel trick} and finding the hyperplane that optimally separates the data. They have been successfully used for differential diagnosis of different types of dementia using post-mortem confirmed subjects \cite{kloppel2008automatic}. Other uses of SVMs on medical images and other high-dimensional data have also been shown \cite{lao2004morphological,fan2005classification,mourao2005classifying,kawasaki2007multivariate}.

Other popular classifiers used in the machine learning field are random forests \cite{ho1995random,breiman2001random}. These work by building a multitude of decision trees on training data, using them to make predictions on the test data and performing majority voting on the predictions of individual trees to select the final labels. Random forests have been demonstrated in medical imaging for diagnosis classification \cite{gray2013random}, image quality transfer \cite{alexander2014image} or myocardium segmentation \cite{lempitsky2009random}. 

Yet another machine learning model popular for time-series predictions is the Long short-term memory (LSTM) network \cite{hochreiter1997long}. LSTMs are a type of recurrent artificial neural network (RNN) that replace the conventional hidden nodes from RNNs with \emph{memory cells}, which ensure that the gradient cannot vanish or explode during training. LSTMs have been applied in medical imaging research for diagnosing between healthy and AD subjects \cite{karlekar2018detecting}, predicting the onset of diseases \cite{razavian2016multi}, predicting diagnoses in electronic health records (EHR) \cite{lipton2015learning}, predicting mortality risk \cite{aczon2017dynamic} as well as several other clinical variables \cite{harutyunyan2017multitask}. 

\subsection{Advantages and Limitations}

Discriminative models such as SVMs, random forests or LSTMs have several advantages. They work for a wide variety of problems, datasets and underlying models and are generally robust to high-dimensional data. Some of them can also be resistant to overfitting, especially when used in conjunction with regularisation and data augmentation techniques. A key advantage of LSTMs in particular is that they are suitable for time-series data, and can be naturally used to predict future biomarker values, although the other models can also be extended to work with continuous data.

These discriminative models also have several limitations. First of all, they generally require labelled data, in the form of a-priori defined clinical categories or stages, which are usually coarse, inaccurate and biased. These limit the temporal resolution of the model. Moreover, it is also harder to interpret what these models learn from the data, which limits their use for understanding the disease process. For some models there is also a lack of mathematical proofs and guarantees regarding their convergence during training, as well as behaviour while making predictions.

\section{Summary}

In Fig. \ref{tab:compDPMs} we show a summary of the main features of data-driven disease progression models, as well as discriminative models. For each model, we show the trajectory shape, indicate whether models incorporated latent subject-specific time-shifts (in terms of intercept or intercept + progression speed), subject-specific trajectories in the form of random effects as well as spatial correlation. For each model, we also indicate the key limitation.

We can observe several key differences between the models. In terms of time-shifts, some models such as the DEM or the network diffusion model do not incorporate any time-shifts, although these could be extended to incorporate such time-shifts. Other models do not model subject-specific trajectories through random effects. Moreover, only spatiotemporal or mechanistic models incorporate correlation between different biomarker measurements. 

In conclusion, over the last few years there have been several models of disease progression that were developed, starting from the early comparisons based on symptomatic groups and moving on to more data-driven approaches and spatiotemporal models. Further work will focus on developing more mechanistic models that enable understanding of the underlying disease process, and can help guide drug development. One example of this is the recent work of \cite{georgiadis2018computational}, which models the dynamics of pathogenic proteins in a neural network and can help understand the effects of such pathogenic proteins in neurodegeneration. However, validation of such models is required through \emph{in vitro} and \emph{in vivo} studies. 

In the following chapters, I will present the application of some of these models to estimate the progression of Posterior Cortical Atrophy (chapter \ref{chapter:pca}), as well as the development of two novel models of disease progression (chapters \ref{chapter:dive} and \ref{chapter:dkt}).


\definecolor{orange0}{rgb}{1, 0, 0}
\definecolor{orange1}{rgb}{0.66, 0.33, 0}
\definecolor{orange2}{rgb}{0.33, 0.66, 0}
\definecolor{orange3}{rgb}{0, 0.7, 0}

\definecolor{green1}{rgb}{0, 0.6, 0}
\definecolor{red1}{rgb}{0.6, 0, 0}
\newcommand{\xmark}{\ding{55}}%

\newcommand{\myyes}{\textcolor{green1}{\Large{\checkmark}}}
\newcommand{\myno}{\textcolor{red1}{\Large{\xmark}}}

\begin{table}
\centering
\footnotesize
\renewcommand{\arraystretch}{1.5}
\begin{tabular}{>{\centering\arraybackslash}p{2.8 cm} | >{\centering\arraybackslash}p{2.3 cm} | >{\centering\arraybackslash}p{1.5 cm} | >{\centering\arraybackslash}p{1.5 cm} | >{\centering\arraybackslash}p{2.0 cm} | >{\centering\arraybackslash}p{2 cm} | >{\centering\arraybackslash}p{2.5 cm}}
 Model & Trajectory shape & \multicolumn{2}{c|}{Subject Time-shifts} & \multirow{2}{2cm}{\centering Subject-specific trajectory} & \multirow{2}{2cm}{\centering Models Biomarker Correlation} & Main Limitation \\
 & & intercept & speed & & \\
 \Xhline{2\arrayrulewidth}
 Event-based Model & step-function & \myyes** & \myno & \myno & \myno & discrete time\\
  \hline
 Differential Equation Model & non-parametric & \myno * & \myno * & \myno & \myno  & univariate\\
  \hline
 Disease Progression Score & sigmoid & \myyes & \myyes & \myno & \myno & sigmoidal trajectory assumption \\
  \hline
 Self-Modelling Regression & non-parametric & \myyes & \myyes & \myyes & \myno & can overfit + identifiability\\
  \hline
 Manifold Model & linear, sigmoid & \myyes & \myyes & \myyes & \myno & user-defined trajectory assumption \\
  \hline
 Voxelwise Model & linear & \myyes & \myyes & \myno & \myyes & linear trajectory assumption\\
  \hline
 Cortical Atrophy Progression Model & linear & \myyes & \myyes & \myyes & \myyes & user-defined trajectory assumption\\
  \hline
 Network diffusion Model & exponential, sigmoidal & \myno * & \myno *&  \myno & \myyes & assumes static connectome + no time-shift\\

\end{tabular}
\caption[Comparison of features of various disease progression models.]{Comparison of features of various disease progression models. By subject-specific trajectories, we exclude deviations from the population-wide trajectory only due to time-shifts. While the models have many limitations, the ones mentioned here are the main ones according to our own opinion. (*) Models cannot be used for subject staging in their basic formulation, but extensions to the model can enable them to estimate subject-specific disease onset and progression speed. Comparison analysis made by me. (**) The subject-specific time-shift in the EBM is discrete and based on cross-sectional data only.}
\label{tab:compDPMs}
\end{table}

\chapter[Longitudinal Neuroanatomical Progression of PCA]{Longitudinal Neuroanatomical Progression of Posterior Cortical Atrophy}
\label{chapter:pca}

This chapter outlines the clinically applied part of my PhD, which focused on modelling the progression of Posterior Cortical Atrophy using already developed methods. The content of this chapter is based on the neuroimaging results from the joint publication below, where I've re-written most of the text for this thesis. I performed all the neuroimaging work: image pre-processing, statistical analysis with EBM and DEM, and the interpretation of the results. The data from table \ref{tab:pcaDemographics} was gathered by Nicholas Firth. Splitting of PCA patients into cognitively-defined subgroups was done by Silvia Primativo. Details in section \ref{sec:pcaParticipants} regarding patient recruitment, patient numbers, clinical diagnosis and pathological confirmation along with image acquisition details from section \ref{sec:pcaImageAcq} were taken from our joint publication.

\section{Publications}

\begin{itemize}
 \item N. C. Firth*, S. Primativo*, R. V. Marinescu*, T. J. Shakespeare, A. Suarez-Gonzalez, M. Lehmann, A. Carton, D. Ocal, I. Pavisic, R. W. Paterson, C. F. Slattery, A. J. M. Foulkes, B. H. Ridha, E. Gil-Néciga, N. P. Oxtoby, A. L. Young, M. Modat, M. J. Cardoso, S. Ourselin, N. S. Ryan, B. L. Miller, G. D. Rabinovici, E. K. Warrington, M. N. Rossor, N. C. Fox, J. D. Warren, D. C. Alexander, J. M. Schott, K. X. X. Yong\^{} and S. J. Crutch\^{}, Longitudinal neuroanatomical and cognitive progression of posterior cortical atrophy, Brain, 2019. (*) joint first authors (\^{}) joint senior authors
 
 In the above manuscript, I preprocessed all the imaging data, performed the modelling and statistical analysis of all the imaging data, and created the figures, tables and diagrams (including statistical tests in the supplementary). I also drafted the section of the results which was related to the imaging results. Other authors recruited patients, collected the data, performed the analysis of cognitive tests, and helped draft the initial version of the manuscript. 

 \item R. V. Marinescu, A. L. Young, Neil P. Oxtoby, N. C. Firth, M. Lorenzi, A. Eshaghi, V. Wottschel, M. J. Cardoso, M. Modat, K. X. X. Yong, S. Primativo, N. C. Fox, M. Lehmann, T. J. Shakespeare, S. J. Crutch, D. C. Alexander, A data-driven comparison of the progression of brain atrophy in Posterior Cortical Atrophy and Alzheimer's disease, AAIC poster, 2016.

 \item R. V. Marinescu, S. Primativo, A. L. Young, N. P. Oxtoby, N. C. Firth, A. Eshaghi, S. Garbarino, J. M. Cardoso, K. Yong, N. C. Fox, M. Lehmann, T. J. Shakespeare, S. J. Crutch, D. C. Alexander, Analysis of the heterogeneity of Posterior Cortical Atrophy: Data-driven Model Predicts Distinct Atrophy Patterns for three different Cognitive Subgroups, AAIC poster, 2017 
\end{itemize}

\section{Introduction}

Posterior Cortical Atrophy (PCA), already described in section \ref{sec:bckPca}, is a progressive neurodegenerative syndrome causing predominantly visuospatial and visuoperceptual impairments \cite{crutch2012posterior}. In order to understand complex disease mechanisms underlying PCA, and design efficient clinical trials for finding treatments of PCA, we need to be able to accurately estimate the temporal progression of atrophy in PCA and contrast it with typical AD (tAD). Previous neuroimaging studies of PCA have shown more atrophy in the superior parietal, occipital and posterior temporal regions as compared to typical AD \cite{lehmann2011cortical, whitwell2007imaging}. However, these studies are all cross-sectional and cannot map the continuous longitudinal progression of the disease. One longitudinal study of PCA \cite{lehmann2012global} showed widespread gray matter loss in both PCA and tAD, but the numbers were small (17 PCA and 16 tAD) and the time interval was short (1 year). Larger longitudinal studies are therefore required to robustly estimate longitudinal progression patterns of PCA as compared to tAD. Moreover, a second aspect that needs to be clarified is the heterogeneity within PCA itself. Some studies have so far reported three dominant subgroups: primary visual (the striate cortex, caudal), parietal (dorsal) and occipito-temporal (ventral) \cite{ross1996progressive,galton2000atypical}. However, evidence for the existence of these groups is mainly limited to individual case reports \cite{ross1996progressive,galton2000atypical} and no previous study looked at the temporal progression of brain atrophy in such subgroups.

The aim of this study is to estimate the progression of MRI brain volumes in PCA as compared to tAD. We used the event-based model (EBM, section \ref{sec:bckEbm}) and the differential equation model (DEM, section \ref{sec:bckDem}) to estimate the progression of brain volumes in 361 individuals (117 PCA, 106 tAD and 138 controls) from three centres in the UK, Spain and US. We also use the event-based model to estimate the progression of atrophy in three cognitively-defined PCA subgroups. Compared to previous studies, our study is the first comprehensive study of atrophy progression in PCA. We also provide the first glimpse into the early progression of atrophy within PCA subgroups.

\section{Methods}

\subsection{Participants}
\label{sec:pcaParticipants}

117 patients with PCA were recruited from three specialist centres: 100 from the Dementia Research Centre (DRC) UK, 9 patients from the University Hospital Virgen del Rocio (HUVR) Memory disorders Unit, Spain and 8 patients from the University of California San Francisco (UCSF) Memory and Aging Center, USA. All PCA participants met two widely-accepted Tang-Wai et al. \cite{tang2004clinical} and Mendez, Ghajarania \& Perryman \cite{mendez2002posterior} criteria. Participants had no clinical features of other neurodegenerative disorders (e.g. visual hallucinations, pyramidal signs), hence fulfilling the criteria for PCA-pure \cite{crutch2017consensus}. 106 tAD patients and 138 healthy controls recruited from the DRC UK were also used for this study. tAD subjects all met criteria for \emph{probable} AD \cite{mckhann2011diagnosis}. Available pathological and molecular analyses for the patients (45/117 = 38\% for PCA, 49/106  = 46\% for tAD) all indicated AD pathology. 

Of all the study participants, 270 had undergone at least one T1 MRI scan and 216 at least one cognitive assessment. Available neuroimaging and neuropsychology data, stratified by the number of visits, are shown in table \ref{tab:pcaDemographics}. PCA, tAD and healthy controls were age-matched (65.44 $\pm$ 7.51 for PCA, 65.67 $\pm$ 7.57 for tAD and 63.13 $\pm$ 5.94 for controls). The gender proportion was as follows: 39\% male for PCA, 62\% male for tAD and 50\% male for controls. PCA and tAD subjects had a similar level of impairment as measured by MMSE scores at first assessment: 20.88 $\pm$ 5.17 for PCA, 19.58 $\pm$ 5.08 for tAD and 29.02 $\pm$ 0.98 for controls. 

\newcommand{\tabwidth}{1cm}

\begin{table}
\centering
\begin{tabular}{c | c c c | c c c}
 & \multicolumn{3}{c|}{\textbf{Imaging}} & \multicolumn{3}{c}{\textbf{Neuropsychology}}\\
\textbf{Visits} & \textbf{Number} & \textbf{Age} & \textbf{Visit Interval} & \textbf{Number} & \textbf{Age} & \textbf{Visit Interval} \\
\multicolumn{7}{c}{}\\
\multicolumn{7}{c}{\textbf{PCA (n=117)}}\\
\hline
All & 89 & 63.52 $\pm$ 6.91 & N/A & 109 & 64.49 $\pm$ 7.54 & N/A \\
≥2 & 46 & 62.11 $\pm$ 6.52 & 1.03 $\pm$ 0.47 & 70 & 63.64 $\pm$ 7.32 & 1.18 $\pm$ 0.48 \\
≥3 & 31 & 62.75 $\pm$ 6.5 & 0.99 $\pm$ 0.47 & 45 & 62.73 $\pm$ 7.26 & 1.15 $\pm$ 0.45 \\ 
≥4 & 15 & 61.46 $\pm$ 4.44 & 0.86 $\pm$ 0.31 & 20 & 63.19 $\pm$ 7.00 & 1.14 $\pm$ 0.40 \\ 
≥5 & 9 & 61.73 $\pm$ 4.06 & 0.81 $\pm$ 0.33 & 7 & 59.44 $\pm$ 4.84 & 1.06 $\pm$ 0.45 \\
≥6 & 2 & 62.35 $\pm$ 1.65 & 0.83 $\pm$ 0.24 & 2 & 57.22 $\pm$ 3.49 & 1.02 $\pm$ 0.35 \\
\multicolumn{7}{c}{}\\
\multicolumn{7}{c}{\textbf{tAD (n=106)}} \\ 
\hline
All & 66 & 66.39 $\pm$ 8.58 & N/A & 58 & 65.68 $\pm$ 7.57 & N/A \\
≥2 & 37 & 66.84 $\pm$ 8.83 & 0.83 $\pm$ 1.46 & 28 & 64.58 $\pm$ 7.08 & 1.35 $\pm$ 0.56 \\ 
≥3 & 21 & 71.0 $\pm$ 6.97 & 0.53 $\pm$ 0.39 & 5 & 66.08 $\pm$ 2.78 & 1.26 $\pm$ 0.43 \\ 
≥4 & 14 & 70.89 $\pm$ 6.33 & 0.47 $\pm$ 0.33 & 0 & N/A & N/A \\
≥5 & 4 & 72.08 $\pm$ 4.81 & 0.49 $\pm$ 0.33 & 0 & N/A & N/A \\ 
≥6 & 1 & 79.9 $\pm$ 0.0 & 0.58 $\pm$ 0.40 & 0 & N/A & N/A \\
\multicolumn{7}{c}{}\\
\multicolumn{7}{c}{\textbf{Controls (n=138)}} \\
\hline
All & 115 & 61.87 $\pm$ 10.43 & N/A & 49 & 63.12 $\pm$ 5.90 & N/A \\
≥2 & 50 & 61 $\pm$ 12.01 & 0.79 $\pm$ 0.66 & 18 & 60.00 $\pm$ 5.87 & 0.91 $\pm$ 0.27 \\
≥3 & 28 & 65.75 $\pm$ 5.96 & 0.66 $\pm$ 0.52 & 0 & N/A & N/A \\
≥4 & 17 & 66.82 $\pm$ 4.88 & 0.45 $\pm$ 0.28 & 0 & N/A & N/A \\
≥5 & 8 & 66.11 $\pm$ 4.83 & 0.44 $\pm$ 0.25 & 0 & N/A & N/A \\
≥6 & 0 & N/A & N/A & 0 & N/A & N/A \\

\end{tabular}
\caption[Demographic details for participants in the PCA study]{Demographic details for participants in the PCA study. Number of participants (n), mean and standard deviation age of participants at baseline visit and mean and standard deviation of visit interval is shown per number of visits. }
\label{tab:pcaDemographics}
\end{table}

\begin{table}
\centering
\begin{tabular}{C{5cm}|c|c|c|c|c|c}
 & \multicolumn{2}{C{3.5cm}|}{\textbf{Vision subgroup (n=30)}} & \multicolumn{2}{C{3.5cm}|}{\textbf{Space subgroup (n=30)}} & \multicolumn{2}{C{3.5cm}}{\textbf{Object subgroup (n=29)}} \\
& n & mean (std) & n & mean (std) & n & mean (std)\\
\hline
MMSE &  29 &     22.38 (5.19) &  28 &    19.54 (4.10) &  29 &     21.90 (5.62) \\
Age &  30 &     64.23 (7.72) &  30 &    63.26 (7.44) &  29 &     65.05 (8.48) \\
Gender (\%male)	&	30 & 16\% &	30 &	46\% &	29 & 41\% \\
PAL &   6 &     15.00 (5.66) &   5 &     5.60 (6.18) &   8 &     10.75 (7.14) \\
Digitspan forwards &  17 &      6.76 (2.86) &  27 &     6.15 (2.46) &  24 &      7.08 (2.36) \\
Digitspan backwards &  16 &      3.88 (1.69) &  28 &     2.89 (1.42) &  23 &      3.78 (2.08) \\
GDA addition &  17 &      1.59 (1.94) &  22 &     0.55 (1.41) &  19 &      1.95 (2.84) \\
GDA subtraction &  17 &      0.76 (1.16) &  22 &     0.23 (0.67) &  19 &      1.84 (2.87) \\
\hline
\textbf{Memory} & & & & & \\
SRMT faces &  18 &     18.78 (4.13) &  25 &    18.88 (4.30) &  20 &     18.10 (3.95) \\
SRMT words &  30 &     21.70 (2.88) &  30 &    19.43 (3.45) &  29 &     20.38 (2.73) \\
\hline
\textbf{Visual processing} & & & & & \\
Shape detection (VOSP) &  28 &     14.43 (3.29) &  29 &    17.17 (2.24) &  28 &     17.18 (2.42) \\
Shape discrimination &  28 &     12.96 (3.18) &  28 &    15.21 (3.45) &  28 &     16.14 (2.68) \\
Crowding &  22 &      6.68 (4.23) &  28 &     9.07 (2.34) &  28 &      8.71 (2.23) \\
\hline
\textbf{Space Perception} & & & & & \\
Number location &  29 &      2.10 (2.66) &  29 &     2.07 (2.69) &  28 &      4.68 (2.83) \\
Dot counting (n correct) &  29 &      4.03 (3.27) &  29 &     3.93 (3.15) &  29 &      6.93 (2.64) \\
A cancellation (time) &  28 &    73.90 (20.24) &  29 &   82.25 (13.95) &  28 &    62.68 (22.65) \\
A cancellation (n misses) &  29 &      5.76 (5.33) &  29 &     4.00 (4.36) &  29 &      3.62 (4.77) \\
\hline
\textbf{Object perception} & & & & & \\
Object decision &  30 &      9.77 (4.51) &  30 &    12.63 (3.95) &  29 &     10.03 (4.13) \\
Fragmented letters &  23 &      3.57 (4.99) &  29 &     6.34 (5.25) &  27 &      5.19 (5.72) \\
Unusual views &  20 &      3.05 (3.97) &  25 &     6.16 (5.03) &  22 &      3.36 (3.95) \\
Usual views &  20 &      9.65 (6.42) &  25 &    15.16 (4.89) &  22 &     12.59 (5.58) \\
\hline
Memory -- visual processing composite score discrepancy &  30 &   -28.90 (16.82) &  30 &   14.26 (20.33) &  29 &    11.93 (14.79) \\
Space -- object composite score discrepancy  &  30 &    -1.62 (26.91) &  30 &   23.06 (16.93) &  29 &   -17.70 (16.30) \\
\end{tabular}
\caption[Baseline population demographics for PCA subgroups]{Baseline population demographics and neuropsychological data for PCA subgroups. For every neuropsychological test, we report the number of participants with available data (n) and the mean and standard deviation of the available measures.}
\label{tab:pcaSubgrDemographics}
\end{table}

For analysing the heterogeneity within PCA, we split the dataset into three groups based on performance on a suite of cognitive tests. For each subject we computed the following composite tests by summing up scores from individual tests:
\begin{itemize}
 \item Early visual processing: shape detection, shape discrimination and crowding
 \item Visuoperceptual processing: object decision, fragmented letters, usual and unusual views.
 \item Visuospatial processing: number location, dot counting and A cancellation (time -- cut off at 90s)
 \item Episodic memory: short recognition memory test (sRMT) for words and faces
\end{itemize}

The score for each of the four categories was computed by standardising each of the sub-scores on a 0-100 scale, corresponding to the minimum and maximum values obtained by the participants, and then taking the average within each category. The subjects were then classified into three groups. The worst 1/3 of subjects (n=30) on the early visual processing tests as compared to the memory tests (i.e. difference between early visual and memory tests) were assigned to the vision subgroup. The remaining 2/3 of participants were split into two groups based on the difference between visuoperceptual and visuospatial tasks: subjects with space $<$ object performance (n=30) were assigned to the space subgroup while remaining subjects (n=29) were assigned to the object subgroup. Of all the PCA subjects selected for the subgroup analysis, only 23 (vision), 21 (space) and 18 (object) had imaging data. Demographics and neuropsychological data of subjects belonging to the PCA subgroups is shown in table \ref{tab:pcaSubgrDemographics}.

\subsection{Image Acquisition and Preprocessing}
\label{sec:pcaImageAcq}

89 PCA, 66 tAD and 115 healthy controls had at least one T1-weighted MRI scan (see Table\ref{tab:pcaDemographics}). A total of five different scanners were used: two 3T Trio (DRC and UCSF), 1.5T Intera (HUVR) and two 1.5 Signa (DRC). The scans had full brain coverage using between 124 and 208 coronal or sagittal slices of 1.0 or 1.5mm in thickness.

Estimation of region-of-interest (ROI) volumes was performed using the Geodesic Information Flow (GIF) \cite{cardoso2015geodesic} based on the Neuromorphometrics atlas \cite{neuromorphometrics}. The atlas produced 144 different brain ROIs across the left and right hemispheres. Segmentation failed for 6 scans belonging to 5 subjects (3 controls, 1 PCA and 1 tAD), which were subsequently removed. A total of 52 brain ROIs were removed (18 were not part of the cerebral cortex, 6 had segmentation errors, 28 were grouped into larger ROIs). After merging the left and right sides of the remaining ROIs, this resulted in a total of 46 ROIs which were further averaged into 8 ROIs corresponding to whole brain, hippocampal, occipital, frontal, entorhinal, parietal and ventricle. GIF-derived ROI volumes were corrected for total intracranial volume (TIV), age, gender, scanner type and site using a general linear model and a one-hot encoding scheme.

\subsection{Statistical Methods}
\label{sec:pcaStaMet}

\subsubsection{The Event-based Model}
\label{sec:pcaStaMetEBM}
For finding the order in which brain regions are affected, we ran the standard Event-Based Model \cite{fonteijn2012event} (section \ref{sec:bckEbm}) independently on estimated volumes from PCA and tAD, using the shared control population in both scenarios. For the EBM event distributions, we chose a Gaussian distribution for likelihood of normal observations, and a uniform distribution for the likelihood of abnormal observations. We further assumed that the control population is well-defined, so we fit the Gaussian distribution directly on the control data. For the uniform distribution, we set the minimum and maximum of the uniform range to be equal to the smaller and largest observed biomarker values. For finding the optimal sequence, we used 25 different starting points and performed greedy ascent for 10,000 biomarker sequences (see section \ref{sec:model_est}). We chose the sequence with the maximum likelihood as the final sequence.

For estimating uncertainty within the EBM sequence, we used MCMC to take 100,000 samples of the event sequence, starting from the maximum likelihood solution. The perturbation rule used is described in detail in section \ref{sec:model_est}. 

\subsubsection{The Differential Equation Model}
\label{sec:pcaStaMetDEM}

\newcommand{\figScale}{0.5}

\begin{figure}
 \centering
 
 \begin{subfigure}{0.47\textwidth}
    \includegraphics[width=\textwidth]{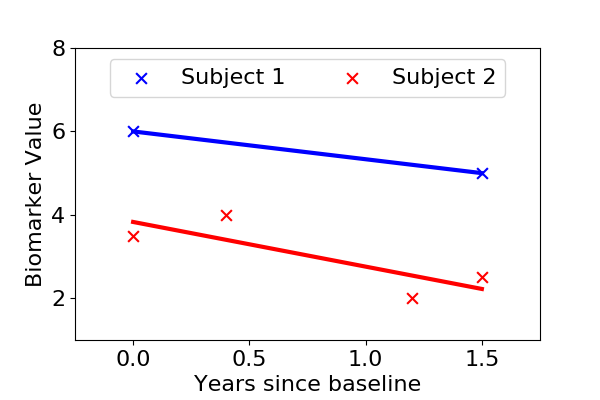}
    \caption{}
    \label{fig:pcaDemDiagramA}
 \end{subfigure}
 \begin{subfigure}{0.47\textwidth}
     \includegraphics[width=\textwidth]{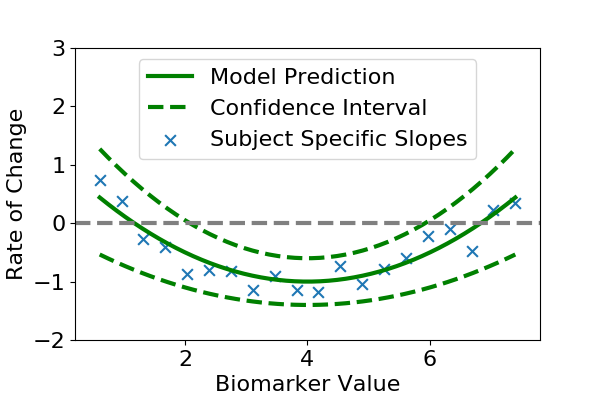}
    \caption{}
    \label{fig:pcaDemDiagramB}
 \end{subfigure}
 \vspace{1em}
 
  \begin{subfigure}{0.47\textwidth}
    \includegraphics[width=\textwidth]{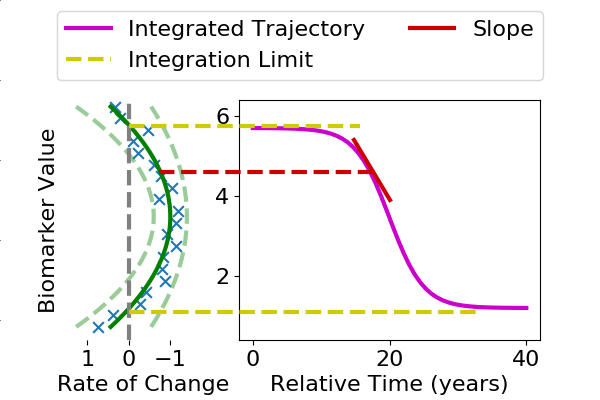}
    \caption{}
    \label{fig:pcaDemDiagramC}
 \end{subfigure}
 \begin{subfigure}{0.47\textwidth}
     \includegraphics[width=\textwidth]{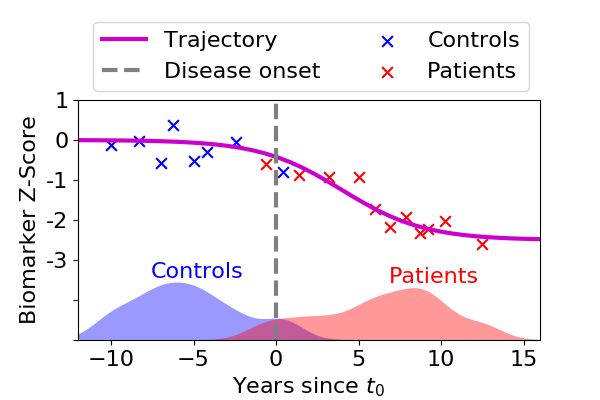}
    \caption{}
    \label{fig:pcaDemDiagramD}
 \end{subfigure}
 \caption[Diagram of the Differential Equation Model]{Diagram of the Differential Equation Model. (a) Subject-specific biomarker rates of change were measured from line of best fit, i.e. line slope. (b) Rate of change model: the slopes of each fitted line were plotted against the average biomarker value of each subject (blue crosses). A non-parametric model (Gaussian Process regression, green line) was then fitted on measurements. (c) Trajectory reconstruction: A line integral was performed on the rate of change model. (d) Anchoring process: to give an absolute time reference, the origin $t_0$ was set as the line that best separates controls from patients, which have been staged along the time axis using their biomarker data. Diagram made by me.}
 \label{fig:pcaDemDiagram}
\end{figure}

For estimating the rate and extent of biomarker decline, we applied the Differential Equation Model \cite{villemagne2013amyloid, oxtoby2018} (section \ref{sec:bckDem}). The methodology is outlined in Fig. \ref{fig:pcaDemDiagram}. The biomarker measurements for each subject were plotted against time since baseline, and a line was fit for each subject independently. The slope of these lines was then used as a measure of the biomarker rate of change (Fig. \ref{fig:pcaDemDiagramA}). The slopes of each fitted line were then plotted against the average biomarker value of each subject (Fig. \ref{fig:pcaDemDiagramB}). A line integral is then performed on the rate of change model (Fig. \ref{fig:pcaDemDiagramC}).  We repeat the DEM fitting for 8 ROIs independently: the four main lobes, whole brain, ventricles, hippocampus and entorhinal cortex. In order to align all images on a common time frame, we staged the controls and patients based on all their data, and then set an absolute time reference $t_0$ as the line that best separated (i.e. maximised the balanced classification accuracy) the controls' and patients' stages.\footnote{The staging of subjects using all their data required an initial trajectory alignment, which we aligned by initially setting $t_0$ to be the mean biomarker value of patients at baseline.}

There were some adaptations that we performed on the DEM model to ensure a good data fit. In the estimation of the rate of change model (Fig. \ref{fig:pcaDemDiagramB}), we did not include the controls, as there was very little change in their biomarker values. We also normalised the average biomarker values to z-scores and standardised the rates of change by dividing them with the average rate of change of all patients. At the line integration step (Fig. \ref{fig:pcaDemDiagramC}), the integration limits were defined as the biomarker values where the corresponding change is zero or the average biomarker value was equal to the minimum or maximum observed in the dataset. 

After the line integration step, we aligned all trajectories on a common time axis through an anchoring process, where we set the time $t_0 = 0$ to correspond to the value of that biomarker in patients at baseline, averaged across all patients. More precisely, we set $f_j(t_0) = avg(X_j)$ where $f_j$ is the trajectory for biomarker $j$ and $X_j$ are the values of biomarker $j$ in tAD/PCA patients at baseline visit. After this initial anchoring, we staged the subjects along their progression and re-set the $t_0 = 0$ to correspond to the threshold that best separated controls from patients (Fig. \ref{fig:pcaDemDiagramD}). After the anchoring process, we converted all biomarker values to z-scores for comparability (Fig. \ref{fig:pcaDemDiagramD}).

To estimate uncertainty in the trajectories, we sampled 20 trajectories from the posterior distribution of the GP, and then anchored them like the mean trajectory. However, the anchoring would've resulted in zero noise interval at the anchor point, so to get realistic confidence intervals we added to each trajectory an extra amount of random noise $N(0, \sigma)$ on the y-axis, with $\sigma$ set to the standard deviation of the biomarker measurements of each subject at baseline visit.

\subsubsection{Statistical Tests}

In order to find out statistically significant differences between the EBM- and DEM-estimated trajectories, we applied several non-parametric statistical tests. 

For EBM results, we tested the effect size of  biomarker $i$ becoming abnormal before another biomarker $j$ both within- and between-group. Within-group differences were assessed using Wilcoxon signed-rank one-tailed tests for all pairs of biomarkers. Between-group (PCA vs tAD) and between-subgroup (space vs object, space vs vision and object vs vision subgroups) differences were assessed using two-tailed Mann-Whitney U tests. We used these non-parametric tests due to non-gaussianity of the data (data is ordinal representing ranks). The reason for using different tests (Wilcoxon vs Mann-Whitney) is because in one case we compare paired samples (two events within the same sequence sample), and in the other unpaired (two events in different sequences, e.g. in a randomly sampled PCA sequence vs a different randomly sampled tAD sequence). We also thinned the MCMC samples (1/100) due to dependence between consecutive samples. 

For DEM results, we tested for differences in estimated biomarker values at different timepoints (-10, 0 and 10 years from $t_0$) both within- and between-groups. For every pair of ROIs, within-group differences were assessed using two-tailed unpaired t-tests. For all ROIs and timepoints, between-group (PCA vs tAD) differences were assessed using similar two-tailed t-tests. For rejecting null hypotheses, we applied Bonferroni-corrected significance thresholds for all tests performed on EBM and DEM results.

\section{Results}
 
\newcommand*{\scaleBrainImg}{0.25}

\newcommand*{\snapLocationPCA}{\pcaPaperFigs/ebmSnapshotsPCA}
\newcommand*{\snapLocationAD}{\pcaPaperFigs/ebmSnapshotsAD} 
\newcommand*{\snapLocationEAR}{\pcaPaperFigs/ebmSnapshotsEAR} 
\newcommand*{\snapLocationPER}{\pcaPaperFigs/ebmSnapshotsPER} 
\newcommand*{\snapLocationSPA}{\pcaPaperFigs/ebmSnapshotsSPA} 

\newcommand*{\snapScale}{0.6} 
\definecolor{light-gray}{gray}{0.6}

\subsection{Progression of PCA and Typical AD}
\label{sec:pcaResPcaAd}

\begin{figure}
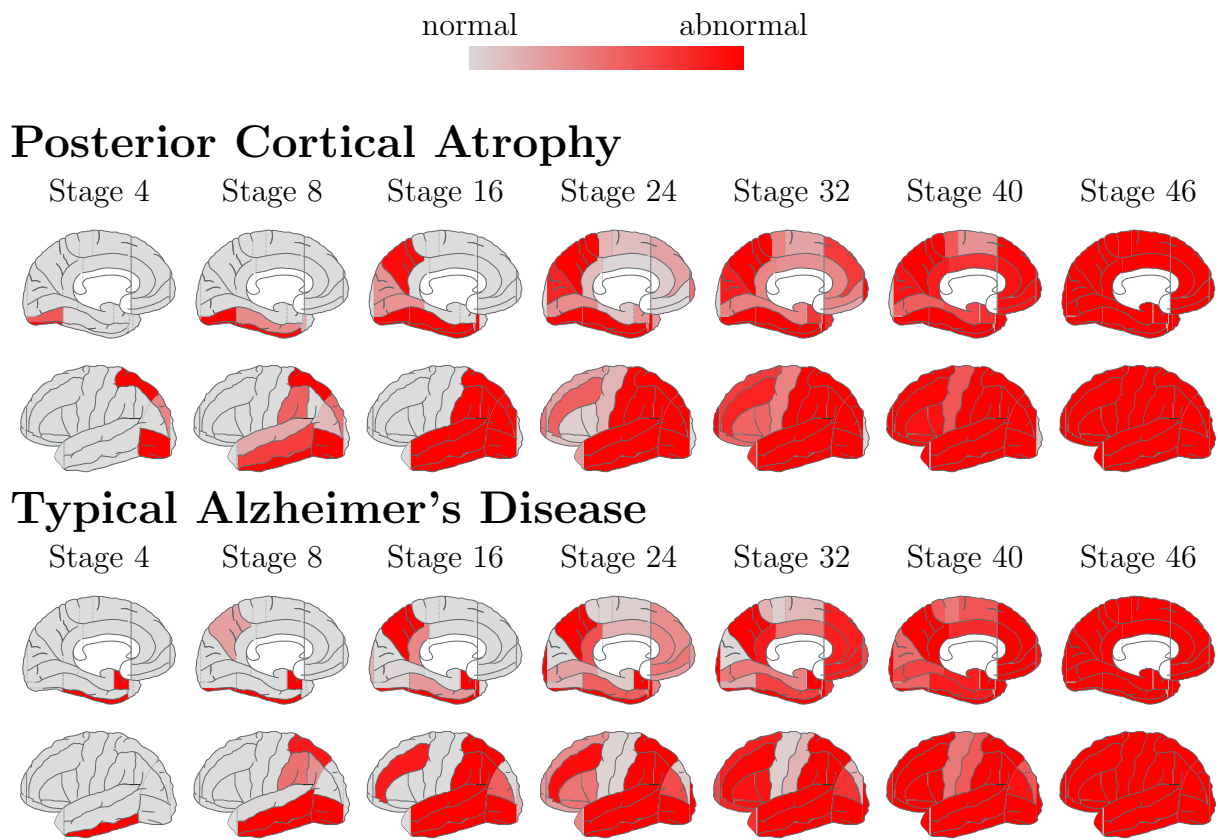

  \begin{subfigure}{\textwidth}
   \centering
  \begin{tikzpicture}[scale=\snapScale,auto,swap]
    \shade[left color=gray!30,right color=red] (0,0) rectangle (6,0.5);
    \node[inner sep=0] (corr_text) at (6,1) {abnormal};
    \node[inner sep=0] (corr_text) at (0,1) {normal};
    \node[inner sep=0] (corr_text) at (0.2,0.5) {};
  \end{tikzpicture}
  \end{subfigure}
  \vspace{1em}
  
    \textbf{\Large{Posterior Cortical Atrophy}}
    
    \begin{tikzpicture}[scale=\snapScale,auto,swap]

    \node (upper_brain) at (0,1.5) { \includegraphics*[scale=\scaleBrainImg,trim=0 0 240 0]{\snapLocationPCA/stage_4-eps-converted-to.pdf}};
    \node (lower_brain) at (0,-1.5) { \includegraphics*[scale=\scaleBrainImg,trim=240 0 0 0]{\snapLocationPCA/stage_4-eps-converted-to.pdf}};
    \node[above=0cm of upper_brain] (stage) {Stage 4};
    
    \end{tikzpicture}
  \hspace{-1.5em}
  ~
    \begin{tikzpicture}[scale=\snapScale,auto,swap]

    \node (upper_brain) at (0,1.5) { \includegraphics*[scale=\scaleBrainImg,trim=0 0 240 0]{\snapLocationPCA/stage_8-eps-converted-to.pdf}};
    \node (lower_brain) at (0,-1.5) { \includegraphics*[scale=\scaleBrainImg,trim=240 0 0 0]{\snapLocationPCA/stage_8-eps-converted-to.pdf}};
    \node[above=0cm of upper_brain] (stage) {Stage 8};
    
    \end{tikzpicture}
  \hspace{-1.5em}
  ~
    \begin{tikzpicture}[scale=\snapScale,auto,swap]

    \node (upper_brain) at (0,1.5) { \includegraphics*[scale=\scaleBrainImg,trim=0 0 240 0]{\snapLocationPCA/stage_16-eps-converted-to.pdf}};
    \node (lower_brain) at (0,-1.5) { \includegraphics*[scale=\scaleBrainImg,trim=240 0 0 0]{\snapLocationPCA/stage_16-eps-converted-to.pdf}};
    \node[above=0cm of upper_brain] (stage) {Stage 16};
    
    \end{tikzpicture}
  \hspace{-1.5em}
  ~
    \begin{tikzpicture}[scale=\snapScale,auto,swap]

    \node (upper_brain) at (0,1.5) { \includegraphics*[scale=\scaleBrainImg,trim=0 0 240 0]{\snapLocationPCA/stage_24-eps-converted-to.pdf}};
    \node (lower_brain) at (0,-1.5) { \includegraphics*[scale=\scaleBrainImg,trim=240 0 0 0]{\snapLocationPCA/stage_24-eps-converted-to.pdf}};
    \node[above=0cm of upper_brain] (stage) {Stage 24};
    
    \end{tikzpicture}
  \hspace{-1.5em}
  ~
    \begin{tikzpicture}[scale=\snapScale,auto,swap]

    \node (upper_brain) at (0,1.5) { \includegraphics*[scale=\scaleBrainImg,trim=0 0 240 0]{\snapLocationPCA/stage_32-eps-converted-to.pdf}};
    \node (lower_brain) at (0,-1.5) { \includegraphics*[scale=\scaleBrainImg,trim=240 0 0 0]{\snapLocationPCA/stage_32-eps-converted-to.pdf}};
    \node[above=0cm of upper_brain] (stage) {Stage 32};
    
    \end{tikzpicture}
  \hspace{-1.5em}
  ~
    \begin{tikzpicture}[scale=\snapScale,auto,swap]

    \node (upper_brain) at (0,1.5) { \includegraphics*[scale=\scaleBrainImg,trim=0 0 240 0]{\snapLocationPCA/stage_40-eps-converted-to.pdf}};
    \node (lower_brain) at (0,-1.5) { \includegraphics*[scale=\scaleBrainImg,trim=240 0 0 0]{\snapLocationPCA/stage_40-eps-converted-to.pdf}};
    \node[above=0cm of upper_brain] (stage) {Stage 40};
    
    \end{tikzpicture}
  \hspace{-1.5em}
  ~
    \begin{tikzpicture}[scale=\snapScale,auto,swap]

    \node (upper_brain) at (0,1.5) { \includegraphics*[scale=\scaleBrainImg,trim=0 0 240 0]{\snapLocationPCA/stage_46-eps-converted-to.pdf}};
    \node (lower_brain) at (0,-1.5) { \includegraphics*[scale=\scaleBrainImg,trim=240 0 0 0]{\snapLocationPCA/stage_46-eps-converted-to.pdf}};
    \node[above=0cm of upper_brain] (stage) {Stage 46};
    
    \end{tikzpicture}
  \hspace{-1.5em}
  
  \textbf{\Large{Typical Alzheimer's Disease}}

    \begin{tikzpicture}[scale=\snapScale,auto,swap]

    \node (upper_brain) at (0,1.5) { \includegraphics*[scale=\scaleBrainImg,trim=0 0 240 0]{\snapLocationAD/stage_4-eps-converted-to.pdf}};
    \node (lower_brain) at (0,-1.5) { \includegraphics*[scale=\scaleBrainImg,trim=240 0 0 0]{\snapLocationAD/stage_4-eps-converted-to.pdf}};
    \node[above=0cm of upper_brain] (stage) {Stage 4};
    
    \end{tikzpicture}
  \hspace{-1.5em}
  ~
    \begin{tikzpicture}[scale=\snapScale,auto,swap]

    \node (upper_brain) at (0,1.5) { \includegraphics*[scale=\scaleBrainImg,trim=0 0 240 0]{\snapLocationAD/stage_8-eps-converted-to.pdf}};
    \node (lower_brain) at (0,-1.5) { \includegraphics*[scale=\scaleBrainImg,trim=240 0 0 0]{\snapLocationAD/stage_8-eps-converted-to.pdf}};
    \node[above=0cm of upper_brain] (stage) {Stage 8};
    
    \end{tikzpicture}
  \hspace{-1.5em}
  ~
    \begin{tikzpicture}[scale=\snapScale,auto,swap]

    \node (upper_brain) at (0,1.5) { \includegraphics*[scale=\scaleBrainImg,trim=0 0 240 0]{\snapLocationAD/stage_16-eps-converted-to.pdf}};
    \node (lower_brain) at (0,-1.5) { \includegraphics*[scale=\scaleBrainImg,trim=240 0 0 0]{\snapLocationAD/stage_16-eps-converted-to.pdf}};
    \node[above=0cm of upper_brain] (stage) {Stage 16};
    
    \end{tikzpicture}
  \hspace{-1.5em}
  ~
    \begin{tikzpicture}[scale=\snapScale,auto,swap]

    \node (upper_brain) at (0,1.5) { \includegraphics*[scale=\scaleBrainImg,trim=0 0 240 0]{\snapLocationAD/stage_24-eps-converted-to.pdf}};
    \node (lower_brain) at (0,-1.5) { \includegraphics*[scale=\scaleBrainImg,trim=240 0 0 0]{\snapLocationAD/stage_24-eps-converted-to.pdf}};
    \node[above=0cm of upper_brain] (stage) {Stage 24};
    
    \end{tikzpicture}
  \hspace{-1.5em}
  ~
    \begin{tikzpicture}[scale=\snapScale,auto,swap]

    \node (upper_brain) at (0,1.5) { \includegraphics*[scale=\scaleBrainImg,trim=0 0 240 0]{\snapLocationAD/stage_32-eps-converted-to.pdf}};
    \node (lower_brain) at (0,-1.5) { \includegraphics*[scale=\scaleBrainImg,trim=240 0 0 0]{\snapLocationAD/stage_32-eps-converted-to.pdf}};
    \node[above=0cm of upper_brain] (stage) {Stage 32};
    
    \end{tikzpicture}
  \hspace{-1.5em}
  ~
    \begin{tikzpicture}[scale=\snapScale,auto,swap]

    \node (upper_brain) at (0,1.5) { \includegraphics*[scale=\scaleBrainImg,trim=0 0 240 0]{\snapLocationAD/stage_40-eps-converted-to.pdf}};
    \node (lower_brain) at (0,-1.5) { \includegraphics*[scale=\scaleBrainImg,trim=240 0 0 0]{\snapLocationAD/stage_40-eps-converted-to.pdf}};
    \node[above=0cm of upper_brain] (stage) {Stage 40};
    
    \end{tikzpicture}
  \hspace{-1.5em}
  ~
    \begin{tikzpicture}[scale=\snapScale,auto,swap]

    \node (upper_brain) at (0,1.5) { \includegraphics*[scale=\scaleBrainImg,trim=0 0 240 0]{\snapLocationAD/stage_46-eps-converted-to.pdf}};
    \node (lower_brain) at (0,-1.5) { \includegraphics*[scale=\scaleBrainImg,trim=240 0 0 0]{\snapLocationAD/stage_46-eps-converted-to.pdf}};
    \node[above=0cm of upper_brain] (stage) {Stage 46};
    
    \end{tikzpicture}
  \hspace{-1.5em}
  
\caption[Atrophy progression in PCA and tAD patients according to the event-based model]{Atrophy progression in PCA and tAD patients according to the event-based model. White regions are within the volume range of healthy controls, while red regions show abnormally low volumes by the corresponding stage, with shading indicating the probability of abnormality. By each stage, a number of biomarkers shaded in red became abnormal. Brain pictures generated using BrainPainter \cite{marinescu2019BrainPainter}}  
\label{fig:pcaSnapshots}
\end{figure}

\begin{figure}
\centering
  \begin{subfigure}{0.7\textwidth}
  \centering
 \textbf{\large{\mbox{Posterior Cortical Atrophy}}}
 \includegraphics[width=1\textwidth,trim=100 30 0 50,clip]{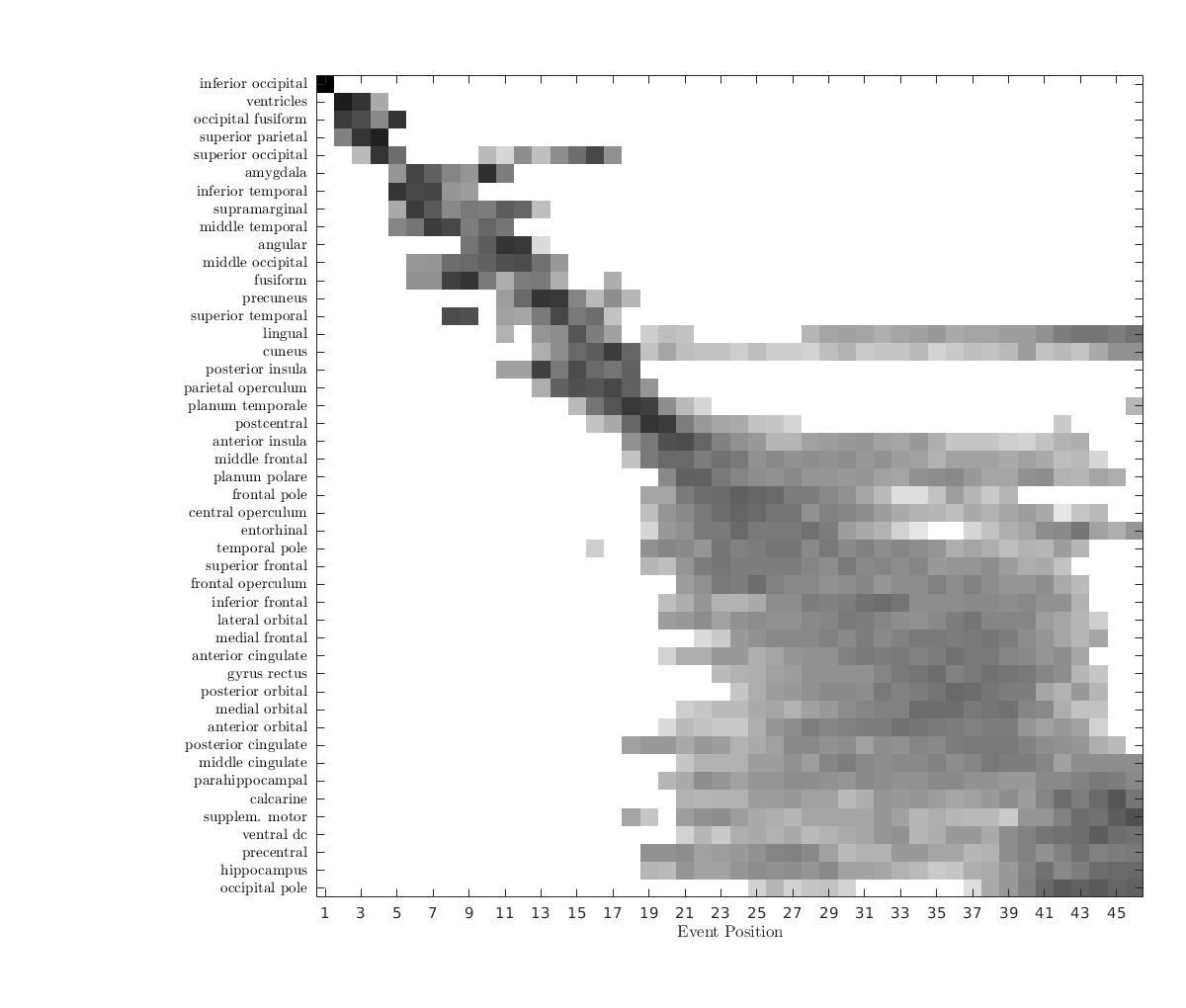} 
 \end{subfigure}
 \hspace{1em}
  \begin{subfigure}{0.7\textwidth}
  \centering
  \textbf{\large{\mbox{Typical Alzheimer's Disease}}}
 \includegraphics[width=1\textwidth,trim=100 30 0 50,clip]{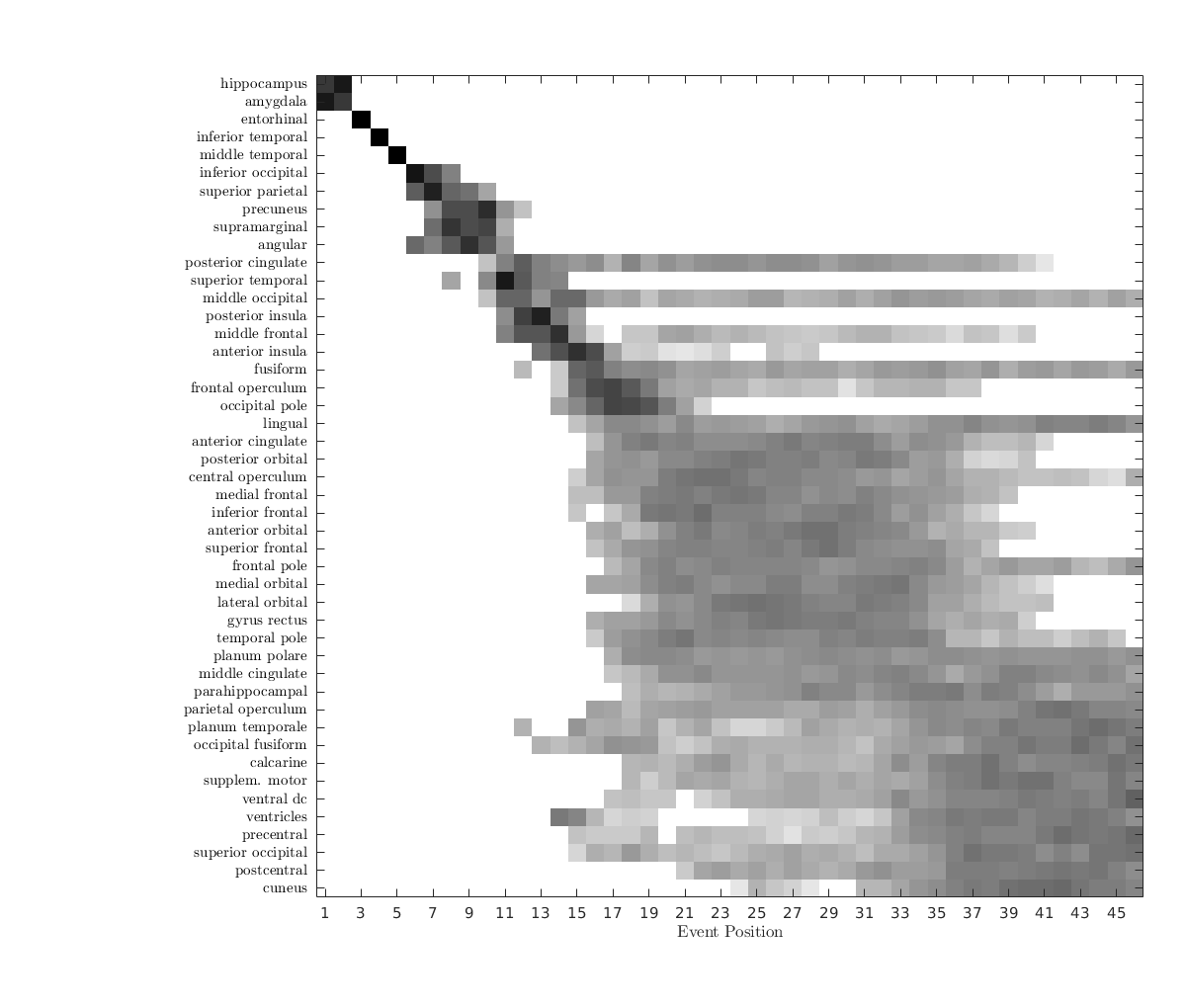}
 \end{subfigure}
 \caption[PCA and tAD positional variance diagrams estimated by the EBM]{Uncertainty in the EBM-estimated atrophy sequences for (top) PCA and (bottom) tAD from Fig \ref{fig:pcaSnapshots}. The ROIs on the Y-axis are ordered according to the timing of abnormality, from early abnormalities on the top to late abnormalities on the bottom. The X-axis shows the position of a biomarker in the abnormality sequence. Each pixel at position $(i,j)$ shows the probability of biomarker $j$ becoming abnormal at position $i$, with darker squares showing higher confidence and whiter squares showing lower confidence. The biomarker orderings are sampled from the EBM posterior distribution.}
 \label{fig:pcaEBMProg}
\end{figure}

Fig. \ref{fig:pcaSnapshots} shows the maximum likelihood progression of atrophy estimated by the EBM, for both PCA and tAD patients. Snapshots of brain atrophy were taken at model stages 4, 8, 16, 24, 32, 40 and 46 (of 46) using the template from Supplementary Fig. \ref{fig:ebmSnapLabels}. Figure \ref{fig:pcaEBMProg} shows the maximum likelihood sequence and the variance in the main sequence. PCA patients show early atrophy in  occipital areas, ventricles and the superior parietal region, while tAD patients show early atrophy in the amygdala, hippocampus and entorhinal cortex, followed by temporal areas. The ordering is largely preserved under bootstrapping (Supplementary Fig. \ref{fig:bootPosVarAllPcaAd}), and supported by statistical testing (Supplementary Fig. \ref{fig:statTestPcaAd}). Differences in abnormality sequences between PCA and tAD are also statistically significant under Bonferroni corrections (Supplementary Figure \ref{fig:ebmStatTestPcaAd}).
 
\begin{figure}
 \centering
\begin{subfigure}{0.8\textwidth}
 \centering
 \includegraphics[width=0.8\textwidth,trim=0 300 0 0,clip]{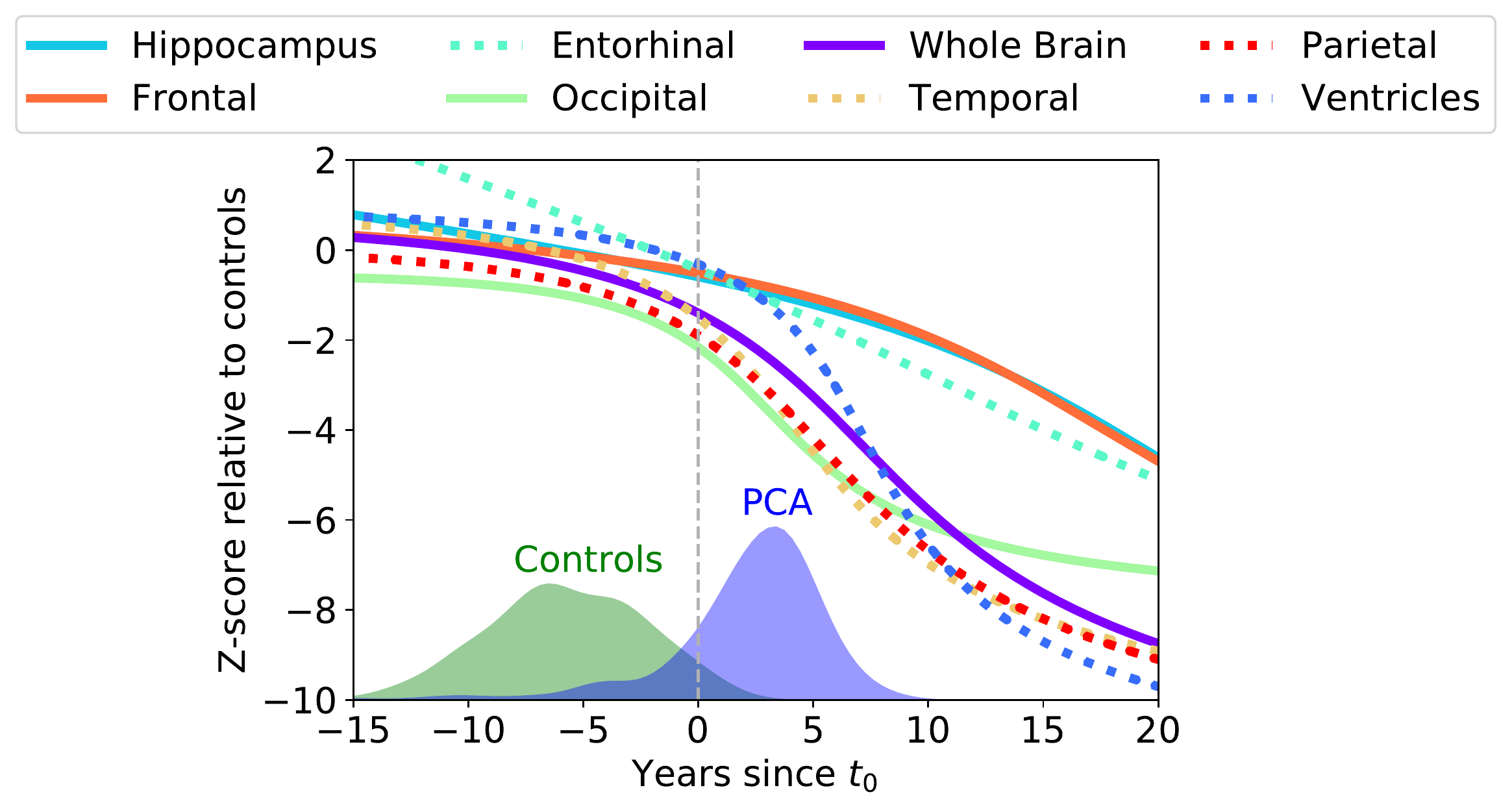}
\end{subfigure}
 
 \begin{subfigure}{0.47\textwidth}
\includegraphics[width=\textwidth,trim=90 0 120 60,clip]{\pcaPaperFigs/trajAlign_600_500PCA.pdf}
 \caption{PCA}
 \label{trajDEMPCA} 
 \end{subfigure}
 \begin{subfigure}{0.47\textwidth}
 \includegraphics[width=\textwidth,trim=90 0 120 60,clip]{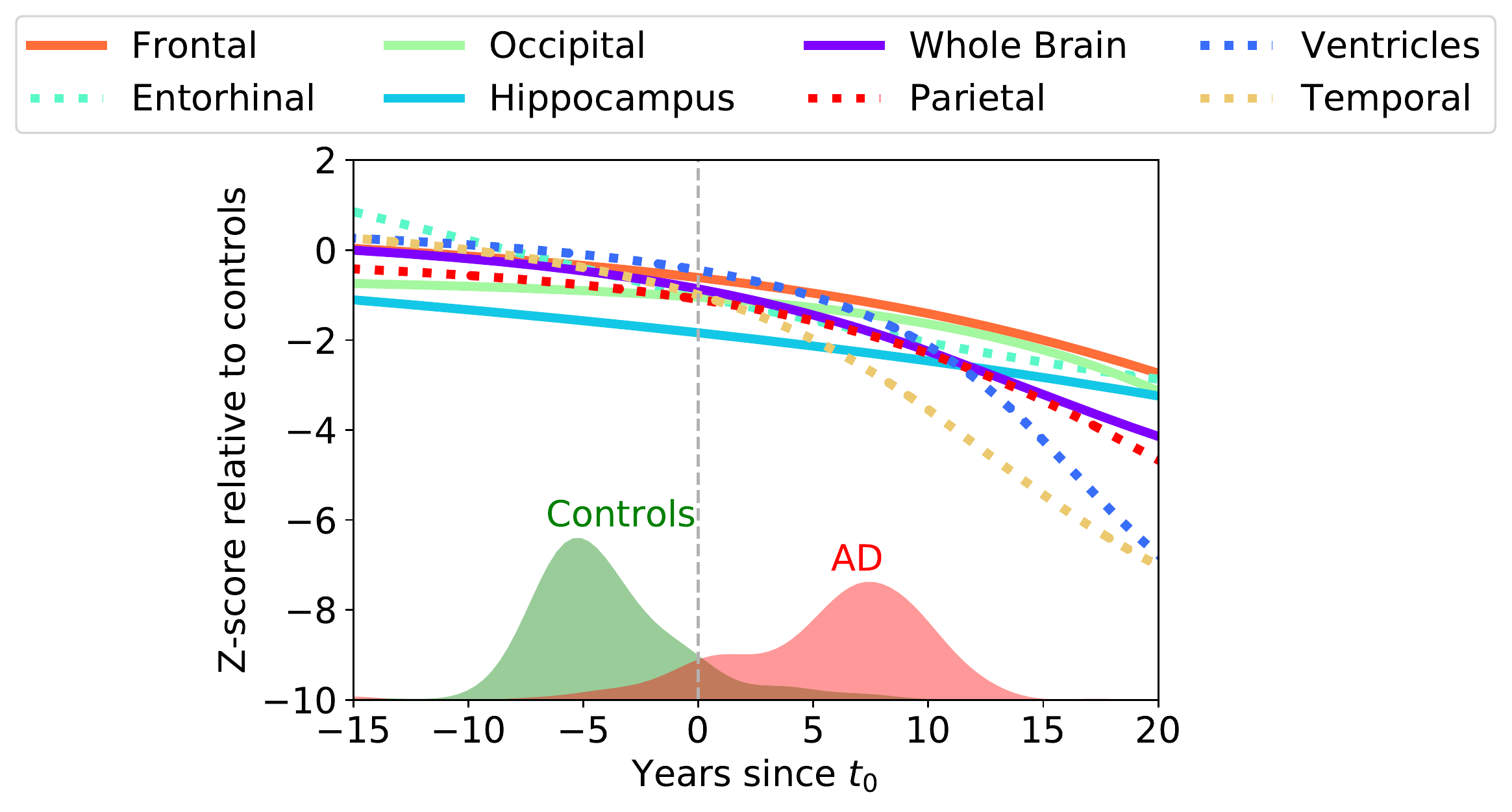}
 \caption{tAD}
 \label{trajDEMAD}
 \end{subfigure}
 \caption[PCA and tAD trajectories estimated by the DEM]{(a-b) Trajectories of different ROI volumes from the differential equation model for (a) PCA progression and (b) tAD progression. The x-axis shows the number of years since $t_0$, and the y-axis shows the z-score of the ROI volume relative to controls. The trajectories of the ventricles have been flipped to aid comparison. Overlayed are histograms of subject stages based on the estimated trajectories.}
 \label{fig:pcaAdDEM}
\end{figure}

\begin{figure}
 \centering
 \includegraphics[width=\textwidth, trim=50 0 50 0]{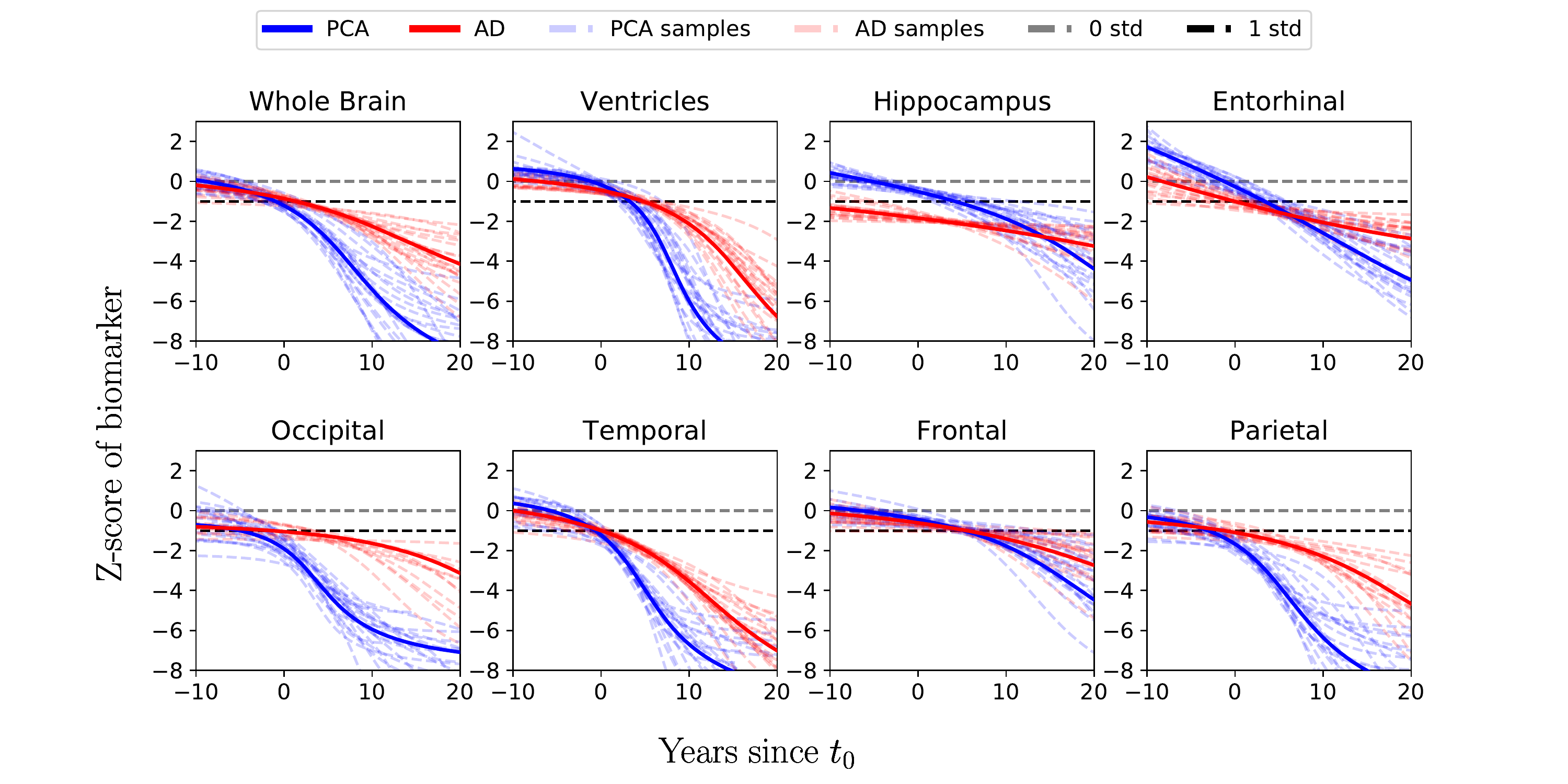}
 \caption[PCA and tAD trajectories aligned in the same space, with samples from the posterior distribution]{Mean trajectories for ROI volumes for PCA and tAD aligned on the same temporal scale with samples from the posterior distribution showing the confidence of the mean trajectory. The axis shows the number of years since $t_0$, and the y-axis shows the z-score of the ROI volume relative to controls. The trajectories for the ventricles have been flipped to aid visual comparison. The 1 std and 0 std horizontal lines represent the limit of 1 and 0 standard deviations away from the mean values of controls.}
 \label{fig:trajDEMPcaAdConf}
\end{figure}

Fig. \ref{fig:pcaAdDEM} shows the DEM-estimated biomarker trajectories for PCA (left) and tAD (right). Confidence estimates of the mean trajectory are also given in Fig. \ref{fig:trajDEMPcaAdConf}. Amongst PCA patients, occipital and parietal atrophy was most evident before $t_0$, and by $t_0$ we also observe considerable atrophy in the temporal lobe. Between $t_0$ and 10 years after $t_0$, we observe a marked increase in the rate of occipital, parietal and temporal atrophy and ventricular expansion. By contrast, hippocampal, entorhinal and frontal atrophy never match the extent of tissue loss in posterior and temporal regions. After 10 years from $t_0$, atrophy rates in occipital, parietal and temporal lobes seem to slow down, but limited data in this time window prevents drawing any clear conclusions. Statistical testing within the PCA cohort also confirms our conclusions -- see Supplementary Tables \ref{tab:demStatTestVolsPcaMinus10}, \ref{tab:demStatTestVolsPca0} and \ref{tab:demStatTestVolsPcaPlus10}.

By contrast, before $t_0$ tAD patients showed most extensive tissue loss in the hippocampus, confirmed by significance tests between hippocampal volume and other regions (p $<$ 4e-05, see Supplementary Figs. \ref{tab:demStatTestVolsAdMinus10} and \ref{tab:demStatTestVolsAd0}). After $t_0$, subsequent rates of change are the highest for temporal atrophy and ventricular expansion. It is of note that within 12 years from $t_0$, model estimates of parietal and ventricular abnormality amongst tAD patients are equivalent to or exceed the relative extent of hippocampal abnormality. Comparing PCA and tAD trajectories directly (Fig. \ref{fig:trajDEMPcaAdConf}), the separation between groups at $t_0$ is greatest in parietal (PCA $>$ tAD, p $<$ 1e-6) and hippocampal (tAD $>$ PCA, p $<$ 1e-22) volumes -- see Supplementary table \ref{tab:demStatTestPcaAd} for full statistical testing. 

\subsection{Progression of PCA Subgroups}
\label{sec:pcaResPcaSub}

\begin{figure}
  \begin{subfigure}{\textwidth}
   \centering
  \begin{tikzpicture}[scale=\snapScale,auto,swap]
    \shade[left color=gray!30,right color=red] (0,0) rectangle (6,0.5);
    \node[inner sep=0] (corr_text) at (6,1) {abnormal};
    \node[inner sep=0] (corr_text) at (0,1) {normal};
    \node[inner sep=0] (corr_text) at (0.2,0.5) {};
  \end{tikzpicture}
  \end{subfigure}

  {\Large \textbf{Vision impairment group}\par}
    \begin{tikzpicture}[scale=\snapScale,auto,swap]

    \node (upper_brain) at (0,1.5) { \includegraphics*[scale=\scaleBrainImg,trim=0 0 240 0]{\snapLocationEAR/stage_1-eps-converted-to.pdf}};
    \node (lower_brain) at (0,-1.5) { \includegraphics*[scale=\scaleBrainImg,trim=240 0 0 0]{\snapLocationEAR/stage_1-eps-converted-to.pdf}};
    \node[above=0cm of upper_brain] (stage) {Stage 1};
    
    \end{tikzpicture}
  \hspace{-1.5em}
  ~
    \begin{tikzpicture}[scale=\snapScale,auto,swap]

    \node (upper_brain) at (0,1.5) { \includegraphics*[scale=\scaleBrainImg,trim=0 0 240 0]{\snapLocationEAR/stage_2-eps-converted-to.pdf}};
    \node (lower_brain) at (0,-1.5) { \includegraphics*[scale=\scaleBrainImg,trim=240 0 0 0]{\snapLocationEAR/stage_2-eps-converted-to.pdf}};
    \node[above=0cm of upper_brain] (stage) {Stage 2};
    
    \end{tikzpicture}
  \hspace{-1.5em}
  ~
    \begin{tikzpicture}[scale=\snapScale,auto,swap]

    \node (upper_brain) at (0,1.5) { \includegraphics*[scale=\scaleBrainImg,trim=0 0 240 0]{\snapLocationEAR/stage_3-eps-converted-to.pdf}};
    \node (lower_brain) at (0,-1.5) { \includegraphics*[scale=\scaleBrainImg,trim=240 0 0 0]{\snapLocationEAR/stage_3-eps-converted-to.pdf}};
    \node[above=0cm of upper_brain] (stage) {Stage 3};
    
    \end{tikzpicture}
  \hspace{-1.5em}
  ~
    \begin{tikzpicture}[scale=\snapScale,auto,swap]

    \node (upper_brain) at (0,1.5) { \includegraphics*[scale=\scaleBrainImg,trim=0 0 240 0]{\snapLocationEAR/stage_4-eps-converted-to.pdf}};
    \node (lower_brain) at (0,-1.5) { \includegraphics*[scale=\scaleBrainImg,trim=240 0 0 0]{\snapLocationEAR/stage_4-eps-converted-to.pdf}};
    \node[above=0cm of upper_brain] (stage) {Stage 4};
    
    \end{tikzpicture}
  \hspace{-1.5em}
  ~
    \begin{tikzpicture}[scale=\snapScale,auto,swap]

    \node (upper_brain) at (0,1.5) { \includegraphics*[scale=\scaleBrainImg,trim=0 0 240 0]{\snapLocationEAR/stage_5-eps-converted-to.pdf}};
    \node (lower_brain) at (0,-1.5) { \includegraphics*[scale=\scaleBrainImg,trim=240 0 0 0]{\snapLocationEAR/stage_5-eps-converted-to.pdf}};
    \node[above=0cm of upper_brain] (stage) {Stage 5};
    
    \end{tikzpicture}
  \hspace{-1.5em}
  ~
    \begin{tikzpicture}[scale=\snapScale,auto,swap]

    \node (upper_brain) at (0,1.5) { \includegraphics*[scale=\scaleBrainImg,trim=0 0 240 0]{\snapLocationEAR/stage_6-eps-converted-to.pdf}};
    \node (lower_brain) at (0,-1.5) { \includegraphics*[scale=\scaleBrainImg,trim=240 0 0 0]{\snapLocationEAR/stage_6-eps-converted-to.pdf}};
    \node[above=0cm of upper_brain] (stage) {Stage 6};
    
    \end{tikzpicture}
  \hspace{-1.5em}
  ~
    \begin{tikzpicture}[scale=\snapScale,auto,swap]

    \node (upper_brain) at (0,1.5) { \includegraphics*[scale=\scaleBrainImg,trim=0 0 240 0]{\snapLocationEAR/stage_7-eps-converted-to.pdf}};
    \node (lower_brain) at (0,-1.5) { \includegraphics*[scale=\scaleBrainImg,trim=240 0 0 0]{\snapLocationEAR/stage_7-eps-converted-to.pdf}};
    \node[above=0cm of upper_brain] (stage) {Stage 7};
    
    \end{tikzpicture}

    \hspace{-1.5em}

    \vspace{-1em}

    {\Large \textbf{Space perception impairment group}\par}
    \begin{tikzpicture}[scale=\snapScale,auto,swap]

    \node (upper_brain) at (0,1.5) { \includegraphics*[scale=\scaleBrainImg,trim=0 0 240 0]{\snapLocationSPA/stage_1-eps-converted-to.pdf}};
    \node (lower_brain) at (0,-1.5) { \includegraphics*[scale=\scaleBrainImg,trim=240 0 0 0]{\snapLocationSPA/stage_1-eps-converted-to.pdf}};
    \node[above=0cm of upper_brain] (stage) {Stage 1};
    
    \end{tikzpicture}
  \hspace{-1.5em}
  ~
    \begin{tikzpicture}[scale=\snapScale,auto,swap]

    \node (upper_brain) at (0,1.5) { \includegraphics*[scale=\scaleBrainImg,trim=0 0 240 0]{\snapLocationSPA/stage_2-eps-converted-to.pdf}};
    \node (lower_brain) at (0,-1.5) { \includegraphics*[scale=\scaleBrainImg,trim=240 0 0 0]{\snapLocationSPA/stage_2-eps-converted-to.pdf}};
    \node[above=0cm of upper_brain] (stage) {Stage 2};
    
    \end{tikzpicture}
  \hspace{-1.5em}
  ~
    \begin{tikzpicture}[scale=\snapScale,auto,swap]

    \node (upper_brain) at (0,1.5) { \includegraphics*[scale=\scaleBrainImg,trim=0 0 240 0]{\snapLocationSPA/stage_3-eps-converted-to.pdf}};
    \node (lower_brain) at (0,-1.5) { \includegraphics*[scale=\scaleBrainImg,trim=240 0 0 0]{\snapLocationSPA/stage_3-eps-converted-to.pdf}};
    \node[above=0cm of upper_brain] (stage) {Stage 3};
    
    \end{tikzpicture}
  \hspace{-1.5em}
  ~
    \begin{tikzpicture}[scale=\snapScale,auto,swap]

    \node (upper_brain) at (0,1.5) { \includegraphics*[scale=\scaleBrainImg,trim=0 0 240 0]{\snapLocationSPA/stage_4-eps-converted-to.pdf}};
    \node (lower_brain) at (0,-1.5) { \includegraphics*[scale=\scaleBrainImg,trim=240 0 0 0]{\snapLocationSPA/stage_4-eps-converted-to.pdf}};
    \node[above=0cm of upper_brain] (stage) {Stage 4};
    
    \end{tikzpicture}
  \hspace{-1.5em}
  ~
    \begin{tikzpicture}[scale=\snapScale,auto,swap]

    \node (upper_brain) at (0,1.5) { \includegraphics*[scale=\scaleBrainImg,trim=0 0 240 0]{\snapLocationSPA/stage_5-eps-converted-to.pdf}};
    \node (lower_brain) at (0,-1.5) { \includegraphics*[scale=\scaleBrainImg,trim=240 0 0 0]{\snapLocationSPA/stage_5-eps-converted-to.pdf}};
    \node[above=0cm of upper_brain] (stage) {Stage 5};
    
    \end{tikzpicture}
  \hspace{-1.5em}
  ~
    \begin{tikzpicture}[scale=\snapScale,auto,swap]

    \node (upper_brain) at (0,1.5) { \includegraphics*[scale=\scaleBrainImg,trim=0 0 240 0]{\snapLocationSPA/stage_6-eps-converted-to.pdf}};
    \node (lower_brain) at (0,-1.5) { \includegraphics*[scale=\scaleBrainImg,trim=240 0 0 0]{\snapLocationSPA/stage_6-eps-converted-to.pdf}};
    \node[above=0cm of upper_brain] (stage) {Stage 6};
    
    \end{tikzpicture}
  \hspace{-1.5em}
  ~
    \begin{tikzpicture}[scale=\snapScale,auto,swap]

    \node (upper_brain) at (0,1.5) { \includegraphics*[scale=\scaleBrainImg,trim=0 0 240 0]{\snapLocationSPA/stage_7-eps-converted-to.pdf}};
    \node (lower_brain) at (0,-1.5) { \includegraphics*[scale=\scaleBrainImg,trim=240 0 0 0]{\snapLocationSPA/stage_7-eps-converted-to.pdf}};
    \node[above=0cm of upper_brain] (stage) {Stage 7};
    
    \end{tikzpicture}
  \hspace{-1.5em}

\vspace{-1em}

    {\Large \textbf{Object perception impairment group}\par}
    \begin{tikzpicture}[scale=\snapScale,auto,swap]

    \node (upper_brain) at (0,1.5) { \includegraphics*[scale=\scaleBrainImg,trim=0 0 240 0]{\snapLocationPER/stage_1-eps-converted-to.pdf}};
    \node (lower_brain) at (0,-1.5) { \includegraphics*[scale=\scaleBrainImg,trim=240 0 0 0]{\snapLocationPER/stage_1-eps-converted-to.pdf}};
    \node[above=0cm of upper_brain] (stage) {Stage 1};
    
    \end{tikzpicture}
  \hspace{-1.5em}
  ~
    \begin{tikzpicture}[scale=\snapScale,auto,swap]

    \node (upper_brain) at (0,1.5) { \includegraphics*[scale=\scaleBrainImg,trim=0 0 240 0]{\snapLocationPER/stage_2-eps-converted-to.pdf}};
    \node (lower_brain) at (0,-1.5) { \includegraphics*[scale=\scaleBrainImg,trim=240 0 0 0]{\snapLocationPER/stage_2-eps-converted-to.pdf}};
    \node[above=0cm of upper_brain] (stage) {Stage 2};
    
    \end{tikzpicture}
  \hspace{-1.5em}
  ~
    \begin{tikzpicture}[scale=\snapScale,auto,swap]

    \node (upper_brain) at (0,1.5) { \includegraphics*[scale=\scaleBrainImg,trim=0 0 240 0]{\snapLocationPER/stage_3-eps-converted-to.pdf}};
    \node (lower_brain) at (0,-1.5) { \includegraphics*[scale=\scaleBrainImg,trim=240 0 0 0]{\snapLocationPER/stage_3-eps-converted-to.pdf}};
    \node[above=0cm of upper_brain] (stage) {Stage 3};
    
    \end{tikzpicture}
  \hspace{-1.5em}
  ~
    \begin{tikzpicture}[scale=\snapScale,auto,swap]

    \node (upper_brain) at (0,1.5) { \includegraphics*[scale=\scaleBrainImg,trim=0 0 240 0]{\snapLocationPER/stage_4-eps-converted-to.pdf}};
    \node (lower_brain) at (0,-1.5) { \includegraphics*[scale=\scaleBrainImg,trim=240 0 0 0]{\snapLocationPER/stage_4-eps-converted-to.pdf}};
    \node[above=0cm of upper_brain] (stage) {Stage 4};
    
    \end{tikzpicture}
  \hspace{-1.5em}
  ~
    \begin{tikzpicture}[scale=\snapScale,auto,swap]

    \node (upper_brain) at (0,1.5) { \includegraphics*[scale=\scaleBrainImg,trim=0 0 240 0]{\snapLocationPER/stage_5-eps-converted-to.pdf}};
    \node (lower_brain) at (0,-1.5) { \includegraphics*[scale=\scaleBrainImg,trim=240 0 0 0]{\snapLocationPER/stage_5-eps-converted-to.pdf}};
    \node[above=0cm of upper_brain] (stage) {Stage 5};
    
    \end{tikzpicture}
  \hspace{-1.5em}
  ~
    \begin{tikzpicture}[scale=\snapScale,auto,swap]

    \node (upper_brain) at (0,1.5) { \includegraphics*[scale=\scaleBrainImg,trim=0 0 240 0]{\snapLocationPER/stage_6-eps-converted-to.pdf}};
    \node (lower_brain) at (0,-1.5) { \includegraphics*[scale=\scaleBrainImg,trim=240 0 0 0]{\snapLocationPER/stage_6-eps-converted-to.pdf}};
    \node[above=0cm of upper_brain] (stage) {Stage 6};
    
    \end{tikzpicture}
  \hspace{-1.5em}
  ~
    \begin{tikzpicture}[scale=\snapScale,auto,swap]

    \node (upper_brain) at (0,1.5) { \includegraphics*[scale=\scaleBrainImg,trim=0 0 240 0]{\snapLocationPER/stage_7-eps-converted-to.pdf}};
    \node (lower_brain) at (0,-1.5) { \includegraphics*[scale=\scaleBrainImg,trim=240 0 0 0]{\snapLocationPER/stage_7-eps-converted-to.pdf}};
    \node[above=0cm of upper_brain] (stage) {Stage 7};
    
    \end{tikzpicture}
  \hspace{-1.5em}

 \begin{subfigure}{0.32\textwidth}
 \centering
 {\footnotesize \textbf{Vision}\par}
 \includegraphics[width=\textwidth]{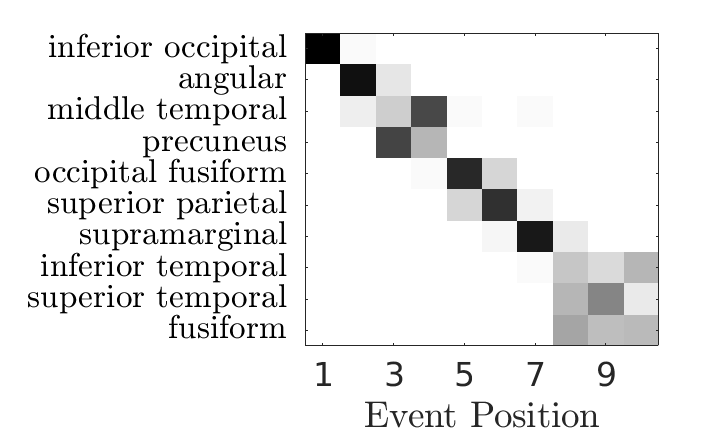}
 \end{subfigure}
  \begin{subfigure}{0.32\textwidth}
  \centering
  {\footnotesize \textbf{Space}\par}
 \includegraphics[width=\textwidth]{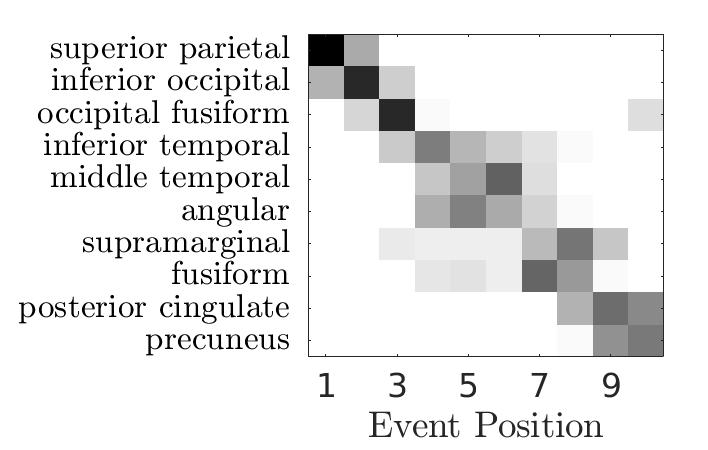}
 \end{subfigure}
  \begin{subfigure}{0.32\textwidth}
  \centering
  {\footnotesize \textbf{Object}\par}
 \includegraphics[width=\textwidth]{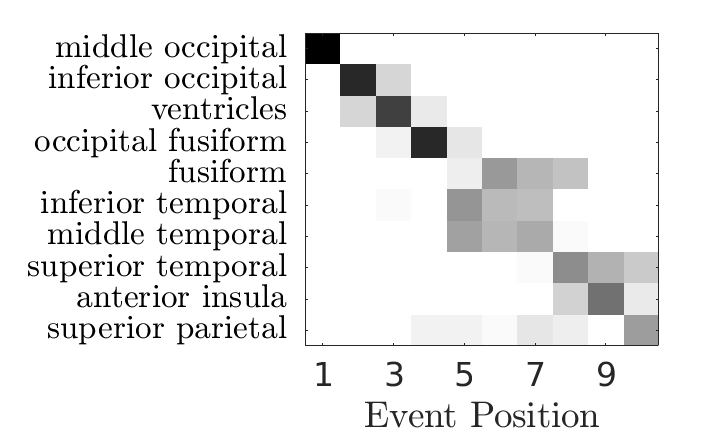}
 \end{subfigure}
 \caption[Early atrophy progression within the three cognitively-defined PCA subgroups]{Early atrophy progression within the three cognitively-defined PCA subgroups, as estimated by the EBM. The top figures shows snapshots of the atrophy patterns for the first 7 stages in the EBM, while the last row shows the uncertainty in the atrophy progression sequence. Brain pictures generated using BrainPainter \cite{marinescu2019BrainPainter}}
 \label{fig:pcaSubgrSnap}
\end{figure}

Fig. \ref{fig:pcaSubgrSnap} shows snapshots of the EBM-estimated atrophy sequence at early stages, in the three cognitively-defined PCA subgroups. The bottom row in the figure shows the uncertainty in the estimated atrophy sequence. The vision subgroup has initial atrophy in the inferior occipital lobe, followed by the angular gyrus, middle temporal, precuneus and superior parietal. The space subgroup shows early atrophy in the superior parietal area (dorsal pattern), followed by inferior occipital and inferior and middle temporal areas. Finally, the object subgroup shows initial atrophy in the middle and inferior occipital areas, with subsequent atrophy in the inferior and middle temporal areas (ventral pattern). Bonferroni-corrected statistically significant differences in atrophy progression have also been observed between PCA subgroups -- see Supplementary Fig. \ref{fig:ebmStatTestPcaSubgroups}. Longitudinal trajectories within PCA subgroups using the DEM could not be estimated due to lack of sufficient data.

\section{Discussion}
\label{sec:pcaDis}

In this work we performed one of the first longitudinal studies of atrophy progression in PCA. Results suggest that in PCA occipital and superior parietal areas are the first to become affected, followed by temporal areas. By 10 years after $t_0$, there seems to be widespread atrophy in the occipital, parietal and temporal areas, as well as ventricular expansion. In contrast, tAD seems to have significant early atrophy in the hippocampus, with subsequent temporal atrophy and ventricular expansion starting 5 years after $t_0$. 

Regarding PCA heterogeneity, our study also provided the first glimpse into the early longitudinal patterns of atrophy within three cognitively defined PCA subtypes. We found early phenotype-specific patterns of atrophy within each cognitively-defined PCA subgroup. These patterns of pathology overlap with the pathways that are hypothesised to be affected within each group: striate cortex for the vision subgroup, dorsal pathway for the space subgroup and ventral pathway for the object subgroup. Nonetheless, among the subgroups there is considerable variability in these patterns as well as spatial overlap, which might suggest that these should not necessarily be interpreted as distinct diseases, but rather that the patients lie on a continuum of phenotypical variation, as suggested by \cite{lehmann2011basic}. 

Our study has several strengths. First of all, the large number of PCA subjects with longitudinal neuroimaging and cognitive data allowed us to perform a robust analysis of PCA atrophy progression. The EBM and DEM methods we used are all data-driven, don't require manual biomarker thresholds and don't rely on diagnostic classes, which are often noisy and biased. Moreover, the ability of the EBM to work with limited cross-sectional data allowed us to estimate the progression of PCA subgroups, which are small and have limited longitudinal data available.  An advantage of the DEM method is its ability to fit continuous, non-parametric biomarker trajectories based on GPs, which makes it suitable for modelling biomarkers whose trajectories have varying shapes.

Nevertheless, our study has several limitations that need to be addressed. First of all, since data was acquired over an extended period of time, not all subjects had CSF, molecular or pathological confirmation for Alzheimer's disease. This can be a problem, as previous studies suggested that at least half of patients who receive a  diagnosis of probable AD actually have other non-AD underlying pathologies \cite{schneider2007mixed,schneider2009neuropathology}. Follow-up studies will need to have a higher proportion of patients with pathological or molecular confirmation. Moreover, the data was acquired in three different centres using different scanners and field strengths, although we adjusted for these covariates. 

The EBM and DEM models that we employed also have several limitations that we acknowledge. First of all, both methodologies assume all subjects follow the same progression sequence. Secondly, the DEM requires longitudinal data, which prevented us from fitting the DEM to the PCA subgroups, who lacked enough longitudinal data. Another assumption made by the EBM is that the control population is well-defined, as we fit the distribution of normal biomarker values directly on the biomarker values of the control population. The EBM also assumes simplistic, step-wise biomarker trajectories that switch from a normal to an abnormal value. With respect to the DEM, the approach requires a reference timepoint, which we took it to be the threshold that best separates the controls from patients after disease staging.  

There are several avenues for future research. Further molecular and pathological confirmation can be obtained for the remaining patients to ensure they all have a reliable diagnosis, which will enable an unbiased estimation of the progression sequence. The EBM and DEM methodologies can be further extended to allow random effects or to fit different progression sequences for different sub-populations in a data-driven way, such as the approach of \cite{young2015multiple}. Information about the rate and extent of atrophy in the PCA subgroups can also be computed after enough data has been acquired. A well-defined control population for the EBM can also be defined by selecting only amyloid-negative subjects or by other types of stratification. The EBM model can be extended to model more complex trajectory shapes, while the DEM can be further extended to a multivariate approach that inherently aligns the biomarker trajectories.

Finally, one of the key directions of future research is to understand the disease mechanisms underlying PCA. To this end, several methods can be used to estimate these mechanisms, such as those based on propagation of pathogenic proteins \cite{raj2012network, georgiadis2018computational} or the architecture of brain networks \cite{zhou2012predicting}. The influence of genetic factors such as Alipoprotein E (APOE) status \cite{schott2006apolipoprotein, snowden2007cognitive} and other factors recently identified \cite{schott2016genetic, schott2006apolipoprotein} from genome-wide association studies also need to be understood. This research will lead the way towards drug development in PCA clinical trials and will allow the selection of robust outcome measures and fine-grained patient stratification in clinical trials in PCA. 

\section{Conclusion}

In this work I performed a statistical analysis of the neuroimaging data from PCA and tAD subjects from the DRC, HUVR and UCSF centres. I pre-processed all the MRI images and applied the event-based model and the differential equation model on the PCA and tAD cohorts, as well as on three cognitively-defined PCA subgroups. The analysis I made gives the first glimpse into the longitudinal progression of atrophy in PCA subjects, and into the early longitudinal patterns of atrophy in the vision, space and object subgroups.

In the following chapter, I will present some novel extensions to the EBM and DEM models that will enable better estimation of the parameters for the EBM and alignment of the biomarker trajectories for the DEM. These improvements can provide a more accurate disease signature, and remove the need for ad-hoc methods of estimating parameters.

\chapter[Novel Extensions to the EBM and DEM]{Novel Extensions to the Event-based Model and Differential Equation Model}
\label{chapter:perf}

\section{Contributions}

In this work I present methodological extensions to the event-based model (EBM) and differential equation model (DEM) and I evaluate their performance compared to the standard implementations. In order to assess differences between these methods more accurately, I also propose novel performance measures based on disease staging consistency and prediction of time elapsed between visits. I formulated and implemented the novel methodologies, and performed their evaluation. I also pre-processed the DRC MRI scans. My colleague Alexandra Young pre-processed the ADNI data. 

\section{Introduction}

Many data-driven disease progression models (DPMs) that have been presented in chapter \ref{chapter:bckDpm} make assumptions about the biomarker data and the model parameters, which limit their usefulness on practical applications. For example, the differential equation model by \cite{villemagne2013amyloid} is univariate, hence it assumes independence across different biomarkers. In order to place biomarker trajectories on the same time frame, in the previous chapter we used a post-hoc anchoring process (see section \ref{sec:pcaStaMetDEM}). This anchoring is inaccurate, as it relies on setting the reference time $t_0$ using biomarker values of a clinical group (i.e. controls or AD). This anchoring process is challenging because of singularities arising from flat trajectories\footnote{The alignment is performed by setting $t_0=0$ so that $f(t_0) = mean(patients)$. However, if the trajectory is flat then there are many points  $t_0$ that match the mean of patients. Even if the trajectory is not fully flat, the measurement noise is amplified by the low slope of $f$.}, and the fact that subjects are at different stages along the disease. Another limitation of some DPMs is that the fitting algorithm assumes independence between different sets of parameters. While this is done in order to ensure computational tractability, this yields inaccurate parameter estimates. In particular, the event-based model parameter estimation procedure proposed by \cite{fonteijn2012event} and \cite{young2014data} assumes that the parameters of the likelihood models for normal and abnormal values are independent of the abnormality sequence. Some better parameter estimation procedures are therefore needed, which can ensure a robust data fit.

The evaluation of the performance of disease progression models is another open problem that has not been addressed so far. While previous studies used accuracy of clinical status predictions\cite{young2014data}, clinical diagnosis is often not reliable without neuropathological confirmation -- one study reported that a clinical diagnosis of \emph{probable AD} has between 70.9--87.3\% accuracy and between 44.3\%--70.8\% specificity. Therefore, performance metrics based on the prediction of clinical diagnosis might not be sufficiently sensitive to differences in the performance of such algorithms. While \cite{fonteijn2012event} computed the number of subjects with increased staging over time -- a performance measure that doesn't rely on clinical diagnosis -- it does not take model uncertainty of staging into account and it is specific to discrete models such as the event-based model. 

In this chapter we suggest novel extensions in the event-based model and differential equation model and propose four novel performance measures for evaluating disease progression models that don't rely on clinical diagnosis. For the event-based model, we devise two novel fitting procedures that perform joint optimisation of the parameters of the normal and abnormal likelihood models, as well as the abnormality sequence. For the differential equation model, we devise a novel data-driven way to align the biomarker trajectories to a common axis by estimating trajectory-specific and subject-specific time shifts. The novel performance measures that we propose exploit uncertainty in the estimated stages and are also suitable for evaluating continuous trajectory models. Using data from the Alzheimer's Disease Neuroimaging Initiative (ADNI) and the Dementia Research Centre (DRC), UK, we show that the novel models generally have better or equal performance compared to standard models. Moreover, we also show that the novel performance measures that we proposed are more sensitive to changes in models than standard measures based on the prediction of diagnosis or conversion status. 

\section{Methods}

\subsection{EBM Extensions}

In this section we outline two novel methods of parameter fitting for the event-based model: a blocked MCMC sampling of the distribution parameters and event ordering (section \ref{sec:simultSampling}), and an Expectation-Maximisation approach (section \ref{sec:ebmEM}). Furthermore, we also present a novel methodology performing a data-driven temporal alignment of the differential equation model trajectories (section \ref{sec:demOptim}). 

\subsubsection{EBM -- Joint MCMC Sampling}
\label{sec:simultSampling}

We present a novel method for fitting the event-based model that jointly optimises the parameters of the normal and abnormal likelihood models using MCMC sampling. The full EBM likelihood model, already been described in Eq. \ref{eq:ebm4}, is:

\begin{equation}
\label{ap:ebmMain}
 p(X|S) = \prod_{i=1}^P \left[ \sum_{k=0}^N p(k) \left( \prod_{j=1}^k p\left(x_{i,s(j)} | E_{s(j)} \right) \prod_{j=k+1}^N p\left(x_{i,s(j)} | \neg E_{s(j)}\right) \right) \right]
\end{equation}
where $x_{ij}$ represents the value of biomarker $j$ from subject $i$  and is informative of event $E_j$ in subject $i$, $P$ is the number of subjects and $N$ is the number of biomarkers. The abnormality sequence $S = [S(1), \dots, S(N)]$ describes the order in which events $E_1, E_2, \dots , E_N$ become abnormal, and models disease progression. 

We now reformulate the EBM likelihood model to explicitly take into account the parameters of the likelihood models for the normal and abnormal biomarker values. This will allow joint optimisation of these parameters, along with sequence parameter $S$. We assume the likelihood models for normal and abnormal biomarker values are Gaussian distributions, i.e. $p(x|E_j) \sim N(\mu^a_j, \sigma^a_j)$ and $p(x|\neg E_j) \sim N(\mu^n_j, \sigma^n_j)$ where $\mu^a_j$ and $\sigma^a_j$ model the distribution of abnormal values for biomarker $j$ (i.e. event $E_j$ occurred), while $\mu^n_j$ and $\sigma^n_j$ model the distribution of normal values for biomarker $j$ (event $E_j$ did not occur). Thus, the full set of parameters that need to be modelled is $\theta = \left[ [\mu^n_j, \sigma^n_j, \mu^a_j, \sigma^a_j]^{j=1 \dots N}, S \right]$. Therefore, the likelihood in equation \ref{ap:ebmMain} can be explicitly written as:

\begin{multline}
\label{ap:ebmExplicit}
 p(X|S, [\mu^n_j, \sigma^n_j, \mu^a_j, \sigma^a_j]^{j=1 \dots N}) = \\ \prod_{i=1}^P \left[ \sum_{k=0}^N p(k) \left( \prod_{j=1}^k p\left(x_{i,S(j)} | \mu^a_{S(j)}, \sigma^a_{S(j)} \right) \prod_{j=k+1}^N p\left(x_{i,S(j)} | \mu^n_{S(j)}, \sigma^n_{S(j)} \right) \right) \right]
\end{multline}

We maximise this likelihood using blocked MCMC sampling, where at each step we only propose parameters for biomarker $j$, i.e. $[\mu^n_j, \sigma^n_j, \mu^a_j, \sigma^a_j]$ along with a new sequence $S_j^{new}$ where only event $E_j$ changed its position. The distribution parameters for the other biomarkers and the ordering of the other events $i \neq j$ are kept the same. This blocked approach can lead to faster convergence because there is strong dependence between parameters corresponding to the same biomarker and between the position of the corresponding event in the sequence. The covariance matrix of the proposal distribution is estimated by taking 100 bootstraps of the dataset and computing the covariance of $[\mu^c, \sigma^c, \mu^p, \sigma^p]$, where $\mu^c$, $\sigma^c$ are the mean and standard deviation of the control group while $\mu^p$, $\sigma^p$ are the mean and standard deviation of the patient group.

\subsubsection{EBM -- Expectation Maximisation}
\label{sec:ebmEM}

The blocked MCMC approach from the previous section can be challenging to implement and execute, due to the difficulty of sampling in a high-dimensional space. The user needs to further tune the covariance matrix in order to get a good acceptance rate, and ensure enough samples are taken in order to exhaustively explore the space of the distribution. To mitigate these issues, we further propose a novel parameter estimation procedure for the EBM based on the Expectation Maximisation (EM) framework \cite{bishop2007pattern}. The EM framework is suitable for estimating parameters of models with discrete latent variables, such as the EBM which has discrete latent variables $k$ representing the subject-specific stages. The EM framework tries to find the parameters $\theta^*$ that maximise the expected log-likelihood of the complete data $\theta^* = \argmax_{\theta} Q(\theta | \theta^{old}) = \argmax_{\theta} \mathbb{E}_{Z|X,\theta^{old}}[log\ p(X,Z|\theta)]$. The key observation to make is that the joint likelihood over the latent variables $Z = [Z_1, ..., Z_N]$ and $X = [X_1, ..., X_N]$ factorises, giving the following form:

\begin{equation}
Q(\theta | \theta^{old}) = \sum_{i=1}^P \sum_{z_i} p(Z_i = z_i|X_i, \theta^{old}) \left[ \sum_{j=1}^{z_i} log\ p(x_{ij}|E_{S(j)}) + \sum_{j=z_i + 1}^N log\ p(x_{ij}| \neg E_{S(j)}) \right]
\end{equation}

We find the maximum for $\mu_k^n$, the mean of $p(x|\neg E_k)$, by solving $\frac{d}{d\mu_k^n}Q(\theta | \theta^{old}) = 0 $. This gives the following update equation for $\mu_k^n$:
\begin{equation}
 \mu_k^n = \sum_{i=1}^P x_{ik} w_i^n
\end{equation}
with weights $w_i^n$ defined as:
\begin{equation}
w_i^n = \frac{p(S^{-1}(k) > Z_i | X, \theta^{old})}{\sum_{i=1}^P \ p(S^{-1}(k) > Z_i | X, \theta^{old})}
\end{equation}
and $p(S^{-1}(k) > Z_i | X, \theta^{old}) = \sum_{l=S^{-1}(k)+1}^{K} p(Z_i = l | X, \theta^{old})$. The full derivation is given in section \ref{sec:appEbmEm}. Similar update rules are derived in the appendix section \ref{sec:appEbmEm} for the other parameters: $\sigma_k^n$, $\mu_k^a$, $\sigma_k^a$. 

Optimising the sequence $S$ in the M-step is intractable, so we use MCMC sampling where at each step of the sampling process we propose a new sequence $S^{new}$, find the optimal distribution parameters for each biomarker given $S^{new}$ using the closed-form EM update rules, and then evaluate the likelihood $Q(\theta | \theta^{old})$. We keep performing a greedy ascent as performed by \cite{fonteijn2012event} until convergence. Although this approach might not guarantee that we truly find the optimal parameters, it still results in an increase of $Q(\theta | \theta^{old})$. This approach, called generalised EM, guarantees that the method will converge to a local maxima \cite{bishop2007pattern}. For parameter initialisation, we use the mean and standard deviation of the control and patient populations. 

\subsection{DEM -- Optimised Trajectory Alignment}
\label{sec:demOptim}

We present a novel extension to the DEM (see section \ref{sec:bckDem}) that aims to place the estimated biomarker trajectories on the same temporal axis, in a data-driven way. The main idea is to use the subjects' data to find optimal subject-specific and trajectory-specific time shifts. Since we are mostly interested in estimating population-level trajectories and to align them on a common time-axis, we do not currently add subject-specific progression speeds and random-effect deviations from the average trajectories.

Let us denote by $X$ our dataset, where $x_{pb}$ is the measurement of biomarker $b$ in patient $p$ and $f_b$ is the shape of the trajectory for biomarker $b$. For every biomarker $b$, we aim to estimate a temporal shift $t_b$ of the trajectory and a measurement noise $\sigma_b$. At the same time. The log-likelihood for the data $X_p$ from patient $p$ can be expressed as:

\beq
p(X_p| t_1, \dots, t_B, \sigma_1, \dots , \sigma_B, z_p ) = \prod_{b=1}^{B}  N(x_{pb}|f_b(z_p-t_b), \sigma_b)
\eeq

where $z_p$ is a latent parameter representing the time-shift of subject $p$. Multiplying by the prior on $z_p$ and summing over all the possible values of $z_p$ we get the marginal:
\beq
p(X_p| t_1, \dots, t_B, \sigma_1, \dots , \sigma_B) = \sum_{z_p} p(Z_p = z_p) \prod_{b=1}^{B}  N(x_{pb}|f_b(z_p-t_b), \sigma_b)
\eeq

Assuming the data from each subject is conditionally independent given $z_p$, we get the full likelihood:
\beq
p(X| t_1, \dots, t_B, \sigma_1, \dots , \sigma_B) = \prod_{p=1}^{P} \sum_{z_p} p(Z_p = z_p) \prod_{b=1}^{B}  N(x_{pb}|f_b(z_p-t_b), \sigma_b)
\eeq

This likelihood can be optimised with any method of choice such as MCMC sampling or gradient methods. We chose to optimise the model using an iterative approach, where for each biomarker $b$ we optimise its trajectory shift $t_b$ conditioned on all the other parameters (Markov blanket), and then estimate its measurement noise $\sigma_b$.

\subsection{Performance Evaluation}
\label{sec:perfEvalMethods}

We compare the performance of the extended EBM and DEM methods against the standard implementations, using novel performance metrics that we propose. In section \ref{sec:stagingConsist}, we present metrics which test staging consistency, i.e. that follow-up stages are greater or equal to baseline stages. We then generalise this concept in section \ref{sec:timeLapse} to test the accuracy of models in predicting the time lapse between two visits of a subject.

\subsubsection{Staging Consistency}
\label{sec:stagingConsist}

The staging consistency metrics test whether subjects' stages at follow-up visits are greater than or equal to the stages at baseline. While such a metric is simple to compute in cases of no uncertainty, we also define a more complex metric that takes staging uncertainty into account. 

Let us consider a set of random variables $z_t^i$ representing the stage of subject $i$ at timepoint $t$, where $i \in [1 \dots N], t \in [1 \dots T_i]$, $N$ being the number of subjects and $T_i$ the number of time points for subject $i$. For most disease progression models, the EBM and DEM included, we can find the posterior $p(z_t^i|X_i, \theta)$, which we will call the staging probabilities. Moreover, let $M^i_t = \argmax_s p(z_t^i = s)$ be the maximum likelihood stage for subject $i$ at time point $t$. The \emph{hard staging consistency} $C_h$ counts the proportion of stages from consecutive visits of every subject where the stage at the later visit must be greater than the stage at the earlier visit. We define $C_h$ as follows:

\begin{equation}
 C_h = \frac{1}{-N +\sum_{i=1}^N T_i} \sum_{i=1}^N \sum_{t=2}^{T_i} \mathbb{I}[M^i_t > M^i_{t-1}] 
\end{equation}
where the element $-N +\sum_{i=1}^N T_i$ is a normalising constant that represents the number of pairs of consecutive visits from all subjects and time points in the dataset. The $C_h$ metric ranges from 0 (no consistent pairs of stages) to 1 (all pairs of stages are consistent).

We further seek to generalise the \emph{hard staging consistency}, in order to make use of the full staging probabilities, instead of using only the maximum likelihood stages. We define the \emph{soft staging consistency} $C_s^i(t_1,t_2)$ for subject $i$ given two time points $t_1$ and $t_2$ as:

\begin{equation}
C_s^i(t_1,t_2) = p(z^i_{t_1} \leq z^i_{t_2}) = \sum_{s \in S} p(z^i_{t_2} = s) p(z^i_{t_1} \leq s) 
\end{equation}
where $S$ is the set of possible stages in the disease progression model. We then define the \emph{soft staging consistency} for the whole population as the mean of subject-specific consistencies for consecutive timepoints:

\begin{equation}
C_s = \frac{1}{-N +\sum_{i=1}^N T_i} \sum_{i=1}^N \sum_{t=2}^{T_i} C_s^i(t_1,t_2) 
\end{equation}

\subsubsection{Time-lapse Prediction}
\label{sec:timeLapse}

Time-lapse prediction is a generalisation of the staging consistency, where the disease progression model needs to predict how much time passed between two visits of the same subject, which is the compared against the true time elapsed. This is only possible for models that estimate continuous latent stage variables, such as the DEM. We define the \emph{hard time-lapse} metric $D_h$ as follows:

\begin{equation}
D_h = \frac{1}{-N +\sum_{i=1}^N T_i} \sum_{i=1}^N \sum_{t=2}^{T_i} \left| \tau(M^i_t) - \tau(M^i_{t-1}) - (a^i_t - a^i_{t-1}) \right|
\end{equation}
where $a^i_t$ is the age of subject $i$ at timepoint $t$, $M^i_t$ is the maximum likelihood stage for subject $i$ at timepoint $t$ and $\tau(M^i_t)$ is the estimated time from onset associated with stage $M^i_t$. The equivalent \emph{soft time-lapse} metric $D_s$, which uses probabilistic staging variables $z_t^i$, is defined as:
\begin{equation}
D_s = \frac{1}{-N +\sum_{i=1}^N T_i} \sum_{i=1}^N \sum_{t=2}^{T_i} \left| \mathbb{E}[\tau(z^i_t) - \tau(z^i_{t-1})] - (a^i_t - a^i_{t-1}) \right|
\end{equation}

\subsection{Data Preprocessing}

\subsection{The Dementia Research Centre Cohort}
\label{sec:dataPrep}

The Dementia Research Centre (DRC), UK cohort contains 89 controls, 74 PCA and 67 tAD subjects that have undergone 1.5T/3T T1-weighted MRI scans,  with an average of 2-3 visits per subject. The demographics of the DRC dataset is given in table \ref{tab:drcDemographics}. More details on the cohort can be found in the PCA longitudinal study from section \ref{sec:pcaParticipants}\footnote{This cohort is a subset of the cohort used in the PCA longitudinal study.}.

\begin{table}[ht]
\centering
 \begin{tabular}{c | c c c}
  Demographics & CN & PCA & AD\\
  \hline
  Number & 89 & 74 & 67\\
  Sex M/F & 33/56 & 28/46 & 35/32 \\
  Age (years) & 61 $\pm$ 11 & 63 $\pm$ 7 & 66 $\pm$ 9\\
  Years from onset & - & 4.5 $\pm$ 2.8 & 4.8 $\pm$ 2.6\\
  Number of visits & 2.8 $\pm$ 2.5 & 2.5 $\pm$ 1.7 & 3.0 $\pm$ 2.7\\
 \end{tabular}
 \caption[Baseline population demographics for DRC data]{Baseline population demographics for the DRC cohort.}
 \label{tab:drcDemographics}
\end{table}

\subsubsection{Image Processing}
The MRI scans were segmented using the Geodesic Information Flows (GIF) algorithm by \cite{cardoso2015geodesic}, which is available as a service at \url{http://cmictig.cs.ucl.ac.uk/niftyweb/}. The atlas that has been used for segmentation is the Neuromorphometrics atlas (provided by Neuromorphometrics, Inc.), which produced 146 different brain ROIs across the left and right hemisphere. All brain volumes have been corrected for total intracranial volume (TIV), age and gender using a general linear model. We summed left and right brain regions together and further selected a subset of 25 ROIs: (a) whole brain; (b) ventricles; (c) 2 subcortical regions: hippocampus and amygdala; (d) 5 occipital regions: inferior, middle and superior occipital and the occipital fusiform and lingual; (e) 5 parietal regions: superior parietal, angular, precuneus, supramarginal and postcentral; (f) 4 temporal regions: inferior, middle and superior temporal along with fusiform; (g) 4 frontal regions: superior, middle and inferior frontal along with precentral; and (h) 3 limbic regions: entorhinal, parahippocampal and posterior cingulate.

\subsection{The Alzheimer's Disease Neuroimaging Initiative Cohort}

In this study we also used the ADNI dataset to evaluate our disease progression models. We used the same biomarker data as previously used by \cite{young2014data}, which included all 285 subjects (Controls, MCI and AD) that had a CSF examination at baseline, cognitive assessment at baseline and MRI scans at baseline and 1 year follow-up. The demographics of the selected subjects is shown in table \ref{tab:adniDemographics}. We also used follow-up imaging, clinical and CSF data at 12- and 24-months after baseline visit. Clinical diagnoses at baseline, 12-, 24- and 36-months were also used for evaluating performance on diagnosis prediction and conversion prediction. The CSF total tau and phosphorylated tau were log-transformed to improve normality. 

\begin{table}[ht]
\centering
 \begin{tabular}{c | c c c}
  Demographics & CN & MCI & AD\\
  \hline
  Number & 92 & 129 & 64\\
  Sex M/F & 48/44 & 82/47 & 34/30\\
  Age (years) & 75 $\pm$ 5 & 73 $\pm$ 7 & 75 $\pm$ 8\\
  Education (years) & 15.6 $\pm$ 2.9 & 15.9 $\pm$ 3 & 15 $\pm$ 3\\
  APOE +/- & 22/70 & 72/57 & 45/19\\
 \end{tabular}
 \caption[Baseline population demographics for the ADNI cohort.]{Baseline population demographics for the ADNI cohort.}
 \label{tab:adniDemographics}
\end{table}

\subsubsection{Image Processing}

FreeSurfer Version 4.3 was used to compute regional volumes of the hippocampus, entorhinal cortex, middle temporal gyrus, fusiform gyrus, ventricles, whole brain and total intracranial volume (TIV) at baseline, 12- and 24-month follow-up. All regional volumes were normalised for each subject by dividing by TIV. Atrophy rates for the whole brain and hippocampus were estimated using the Boundary Shift Integral (BSI) (\cite{freeborough1997boundary}) using the scans at baseline and 12-months follow-up. In particular, volume change for the whole brain was measured using the KN-BSI method (\cite{leung2010robust}) and for hippocampus using the MAPS-HBSI method (\cite{leung2010automated}). 

We used the same biomarker set as the one used by \cite{young2014data}, which included 14 biomarkers in total: (a) three CSF biomarkers: amyloid-$\beta_{1-42}$, phosphorylated tau and total tau; (b) 3 cognitive tests: Alzheimer's Disease Assessment Scale - Cognitive Subscale (ADAS-Cog), Rey Auditory Verbal Learning Test (RAVLT) and the Mini-Mental State Examination (MMSE); (c) six regional brain volumes: whole brain, ventricles, hippocampus, entorhinal, middle temporal gyrus and fusiform gyrus; (d) rates of atrophy for two ROIs: hippocampus and whole brain.

\section{Results}
\label{sec:perfRes}

We tested all novel EBM and DEM methods, along with their standard implementations. We evaluated each model using the staging consistency and time-lapse metrics, using data from the DRC and ADNI datasets. On the DRC dataset, we also evaluated the models with respect to diagnosis prediction, while on ADNI we evaluated them based on prediction of conversion from healthy controls to mild cognitive impairment (MCI) and from MCI to Alzheimer's disease.

\subsection{DRC Results}
\label{sec:perfResDrc}

We ran all the standard and novel methods for the EBM and DEM described above. For the EBM joint MCMC sampling method, we took 100,000 samples who had an acceptance rate in the range of 0.29-0.33 (min-max) across all cross-validation folds, which suggested a good mixing, while the sample autocorrelation was in the range of 0.86-0.93. In order to get the acceptance rate to this reasonable rate, we estimated the covariance matrix  of the MCMC proposal distribution based on the covariance of maximum likelihood parameter estimates on bootstrapped subsets of the full dataset. For the EBM-EM method, we performed 20 iterations as we noticed that the method converged after a maximum of 3-4 iterations.

In table \ref{tab:drcStagingResPCA} we show the staging-based metrics for the PCA cohort. Each entry shows the mean and standard deviation of the metric calculated over 10 cross-validation folds. Similar results are shown for the AD cohort in table \ref{tab:drcStagingResAD}. In table \ref{tab:drcDiagRes} we show the models' balanced accuracy in diagnosis prediction on the DRC cohort. For reference, we also show similar results using a standard Support Vector Machine (SVM) classifier \cite{vapnik2006estimation}. The SVM classifier was trained with a linear kernel using sequential minimal optimisation \cite{platt1998sequential} and a box-constraint parameter $C=1$. In each entry we show the mean and standard deviation of the balanced classification accuracy across the 10 cross-validation folds. 

\begin{table}[ht]
\centering
 \begin{tabular}{c | c | c | c | c}
  Model & \multicolumn{2}{c |}{Staging Consistency} & \multicolumn{2}{c}{Time-lapse}\\
  & Hard & Soft & Hard & Soft\\
  \multicolumn{5}{c }{Event-based Model}\\
  \hline
  EBM - Standard & 0.88 $\pm$ 0.12 & 0.66 $\pm$ 0.09 & - & -\\ 
  EBM - MCMC & 0.96 $\pm$ 0.06 & 0.70 $\pm$ 0.06  & - & -\\
  EBM - EM & 0.95 $\pm$ 0.10 & 0.68 $\pm$ 0.11 & - & -\\
  \multicolumn{5}{c }{\textbf{}}\\
  \multicolumn{5}{c }{Differential Equation Model}\\
  \hline
  DEM - Standard & 0.94 $\pm$ 0.06 & 0.95 $\pm$ 0.05 & 0.54 $\pm$ 0.31 & 0.52 $\pm$ 0.29\\
  DEM - Optimised & 0.95 $\pm$ 0.05 & 0.95 $\pm$ 0.04 & 0.56 $\pm$ 0.28 & 0.52 $\pm$ 0.27\\
  
 \end{tabular}
 \caption[Model performance according to staging-based metrics on PCA subjects from the DRC cohort]{Model performance according to staging-based metrics on PCA subjects from the DRC cohort. The mean and standard deviations are calculated for each testing set in 10-fold cross-validation.}
 \label{tab:drcStagingResPCA}
\end{table}

\begin{table}[ht]
\centering
 \begin{tabular}{c | c | c | c | c}
  Model & \multicolumn{2}{c |}{Staging Consistency} & \multicolumn{2}{c}{Time-lapse}\\
  & Hard & Soft & Hard & Soft\\
  \multicolumn{5}{c }{Event-based Model}\\
  \hline
  EBM - Standard & 0.91 $\pm$ 0.16 & 0.71 $\pm$ 0.07 & - & -\\
  EBM - MCMC & 0.96 $\pm$ 0.07 & 0.76 $\pm$ 0.10 & - & -\\
  EBM - EM & 0.99 $\pm$ 0.01 & 0.72 $\pm$ 0.07 & - & -\\
  \multicolumn{5}{c }{\textbf{}}\\
  \multicolumn{5}{c }{Differential Equation Model}\\
  \hline
  DEM - Standard & 0.87 $\pm$ 0.10 & 0.88 $\pm$ 0.08 & 0.72 $\pm$ 0.91 & 0.67 $\pm$ 0.92\\
  DEM - Optimised & 0.87 $\pm$ 0.10 & 0.88 $\pm$ 0.08 & 0.74 $\pm$ 0.92 & 0.69 $\pm$ 0.92\\
  
 \end{tabular}
 \caption[Model performance according to staging-based metrics on typical AD subjects from the DRC cohort.]{Model performance according to staging-based metrics on typical AD subjects from the DRC cohort. The mean and standard deviations are calculated for each testing set in 10-fold cross-validation.}
 \label{tab:drcStagingResAD}
\end{table}

\begin{table}[H]
\centering
 \begin{tabular}{c | c c c}
  Model & PCA vs AD &  Controls vs PCA & Controls vs AD\\
  \multicolumn{4}{c }{Event-based Model}\\
  \hline
  EBM - Standard & 0.72 $\pm$ 0.13 & 0.95 $\pm$ 0.05 & 0.90 $\pm$ 0.06\\
  EBM - MCMC & 0.79 $\pm$ 0.09 & 0.94 $\pm$ 0.06 & 0.90 $\pm$ 0.05\\
  EBM - EM & 0.80 $\pm$ 0.07 & 0.95 $\pm$ 0.05 & 0.87 $\pm$ 0.05\\
  \multicolumn{4}{c }{\textbf{}}\\
  \multicolumn{4}{c }{Differential Equation Model}\\
  \hline
  DEM - Standard & 0.81 $\pm$ 0.07 & 0.95 $\pm$ 0.05 & 0.90 $\pm$ 0.11\\
  DEM - Optimised & 0.82 $\pm$ 0.09 & 0.93 $\pm$ 0.06 & 0.88 $\pm$ 0.14\\
  \multicolumn{4}{c }{\textbf{}}\\
  \multicolumn{4}{c }{Support Vector Machine}\\
  \hline
  SVM & 0.79 $\pm$ 0.14 & 0.91 $\pm$ 0.06 & 0.88 $\pm$ 0.07\\
  
 \end{tabular}
 \caption[Model performance at diagnosis prediction on DRC data.]{Model performance at diagnosis prediction on the DRC cohort. Each entry shows the mean and standard deviation of the balanced accuracy across the cross-validation folds. }
 \label{tab:drcDiagRes}
\end{table}

\subsection{ADNI Results}
\label{sec:perfResAdni}

In table \ref{tab:adniStagingRes} we show the staging-based performance results of the progression models on the ADNI dataset. As with the DRC results, for each metric we show its mean and standard deviation over the 10 cross-validation folds. In table \ref{tab:adniConvPredRes} we also evaluated the models on how well they predict conversion from MCI to AD at 12-months, 24-months and 36-months from baseline visit. We did not compute results for prediction of conversion status in controls due to small and very imbalanced datasets.

\begin{table}[H]
\centering
 \begin{tabular}{c | c c | c c}
  Model & \multicolumn{2}{c |}{Staging Consistency} & \multicolumn{2}{c}{Time-lapse}\\
  & Hard & Soft & Hard & Soft\\
  \multicolumn{5}{c }{Event-based Model}\\
  \hline
  EBM - Standard & 0.83 $\pm$ 0.07 & 0.76 $\pm$ 0.05 & - & -\\ 
  EBM - MCMC & 0.84 $\pm$ 0.05 & 0.76 $\pm$ 0.06 & - & -\\
  EBM - EM & 0.84 $\pm$ 0.08 & 0.74 $\pm$ 0.06 & - & -\\
  \multicolumn{5}{c }{\textbf{}}\\
  \multicolumn{5}{c }{Differential Equation Model}\\
  \hline
  DEM - Standard & 0.87 $\pm$ 0.05 & 0.83 $\pm$ 0.08 & 0.85 $\pm$ 0.17 & 0.85 $\pm$ 0.16\\
  DEM - Optimised & 0.87 $\pm$ 0.05 & 0.84 $\pm$ 0.07 & 0.86 $\pm$ 0.15 & 0.86 $\pm$ 0.16\\
 \end{tabular}
 \caption{Model performance according to staging metrics on ADNI data.}
 \label{tab:adniStagingRes}
\end{table}


\begin{table}[H]
\centering
 \begin{tabular}{c | p{2.5cm} p{2.5cm} p{2.5cm}}
  Model & \multicolumn{3}{c}{Duration between baseline and follow-up}\\
  
  & 12 months & 24 months & 36 months\\
  \multicolumn{4}{c }{Event-based Model}\\
  \hline
  EBM - Standard & 0.69 $\pm$ 0.14 & 0.64 $\pm$ 0.11 & 0.72 $\pm$ 0.15\\
  EBM - MCMC & 0.66 $\pm$ 0.14 & 0.63 $\pm$ 0.10 & 0.74 $\pm$ 0.14\\
  EBM - EM & 0.69 $\pm$ 0.15 & 0.63 $\pm$ 0.10 & 0.76 $\pm$ 0.15\\
  \multicolumn{4}{c }{\textbf{}}\\
  \multicolumn{4}{c }{Differential Equation Model}\\
  \hline
  DEM - Standard & 0.73 $\pm$ 0.13 & 0.72 $\pm$ 0.14 & 0.70 $\pm$ 0.13\\
  DEM - Optimised & 0.64 $\pm$ 0.11 & 0.69 $\pm$ 0.12 & 0.75 $\pm$ 0.14\\
  \multicolumn{4}{c }{\textbf{}}\\
  \multicolumn{4}{c }{Support Vector Machine}\\
  \hline
  SVM & 0.68 $\pm$ 0.15 & 0.70 $\pm$ 0.10 & 0.77 $\pm$ 0.08\\
  
 \end{tabular}
 \caption{Model performance at prediction of conversion from MCI to AD on ADNI data.}
 \label{tab:adniConvPredRes}
\end{table}

\section{Discussion}
\label{sec:perfDis}

\subsection{Model Performance on DRC cohort}
\label{sec:perfDisDrc}

In the PCA cohort, we notice that the extended EBM methods show better results compared to the standard EBM method, whereas the extended DEM method has equal performance compared to the standard method. When comparing EBM vs DEM models, most EBM models perform as well as the DEM models in terms of \emph{hard staging consistency} but relatively worse in \emph{soft staging consistency}. There is also a drop in EBM staging consistency when moving from the \emph{hard} to the \emph{soft staging consistency}, which can be explained by the discrete nature of the EBM and by the simplistic biomarker trajectories, effectively modelled as step-functions, which can result in significant staging uncertainty.

In the AD cohort we again find that the novel EBM methods show improvements over standard methods, while there is no significant difference between the novel and standard DEM methods. When comparing EBMs vs DEMs, we notice that the EBM models actually perform better in terms of \emph{hard-}, but worse in \emph{soft-staging consistency}. This could again be due to overly simplistic EBM trajectories that might not offer a good fit to the data.

In the diagnosis prediction tasks, most disease progression models have similar performance, with only the Standard EBM having a low performance in the PCA vs AD test. The SVM classifier has slightly worse results compared to the disease progression models for the Controls vs PCA task, but similar results for the other tasks.  

\subsection{Model Performance on ADNI cohort}
\label{sec:perfDisAdni}

In the ADNI cohort, we notice that the extended EBM and DEM methods have similar performance to the standard methods. There is again a drop in EBM performance on the soft consistency metric as compared to the hard consistency. The fact that there is no improvement in ADNI data between the novel methods and the standard methods suggests that the standard methods already offered a good fit on this dataset, and further that the ADNI dataset has different characteristics compared to the DRC dataset. We attribute this to the fact that the biomarkers present in the ADNI dataset were multimodal and included both early-stage molecular markers as well as late-stage cognitive tests, which enabled even the standard models to robustly estimate the subjects' disease stages.

The results on conversion prediction in ADNI show that all models have a broadly similar performance at this task. However, a few clear differences can be noticed in some models. The model with the best performance at 12-months and 24-months conversion prediction is the DEM with standard trajectory alignment, while at 36-month conversion the SVM and the novel EBM and DEM methods perform the best. The fact that different models have different performance at different durations-of-conversion suggests different models have better fits on certain time-frames of the disease time course.

\subsection{Staging-based Metrics}

The staging-based performance measures can pick up differences in the performance of the models in both DRC and ADNI datasets, in particular between different classes of models such as EBM vs DEM. This is most clear with the soft staging consistency, which penalises the EBM more than the DEM. This might be the case because the DEM constructs continuous non-parametric biomarker trajectories which might give a better fit than the simplistic step-wise trajectories of the EBM. 

The staging consistency metrics can also pick up differences across fitting procedures within the same model, such as in the case of EBMs. On the other hand, there are generally no statistically significant differences in the DEM between the standard versus optimised alignment, probably due to the fact that the standard alignment is already good enough for the two datasets that we tested them on.

There are some differences between the results on the \emph{hard-} vs \emph{soft-staging consistency} metrics. In particular, the \emph{soft-staging consistency} penalised the EBM models more than the DEM models, probably due to increased staging uncertainty in the EBM models. In terms of time-lapse, there were no significant differences between the \emph{hard} and the \emph{soft} versions of this metric.

\subsection{Diagnosis Prediction Metrics}

We found that the performance metrics which are based on diagnosis or conversion prediction are less able to discriminate between different types of models or fitting procedures. Moreover, there was a lot of variability in the values of these performance metrics across folds and also in the model rankings across experiments, especially in the ADNI cohort, which made it hard to identify an overall best model. The variability of the diagnosis prediction metrics can be attributed to inaccuracies and biases in the diagnostic labels, and to the heterogeneity present in these diseases.

\section{Summary}
\label{sec:perfSum}

In this work we presented several extensions of the EBM and the DEM. We further devised performance metrics that measure the accuracy of the predicted subject stages and clinical diagnosis. We evaluated the new methodologies on data from two distinct diseases (PCA vs tAD), and on two independent datasets (ADNI and DRC). Our results show that in many situations the novel EBM and DEM fitting methods show improvements with respect to our performance metrics compared to the standard versions. 

\subsection{Limitations and Future Work}
\label{sec:perfSumLim}

The performance metrics we used for evaluation have certain inherent limitations which might limit their use for some disease progression models. For example, the staging consistency metrics are prone to cheating using a specially crafted model that can get perfect consistency if it simply assigns the same stage to all subjects. However, the time-difference metric does not suffer from this problem, due to the fact that the model needs to predict precisely the time that passed between different visits to the clinic. 

The staging consistency metrics are based on the idea that the biomarkers evolve monotonically as the disease progresses. However, this is not the case with some neurological disorders such as multiple sclerosis (MS), where a patient can have an attack (relapse) followed by a period of steady recovery (remission). For such ``non-monotonic`` diseases, other performance measures based on biomarker predictions would be more suitable, such as the ones used in the TADPOLE Challenge \cite{marinescu2018tadpole}. 

One limitation of the time difference metrics is that they require the disease progression model to estimate the time from onset for every stage. This is not normally modelled in discrete models such as the EBM or other methods based on Markov chains (e.g. \cite{sukkar2012disease}). However, it might be possible to extend these discrete models in order to estimate time since onset for each of the states.

While we have tested these models only on the DRC and ADNI datasets, their performance might be different on other datasets with different types of neurodegenerative diseases and biomarker data. Future work should include validation on other datasets, including well-phenotyped datasets of autosomal-dominant Alzheimer's disease or genetic frontotemporal dementia. Moreover, the models should be tested also on other types of biomarkers, such as non-MRI imaging biomarkers, molecular measurements from cerebro-spinal fluid or cognitive tests.

Future work can also include performance evaluation of these models on simulated datasets, in presence of ground truth. This will enable the detection of more subtle differences in these methodologies, which might not be detectable in patient datasets due to inherent measurement noise and disease heterogeneity.

\section{Conclusion}
\label{sec:perfCon}

In this chapter I presented methodological extensions in the EBM and DEM, and evaluated their performance based on a set of performance measures, some of which I proposed. Future work will focus on evaluating other types of disease progression models presented in chapter \ref{chapter:bckDpm}, or on devising more sensitive performance metrics, for evaluation on both simulated data as well as patient datasets.

In the next chapter, I will present DIVE: a novel disease progression model that can estimate fine-grained spatial patterns of brain pathology, and estimate latent subject-specific time-shifts. Such a model overcomes a some limitations of the EBM and DEM models, which do not take spatial correlation into account and assume a pre-defined ROI atlas. DIVE can also help us better understand underlying disease mechanisms by studying the overlap between spatial patterns of pathology and brain connectomes.

\chapter[DIVE: A Spatiotemporal Progression Model of Brain Pathology]{DIVE: A Spatiotemporal Progression Model of Brain Pathology in Neurodegenerative Disorders}
\label{chapter:dive}

In this chapter I present DIVE, a novel spatiotemporal model of disease progression that I estimates fine-grained spatial patterns of atrophy in the brain. I did the entire work: model development, mathematical derivations (Supplementary section \ref{sec:diveAppendix}), image pre-processing and analysis and wrote the manuscripts. Arman Eshaghi showed me how to perform the image processing with Freesurfer. Daniel Alexander and Marco Lorenzi offered suggestions with modelling and experiment design. All collaborators gave me feedback on the manuscripts.

\section{Publications}
\begin{itemize}
 \item R. V. Marinescu, A. Eshaghi, M. Lorenzi, A. L. Young, N. P. Oxtoby, S. Garbarino, T. J. Shakespeare, S. J. Crutch and D. C. Alexander, A Vertex Clustering Model for Disease Progression: Application to Cortical Thickness Images, Information Processing in Medical Imaging, 2017
 \item R. V. Marinescu, A. Eshaghi, M. Lorenzi, A. L. Young, N. P. Oxtoby, S. Garbarino, S. J. Crutch, D. C. Alexander, DIVE: A spatiotemporal progression model of brain pathology in neurodegenerative disorders, NeuroImage, 2019.
\end{itemize}

\section{Introduction}
\label{sec:diveInt}

Current image-based disease progression models, such as those presented in section \ref{sec:bckSca}, estimate the evolution of the disease using a small set of biomarkers corresponding to pre-defined regions-of-interest (ROI). This ROI parcellation is usually coarse and doesn't allow one to find spatially dispersed patterns of atrophy. While spatiotemporal longitudinal models have already been demonstrated \cite{derado2010modeling, hyun2016stgp, lorenzi2015efficient}, these models regress against pre-defined sets of covariates such as age, time since baseline or clinical markers. This is problematic because, age-based alignment of subjects assumes all subjects have the same age of disease onset, while for time since baseline, its relationship with disease onset is unknown. Similarly, clinical markers are noisy, biased, suffer from floor/ceiling and training effects, are not sensitive in pre-symptomatic phases, and have low test-retest reliability \cite{johnson2012brain}.  Recently, some spatiotemporal models that estimate subject-specific time-shifts have been developed \cite{bilgel2016multivariate,koval2017statistical}. However, these models generally cannot recover dispersed and disconnected pathological patterns, because they assume voxel measurements correlate based on spatial distance, either through a distance function or distance from control points. However, spatially dispersed pathological patterns have been observed in AD and related dementias and are hypothesised to appear due to the interaction of pathology with brain networks \cite{seeley2009neurodegenerative}. Discovering such fine-grained patterns could allow one to understand underlying mechanisms of pathology propagation along these networks. However, a spatiotemporal disease progression model that allows recovery of dispersed and disconnected atrophy patterns present in AD, is not currently available. 

In this work, we present DIVE: Data-driven Inference of Vertexwise Evolution. DIVE is a novel disease progression model with single vertex resolution that makes only weak assumptions on spatial correlation. In contrast to approaches which model temporal trajectories for a small set of biomarker measures based on a priori defined ROIs, DIVE models temporal trajectories for each vertex on the cortical surface. DIVE combines unsupervised learning and disease progression modelling to identify clusters of vertices on the cortical surface that show a similar trajectory of brain pathology over a particular patient cohort. This formulation enables us to estimate a fine-grained spatial distribution of pathology and also provides a novel parcellation of the brain based on temporal change. 

We first test DIVE on synthetic data and show that the model can recover known biomarker trajectories and time-shifts. We then demonstrate the model on both MRI and PET data from two cohorts: the Alzheimer's Disease Neuroimaging Initiative (ADNI) and the Dementia Research Centre (DRC), UK. We use the model to reveal spatiotemporal patterns of pathology to a much finer resolution than previous models and demonstrate the ability to assign subjects to stages that predict progression. Finally, we validate DIVE in terms of how robust are the estimated pathology patterns and how well the disease progression scores correlate with cognitive tests. Code for DIVE is available online: \url{https://github.com/mrazvan22/dive}.

\section{Methods}
\label{sec:diveMet}

In this section we describe the mathematical formulation of DIVE (section \ref{sec:diveMod}), then we show how to fit the model using Expectation Maximisation (section \ref{sec:diveFit}) and we describe further implementation details of the algorithm (section \ref{sec:diveImplem}). Afterwards, we outline the synthetic data-generation process (section \ref{sec:diveSimulations}) for testing the model in the presence of ground truth, as well as the pipeline for pre-processing the ADNI and DRC datasets (section \ref{sec:diveDataAcquis}).

\begin{figure}
 \centering
 \includegraphics[width=\textwidth]{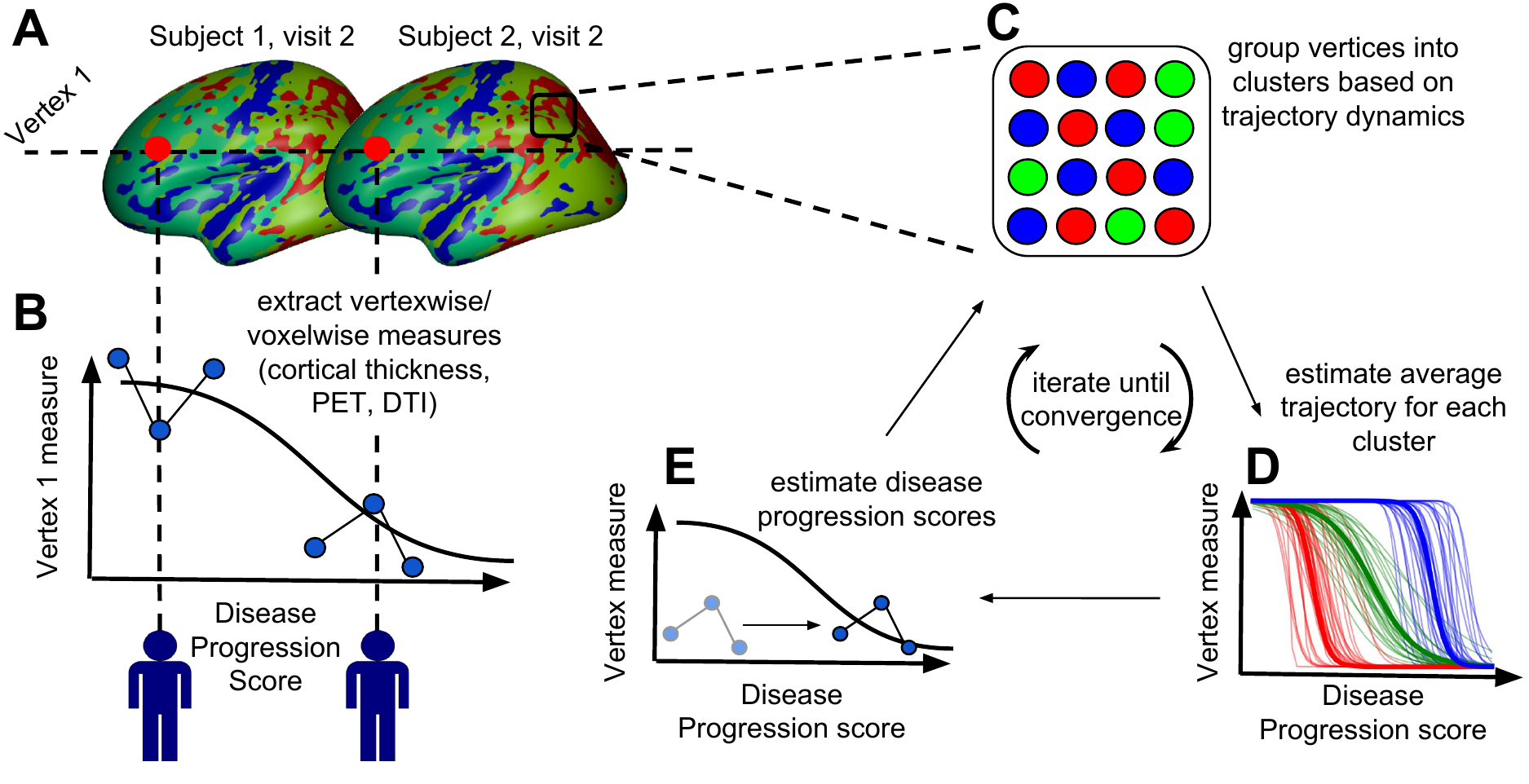}
 \caption[Diagram of the proposed DIVE model.]{Diagram of the proposed DIVE model. DIVE assumes that biomarkers of pathology (e.g. cortical thinning) can be measured at many vertices (i.e. locations) on the cortical surface (A), where each vertex has a distinct trajectory of change during disease progression (B). In (B), each individual has measurements for vertex 1 at three visits. DIVE assigns to every cortical vertex one of a small set of temporal trajectories describing the change in some image-based measurement (e.g. cortical thickness, amyloid PET, DTI fractional anisotropy measures) from beginning to end of the disease progression. The estimation process simultaneously estimates the set of clusters, the trajectory defining each cluster, and the position of each subject along the trajectories, which are defined on a common timeline. The process iterates assignment of each vertex to clusters (red, green and blue in this diagram) (C), estimation of the trajectory in each cluster (D) and estimation of the disease progression score (location along trajectory) for  each subject (E), all within an Expectation-Maximisation framework, until convergence. In particular, (E) shows how the disease progression score, which is initially set to the individual's age, converges to the disease stage of the subject. Diagram made by me. }
 \label{fig:diveDiagram}
\end{figure}

\subsection{DIVE Model}
\label{sec:diveMod}

Figure \ref{fig:diveDiagram} illustrates the DIVE aims and implementation. DIVE input measures are vertexwise or voxelwise biomarker measures in the brain (Fig \ref{fig:diveDiagram}\textcolor{blue4}{A}), such as cortical thickness or amyloid load. A vertex is a location on the cortical surface at which a biomarker of pathology is quantifiable (e.g. cortical thickness). For each vertex on the cortical surface (or voxel in the 3D brain volume), we estimate a unique trajectory along the disease progression timeline (Fig \ref{fig:diveDiagram}\textcolor{blue4}{B}), while also estimating subject/visit-specific disease progression scores (i.e. disease stages). We do that by grouping vertices with similar biomarker trajectories into clusters (Fig \ref{fig:diveDiagram}\textcolor{blue4}{C}), and we estimate a representative trajectory for every cluster (Fig \ref{fig:diveDiagram}\textcolor{blue4}{D}). Each trajectory is a function of subject-/visit-specific disease progression scores (DPS) (Fig \ref{fig:diveDiagram}\textcolor{blue4}{E}). The DPS depends linearly on the time since baseline visit, but with subject-specific slope and intercept.

\subsection{Modelling Subject-specific Parameters}

The disease progression score $s_{ij}$ for subject $i$ at visit $j$ is a latent variable denoting the current disease stage of the subject at this visit. It  defined as a linear transformation of time since baseline measurement $t_{ij}$ (in years):

\begin{equation}
\label{eq:dps_vwdpm}
 s_{ij} = \alpha_i t_{ij} + \beta_i
\end{equation}
where $\alpha_i$ and $\beta_i$ represent the speed of progression and time shift (i.e. disease onset) of subject $i$ respectively. 

\subsection{Modelling Biomarker Trajectory for a Single Vertex}

DIVE assumes that the biomarker measure at each vertex on the cortical surface follows a sigmoidal trajectory $f(s ; \theta)$ over the disease progression score $s$ and with parameters $\theta$. We choose a parametric sigmoid function because it is a parsimonious parametric model that offers better fit compared to linear models, is monotonic, and can account for floor and ceiling effects \cite{caroli2010dynamics, sabuncu2011dynamics}. We also assume that vertices are grouped into $K$ clusters and we model a unique trajectory for each cluster $k \in {1, ... , K}$, which will be referred to as cluster trajectories. The sigmoidal function $f(s; \theta_k)$ for cluster $k$ is defined as: 

\begin{equation}
\label{eq:dps_vwdpm2}
 f(s;\theta_k) = \frac{a_k}{1+exp(-b_k(s-c_k))} + d_k
\end{equation}
where $s$ is the disease progression score from Eq. \ref{eq:dps_vwdpm} and $\theta_k = [a_k, b_k, c_k, d_k]$ are parameters controlling the shape of the trajectory -- $d_k$ and $d_k + a_k$ represent the lower and upper limits of the sigmoidal function, $c_k$ represents the inflection point and $a_k b_k/4$ represents the slope at the inflection point. 

For a given subject $i$ at visit $j$, the value $V_l^{ij}$ of its biomarker measurement at vertex $l$ is a random variable that has an associated discrete latent variable $Z_l \in [1, ... , K]$ denoting the cluster it was generated from. The value of $V_l^{ij}$ given that it was generated from cluster $Z_l$ can be modelled as:

\begin{equation}
\label{eq:dps_vwdpm3}
 p(V_l^{ij} | \alpha_i, \beta_i, \theta_{Z_l}, \sigma_{Z_l}, Z_l) = N(V_l^{ij} | f(\alpha_i t_{ij} + \beta_i | \theta_{Z_l}), \sigma_{Z_l})
\end{equation}
where $N(V_l^{ij} | f(\alpha_i t_{ij} + \beta_i | \theta_{Z_l}), \sigma_{Z_l})$ represents the probability density function (pdf) of the normal distribution that models the measurement noise along the sigmoidal trajectory of cluster $Z_l$, having variance $\sigma_{Z_l}$. Next, we assume the measurements from different subjects are independent, while the measurements from the same subject $i$ at different visits $j$ are linked using the disease progression score from equation \ref{eq:dps_vwdpm}. Moreover, we also assume a uniform prior on $Z_l$. This gives the following model:

\begin{equation}
\label{eq:dps_vwdpm4}
 p(V_l, Z_l | \alpha, \beta, \theta, \sigma) = \prod_{(i,j) \in I} N(V_l^{ij} | f(\alpha_i t_{ij} + \beta_i | \theta_{Z_l}), \sigma_{Z_l})
\end{equation}
where $I = {(i,j)}$ represents the set of all the subjects $i$ and their corresponding visits $j$. Furthermore, $V_l = [V_l^{ij} | (i,j) \in I]$ is the 1D array of all the values for vertex $l$ across every subject and corresponding visit. Vectors $\alpha = [\alpha_1, \dots, \alpha_S]$ and $\beta = [\beta_1, \dots, \beta_S]$, where $S$ is the number of subjects, denote the stacked parameters for the subject shifts. If a subject $i$ has multiple visits, these visits share the same parameters $\alpha_i$ and $\beta_i$. Vectors $\theta = [\theta_1, \dots, \theta_K]$ and $\sigma = [\sigma_1, \dots, \sigma_K]$, with $K$ being the number of clusters, represent the stacked parameters for the sigmoidal trajectories and measurement noise specific to each cluster.

Due to our main motivation of modelling population trajectories and in order to ensure robustness and identifiability, we did not add random effects to the trajectories of specific subjects.

\subsection{Modelling Biomarker Trajectories for all Vertices}

So far we have a model for only one vertex on the brain surface. We therefore extend the formulation to all the vertices by assuming all these vertex measurements are spatially independent, giving the complete data likelihood:

\begin{equation}
\label{eq:dps_vwdpm5}
 p(V, Z | \alpha, \beta, \theta, \sigma) = \prod_l^L \prod_{(i,j) \in I} N(V_l^{ij} | f(\alpha_i t_{ij} + \beta_i | \theta_{Z_l}), \sigma_{Z_l})
\end{equation}
where $V = [V_1, \dots, V_L]$, $Z = [Z_1, \dots, Z_L]$, $L$ being the total number of vertices on the cortical surface. The formulation so far assumes spatial independence between measurements in different vertices, but in section \ref{sec:diveSpatialCorr} the model is extended to capture spatial correlations. The full joint distribution is given by:

\begin{equation} \label{eq21}
  p(V, Z, \alpha, \beta, \theta, \sigma) = p(V, Z | \alpha, \beta, \theta, \sigma) p(\alpha, \beta, \theta, \sigma)
\end{equation}
where $p(\alpha, \beta, \theta, \sigma)$ is an informative prior on the model parameters defined as follows: 
\begin{equation} \label{eq1}
\begin{split}
 p(V, Z, \alpha, \beta, \theta, \sigma) & = \prod_{i} p(\alpha_i) p(\beta_i)\\
 p(\alpha_i) & \sim \Gamma(\alpha_{shape}, \alpha_{rate})\\
 p(\beta_i) & \sim N(\beta_{mean}, \beta_{std})\\
\end{split}
\end{equation}
where $\alpha_{shape}$, $\alpha_{rate}$, $\beta_{mean}$, $\beta_{std}$ are a-priori defined hyperparameters. The informative priors on the subject-specific parameters help ensure model identifiability, as the model otherwise has two extra degrees of freedom. Such informative priors on $\alpha_i$ and $\beta_i$ also help deal with singularities in the objective functions of $\alpha_i$ and $\beta_i$ when the biomarker trajectories are flat.

We get the final model log likelihood for incomplete data by marginalising over the latent variables $Z$:
\begin{equation}
\label{eq:dps_vwdpm6}
 p(V|\alpha, \beta, \theta, \sigma) = \prod_{l=1}^L \sum_{k=1}^K p(Z_l = k) \prod_{(i,j) \in I} N(V_l^{ij} | f(\alpha_i t_{ij} + \beta_i | \theta_k), \sigma_k)
\end{equation}

Throughout the article, we will use the shorthand $z_{lk} = p(Z_l = k)$.

\subsection{Modelling Spatial Correlation}
\label{sec:diveSpatialCorr}

The version of the model presented so far assumes spatial independence between vertex measurements. However, the regional organisation of the cortex suggests we would expect spatial correlation\footnote{By correlation here we mean that these vertex measurements are not statistically independent} of the vertex measurements. More precisely, measures of cortical thickness or other modalities are often similar in neighbouring vertices on the cortical surface and likely belong to the same cluster. DIVE can be easily extended to include mild spatial constraints on the correlation of vertex measurements via a Markov Random Field (MRF), which encourages neighbouring vertices to have the same corresponding cluster. We hypothesise that incorporating such constraints should reduce the effects of noise and produce a more stable clustering. However, this does not model correlation between the actual vertex values, but only between the latent variables $Z_l$, i.e. the cluster membership of each vertex. The MRF thus has the advantage of not requiring the use of huge covariance matrices, which are otherwise needed if we want to model correlation of vertex values directly. Moreover, in contrast to previous methods that use correlation based on spatial distance \cite{bilgel2016multivariate,koval2017statistical}, we use neighbourhood correlations, which allow us to estimate fine-grained spatial patterns of pathology. With the MRF, the full-data likelihood function of the model now becomes:

\begin{equation}
 p(V, Z | \alpha, \beta, \theta, \sigma, \lambda) = \prod_l^L \left[ \prod_{(i,j) \in I} N(V_l^{ij} | f(\alpha_i t_{ij} + \beta_i | \theta_{Z_l}), \sigma_{Z_l}) \prod_{l_2 \in N_l} \Psi (Z_{l}, Z_{l_2}) \right]
\end{equation}

where $\Psi(Z_l, Z_{l2})$ is a clique term representing the likelihood of a neighbouring vertex $l_2$ to have similar label with vertex $l$. The formula for the clique term is:

\begin{equation}
 \Psi (Z_{l}=k, Z_{l_2}=k_2) = 
 \begin{cases}
  exp(g(\lambda)) & \text{if } k = k_2\\
  exp(-h(\lambda)) & \text{otherwise}
 \end{cases}
\end{equation} 

where $\lambda$ is a parameter controlling how much to penalise neighbouring vertices that belong to distinct clusters, and $g$ and $h$ are positive, monotonic functions over the $\lambda>0$ range. We choose $g(\lambda)=\lambda$ and $h(\lambda)=\lambda^2$, which results in a concave objective function for $\lambda$, ensuring that it can later be optimised (see M-step).

Therefore, the model parameters that need to be estimated are $M = [\alpha, \beta, \theta, \sigma, \lambda]$ where $\alpha$ and $\beta$ are the subject specific shifting parameters, $\theta$ and $\sigma$ are the cluster specific trajectory and noise parameters and $\lambda$ is the clique parameter denoting the penalisation of spatially non-smooth assignments of latent variables $Z$. 

\subsection{Fitting the Model using Generalised Expectation-Maximisation}
\label{sec:diveFit}

We choose to fit our model using Expectation-Maximisation (EM), because it offers a fast convergence given the large number of parameters that need to be estimated and the huge dimensionality of relevant datasets (e.g. 1973 subjects x 163,842 vertices in ADNI). In the next two sections we outline the E-step and M-step. While both of these steps have no closed-form solution, we will solve them using numerical optimisation, which only results in an increase in the objective function at each iteration. However, the EM algorithm is still guaranteed to converge, and this approach is called Generalised EM \cite{bishop2007pattern}.

Algorithm \ref{fig:algo_vwdpm} shows the model fitting procedure using the EM algorithm. The procedure first initialises (line 1) some parameters required to start the EM algorithm: the subject parameters $\alpha$ and $\beta$ and the latent parameters $z_{lk}$ which represent the assignment of vertices to clusters. In the M-step, the method updates the trajectories of each cluster (lines 4-6), the subjects-specific parameters (line 9) and the clique penalty term $\lambda$ (line 17). In the E-step, the method computes $z_{lk}$ (line 18) using previously defined functions that compute $z_{lk}$ given a fixed $\lambda$ (line 14).

\begin{figure}
\begin{algorithm}[H]
 Initialise $\alpha^{(0)}$, $\beta^{(0)}$, $z_{lk}^{(0)}$ \\
  \While{$\theta$, $\sigma$, $\alpha$, $\beta$ or $z_{lk}$ not converged}{
   \tcp*[l]{M-step 1: For each cluster, optimise its trajectory}
    \For{$k=1$ to $K$}{
      ${\theta_k^{(u)} = \argmin_{\theta_k} \sum_{l=1}^L z_{lk}^{(u-1)} \sum_{(i,j) \in I} (V_l^{ij} - f(\alpha_i^{(u-1)} t_{ij} + \beta_i^{(u-1)} | \theta_k))^2  - log\ p(\theta_k)}$\\
      $\theta_k^{(u)} = \mbox{make\_identifiable}(\theta_k^{(u)})$\\
      ${ \left(\sigma_k^{(u)}\right)^2 = \frac{1}{|I|} \sum_{l=1}^L z_{lk}^{(u-1)} \sum_{(i,j) \in I} (V_l^{ij} - f(\alpha_i^{(u-1)} t_{ij} + \beta_i^{(u-1)} | \theta_k^{(u)}))^2 - log\ p(\sigma_k)}$
    }
     \tcp*[l]{M-step 2: For each subject, optimise its time shift $\alpha_i$ and progression speed $\beta_i$}
    \For{$i=1$ to $S$}{
    \footnotesize
      ${\alpha_i^{(u)}, \beta_i^{(u)} = \argmin_{\alpha_i, \beta_i}  \left[ \sum_{l=1}^L \sum_{k=1}^K \frac{z_{lk}^{(u-1)}}{2 \left(\sigma_k^{(u)}\right)^2 } \sum_{j \in I_i} (V_l^{ij} - f(\alpha_i t_{ij} + \beta_i | \theta_k^{(u)}))^2\right] - log\ p(\alpha_i, \beta_i)}$
    \normalfont
    }
   \tcp*[l]{E-step 1: Define functions $\zeta_{lk}(\lambda)$ computing $z_{lk}$, the probability of vertex $l$ being assigned to cluster $k$, given fixed $\lambda$}
  \For{$l = 1$ to $L$}{
    \For{$k = 1$ to $K$}{
      \tcp*[l]{Pre-compute data fit terms $D_{lk}$}
      $D_{lk} = -\frac{1}{2}log\ (2 \pi \left(\sigma_k^{(u)}\right)^2) |I| - \frac{1}{2\left(\sigma_k^{(u)}\right)^2} \sum_{i,j \in I} (V_l^{ij} - f(\alpha_i^{(u)} t_{ij} + \beta_i^{(u)} | \theta_k^{(u)}))^2$\\
      $ \zeta_{lk}(\lambda) \approx exp \left( D_{lk} +   \sum_{l_2 \in N_l} log\ \left[ exp(-\lambda^2) + z_{l_2k}^{(u-1)} (exp(\lambda) - exp(-\lambda^2)) \right] \right) $

    }
  }
  \tcp*[l]{M-step 3: optimise clique term $\lambda$ using above definitions in E-step 1}
$ \lambda^{(u)} = \argmax_{\lambda}\ \sum_{l=1}^L \sum_{k=1}^K \zeta_{lk}(\lambda) \left[  D_{lk} \  + \lambda \sum_{l_2 \in N_l}  \zeta_{l_2 k}(\lambda)\  -\lambda^2 \sum_{l_2 \in N_l} (1- \zeta_{l_2 k}(\lambda))  \right]  $\\
  
  \tcp*[l]{E-step 2: Compute next $z_{lk}$ using the best $\lambda$}
  $z_{lk}^{(u)} = \zeta_{lk}(\lambda^{(u)})$
  }

  $\alpha_i^{(u)} = \frac{\alpha_i^{(u)}}{\sigma_N}, \beta_i^{(u)} = \frac{\beta_i^{(u)} - \mu_N}{\sigma_N}$   \tcp*[l]{Re-scale subject shifts}
\end{algorithm}
\caption[The DIVE parameter estimation algorithm.]{The DIVE parameter estimation algorithm. The algorithm, based on Expectation-Maximisation, iteratively optimises the assignment of vertices to clusters (E-step) and the parameters for the biomarker trajectories and subject time-shifts (M-step). }
\label{fig:algo_vwdpm}
\end{figure}
\subsubsection{E-step}

In the Expectation step, at iteration $u$ we seek an estimate of $p(Z | V, M^{(u-1)})$, given the current estimates of the parameters $M^{(u-1)} =[ \theta_k^{(u-1)}, \sigma_k^{(u-1)}, \alpha_i^{(u-1)}, \beta_i^{(u-1)}, \lambda_i^{(u-1)}]$. We perform this using Iterated Conditional Modes \cite{bishop2007pattern}, which performs coordinate-wise gradient ascent. This works by conditioning the clique terms Z on the values of Z from the previous iterations.  This approximation gives the following factorisable likelihood:

\begin{equation}
\label{eq:e_approx}
 p(Z | V, \Mu^{(u-1)}) \approx \prod_l^L \mathbb{E}_{Z_{N_l}^{(u-1)}|V_l, M} \left[ p(Z_l|V_l, \Mu, Z_{N_l}^{(u-1)}) \right]
\end{equation}

The factorised form allows for tractable computation and memory storage of $p(Z)$. Let $z_{lk}(u) = p(Z_l = k | V_l, M^{(u-1)},Z^{(u-1)})$. After simplifications we reach the following update rule:

\begin{equation}
\label{eq:e-step}
\begin{split}
 log\ z_{lk}^{(u)} \propto D_{lk}+ \left[ \sum_{l_2 \in N_l} log\ \left[ exp(-\lambda^2) + z_{l_2k}^{(u-1)} (exp(\lambda) - exp(-\lambda^2)) \right] \right]
\end{split}
\end{equation}
where the data-fit term $D_{lk}$ has the following form:

\begin{equation}
\label{eq:e-step_Dlk}
D_{lk} = -\frac{log\ (2 \pi \sigma_k^2) |I|}{2} - \sum_{i,j \in I}  \frac{1}{2\sigma_k^2}(V_l^{ij} - f(\alpha_i t_{ij} + \beta_i | \theta_k))^2 
\end{equation}

The full derivation is given in Supplementary section \ref{sec:diveEmDerivAppendix}. In order to enable optimisation over $\lambda$, a final modification of this step is performed, by considering $z_{lk}$ to be functions $\zeta_{lk}(\lambda)$ over $\lambda$. This results in the update equation from Alg. \ref{fig:algo_vwdpm}, line 18 which is based on pre-defined terms on lines 13-14.

\subsubsection{M-step}

In the Maximisation step we try to estimate the model parameters $M = (\alpha, \beta, \theta, \sigma, \lambda)$ that maximise $E_{Z|V,M^{(u-1)}}[log\ p(V,Z|M)]$. We cannot simultaneously optimise all 5 sets of parameters, so we optimise them independently. In order to get the update rule for the trajectory parameters $\theta_k$ corresponding to cluster $k$ we need to maximise the expected log likelihood with respect to $\theta_k$. The key observation here is that if we assume fixed $\alpha$, $\beta$ and $Z$, then the trajectory parameters $\theta_k$ for every cluster $k$ are conditionally independent, i.e. $\theta_k \ci \theta_m | (Z, \alpha, \beta, \sigma)\ \forall\ (k, m)$, $k \neq m$. This allows us to maximise every $\theta_k$ independently using the following equation:

\begin{equation}
 \theta_k = \argmax_{\theta_k} \sum_{z_1,\dots, z_L}^K p(Z | V, \Mu^{(u-1)})\ log \left[ \prod_{l=1}^L \prod_{(i,j) \in I} N(V_l^{ij} | f(\alpha_i t_{ij} + \beta_i | \theta_{z_l}), \sigma_{z_l}) \right] + log\ p(\theta_k)
\end{equation}

A similar observation of conditional independence can also be observed for the latent variables $Z$. This allows us to decompose the joint distribution over $Z$, and after expanding the noise model we reach the optimisation problem from Alg. \ref{fig:algo_vwdpm}, line 4. See Supplementary section \ref{sec:diveEmDerivAppendix} for full derivation. This does not have a closed-form solution, so we use numerical optimisation for finding $\theta_k$ that maximises the equation from Alg. \ref{fig:algo_vwdpm}, line 4.

A similar equation, yet in closed form, is also obtained for $\sigma_k$ (Alg. \ref{fig:algo_vwdpm}, line 6).  After estimating $\theta$ and $\sigma$ for every cluster, we use the new values to estimate the subject specific parameters $\alpha$ and $\beta$. For every subject $i$, we maximise the expected log likelihood with respect to $\alpha_i$, $\beta_i$ independently, and after simplifications we obtain the update rule from Alg. \ref{fig:algo_vwdpm}, line 9, which is again solved using numerical optimisation. For the numerical optimisation of $\theta$ we used the Nelder-Mead method for its robustness, while for $\alpha$ and $\beta$ we used the second-order Broyden--Fletcher--Goldfarb--Shanno algorithm due to fast convergence. 

The large dimensionality of the dataset (around 163,428 vertices x 400 subjects x 4 timepoints each) makes model fitting extremely difficult from a computational perspective. Initial optimisation on a smaller subset of around 100 ADNI subjects took around 30h. However, we achieved a significant speed-up in the evaluation of objective functions by computing a $z_{lk}$-weighted average of vertex measurements within each cluster (see Appendix section \ref{sec:appDivFas}). This resulted in a final convergence time of around 4-6h depending on the size of the dataset, using an Intel Xeon E3-1271 @ 3.60GHz CPU. Regarding memory requirements, loading into memory around 1600 and fitting the model required around 12GB of RAM. However, we dropped it down by a factor of x4 by using small 16-bit floating representations for the vertexwise biomarkers.

For optimising $\lambda$, we again try to optimise in the M-step the expected full data likelihood under the $Z$ estimates from the previous iteration: 

\begin{equation}
\lambda^{(u)} = \argmax_{\lambda} E_{p(Z|V, M^{(u-1)}, \lambda, Z^{(u-1)})}[log\ p(V,Z|M^{(u-1)})]
\end{equation}

We simplify the above equation by expanding the likelihood model and approximating the joint over $Z$ with the product of the marginals $z_{lk}$ over all vertices $l$. This results in the update equation from Alg. \ref{fig:algo_vwdpm} line 17 -- see appendix for full derivation. In this final equation we also replaced $z_{lk}$ with a function $\zeta_{lk}(\lambda)$ over $\lambda$, which updates $z_{lk}$ based on the current value of $\lambda$ being evaluated. This is done to increase convergence, as latent variables $z_{lk}$ are highly coupled with the value of $\lambda$ being evaluated.

\subsection{Implementation Details}
\label{sec:diveImplem}

\subsubsection{Parameter Initialisation and Priors}

Before starting the fitting process, we need to initialise $\alpha$, $\beta$ and the clustering probabilities $z_{lk}$ (Alg. \ref{fig:algo_vwdpm}, line 1). We set $\alpha_i$ and $\beta_i$ to be 1 and 0 respectively for each subject, which sets the initial disease progression score to the time since baseline of the subject at the clinical visit. We initialise $z_{lk}$ using k-means clustering of the vectors $V_l$. We also initialise hyperparameters $\alpha_{shape}=16e4$, $\alpha_{rate}=16e4$, $\beta_{mean} = 0$ $\beta_{std} = 0.1$, which work well in practice as they result in realistic ranges for $\alpha_i$ and $\beta_i$ of around [0.3, 3] and [-15,15] respectively. The reason why we need to give such large numbers of 16e4 is because there are many vertex measurements ($>$ 100,000) that each drag the subject to an extremity if most values are above/below the population curve. This can be avoided in the future by adding subject-specific random effects to the population trajectory.

As already explained in \cite{jedynak2012computational}, the sigmoid parameters $\theta_k$ are not identifiable because $f(t;a_k,b_k, c_k, d_k) = f(t;-a_k,-b_k, c_k, a_k + d_k)$. We thus need to apply the following transformation on line 5 of Alg. \ref{fig:algo_vwdpm}: if $b_k^{(u)} < 0$ then $a_k^{(u)} = - a_k^{(u)}; b_k^{(u)} = - b_k^{(u)}; d_k^{(u)} = d_k^{(u)} - a_k^{(u)}$. This ensures model identifiability and is performed at every iteration. 

\subsubsection{Estimating the Optimal Number of Clusters}

The EM procedure needs to specify a-priori the number of clusters to fit on the data. We optimise the number of clusters $K$ using Akaike Information Criterion (AIC), which we found to better agree with ground truth in simulations than other information criteria such as the Bayesian Information Criteria (BIC). The number of parameters of the fitted model is 5$K$+2$S$+1, where $S$ is the number of subjects. Note that $z_{lk}$ are not included as parameters of the model because they are latent variables that are marginalised (see Eq. \ref{eq:dps_vwdpm6}). We repeat the fitting procedure for each $K$ from 2 to 100 clusters and select the $K$ that minimises the AIC.

\subsection{Simulation Experiments}
\label{sec:diveSimulations}

\subsubsection{Motivation}

Initial assessment of DIVE performance uses synthetic data, where we know the ground truth. The aim is to explore how accurately we can recover ground truth parameters as the problem becomes harder in three different scenarios:
\begin{itemize}
 \item Scenario 1: as the number of clusters increases, evaluate how well DIVE can estimate the correct number of clusters using AIC and BIC
 \item Scenario 2: as the trajectories become more similar, test how well we can recover the assignment of vertices to clusters and the DIVE parameters
 \item Scenario 3: same as Scenario 2, but for decreasing number of subjects
\end{itemize}

\subsubsection{Synthetic Data Generation}

We first designed a basic simulation, which the model should be able to fit well since the trajectories were designed to be well separated and enough subject data was generated along the disease time course. The data in the basic simulation was generated as follows: 

\begin{enumerate}
 \item Sampled baseline age $a_{i1}$ and shift parameters $\alpha_i$, $\beta_i$ for 300 subjects with 4 timepoints (each timepoint 1 year apart), with $a_{i1} \sim U(40,80)$, $\alpha_i \sim \Gamma(6.25, 6.25)$, $\beta_i \sim N(0, 10)$. Time since baseline has been obtained for every visit $j$ of subject $i$ as follows: $t_{ij} = a_{ij} - a_{i1}$.
 \item Generated three sigmoids with different (slope, centre) parameters: [(-0.1, -15), (-0.1, 2.5), (-0.1, 20)] (Fig. \ref{diveResSynthA}, red lines). Upper and lower limits have been set to 1 and 0 respectively.
 \item randomly assign every vertex $l \in \{1, \dots, L\}$, where $L = 1000$, to a cluster $a[l] \in \{1,2,3\}$
 \item Sampled a set of $L$ perturbed trajectories $\theta_l$ from each of the original trajectories, one for each vertex (Fig. \ref{diveResSynthA}, gray lines) using covariance matrix $C_{\theta} = diag([0, 2b_k/15,  11.6, 0])$.
 \item Sampled subject data for every vertex $l$ from its corresponding perturbed trajectory $\theta_l$ with noise standard deviation $\sigma_l = 1$
\end{enumerate}

From the basic simulation, we generated synthetic data for each of the three scenarios by varying one parameter at a time and kept the other parameters constant, having the same values as in the basic simulation. We varied the following parameters:
\begin{itemize}
 \item Scenario 1: number of clusters - 2, 3, 5, 10, 15, 20, 30 and 40. The cluster centres were spread evenly across a fixed total DPS range where the data was available. 
 \item Scenario 2: distance between trajectory centres (as proportion of total DPS range sampled) -- 0.33, 0.30, 0.23, 0.17, 0.10, 0.07, 0.03 and 0.02 
 \item Scenario 3: number of subjects - 300, 200, 100, 50, 35, 20, 10 and 5
\end{itemize}

\subsubsection{Model Fitting and Evaluation}

Since there was no spatial information in the data generation procedure, we used DIVE without the MRF extension. For Scenario 1, we estimated using AIC and BIC the optimal number of clusters. For Scenarios 2 and 3, after fitting the parameters of DIVE, we calculated the agreement between the final clustering probabilities $p(Z_l)$ and the true clustering assignments $a[l]$. This agreement, which we will call the clustering agreement, is defined as $\aleph = max_{\tau} (1/L) \sum_{l=1}^L p(Z_l = \tau(a[l]))$, where $\tau$ is any permutation of cluster labels. We also computed the error in the DPS estimation (sum of squared differences, SSD) and trajectory estimation (SSD between predicted trajectory and true trajectory at DPS points of every subject visit). 

\subsection{Data Acquisition and Pre-processing}
\label{sec:diveDataAcquis}

Data used in this work were obtained from the Alzheimer's Disease Neuroimaging Initiative (ADNI) database (\url{adni.loni.usc.edu}) and from the Dementia Research Centre, UK. For ADNI, we downloaded all T1 MR images that have undergone gradient warping, intensity correction, and scaling for gradient drift. We included subjects that had at least 3 scans, to ensure we get a robust estimate of the subject specific parameters. This resulted in 138 healthy controls, 235 subjects with mild cognitive impairment (MCI) and 81 subjects with Alzheimer's disease. 

We also downloaded all AV45 PET images from ADNI that were fully pre-processed, having the following tag: \emph{Co-reg, Avg, Std Img and Vox Siz, Uniform Resolution}. This meant that the images were co-registered, averaged across the 6 five-minute frames, standardised with respect to the orientation and voxel size and smoothed to produce a uniform resolution of 8mm full-width/half-max (FWHM). 

The DRC dataset consisted of T1 MRI scans from 31 healthy controls, 32 PCA and 23 typical AD subjects with at least 3 scans each and an average of 5.26 scans per subject. All PCA patients fulfilled both Tang-Wai \cite{tang2004clinical} and Mendez \cite{mendez2002posterior} criteria based on clinical review. The typical AD patients all met the criteria for probable Alzheimer's disease \cite{dubois2007research,dubois2010revising}. 

Given that the ADNI and DRC datasets contained subjects with different modalities or diseases, we ran DIVE independently on the following four cohorts (see Table \ref{tab:divePcaDemogr} for demographics): 
\begin{enumerate}
 \item ADNI MRI: controls, MCI and tAD subjects from ADNI (cortical thickness data) 
 \item DRC tAD: tAD subjects and controls from the DRC dataset (cortical thickness data)
 \item DRC PCA: PCA subjects and controls from the DRC dataset (cortical thickness data)
 \item ADNI PET: AV45 scans from ADNI containing subjects with following diagnoses: healthy controls, subjective memory complaints, early MCI, late MCI and Alzheimer's disease.
\end{enumerate}

\begin{table}
  \centering
  \begin{tabular}{c | c | c | C{3.5cm} | C{3.5cm} } 
  \textbf{Cohort} & \textbf{Diagnosis} & \textbf{Number of Subjects} & \textbf{Number of Scans} & \textbf{Age at baseline (years)}\\
  \hline
  ADNI & Controls & 138 & 4.3 & 76.3\\ 
  MRI & MCI & 235 & 4.6 & 74.8\\ 
  & AD & 81 & 3.5 & 75.8\\ 
  \hline
  DRC & Controls & 31 & 5.0 & 66.3\\ 
  tAD & AD & 24 & 5.4 & 71.2\\ 
  \hline
  DRC & Controls & 31 & 5.0 & 66.3\\ 
  PCA & PCA & 32 & 4.1 & 62.6\\ 
  \hline
  & Controls & 141 & 2.4 & 85.5\\ 
  ADNI & SMC & 27 & 2.0 & 86.1\\ 
  PET & EMCI & 149 & 2.4 & 85.6\\
  & LMCI & 104 & 2.4 & 86.0\\
  & AD & 12 & 2.0 & 87.3\\
  \end{tabular}
  \caption[Demographics of the four cohorts from ADNI and DRC]{Demographics of the four cohorts used in our analysis. ADNI MRI and the DRC cohorts were used for the cortical thickness analysis, while ADNI PET was used for the PET AV45 analysis. MCI -- mild cognitive impairment, SMC - subjective memory complaints, EMCI -- early MCI, LMCI -- late MCI.}
 \label{tab:divePcaDemogr}
\end{table}

\subsubsection{MRI Preprocessing}

On both datasets, in order to extract reliable cortical thickness measures, we ran the Freesurfer longitudinal pipeline \cite{reuter2012within}, which first registers the MR scans to an unbiased within-subject template space using inverse-consistent registration. The longitudinally registered images were then registered to the average Freesurfer template. No further smoothing was performed on these images (FWHM level of zero mm). From these template-registered volumetric images, cortical thickness measurements were computed at each vertex (i.e. point) on an average 2D cortical surface manifold. For each vertex we averaged the thickness levels from both hemispheres in order to later ease visualisation and to obtain a smaller representation of the input data. Each of the final images had a resolution of 163,842 vertices on the cortical surface. 

Finally, we standardised the data by computing Z-scores for each vertex with respect to the values of that vertex in the control population. This normalisation step ensures that the model will not be affected by different thicknesses of the cortex at various locations on the cortical surface. This step is specific for MRI cortical thickness data, and might not be necessary for other modalities (e.g. PET). 

\subsubsection{PET Preprocessing}

We computed amyloid standardised uptake value ratio (SUVR) levels using the PetSurfer pipeline \cite{greve2014cortical,greve2016different}, which is available with Freesurfer version 6. The PetSurfer pipeline first registers the PET image with the corresponding MRI scan, then applies Partial Volume Correction, and finally resamples the voxelwise SUVR values onto the cortical surface. While the final images also had a resolution of 163,842 vertices, the PET data we obtained from ADNI was inherently more smooth than the MRI cortical thickness data (8mm FWHM). We did not standardise the SUVR values like we did for cortical thickness, due to the fact that we did not observe different uptake based on anatomy within the control population.

\subsubsection{The MRF Neighbourhood Graph}

We estimated the MRF neighbourhood graph based on a Freesurfer triangular mesh for the fsaverage template. Each vertex was a triangle on the brain surface estimated with Freesurfer, and we connected the vertices if the corresponding triangles had a shared edge. For the MRF neighbourhood graph, we used a 3rd degree neighbourhood structure, meaning that two vertices were considered neighbours if the shortest path between them was not higher than 3.

\section{Results}
\label{sec:diveResults}

\subsection{Results on Synthetic Data}
\label{sec:diveResultsSynth}

\begin{figure}
\centering
\begin{subfigure}[b]{0.7\textwidth}
\includegraphics[width=1\textwidth,trim=30 0 60 0,clip]{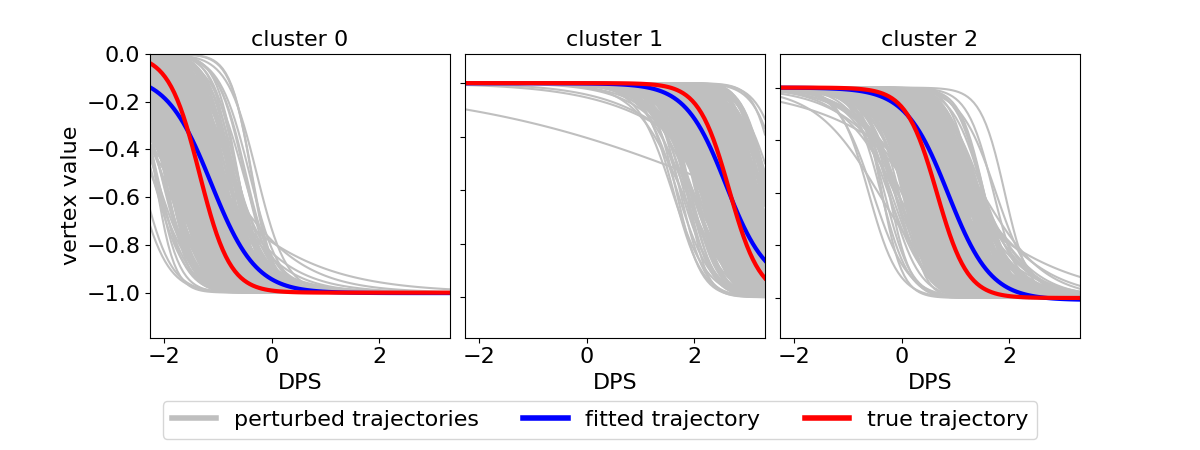}
\caption{}
\label{diveResSynthA}
\end{subfigure}
\begin{subfigure}[b]{0.25\textwidth}
\includegraphics[width=1\textwidth,trim=0 0 0 0,clip]{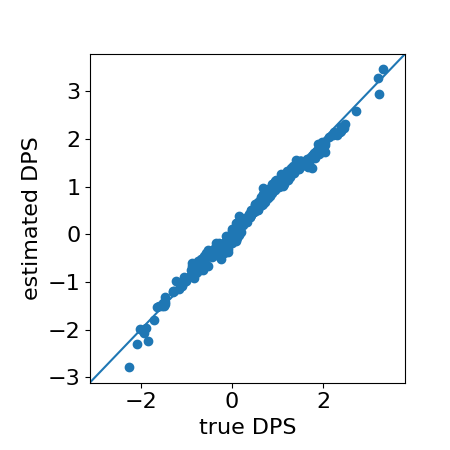}
\vspace{0.4em}
\caption{}
\label{diveResSynthB}
\end{subfigure}
\begin{subfigure}[b]{0.32\textwidth}
\includegraphics[width=1\textwidth]{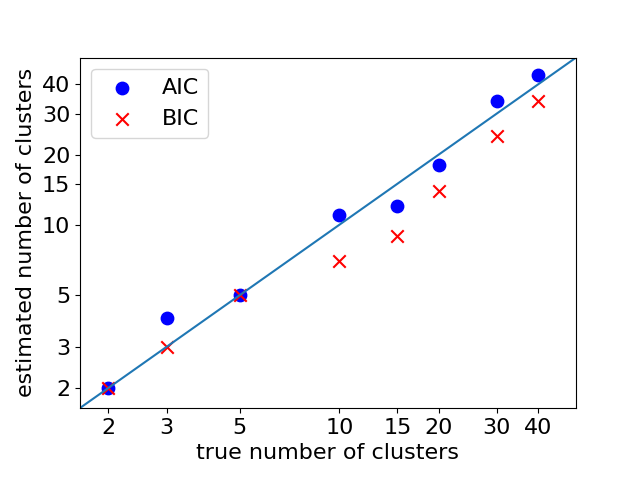}
\caption{}
\label{diveResSynthC}
\end{subfigure}
\begin{subfigure}[b]{0.32\textwidth}
\includegraphics[width=1\textwidth]{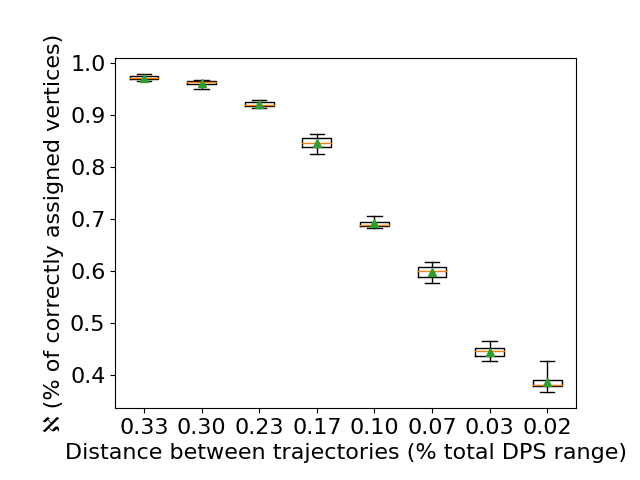}
\caption{}
\label{diveResSynthD}
\end{subfigure}
\begin{subfigure}[b]{0.32\textwidth}
\includegraphics[width=1\textwidth]{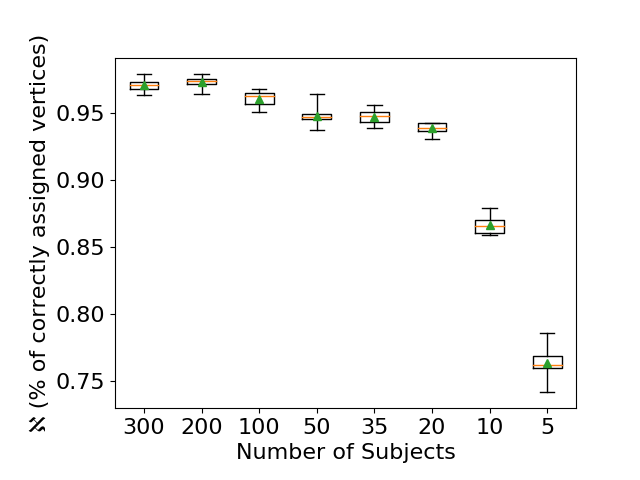}
\caption{}
\label{diveResSynthE}
\end{subfigure}
\caption[DIVE Simulation Results]{(a-b) Results for the basic simulation, where trajectories are relatively well separated. (a) Reconstructed temporal trajectories (blue) plotted against the true trajectories (red). The x-axis shows the disease progression score (DPS), while the y-axis shows the biomarker values of the vertices. (b) Estimated subject-specific DPS scores compared to the true scores. (C-E) Simulation results for the three scenarios: (c) increasing number of clusters, (d) trajectories becoming similar and (e) decreasing number of subjects. On the x-axis we show the variable that was changing within the scenario (e.g. number of clusters), while on the y-axis we show the agreement measure $\aleph$, representing the percentage of vertices that were assigned to the correct cluster.}
\label{diveResSynth}
\end{figure}

In the basic simulation, we obtained a clustering agreement $\aleph$ of 0.97, which suggests that almost all vertices were assigned to the correct cluster. Fig. \ref{diveResSynthA} shows the original trajectories and the recovered trajectories using our model, plotted against the disease progression score on the x-axis and the vertex value on the y-axis. In Fig. \ref{diveResSynthB} we plotted the recovered DPS of each subject along with the true DPS. The results for the three scenarios are shown in Figs. \ref{diveResSynthC}-\ref{diveResSynthE}. In Fig. \ref{diveResSynthC}, we show for Scenario 1 the estimated number of clusters against the true number of clusters using both AIC and BIC criteria. In Figs. \ref{diveResSynthD}-\ref{diveResSynthE} we show the distributions for $\aleph$ in Scenarios 2 and 3 as the problem becomes harder in each successive step.

The results show that, in a simple experiment where the trajectories are well separated, DIVE can very accurately estimate which clusters generated each vertex. Moreover, the recovered trajectories and DPS scores are close to the true values. The results of Scenario 1 also suggest that both AIC and BIC are effective at estimating the correct number of known clusters, with AIC having slightly better performance than BIC for larger numbers of clusters. On the other hand, the results of the stress test scenarios 2 and 3 show that performance measure $\aleph$ drops when the trajectories become very similar with each other or when the number of subjects decreases. This happens because small differences in trajectories are hard to detect in the presence of measurement noise, while a small number of subjects doesn't provide enough data to accurately estimate the parameters. Similar decreases in performance for scenarios 2 and 3 are observed also for other measures, such as the error in recovered trajectories or DPS scores (Supplementary Fig \ref{diveTrajError}).

\subsection{Results with ADNI and DRC Datasets}
\label{sec:diveResAdniDrc}

\subsubsection{Initial Hypotheses}

Using ADNI and DRC datasets, we aim to recover the spatial distribution of cortical atrophy and amyloid pathology, as well as the rate and timing of these pathological processes. In particular, we hypothesise that these spatial patterns of pathology and their evolution will be: 
\begin{itemize}
 \item similar on two independent typical AD datasets: ADNI and DRC
 \item different on distinct diseases: tAD vs PCA 
 \item different in distinct modalities: cortical thickness from MRI vs amyloid load from AV45 PET.
\end{itemize}

\subsubsection{Results}

The optimal number of clusters, as estimated with AIC, was three for the ADNI MRI dataset, three for the DRC tAD dataset, five for the DRC PCA dataset and eighteen for the ADNI PET dataset. Fig. \ref{diveClustAdniMri} (left) shows the results from the ADNI MRI dataset, where in the left image we coloured the vertices on the cortical surface according to the cluster they most likely belong to. We assigned a colour for each cluster (both the brain figures on the left and the trajectory figures on the right) according to the extent of pathology of its corresponding trajectory at a DPS score of 1. The cluster colours range from red (severe pathology) to blue (moderate pathology). In Fig. \ref{diveClustAdniMri} (right), we show the resulting cluster trajectories with samples from the posterior distribution of each $\theta_k$. Similar results are shown for the other three datasets: the DRC tAD dataset (Fig. \ref{diveClustDrcAd}), DRC PCA dataset (Fig. \ref{diveClustDrcPca}) and the ADNI PET dataset (Fig. \ref{diveClustAdniPet}).

We notice that in tAD subjects using the ADNI datasets (Fig. \ref{diveClustAdniMri}), there is more severe cortical thinning mainly in the inferior temporal lobe (red cluster), with disperse atrophy also in parietal and frontal regions (green cluster), with relative sparing of the inferior frontal and occipital lobes. In tAD subjects from the DRC dataset (Fig. \ref{diveClustDrcAd}), we see a relatively similar pattern, however with more pronounced atrophy in the supramarginal cortex (red cluster) compared to ADNI. This could be due to the younger ages of controls and tAD subjects in the DRC dataset as compared to ADNI. The spatial distribution of cortical thinning found with DIVE resembles results from previous longitudinal studies such as \cite{dickerson2008cortical,singh2006spatial}. However, in contrast to these approaches, our model gives insight into the timing and rate of atrophy and is also able to stage subjects across the disease time course. We also find that the cluster trajectories in the DRC tAD dataset have similar dynamics to the ADNI MRI dataset, although they show a clearer separation between each other.

In the PCA subjects (Fig. \ref{diveClustDrcPca}), we find that atrophy is mainly focused on the posterior part of the brain, with limited spread in the motor cortex, anterior temporal and frontal areas. This posterior-focused pattern of atrophy is different from the one found in the tAD datasets, and agrees with previous findings in the literature \cite{crutch2012posterior,lehmann2011cortical}.  However, as opposed to the results from \cite{lehmann2011cortical} which showed posterior regions uniformly affected, we notice that there are two clusters within the posterior region with different pathology dynamics, with the superior parietal and supramarginal areas affected more than the remaining posterior regions. This might be attributable to DIVE's ability to model subjects' disease onset and progression speed, along with non-linear cortical thinning dynamics, other differences due to the different subjects analysed, and the merging of left and right hemispheres could also give such differences.

In ADNI PET (Fig. \ref{diveClustAdniPet}) we see that the regions with the highest amyloid uptake are more spatially continuous, comprising the precuneus and anterior frontal areas. On the other hand, the anterior-superior temporal gyrus shows the least uptake of amyloid. This result closely matches the result by \cite{bilgel2016multivariate}, which used a completely different dataset and modelling technique. These results using AV45 PET are also noticeably different from results using cortical thickness (e.g. Fig. \ref{diveClustAdniMri}), which have more high-frequency patterns and only give 3-5 optimal clusters instead of 20. The “layers of clusters” starting from the precuneus and frontal lobes, which range from severe to less severe atrophy, suggest a continuum of variation in vertex trajectories in the case of the PET dataset (Fig \ref{diveClustAdniPet}, right). These trajectories all start with a low amyloid SUVR, between 0 and 0.25, but in late stages the trajectories for some clusters such as cluster 0 can reach an SUVR of 1.5. The reason for seeing this continuum might be because the PET images have a much lower resolution than MR images and were smoothed by ADNI during the pre-processing steps. 

\newcommand{\scalingFactor}{1.2}
\newcommand{\scalingFactorSubfigBrain}{0.35}

\newcommand{\gradLimLeft}{-1.6}
\newcommand{\gradLimRight}{1.6}

\newcommand{\scalingFactorBrains}{0.75}
\newcommand{\scalingFactorTraj}{1.05}

\newcommand{\typeOfBrainColoring}{atrophyExtent}

\definecolor{barGreen}{rgb}{0.4,1,0.4}

\begin{figure}
  \centering
  \vspace{-1em}

  \begin{subfigure}[b]{0.45\textwidth}
   \centering
  \begin{tikzpicture}[scale=\scalingFactor]
    \shade[left color=red,right color=yellow] (\gradLimLeft,2.5) rectangle (-0.8,2.75);
    \shade[left color=yellow,right color=barGreen] (-0.8,2.5) rectangle (0,2.75);
    \shade[left color=barGreen,right color=cyan] (0,2.5) rectangle (0.8,2.75);	
    \shade[left color=cyan,right color=blue] (0.8,2.5) rectangle (\gradLimRight,2.75);   
    \node[inner sep=0] (corr_text) at (\gradLimLeft,3) {severe pathology};
    \node[inner sep=0] (corr_text) at (\gradLimRight,3) {moderate pathology};
  \end{tikzpicture}
  \vspace{1em}
  \end{subfigure}
  

  \begin{subfigure}[b]{\textwidth}
   \centering
  \includegraphics[width=\scalingFactorSubfigBrain \textwidth,trim=0 0 0 20,clip]{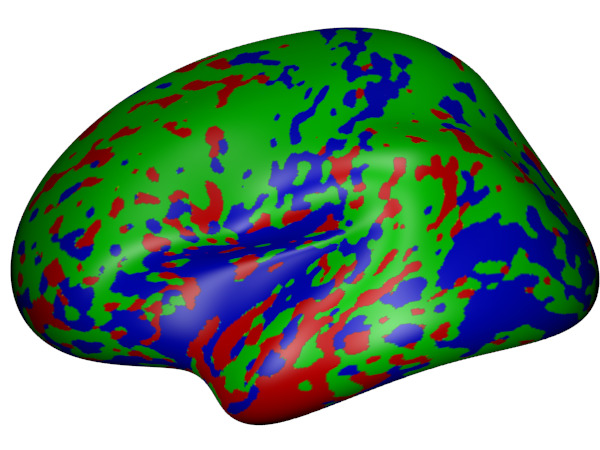} \includegraphics[width=\scalingFactorSubfigBrain \textwidth,trim=0 10 0 30,clip]{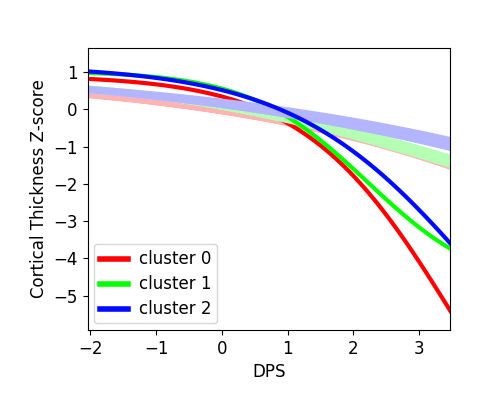}
    \caption{ADNI MRI}
    \label{diveClustAdniMri}
  \end{subfigure}

  \begin{subfigure}[b]{\textwidth}
   \centering
  \includegraphics[width=\scalingFactorSubfigBrain \textwidth,trim=0 0 0 20,clip]{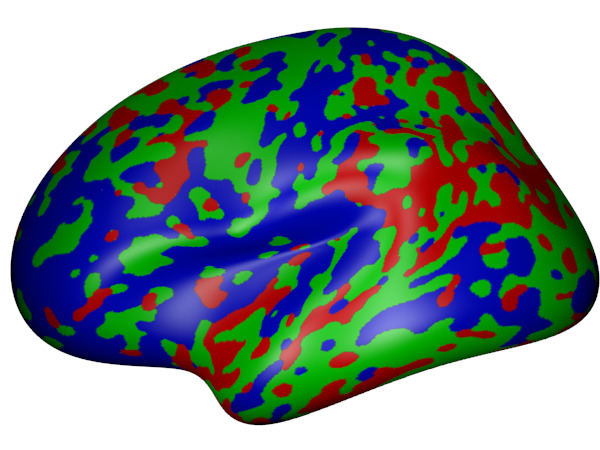} \includegraphics[width=\scalingFactorSubfigBrain \textwidth,trim=0 10 0 30,clip]{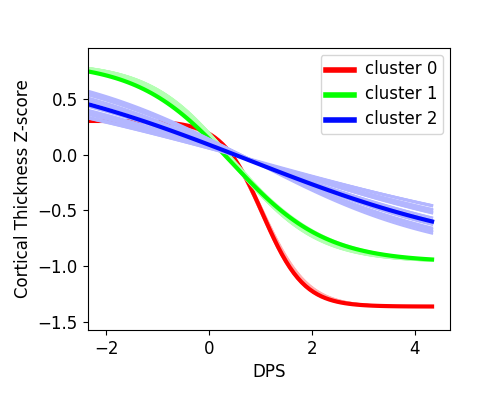}
    \caption{DRC tAD}
    \label{diveClustDrcAd}
  \end{subfigure}
  
  \begin{subfigure}[b]{\textwidth}
   \centering
  \includegraphics[width=\scalingFactorSubfigBrain \textwidth,trim=0 0 0 20,clip]{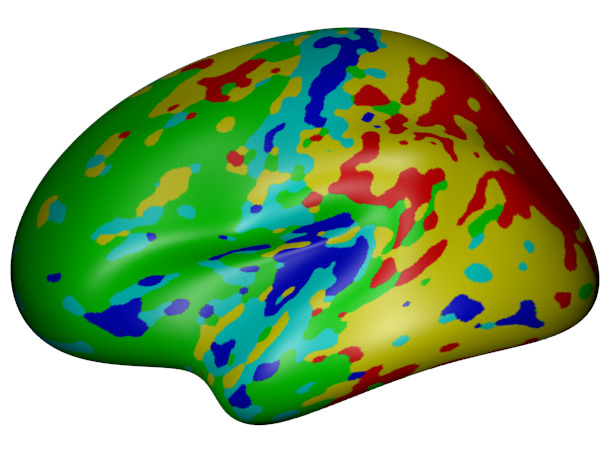} \includegraphics[width=\scalingFactorSubfigBrain \textwidth,trim=0 10 0 30,clip]{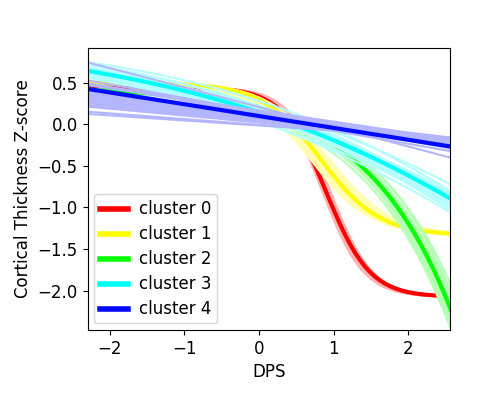}
    \caption{DRC PCA}
    \label{diveClustDrcPca}
  \end{subfigure}
  
  \begin{subfigure}[b]{\textwidth}
   \centering
  \includegraphics[width=\scalingFactorSubfigBrain \textwidth,trim=0 0 0 20,clip]{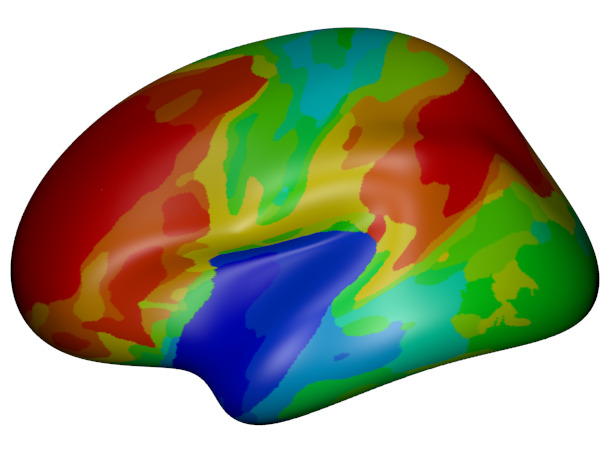} \includegraphics[width=\scalingFactorSubfigBrain \textwidth,trim=0 10 0 30,clip]{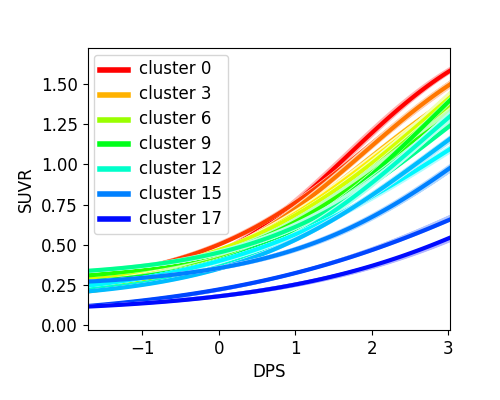}
    \caption{ADNI PET}
    \label{diveClustAdniPet}
  \end{subfigure}
  

  \caption[DIVE Results on ADNI and DRC cohorts]{(left column) DIVE estimated clusters (left column) and corresponding disease progression trajectories (right column) on four datasets: (a) ADNI MRI (b) DRC tAD (c) DRC PCA and (d) ADNI PET. We coloured each cluster according to the extent of pathology (cortical thickness or amyloid uptake) at DPS=1.}
  \label{diveClustTrajAll}
\end{figure}
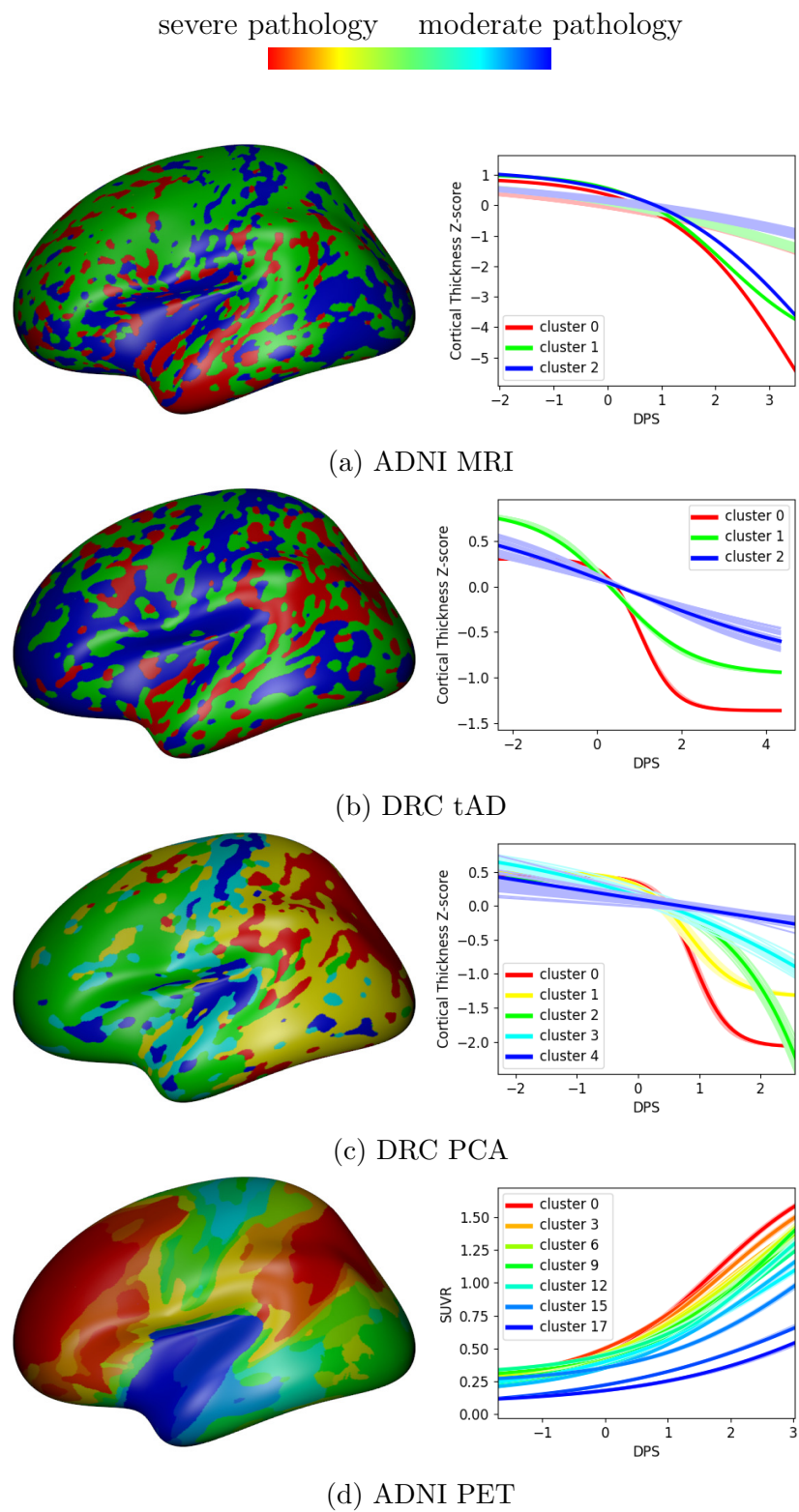

\subsection{Model Evaluation}
\label{sec:diveEval}

\subsubsection{Motivation}
\label{sec:diveEvalMotiv}

We further tested the robustness and validity of the model as follows: 
\begin{itemize}
 \item Robustness in parameter estimation: test whether similar spatial clustering is estimated for different subsets of the data
 \item Clinical validity of DPS scores: test whether the subject disease progression scores, based purely on MRI or PET data, correlate with cognitive tests such as Clinical Dementia Rating Scale - Sum of Boxes (CDRSOB), Alzheimer's Disease Assessment Scale - Cognitive (ADAS-COG), Mini-Mental State Examination (MMSE) and Rey Auditory and Verbal Learning Test (RAVLT).
 \item Comparison with other models: to evaluate the benefit of estimating fine-grained patterns of pathology in DIVE, as well as latent time shifting of subjects, we compared the performance of DIVE with a region-of-interest based method \cite{jedynak2012computational} and a no-staging method that doesn't estimate subject time shifts. See Supplementary Section \ref{sec:diveCompAppendix} for precise specifications.
\end{itemize}

\subsubsection{Evaluation Procedure}

For all scenarios, we ran 10-fold cross-validation (CV) on the ADNI MRI dataset. At each fold we fit the model using 3 clusters, since this was the optimal number of clusters found previously on the entire dataset. The trained model was then used to estimate the DPS of the test subjects. 

For the performance comparison of DIVE with other models, we compute two performance metrics: (1) between-subject correlation of the models' estimated DPS values with cognitive tests; we estimated a unique DPS for every subject and every visit, which we then matched with the corresponding cognitive tests at that subject's visit and (2) prediction root mean squared error (RMSE) between the predicted vertex-wise values and actual measurements, averaged over all subjects and all locations on the brain; to evaluate these predictions, for every subject we use the first n-1 scans for training and the last scan for testing the prediction.

\subsubsection{Evaluation Results}

\newcommand{\outFoldADNICVbrains}{images/vwdpm/crossvalid/adniThavgFWHM0Initk-meansCl3Pr0Ra1_VWDPMMean}
\newcommand{\outFoldADNIPetCVbrains}{figures/validAdniPET/brainAtrophyExtent}

\newcommand{\adniThickCVExpName}{adniThInitk-meansCl3Pr1Ra1_VDPM_MRF}
\newcommand{\adniThickCVFolder}{\voxDpmFolder/resfiles/cogCorr/\adniThickCVExpName}
\newcommand{\trimModelValidTop}{0}

\begin{figure}
 \centering
\foreach \f in {0,1,2,3,4,5,6,7,8,9}
{
\begin{subfigure}[b]{0.185\textwidth}
\centering
  \FPeval{\faddOne}{clip(\f+1)}
  f = \faddOne \\
  \includegraphics[width=\textwidth, trim=0 0 0 \trimModelValidTop]{\adniThickCVFolder/f\f/atrophyExtent_cogCorr_\adniThickCVExpName_f\f}
\end{subfigure}
}
\vspace{1em}

\begin{subfigure}[b]{0.223\textwidth}
\centering
f=1
\includegraphics[width=\linewidth, trim=24 0 35 20,clip]{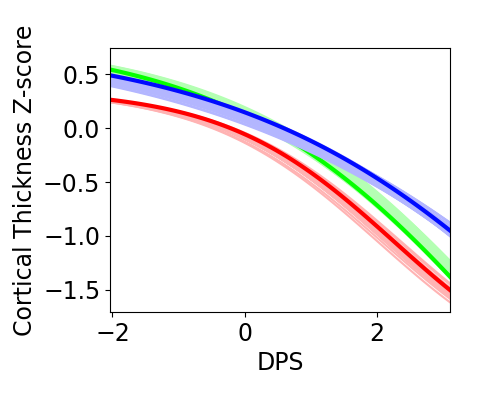}
\end{subfigure}
\hfill
\begin{subfigure}[b]{0.185\textwidth}
\centering
f=2
\includegraphics[width=\linewidth, trim=75 0 35 20,clip]{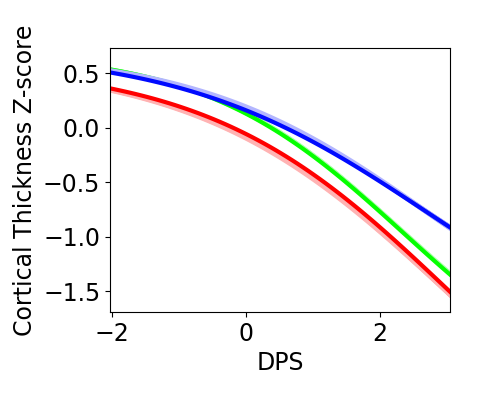}
\end{subfigure}
\hfill
\begin{subfigure}[b]{0.185\textwidth}
\centering
f=3
\includegraphics[width=\linewidth, trim=75 0 35 20,clip]{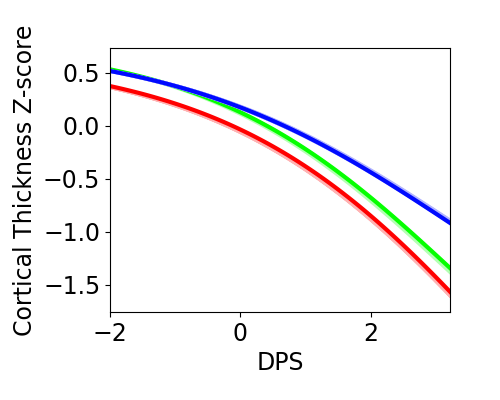}
\end{subfigure}
\hfill
\begin{subfigure}[b]{0.185\textwidth}
\centering
f=4
\includegraphics[width=\linewidth, trim=75 0 35 20,clip]{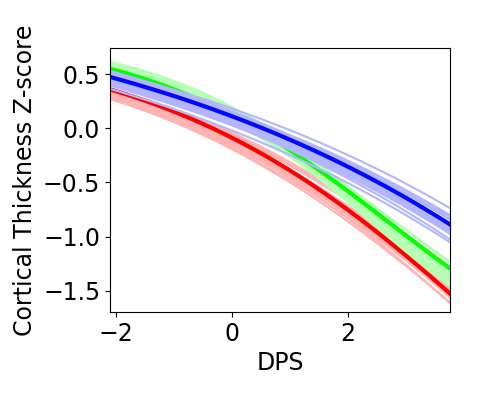}
\end{subfigure}
\hfill
\begin{subfigure}[b]{0.185\textwidth}
\centering
f=5
\includegraphics[width=\linewidth, trim=75 0 35 20,clip]{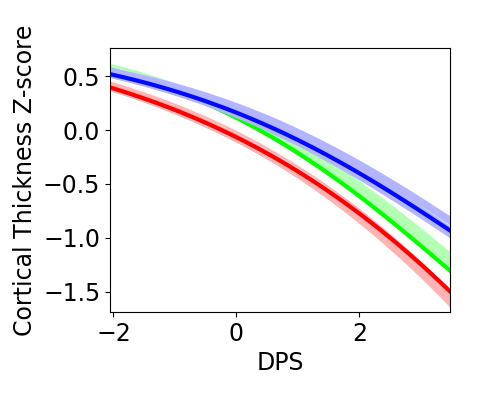}
\end{subfigure}
\hfill

\begin{subfigure}[b]{0.223\textwidth}
\centering
f=6
\includegraphics[width=\linewidth, trim=24 0 35 20,clip]{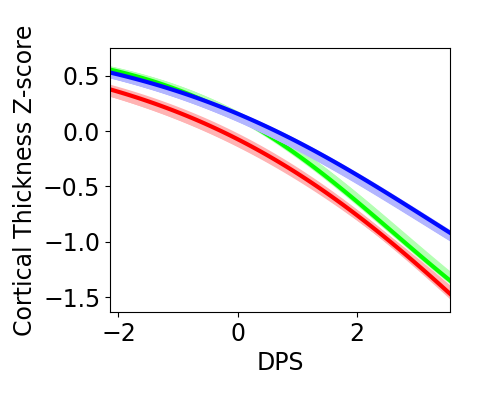}
\end{subfigure}
\hfill
\begin{subfigure}[b]{0.185\textwidth}
\centering
f=7
\includegraphics[width=\linewidth, trim=75 0 35 20,clip]{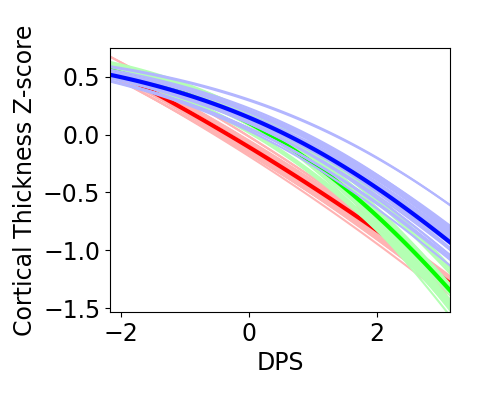}
\end{subfigure}
\hfill
\begin{subfigure}[b]{0.185\textwidth}
\centering
f=8
\includegraphics[width=\linewidth, trim=75 0 35 20,clip]{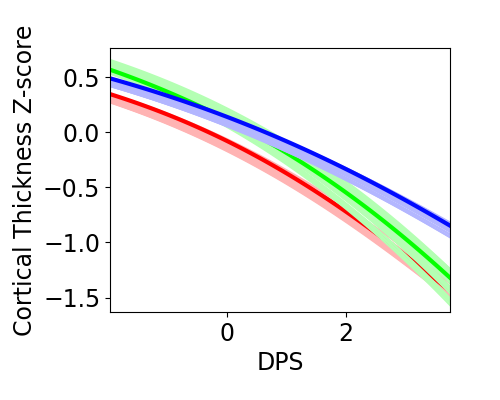}
\end{subfigure}
\hfill
\begin{subfigure}[b]{0.185\textwidth}
\centering
f=9
\includegraphics[width=\linewidth, trim=75 0 35 20,clip]{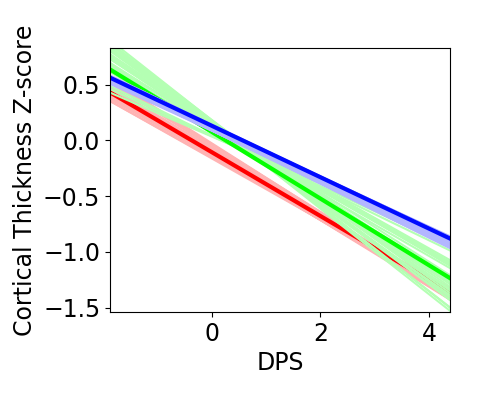}
\end{subfigure}
\hfill
\begin{subfigure}[b]{0.185\textwidth}
\centering
f=10
\includegraphics[width=\linewidth, trim=75 0 35 20,clip]{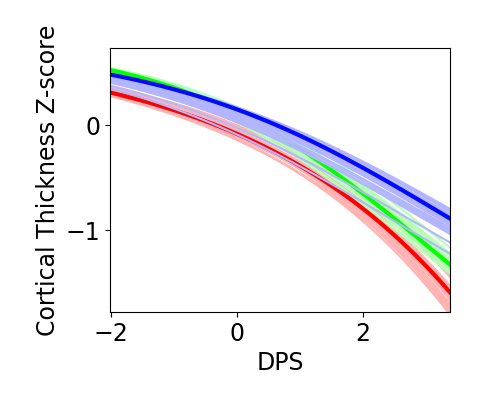}
\end{subfigure}
\hfill

%
%
\caption[DIVE estimated clusters and trajectories over the 10 cross-validation folds]{(top) Clusters estimated from 10-fold cross-validation training sets on the ADNI MRI dataset. (bottom) Estimated trajectories for each fold. }
\label{fig:diveClustTrajCV}
\end{figure}

Fig. \ref{fig:diveClustTrajCV} shows the brain clusters and corresponding trajectories, estimated for all the cross-validation folds after fitting the model on the training data. The clusters have been coloured using a similar colour scheme as in Fig. \ref{diveClustTrajAll}. In Fig \ref{fig:diveCogCorr} we show scatter plots of the DPS scores with clinical measures such as CDRSOB, ADAS-COG, MMSE and RAVLT.

\newcommand{\figFont}{\normalfont}
\newcommand{\pValFont}{\footnotesize}

\newcommand{\cogCorrScatterFold}{\voxDpmFolder/resfiles/adniThInitk-meansCl3Pr1Ra1_VDPM_MRF}

\begin{figure}[h]
  \begin{subfigure}{0.245\textwidth}
    \centering
    \hspace{1.5em}\figFont{CDRSOB}\\ 
    \hspace{1.5em}\pValFont{($\rho = 0.37$, $p < 1e-65$)}
    \includegraphics[width=1.1\textwidth]{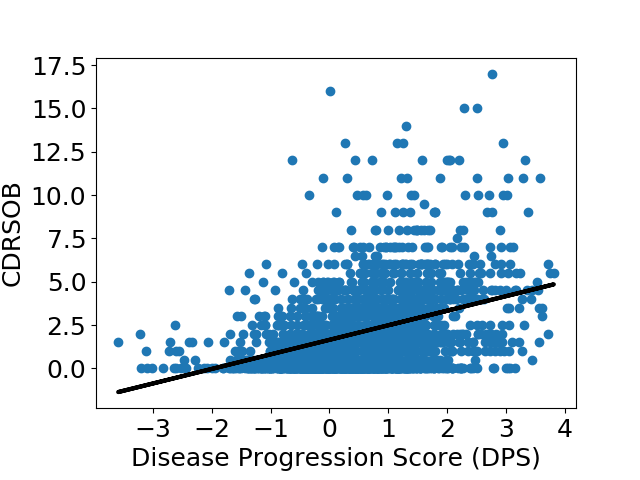}
  \end{subfigure}
  \begin{subfigure}{0.245\textwidth}
    \centering
    \hspace{1.5em}\figFont{ADAS-COG}\\ 
    \hspace{1.5em}\pValFont{($\rho = 0.37$, $p < 1e-64$)}
    \includegraphics[width=1.1\textwidth]{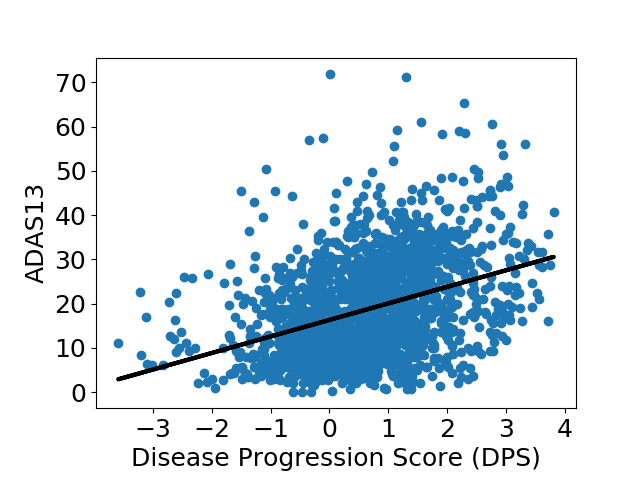}
  \end{subfigure}
    \begin{subfigure}{0.245\textwidth}
    \centering
    \hspace{1.4em}\figFont{MMSE}\\ 
    \hspace{1.4em}\pValFont{($\rho = -0.36$, $p < 1e-63$)}
    \includegraphics[width=1.1\textwidth]{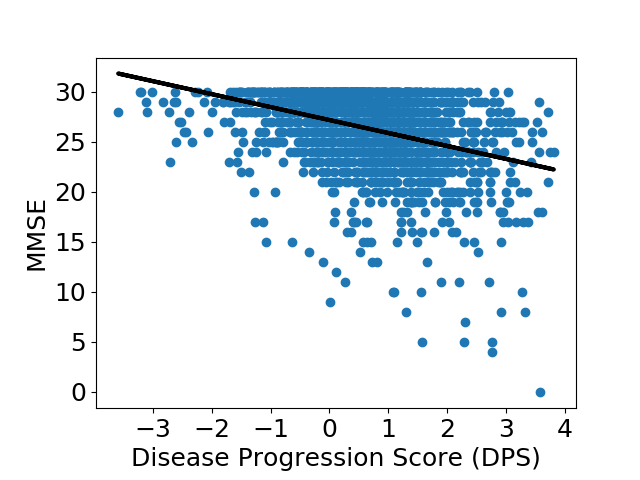}
  \end{subfigure}
    \begin{subfigure}{0.245\textwidth}
    \centering
    \hspace{1.4em}\figFont{RAVLT}\\ 
    \hspace{1.4em}\pValFont{($\rho = -0.32$, $p < 1e-49$)}
    \includegraphics[width=1.1\textwidth]{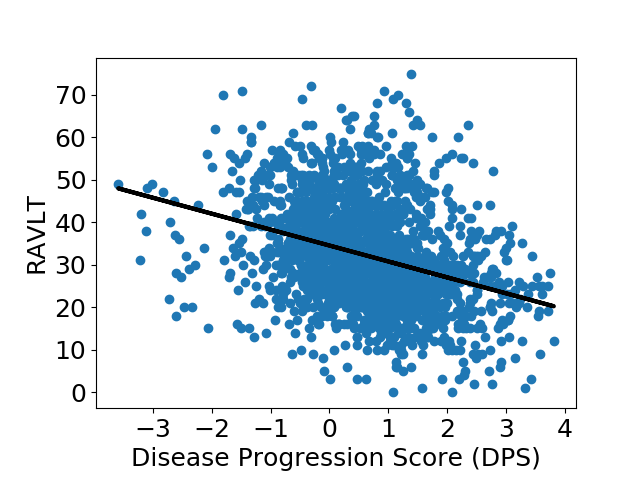}
  \end{subfigure}
  \caption[Scatter plot of DIVE-derived DPS scores vs cognitive tests]{Scatter plots of the DPS scores estimated from the ADNI MRI dataset, plotted against four cognitive tests: CDRSOB, ADAS-COG, MMSE and RAVLT. For each cognitive test we also report the Pearson correlation coefficient and p-value. The disease progression scores, computed only based on MRI cortical thickness data, correlate with these cognitive measures, suggesting that the DPS scores are clinically meaningful. }
  \label{fig:diveCogCorr}
\end{figure}

\begin{table}[H]
\centering
\begin{footnotesize}
 \begin{tabular}{c | c c c c | c}
  Model & CDRSOB ($\rho$) & ADAS13 ($\rho$) & MMSE ($\rho$) & RAVLT ($\rho$) & Prediction (RMSE)\\
  \hline 
DIVE & 0.37 $\pm$ 0.09 & 0.37 $\pm$ 0.10 & 0.36 $\pm$ 0.11 & 0.32 $\pm$ 0.12 & 1.021 $\pm$ 0.008 \\
ROI-based model & 0.36 $\pm$ 0.10 & 0.35 $\pm$ 0.11 & 0.34 $\pm$ 0.13 & 0.30 $\pm$ 0.13 & 1.019 $\pm$ 0.010 \\
No-staging model & *0.09 $\pm$ 0.06 & *0.03 $\pm$ 0.09 & *0.05 $\pm$ 0.06 & *0.02 $\pm$ 0.06 & *1.062 $\pm$ 0.024 \\

 \end{tabular}
 \end{footnotesize}
 \caption[Performance evaluation of DIVE and two simplified models on the ADNI MRI dataset]{Performance evaluation of DIVE and two simplified models on the ADNI MRI dataset using 10-fold cross-validation. In the middle four columns, we show between-subject correlations between the DPS scores and several cognitive tests: CDRSOB, ADAS-Cog13, MMSE and RAVLT. The last column shows the prediction error (RMSE) of cortical thickness values from follow-up scans. (*) Statistically significant differences between the model and DIVE, Bonferroni corrected for multiple comparisons.}
 \label{tab:divePerfEval}
\end{table}

The results in Fig. \ref{fig:diveClustTrajCV} demonstrate that DIVE is robust in cross-validation, as the estimated clusters and trajectory parameters are all similar across folds. The average Dice score overlap across the 10-folds range were 0.77, 0.76 and 0.90 for clusters 0, 1 and 2 respectively. The DIVE-derived DPS scores, which were estimated purely based on MRI data, are also clinically relevant as they correlate with cognitive tests (Fig. \ref{fig:diveCogCorr}). 

The performance of DIVE in terms of subject staging and biomarker prediction also compares favourably with simpler no-staging and ROI-based models (Table \ref{tab:divePerfEval}). Results show that DIVE has comparable performance to the ROI-based model, both in terms of subject staging and cortical thickness prediction. The fact that DIVE has similar performance to a simpler model which has less parameters is evidence that the estimated patterns are meaningful. Moreover, DIVE offers qualitative insight into the fine-grained spatial patterns of pathology and their temporal progression. Furthermore, the No-staging model performs significantly worse than DIVE, both in terms of subject staging and for biomarker prediction. This suggests that, when modelling progression of AD, it is important to account for the fact that patients are at different stages along the disease time-course.

\section{Discussion}
\label{sec:diveDis}

\subsection{Summary and Key Findings}

We presented DIVE, a spatiotemporal model of disease progression that clusters vertex- or voxel-wise measures of pathology in the brain based on similar temporal dynamics. The model highlights, for the first time, groups of cortical vertices that exhibit a similar temporal trajectory over the population. The model also estimates the temporal shift and progression speed for every subject. We applied the model on cortical thickness vertex-wise data from three MRI datasets (ADNI, DRC tAD and DRC PCA), as well as an amyloid PET dataset (ADNI). Our model found qualitatively similar patterns of cortical thinning in tAD subjects using the two independent datasets (ADNI and DRC). Moreover, it also found different patterns of pathology dynamics on two distinct diseases (tAD and PCA) and on different types of data (PET and MRI-derived cortical thickness). Finally, DIVE also provides a new way to parcellate the brain that is specific to the temporal trajectory of a particular disease, and enables staging of individuals at risk of disease, which can potentially help stratification in clinical trials.

The characteristics of the subjects' data used for training can affect the DIVE output. For instance, in cortical thinning analyses we standardised the data with respect to controls, which might have already shown cortical thinning due to early pathology. This can be mitigated through enrichment of the control population to amyloid-negative individuals. DIVE also relies on subjects spanning the entire disease progression, so inclusion of subjects in middle stages is recommended for a robust estimation of trajectories and spatial patterns. To reliably estimate the subject-specific time shift and progression speed, multiple follow-up scans are required. We mitigated this by using only subjects with at least three scans, and further placing informative priors on these parameters. 

The DIVE-estimated spatial patterns are patchier in MRI compared to PET scans, which had lower resolution and were smoothed a-priori. However, we believe MRI images should not instead smoothed a-priori, as the spatial correlation mechanism within DIVE enables it to automatically remove high-frequency patterns from MRI that are not meaningful.  Moreover, such a-priori smoothing could potentially loose dispersed patterns of pathology that arise due to underlying disruption of brain networks.

\subsection{Limitations and future work}

DIVE has some limitations that can be addressed. First, we assumed that cluster trajectories follow sigmoidal shapes, which is not the case for many types of biomarkers in ADNI which do not plateau in later stages. The assumption of sigmoidal trajectories can be avoided using non-parametric curves such as Gaussian Processes \cite{lorenzi2017disease}, which would be straightforward to incorporate into the DIVE framework. To get a reliable estimate of the subject-specific parameters, we only tested DIVE on balanced datasets, where subjects had at least three scans. However, DIVE can also be applied to less balanced datasets, by setting stronger priors on these parameters or even fixing the progression speed for every subject to 1. Another limitation of the model is that it assumes all subjects follow the same disease progression pattern, which might not be the case in heterogeneous datasets such as ADNI or DRC. This can be a concern, as there might be a pattern of pathology that occurs in a small set of subjects. However, DIVE can be extended to account for heterogeneity in the datasets by modelling subject-specific trajectories using random effects, or different progression dynamics for distinct subgroups, using unsupervised learning methods like the SuStaIn model by \cite{young2018uncovering}. While SuStaIn, just like DIVE, estimates clusters and trajectories within the dataset, the clusters in SuStaIn are made of subjects with similar disease progression, while the clusters in DIVE are made of vertices with similar progression. Future work could combine clustering along both subjects and vertices simultaneously to estimate disease subtypes with distinct spatiotemporal dynamics at the vertexwise level.

There are several potential future applications of DIVE. One of the advantages of DIVE is that it can be used to study the link between disconnected patterns of brain pathology and connectomes extracted from diffusion tractography or functional MRI (fMRI). Such an analysis would enable further understanding of the exact underlying mechanisms by which the brain is affected by the disease. Our model, which can estimate fine-grained spatial patterns of pathology, is more suitable than standard ROI-based methods for studying the link between pathology and these structural or functional connectomes, because white matter or functional connections have a fine-grained and spatially-varying distribution of endpoints on the cortex.

Apart from studying the link with brain connectomes, there are other potential applications for DIVE. While we only applied it to vertexwise data, the model can also be applied to study voxelwise data. Moreover, DIVE can be applied to other modalities or types of data, including FDG PET, tau PET, DTI or Jacobian compression maps from MRI. Moreover, the model can also be extended to cluster points on the brain surface according to a more complex disease signature, that can be made of two or more biomarkers. For example, using our cortical thickness and amyloid PET datasets from ADNI, we could have clustered points on the brain based on both modalities simultaneously. Such complex disease signatures can offer important insights into the relationships between different modalities and underlying disease mechanisms.

DIVE is a spatiotemporal model that can be used for accurately predicting and staging patients across the progression timeline of neurodegenerative diseases. The spatial patterns of pathology can also be used to test mechanistic hypotheses which consider AD as a network vulnerability disorder. All these avenues can help towards disease understanding, patient prognosis, as well as clinical-trials for assessing efficacy of a putative treatment for slowing down cognitive decline.

\section{Conclusion}
\label{sec:diveConclusion}

In this chapter I developed DIVE, a spatiotemporal model of disease progression that estimates fine-grained spatial patterns of brain pathology, while simultaneously placing subjects optimally on a disease time axis. I applied it to two typical AD MRI datasets (ADNI and DRC), one dataset of PCA patients, and one typical AD PET dataset. I also tested the robustness of the method in simulations, under cross-validation, and I've also compared its performance to simpler feature-based models.

In the next chapter, I will present another model, DKT, that can transfer information across different types of dementias in order to estimate the progression of rare dementias from limited, unimodal datasets.

\chapter{Disease Knowledge Transfer across Neurodegenerative Diseases}
\label{chapter:dkt}

\newcommand{\expFld}{./jointModellingDisease/resfiles/tad-drcTiny_JMD}
\newcommand{\jmdFld}{./jointModellingDisease}

\section{Contributions}

In this chapter I present Disease Knowledge Transfer (DKT), a novel method for transferring biomarker information between related neurodegenerative diseases. I performed the mathematical modelling, implementation of the DKT method, data pre-processing, statistical analysis and model validation. The TADPOLE dataset has been assembled by myself and Neil Oxtoby, with suggestions from the EuroPOND team. The PCA dataset was acquired by the Dementia Research Centre, UK. 

While the original DKT implementation relied on a non-parametric GP disease progression model by Marco Lorenzi \cite{lorenzi2017disease} as a building block, for this thesis I chose a simpler parametric model, due to the complexity of fitting hierarchical, non-parametric, latent-space models.

\section{Publications}
\begin{itemize}
 \item R. V. Marinescu, M. Lorenzi, S. B. Blumberg, P. Planell-Morell, A. L. Young, N. P. Oxtoby,  A. Eshaghi, K. X. X. Yong, S. Crutch, D. C. Alexander, arXiv, 2019.
\end{itemize}

\section{Introduction}
\label{sec:dktInt}

The estimation of accurate biomarker signatures in Alzheimer's disease (AD) and related neurodegenerative diseases is crucial for understanding underlying disease mechanisms, predicting subjects' progressions, and selecting the right subjects in clinical trials. As a result, data-driven disease progression models (chapter \ref{chapter:bckDpm}) were proposed that reconstruct long term biomarker signatures from collections of short term individual measurements. When applied to large datasets of typical AD, disease progression models have shown important benefits in understanding the earliest events in the Alzheimer's disease cascade \cite{iturria2016early, young2014data}, the heterogeneity of AD \cite{young2018uncovering}, helped discover novel genes involved in AD \cite{scelsi2018genetic} and they showed improved predictions over standard approaches \cite{oxtoby2018}. However, by necessity these models require large datasets -- in addition they must be both multimodal and longitudinal. Such data is not available in rare neurodegenerative diseases. In particular, most datasets for rare neurodegenerative diseases come from local clinical centres, are unimodal (e.g. MRI only) and limited both cross-sectionally and longitudinally -- this makes the application of disease progression models extremely difficult.  Moreover, such a model estimated from common diseases such as typical AD may not generalise to specific variants. For example, in Posterior Cortical Atrophy -- a neurodegenerative syndrome causing visual disruption -- posterior regions such as the occipital lobe and superior parietal regions are affected early, instead of the hippocampus and temporal regions that are affected early in typical AD. 

The problem of limited data in medical imaging has so far been addressed through transfer learning methods. Such techniques have been successfully used to improve the accuracy of AD diagnosis \cite{hon2017towards, cheng2017multi} or prediction of MCI conversion \cite{cheng2015domain}, but have two key limitations. First, they use deep learning or other machine learning methods, which are not interpretable and don't allow us to understand underlying disease mechanisms that are either specific to rare diseases, or shared across related diseases. Secondly, these models cannot be used to forecast the future evolution of subjects at risk of dementia, which is important for selecting the right subjects in clinical trials. 

We propose Disease Knowledge Transfer (DKT), a generative joint model that estimates continuous multimodal biomarker progressions for multiple neurodegenerative diseases simultaneously -- including rare neurodegenerative diseases -- and which inherently performs transfer learning between the modelled phenotypes. This is achieved by exploiting biomarker relationships that are shared across diseases, whilst accounting for differences in the spatial distribution of brain pathology. DKT is interpretable, which allows us to understand underlying disease mechanisms, and can also predict the future evolution of subjects at risk of diseases. We apply DKT on Alzheimer's variants and demonstrate its ability to predict non-MRI trajectories for patients with Posterior Cortical Atrophy, in lack of such data. This is done by fitting DKT to two datasets simultaneously: (1) the TADPOLE Challenge \cite{marinescu2018tadpole} dataset containing subjects from the Alzheimer's Disease Neuroimaging Initiative (ADNI) with MRI, FDG-PET, DTI, AV45 and AV1451 scans and (2) MRI scans from patients with Posterior Cortical Atrophy from the Dementia Research Centre (DRC), UK. We first show that the estimated non-MRI trajectories for PCA subjects are plausible as they agree with previous literature findings. We finally validate DKT on three datasets: 1) simulated data with known ground truth, 2) TADPOLE sub-populations with different progressions and 3) 20 DTI scans from controls and PCA patients from the DRC, showing it yields favourable performance compared to standard approaches. Code for DKT is available online: \url{https://github.com/mrazvan22/dkt}.

\begin{figure}[h]
 \centering
 \includegraphics[width=1\textwidth,trim=0 0 0 0,clip]{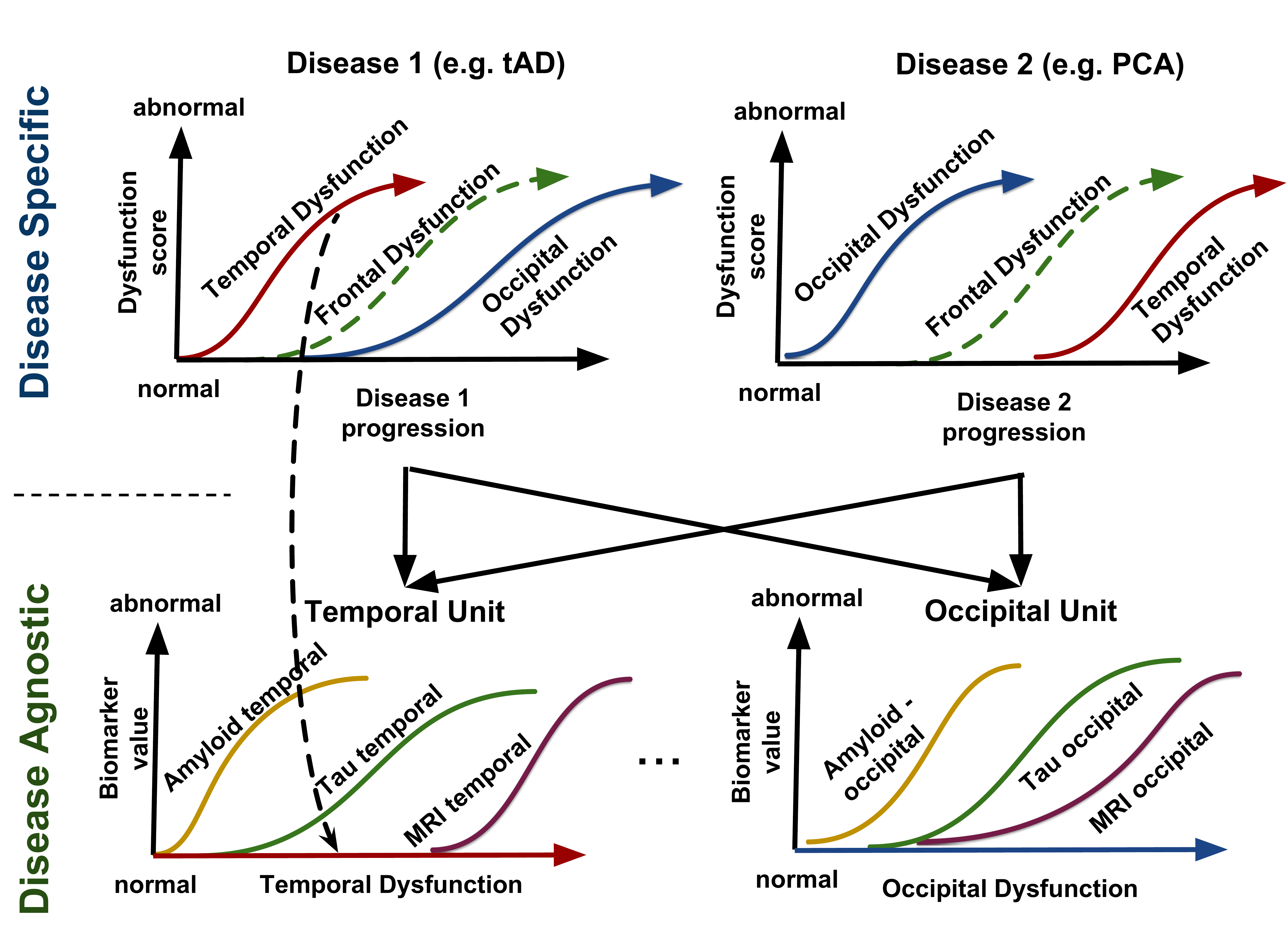}
 \caption[Diagram of the proposed framework for joint modelling of multiple diseases.]{Diagram of the proposed framework for joint modelling of multiple diseases. We assume that each disease can be modelled as the evolution of abstract dysfunctionality scores (Y-axis, top row), each one related to different brain regions. Each region-specific dysfunctionality score then further models (X-axis, bottom row) the progression of several modality-specific biomarkers within that same region. For instance, the temporal dysfunction, modelled as a biomarker in the disease specific model (top row), is the X-axis in the disease agnostic model (temporal unit, bottom row), which aggregates together abnormality from amyloid, tau and MR imaging within the temporal lobe. The biomarker correlations within the bottom units are assumed to be disease agnostic and shared across all diseases modelled. Disease knowledge transfer can then be achieved via the disease-agnostic units.}
 \label{fig:diagram}
\end{figure}

\section{Methods}
\label{sec:dktMet}

\subsection{DKT Framework}
\label{sec:dktMethFramework}
\newcommand{\lp}{\lambda_{d_i}^{\psi(k)}}
\newcommand{\lpuu}{\lambda_{d_i}^{\psi(k),(u)}}
\newcommand{\lpum}{\lambda_{d_i}^{\psi(k),(u-1)}}

Fig. \ref{fig:diagram} shows the overall diagram of our proposed framework for joint modelling of diseases. We assume that the progression of each disease (X-axis, top row) can be modelled as the evolution of abstract dysfunctionality scores, each one related to different brain regions (top row). Each dysfunctionality score is then modelled as the progression of several biomarkers within that same region, but acquired using different noninvasive imaging modalities (bottom row). Each group of biomarkers in the bottom row will be called a \emph{functional unit}, because the correlations between biomarkers are related through common "function" in a disease--agnostic way, since they are related to the same underlying brain region. Biomarker groupings into functional units are defined a-priori. We choose to model the correlations within each unit using the disease progression model (DPM) by Jedynak et al. \cite{jedynak2012computational}, but any other DPM can also be used. The DPM allows us to reconstruct unit-specific \emph{dysfunction} progression manifolds (bottom row, X axis), which can be used for staging subjects. Finally, we use the same DPM to express the progression within each disease (Figure 1, top) in terms of the dysfunction scores estimated within each functional unit. More precisely, the X-axis dysfunction scores from the functional units become Y-axis measurements in the disease specific models.

The model has a concise mathematical formulation. We assume a set of given biomarkers measurements $Y = [y_{ijk} | (i,j,k) \in \Omega]$ for subject $i$ at visit $j$ in biomarker $k$, where $\Omega$ is defined as the set of available biomarker measurements, since subjects can have missing data at various visits. We assume that each subject $i$ at each visit $j$ has an underlying disease stage $s_{ij} = \beta_i + m_{ij}$, where $m_{ij}$ represents the months since baseline visit for subject $i$ at visit $j$ and $\beta_i$ represents the time shift of subject $i$. We further denote by $\theta_k$ the parameters used to represent the trajectory for biomarker $k \in K$ within its functional unit $\psi(k)$, where $\psi$: \{1, ..., K\} $ \rightarrow \Lambda$ is a function that maps each biomarker $k$ to a unique functional unit $l \in \Lambda$, where $\Lambda$ is the set of functional units. Moreover, we denote by $\lambda_d^l$ the parameters for the trajectory of the dysfunction score corresponding to functional unit $l \in \Lambda$ in the space of disease $d$. These definitions allow us to formulate the likelihood for a single measurement $y_{ijk}$ as follows:

\begin{equation}
 p(y_{ijk}|\theta_k, \lp, \beta_i, \epsilon_k) = N(y_{ijk}| g(f(\beta_i + m_{ij}); \lp; \theta_k), \epsilon_k)
\end{equation}
where $g(\ .\ ; \theta_k)$ represents the trajectory of biomarker $k$ within functional unit $\psi(k)$ and $f(\ .\ ; \lambda_{d_i}^{\psi(k)})$ represents the trajectory of the functional unit $\psi(k)$ within the space of disease $d_i$. To be precise, $d_i \in \mathbb{D}$ represents the index of the disease space where subject $i$ belongs, where $\mathbb{D}$ is the set of all diseases modelled. For example, MCI and tAD subjects from ADNI as well as tAD subjects from the DRC cohort can all be assigned $d_i=1$, while PCA subjects from the DRC dataset can be assigned $d_i=2$. Healthy controls can be assigned to either disease space, although a more precise assignment would take molecular biomarkers into account. Variable $\epsilon_k$ denotes the variance of measurements for biomarker $k$. 

We extend the above model to multiple subjects, visits and biomarkers to get the full model likelihood:
\begin{equation}
 p(\boldsymbol{y}|\theta, \lambda, \beta , \epsilon) = \\ \prod_{(i,j,k) \in \Omega} p(y_{ijk}|\theta_k, \lp, \beta_i) 
\end{equation}

where $\boldsymbol{y} = [y_{ijk} | \forall (i,j,k) \in \Omega ]$ is the vector of all biomarker measurements, while $\boldsymbol{\theta} = [\theta_1, ..., \theta_K]$ represents the stacked parameters for the trajectories of biomarkers in functional units, $\boldsymbol{\lambda} = [\lambda_d^{l}|l \in \Lambda, d \in \mathbb{D}]$ are the parameters of the dysfunctionality trajectories within the disease models, $\boldsymbol{\beta} =[\beta_1, ..., \beta_N]$ are the subject-specific time shifts and $\boldsymbol{\epsilon} = [\epsilon_k | k \in K]$  estimates biomarker measurement noise. Although we assumed independence across different subjects, biomarker measurements and visits are linked using the latent time-shift $\beta_i$ for each subject. The parameters of the model that need to be estimated are $[\boldsymbol{\theta}, \boldsymbol{\lambda}, \boldsymbol{\beta}, \boldsymbol{\epsilon}]$. For model simplicity, we did not account for inter-individual variability other than that expressed by the time-shift $\beta_i$, although this could be extended in future work.

\subsection{Modelling Biomarker Trajectories}
\label{sec:dktBiomkTraj}

So far we defined the DKT framework using generic models $g(\ .\ ; \theta_k)$ and $f(\ .\ ; \lp)$ for the biomarker trajectories within the functional units and the disease models. Now we choose to implement the $f$ and $g$ models as parametric sigmoidal curves, to enable a robust optimisation and because these models account for the floor and ceiling effects normally observed in AD biomarkers \cite{sabuncu2011dynamics,caroli2010dynamics}. The sigmoidal model for $f$ is defined as:

\begin{equation}
 f(s;\theta_k) = \frac{a_k}{1+exp(-b_k(s-c_k))} + d_k
\end{equation}

where $s$ is the disease progression score of a subject and $\theta_k = [a_k, b_k, c_k, d_k]$ are parameters controlling the shape of the trajectory for biomarker $k$: $d_k$ and $d_k + a_k$ represent the lower and upper limits of the sigmoidal function, $c_k$ represents the inflection point and $a_k b_k/4$ represents the slope at the inflection point. A similar model is used also for $g$. 

\subsection{Parameter Estimation}

\newcommand{\uu}{^{(u)}}
\newcommand{\um}{^{(u-1)}}

We estimate the model parameters using a two-stage approach. In the first stage, we perform belief propagation within each functional unit and then within each disease model. Each functional unit and disease model is assumed to be an independent disease progression model that we fit by alternatively optimising the fit of biomarker trajectories and subject-specific time-shifts, using the approach described in \cite{jedynak2012computational}. At this stage we assume the existence of a latent variable $\beta_i^{\psi(k)} = f(\beta_i + m_{ij}; \lp)$ representing the dysfunctionality score of subject $i$ within the functional unit $\psi(k)$, which represents a time-shift within that functional unit.

In the second stage we jointly optimise across all functional units and disease models using loopy belief propagation. An overview of the algorithm is given in Figure \ref{fig:dktAlgo}. Given the initial parameters estimated from the first stage (line 1), the algorithm continuously updates the biomarker trajectories within the functional units (lines 4-5), dysfunctionality trajectories (line 9) and subject-specific time shifts (line 13) until convergence. The cost function for all parameters is nearly identical, the main difference being the measurements $(i,j,k)$ over subjects $i$, visits $j$ and biomarkers $k$ that are selected for computing the measurement error. For estimating the trajectory of biomarker $k$ within functional unit $\psi(k)$, measurements are taken from $\Omega_k$ representing all measurements of biomarker $k$ from all subjects and visits. For estimating the dysfunctionality trajectories,  $\Omega_{d,l}$ represents the measurement indices from all subjects with disease $d$ (i.e. $d_i = d$) and all biomarkers $k$ that belong to functional unit $l$ (i.e. $\psi(k) = l$). Finally, $\Omega_i$ (line 13) represents all measurements from subject $i$, for all biomarkers and visits. 

The algorithm we proposed in Figure \ref{fig:dktAlgo} has a complexity of $O(I*S)$, where $S$ is the number of subjects in the dataset and $I$ is the number of iterations until convergence. In practice, convergence is achieved after around 10-15 iterations, which takes around 1h on a Xeon CPU E5-2680 @ 2.5GHz.

\begin{figure}
\begin{algorithm}[H]
 Initialise $\boldsymbol{\theta}^{(0)}$, $\boldsymbol{\lambda}^{(0)}$, $\boldsymbol{\beta}^{(0)}$\\
  \While{$\boldsymbol{\theta}$, $\boldsymbol{\lambda}$, $\boldsymbol{\beta}$ not converged}{
   \tcp*[l]{Estimate biomarker trajectories (disease agnostic)}
    \For{$k=1$ to $K$}{
      ${\theta_k\uu = \argmin_{\theta_k} \sum_{(i,j) \in \Omega_k} \left[y_{ijk} - g\left(f(\beta_i\um + m_{ij}; \lpum) ; \theta_k\right) \right]^2  - log\ p(\theta_k)}$\\
      ${\epsilon_k\uu = \frac{1}{|\Omega_k|} \sum_{(i,j) \in \Omega_k}    \left[y_{ijk} - g\left(f(\beta_i\um + m_{ij}; \lpum) ; \theta_k\uu \right) \right]^2 }$\\
    }
     \tcp*[l]{Estimate dysfunctionality trajectories (disease specific)} 
    \For{$d=1 \in \mathbb{D}$}{
      \For{$l=1 \in \Lambda$}{
        ${\lambda_{d}^{l, (u)} = \argmin_{\lambda_{d}^{l}} \sum_{(i,j,k) \in \Omega_{d,l}} \left[y_{ijk} - g\left(f(\beta_i\um + m_{ij}; \lambda_{d}^{l}) ; \theta_k\uu 
        \right) \right]^2  - log\ p(\lambda_{d}^{l})}$\\
      }
    }
    \tcp*[l]{Estimate subject-specific time shifts} 
    \For{$i=1 \in [1, \dots, S]$}{
      ${\beta_i\uu = \argmin_{\beta_i} \sum_{(j,k) \in \Omega_i} \left[y_{ijk} - g\left(f(\beta_i + m_{ij}; \lpuu) ; \theta_k\uu
      \right) \right]^2  - log\ p(\beta_i)}$\\
    }
}
\end{algorithm}
\caption[The algorithm for estimating the DKT parameters]{The algorithm for estimating the DKT parameters. The algorithm successively updates the biomarker trajectories within the functional units (disease agnostic models), dysfunctionality trajectories (disease specific) and subject-specific time shifts until convergence.}
\label{fig:dktAlgo}
\end{figure}

\subsection{Synthetic Experiment}
\label{sec:dktMetSyn}

We first test DKT on synthetic data, in order to assess the performance when ground truth is known. We generate synthetic data from two diseases as follows:
\begin{itemize}
 \item[] \textbf{Disease model}
 \item We define two functional units $l_0$ and $l_1$ and 6 biomarkers $k_0-k_5$, which we allocate to functional units as follows: $l_0:\{k_0, k_2, k_4\}$, $l_1: \{k_1, k_3, k_5\}$. Within their units, we define the trajectory of each biomarker as a sigmoidal curves with the following $\theta_k$ parameters:
 \begin{itemize}
  \item functional unit $l_0$: $\theta_0 = (1,5,0.2,0)$, $\theta_2 = (1,5,0.55,0)$ and $\theta_4 = (1,5,0.9,0)$ 
  \item functional unit $l_1$: $\theta_1 = (1,10,0.2,0)$, $\theta_3 = (1,10,0.55,0)$ and $\theta_5 = (1,10,0.9,0)$ 
 \end{itemize}
 \item We define two synthetic diseases, "synthetic AD" ($d=0$) and "synthetic PCA" ($d=1$). For each disease $d$, each functional unit $l$ has a distinct dysfunctionality trajectory defined as a sigmoidal curve with parameters $\lambda_d^l$ as follows: 
 \begin{itemize}
  \item "synthetic AD" disease: $\lambda_0^0 = (1, 0.3, -4, 0)$  and $\lambda_0^1 = (1, 0.2, 6, 0)$.
  \item "synthetic PCA" disease: $\lambda_1^0 = (1, 0.3, 6, 0)$ and $\lambda_1^1 = (1, 0.2, -4, 0)$.
 \end{itemize}

 \item[] \textbf{Subject model}
 \item We generated time-shifts $\beta_i$ for 100 subjects (disease $d_0$) and 50 subjects (disease $d_1$) based on a uniform distribution with ranges $(-13, 10)$ years before/after disease onset. 
 \item Within each disease, we generated the subjects' diagnosis (controls/patients) based on an exponential likelihood model with mean -4.5 (controls)/4.5 (patients) years before/after disease onset. 
 \item For each subject and each biomarker, we generated data for four consecutive visits, each visit one year apart, using a noise standard deviation of 0.05.
\end{itemize}

These trajectory and subject parameters were chosen to mimic the TADPOLE and DRC cohorts, described below. Before fitting DKT on the synthetic dataset, we discarded the data from biomarkers $k_0$, $k_1$, $k_4$ and $k_5$ for all subjects within the synthetic PCA cohort, to simulate the lack of multimodal data in these subjects. Remaining biomarkers $k_2$ and $k_3$, for which data was still available in the synthetic PCA cohort, are assumed to be of the same modality (e.g. MRI volume) but to represent measurements from different brain regions (e.g. temporal and occipital).

\subsection{Data Acquisition and Preprocessing}

We trained DKT on ADNI data from the TADPOLE challenge \cite{marinescu2018tadpole}, since it contained a large number of multimodal biomarkers already pre-processed and aggregated into one table. From the TADPOLE dataset we selected a subset of 230 subjects which had at least one FDG PET, AV45, AV1451 or DTI scan. Most subjects also had MRI scans and cognitive tests. In order to model another disease, we further included 76 PCA subjects from the DRC in the training set, along with 67 tAD and 87 age-matched controls, all of which only had MRI scans. 

For both datasets, volumetric measures for each subject have been obtained using the Freesurfer software. For FDG, AV45 and AV1451 PET, we used already extracted SUVR measures from ADNI. For DTI, we used fractional anisotropy (FA) measures from white-matter regions adjacent to each lobe. For every lobe, we averaged the biomarker values for regions of interest within each lobe and regressed out the following covariates: age, gender, total intracranial volume (TIV) and dataset (ADNI vs DRC dataset). Finally, we normalised the biomarker values to lie within the [0,1] range. 

For validating DKT's performance at predicting missing biomarkers in PCA, we used a separate test set of DTI scans from the DRC controls and PCA subjects. As this validation set was acquired at a centre different from ADNI and on different scanners, we matched the FA mean and standard deviation of the DRC controls to be equal to the FA mean and standard deviation of the ADNI controls. No DTI data from PCA subjects was exposed to the algorithm at training time.

\section{Results}
\label{sec:dktRes}

\subsection{Synthetic Results}
\label{sec:dktResSyn}

\begin{figure}
\includegraphics[width=\textwidth]{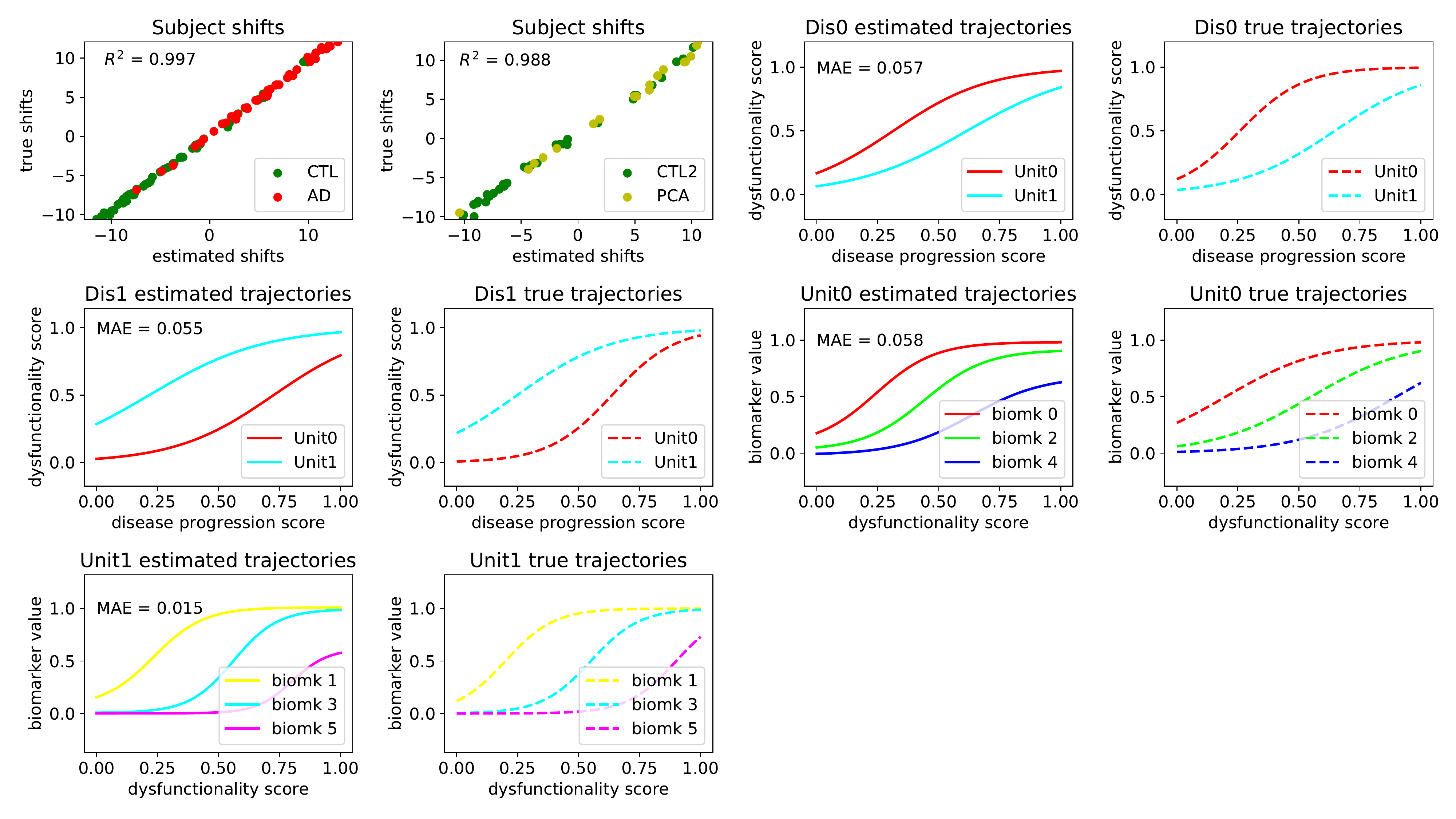}
 \caption[DKT Simulation Results - Comparison between true and DKT-estimated biomarker trajectories and subject time-shifts.]{Comparison between true and DKT-estimated subject time-shifts and biomarker trajectories. (top-left) Scatter plots of the true shifts (y-axis) against estimated shifts (x-axis), for the 'synthetic AD' (left) and 'synthetic PCA' (right) diseases. We also show the DKT-estimated and true trajectories of the functional units within the 'synthetic AD' disease (top-right) and the 'synthetic PCA' disease (middle-left). For these figures, the x-axis measures the normalised disease progression score $s_i$ while the y-axis measures the dysfunctionality scores $f(s_i;\lambda_d^l)$. Finally, we also show the biomarker trajectories within unit 0 (middle-right) and unit 1 (bottom), where the x-axis represents the dysfunctionality scores $f(s_i;\lambda_d^l)$ and the y-axis represents the biomarker value.}
 \label{fig:dktSynthTrajCompTrue}
\end{figure}

\begin{figure}
\includegraphics[width=\textwidth]{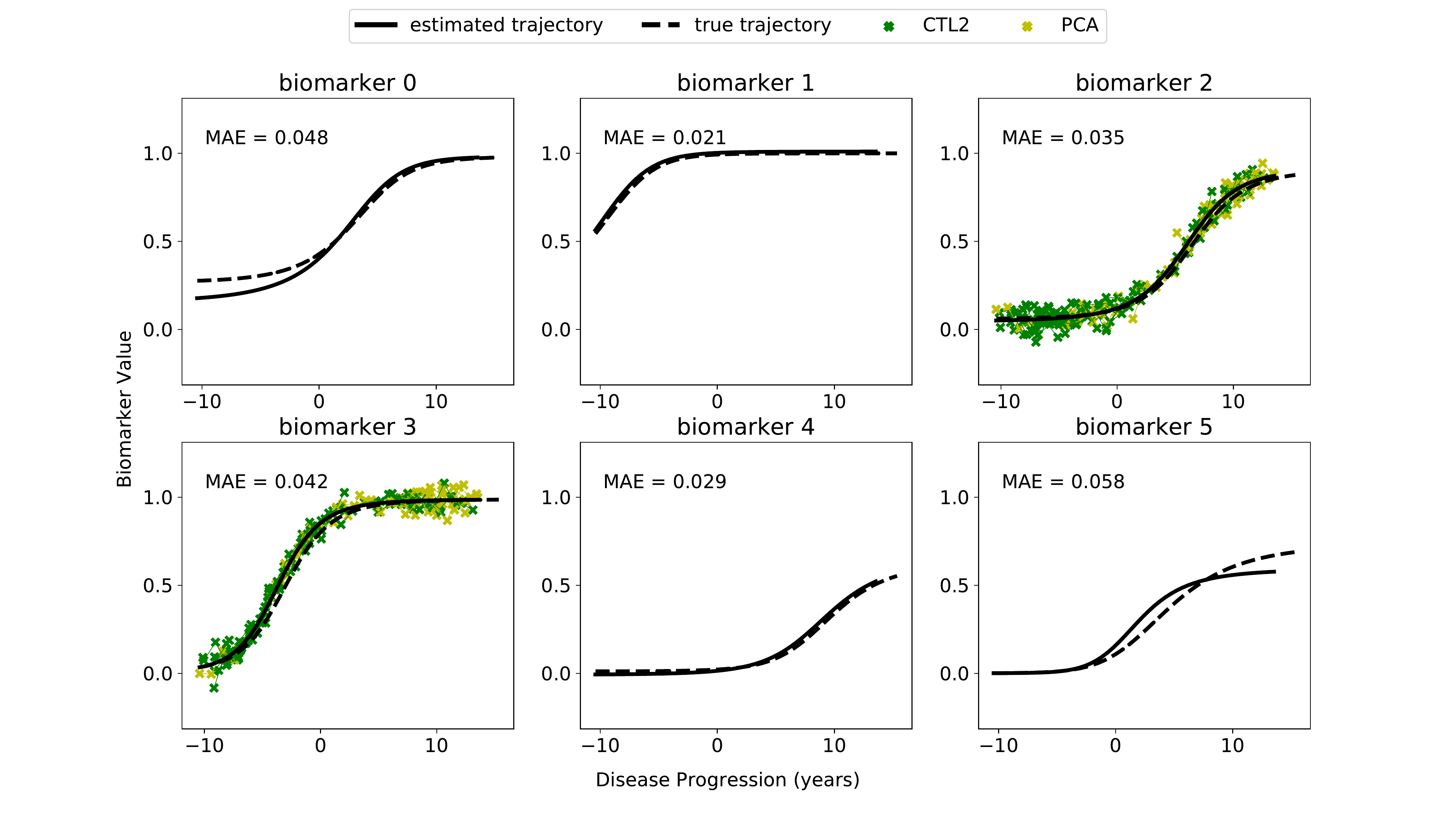}
 \caption[Estimated biomarker trajectories for the "synthetic PCA" disease, plotted alongside true trajectories]{Estimated biomarker trajectories for the "synthetic PCA" disease, plotted alongside true trajectories. Estimation of the trajectories in biomarkers 0,1,4 and 5 has been done without any data from the "synthetic PCA" disease, only based on the disease-agnostic correlations with biomarkers 2 and 3.}
 \label{fig:dktSynthTrajDrcSpace}
\end{figure}

Fig. \ref{fig:dktSynthTrajCompTrue} shows the true and estimated subject shifts and trajectories for each functional unit $l$ and biomarker $k$. In the top-left figures we show scatter plots of the true shifts (y-axis) against estimated shifts (x-axis), for the 'synthetic AD' and 'synthetic PCA' diseases. On the top-right and middle-left figures, we show the trajectories of the functional units within disease $d=0$ (synthetic AD) and $d=1$ (synthetic PCA). In the middle-right and bottom-left figures, we show the biomarker trajectories within units $l_0$ and $l_1$. In Figure \ref{fig:dktSynthTrajDrcSpace}, we show the corresponding trajectories of PCA patients, which as opposed to Fig. \ref{fig:dktSynthTrajCompTrue}, are plotted directly against the time-shifts, as it is normally done in a classical disease progression model. We also show the true trajectories and the data of the synthetic PCA cohort.

The results in Fig. \ref{fig:dktSynthTrajCompTrue} suggest that the DKT-estimated trajectories match closely (mean absolute error, MAE $<$ 0.058) with the true trajectories, for both the unit-trajectories within the disease-specific models and the biomarker trajectories within the disease-agnostic models. Moreover, the subject time-shifts are very close ($R^2$ $>$ 0.98) to the true time-shifts. When plotted directly against the disease space, the estimated PCA trajectories also match the true trajectories, even when there is a complete lack of such data (Fig. \ref{fig:dktSynthTrajDrcSpace}, biomarkers 0,1,4 and 5). There are however small errors in  biomarkers 0 and 5 which are due to measurement noise (confirmed by experiments with smaller noise level -- not shown here). The equivalent trajectories estimated for the synthetic AD cohort also show very good agreement with the true trajectories (Fig. \ref{fig:dktSynthTrajADSpace}).


\subsection{Results on TADPOLE and DRC Datasets}
\label{sec:dktResTadDrc}

\begin{figure}
\centering
\begin{subfigure}{\textwidth}
\centering
\includegraphics[width=1\textwidth, trim=90 20 110 0, clip]{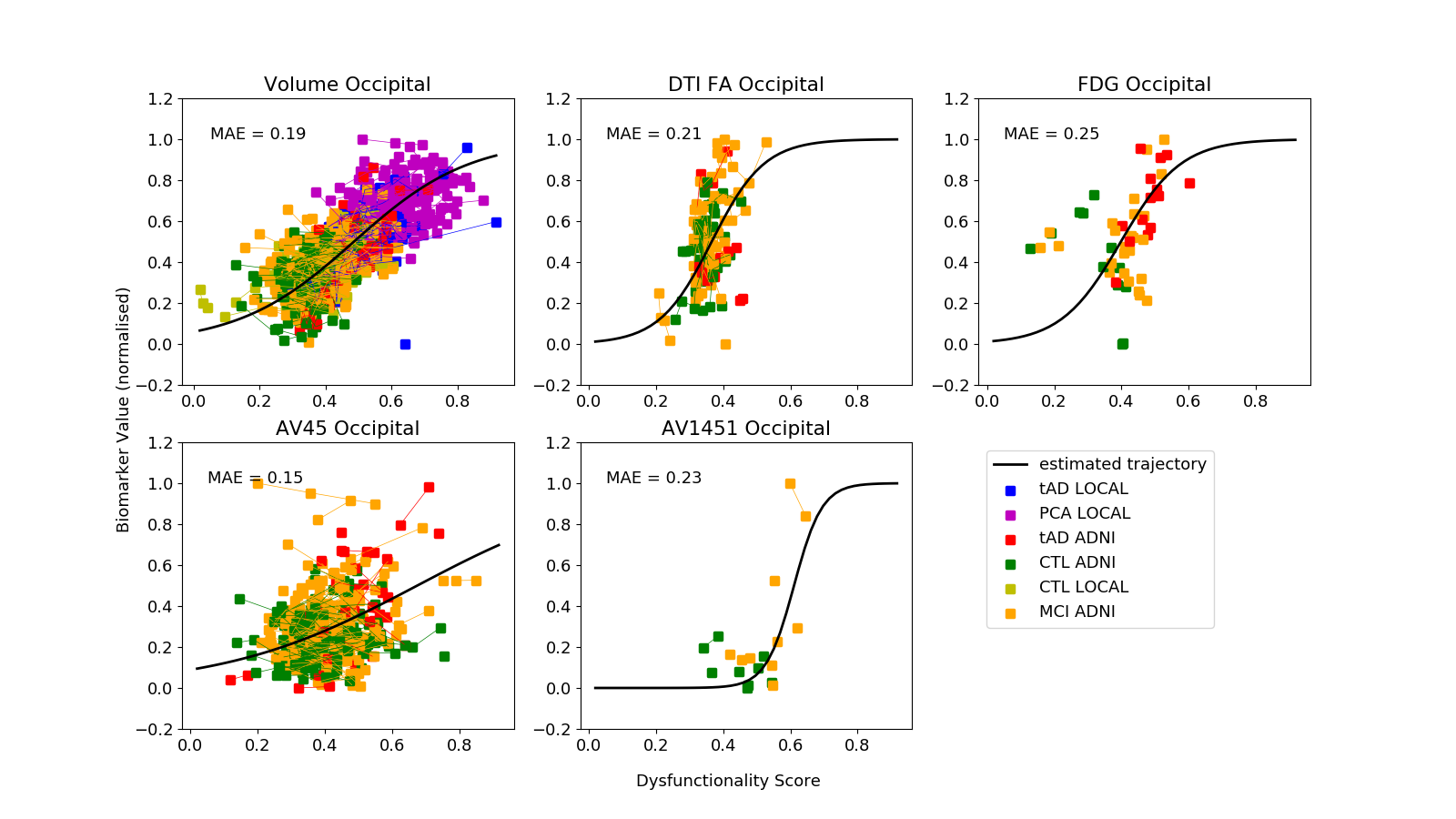}
\caption{Occipital unit}
\label{fig:dktFitUnit}
\end{subfigure}
\vspace{2em}

\begin{subfigure}{0.47\textwidth}
\centering
\includegraphics[width=1\textwidth, trim=0 0 0 20, clip]{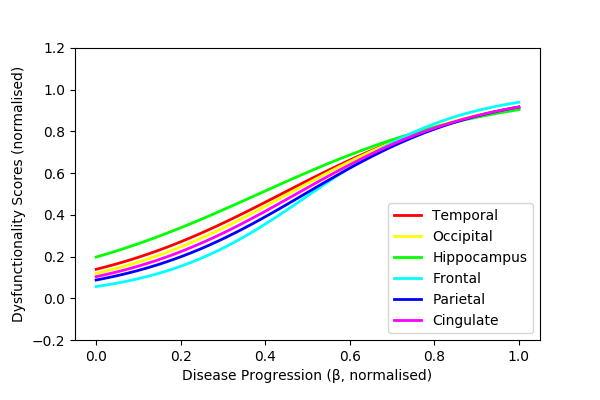}
\caption{tAD}
\label{fig:dktFitAD}
\end{subfigure}
\begin{subfigure}{0.47\textwidth}
\centering
\includegraphics[width=1\textwidth, trim=0 0 0 20, clip]{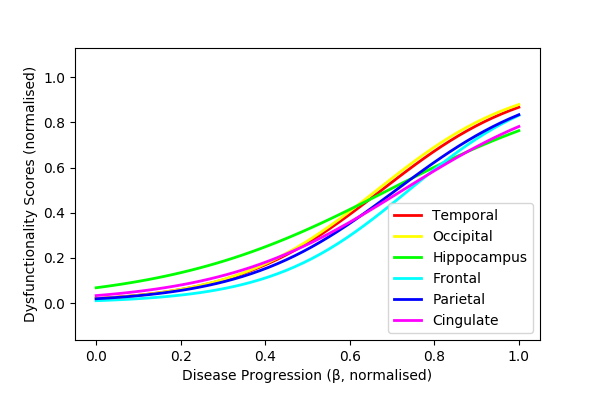}
\caption{PCA}
\label{fig:dktFitPCA}
\end{subfigure}
\caption[DKT results - biomarker trajectories in the occipital unit and dysfunctionality scores for tAD and PCA]{(a) DKT-estimated biomarker trajectories in the occipital functional unit. Subject data from ADNI and our local DRC cohort are also shown. The X-axis, defined as the occipital dysfunctionality score, represents the time-shifts (in months) of each subject. (b-c) Progression of DKT-estimated dysfunctionality scores for (b) typical AD and (c) PCA.}
\label{fig:pcaTadDisSpace}
\end{figure}

Fig. \ref{fig:dktFitUnit} shows the estimated biomarker trajectories within the \emph{occipital unit} plotted over the dysfunction scores, along with samples from the model posterior and aligned subject data. The X-axis shows the dysfunctionality scores within the occipital unit, which represent estimated time-shifts, in months, from an arbitrary reference X=0, while the Y-axis shows biomarker values normalised to [0,1] range. The model shows an unbiased data fit (Fig. \ref{fig:dktFitUnit}), and we can observe most PCA subjects having abnormal occipital volumes, thus leading to high occipital dysfunctionality scores, in line with the current understanding of PCA as affecting posterior regions \cite{crutch2012posterior}. We also show the progression of dysfunctionality scores over the disease stage for typical AD (Fig \ref{fig:dktFitAD}) and PCA (Fig \ref{fig:dktFitPCA}). In typical AD, we see that hippocampal dysfunction becomes abnormal earliest, while PCA shows early hippocampal dysfunction, which is later exceeded by the dysfunction in the occipital, temporal and parietal regions, which are known to be affected in PCA \cite{crutch2012posterior,Baron2001}. 

\begin{figure}
 \includegraphics[width=\textwidth, trim=0 20 0 0, clip]{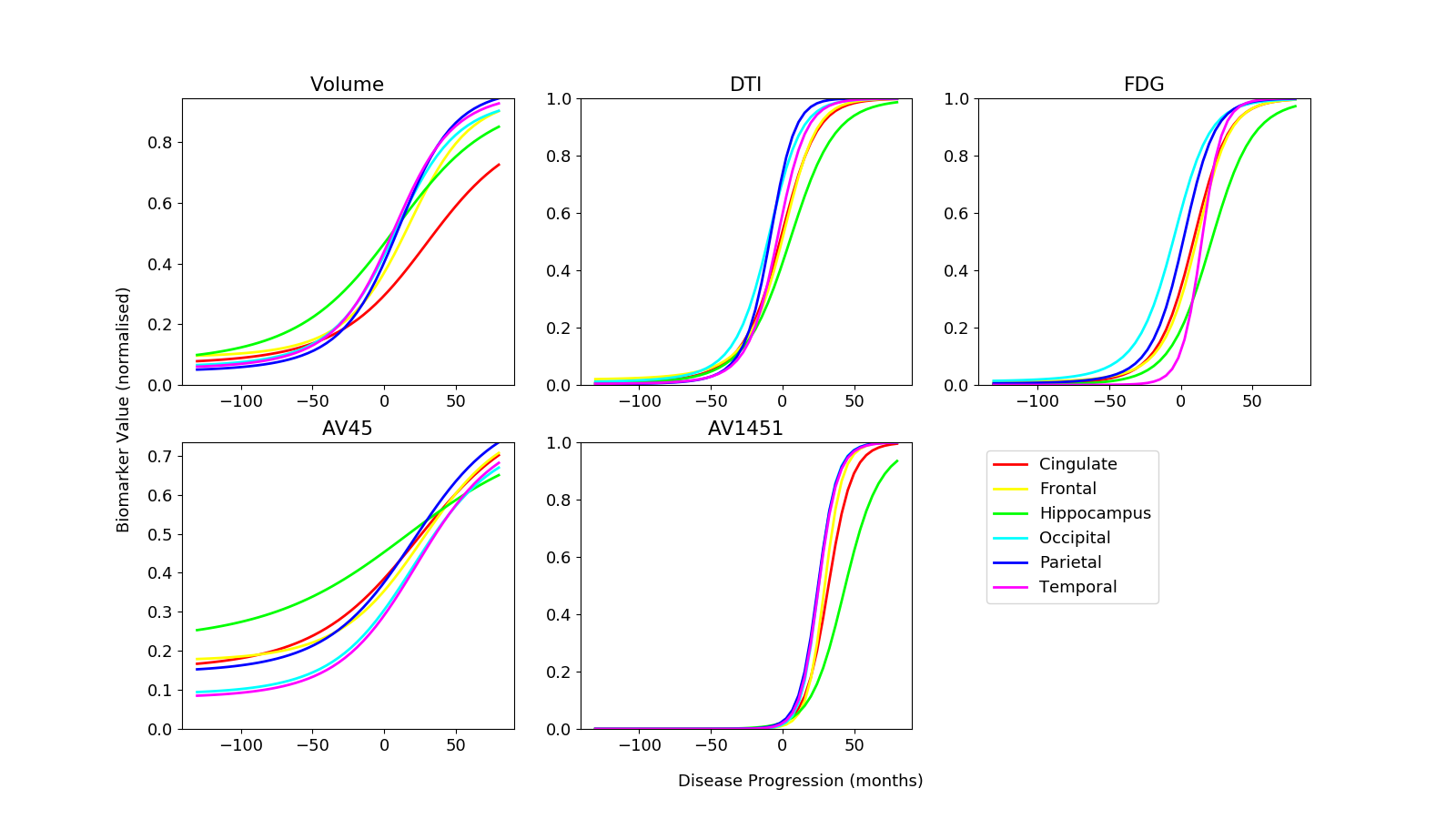}
 \caption[Estimated multi-modal trajectories for the PCA cohort.]{Estimated multi-modal trajectories for the PCA cohort. The only data that were available were the MRI volumetric data. The dynamics of the other biomarkers has been inferred by the model using data from typical AD, and taking into account the different spatial distribution of pathology in PCA as compared to typical AD.}
 \label{fig:PCAtrajByModality}
\end{figure}

In Fig. \ref{fig:PCAtrajByModality}, we plot the inferred biomarker trajectories for PCA directly across the disease progression. We do this for five different modalities: MRI Volumes, DTI, FDG, AV45 and AV1451. The results again recapitulate known patterns in PCA, where posterior regions are predominantly affected in all modalities. However, for MRI volumes and AV45, we also see early abnormalities, which we attribute to the models underestimating the biomarker measurement noise.

\section{Validation on DTI Data in PCA}
\label{sec:dktResVal}

We further validated DKT by predicting unseen DTI data from two patient datasets:
\begin{itemize}
 \item TADPOLE subjects with a different progression from the training subjects
 \item A separate test set of 20 DTI scans from controls and PCA patients from our own cohort.
\end{itemize}

To split TADPOLE into subgroups with different progression, we used the SuStaIn model by \cite{young2018uncovering}, which resulted into three subgroups: hippocampal, cortical and subcortical, with prominent early atrophy in the hippocampus, cortical and subcortical regions respectively. To evaluate prediction accuracy, we computed the rank correlation between the DKT-predicted biomarker values and the measured values in the test data. We compute the rank correlation instead of mean squared error as it is not susceptible to systemic biases of the models when predicting "unseen data" in a certain disease. We also compared the performance of DKT at predicting unseen data with four other models: 
\begin{itemize}
 \item \emph{Latent stage model}: a sigmoidal based disease progression model, as described in \cite{jedynak2012computational}. This model assumes all tAD and PCA subjects follow the same progression.
 \item \emph{Multivariate}: A multivariate Gaussian Process model with RBF kernel that predicts a DTI ROI marker from multiple MRI markers.
 \item \emph{Spline}: a univariate cubic spline regression model that predicts the DTI biomarker based on the corresponding MRI biomarker, independently for each region.
 \item \emph{Linear}: Same as above but linear model instead of spline.
\end{itemize}

Validation results are shown in Table \ref{sec:dktPerfMetrics}, for hippocampal to cortical TADPOLE subgroups, as well as PCA subjects from our DRC cohort. When predicting missing DTI markers from the TADPOLE cortical subgroup as well as PCA subjects, the DKT correlations are generally high for the cingulate, hippocampus and parietal, and lower for the frontal lobe. DKT further shows favourable performance compared to the other models, due to its ability to disentangle the progressions of each disease separately. In particular, it shows the best results for DTI FA prediction in the parietal and temporal lobes on both datasets and similar performance to the latent-stage model on the PCA dataset for the cingulate, frontal and hippocampal (differences here are not statistically significant). Due to the challenging problem of predicting unseen data in these diseases/subtypes, notice how the models yield bad predictions for the occipital lobe (negative correlations), most likely due to overfitting.

\newcommand{\cw}{c}

\begin{table}
\centering
\fontsize{9}{12}\selectfont
\begin{tabular}{c | c c c c c c}
\textbf{Model} & \textbf{Cingulate} & \textbf{Frontal} & \textbf{Hippocam.} & \textbf{Occipital} & \textbf{Parietal} & \textbf{Temporal}\\
& \multicolumn{6}{c}{\textbf{TADPOLE subgroups: Hippocampal subgroup to Cortical subgroup}}\\
DKT (ours) &      0.56 $\pm$ 0.23 &    \textbf{0.35 $\pm$ 0.17} &        \textbf{0.58 $\pm$ 0.14} &     -0.10 $\pm$ 0.29 &     \textbf{0.71 $\pm$ 0.11} &     \textbf{0.34 $\pm$ 0.26} \\
Latent stage &      0.44 $\pm$ 0.25 &    0.34 $\pm$ 0.21 &       0.34 $\pm$ 0.24* &     \textbf{-0.07 $\pm$ 0.22} &     0.64 $\pm$ 0.16 &    0.08 $\pm$ 0.24* \\
Multivariate &      \textbf{0.60 $\pm$ 0.18} &   0.11 $\pm$ 0.22* &       0.12 $\pm$ 0.29* &     -0.22 $\pm$ 0.22 &   -0.44 $\pm$ 0.14* &   -0.32 $\pm$ 0.29* \\
Spline &    -0.24 $\pm$ 0.25* &  -0.06 $\pm$ 0.27* &        0.58 $\pm$ 0.17 &     -0.16 $\pm$ 0.27 &    0.23 $\pm$ 0.25* &    0.10 $\pm$ 0.25* \\
Linear &    -0.24 $\pm$ 0.25* &   0.20 $\pm$ 0.25* &        0.58 $\pm$ 0.17 &     -0.16 $\pm$ 0.27 &    0.23 $\pm$ 0.25* &    0.13 $\pm$ 0.23* \\
& \multicolumn{6}{c}{\textbf{typical Alzheimer's to Posterior Cortical Atrophy}}\\
DKT (ours) &    0.77 $\pm$ 0.11 &    0.39 $\pm$ 0.26 &      0.75 $\pm$ 0.09 &    0.60 $\pm$ 0.14 &    \textbf{0.55 $\pm$ 0.24} &    \textbf{0.35 $\pm$ 0.22} \\
Latent stage &    \textbf{0.80 $\pm$ 0.09} &    \textbf{0.53 $\pm$ 0.17} &      \textbf{0.80 $\pm$ 0.12} &    0.56 $\pm$ 0.18 &    0.50 $\pm$ 0.21 &    0.32 $\pm$ 0.24 \\
Multivariate &   0.73 $\pm$ 0.09 &   0.45 $\pm$ 0.22  &    0.71 $\pm$ 0.08 & -0.28 $\pm$ 0.21* &  0.53 $\pm$ 0.22  &  0.25 $\pm$ 0.23* \\
Spline &   0.52 $\pm$ 0.20* &  -0.03 $\pm$ 0.35* &     0.66 $\pm$ 0.11* &   0.09 $\pm$ 0.25* &    0.53 $\pm$ 0.20 &   0.30 $\pm$ 0.21* \\
Linear &   0.52 $\pm$ 0.20* &    0.34 $\pm$ 0.27 &     0.66 $\pm$ 0.11* &    \textbf{0.64 $\pm$ 0.17} &    0.54 $\pm$ 0.22 &   0.30 $\pm$ 0.21* \\
\end{tabular}
\vspace{0.5em}
\caption[Performance evaluation of DKT and other models]{Performance evaluation of DKT and four other statistical models of decreasing complexity. We show the rank correlation between predicted biomarkers and measured biomarkers in (top) TADPOLE subgroups -- information transfer from hippocampal subgroup to cortical subgroup -- and (bottom) PCA. (*) Statistically significant difference in the performance of DKT vs the other models, based on a two-tailed t-test, Bonferroni corrected.}
\label{sec:dktPerfMetrics}
\end{table}

\section{Discussion}
\label{sec:dktDis}

We presented DKT, a framework that enables, for the first time, joint modelling of biomarker progressions in multiple neurodegenerative diseases simultaneously. The framework allows the inference of biomarker trajectories in rare diseases, for which there is not enough data to allow estimation of such trajectories, and accounts for a different spatial distribution of pathology between distinct types of dementia. This further enables us to understand the complex mechanisms of rare diseases, as well as mechanisms shared between different types of related diseases.

We provided an example implementation of DKT using specific models of the biomarker trajectories, measurement noise and link function (the disease progression score). However, DKT should be considered as a general framework for joint modelling of biomarker trajectories within different diseases simultaneously. The actual implementation of DKT can thus be extended to use non-parametric trajectories, or more complex link functions that estimate not only subject time-shifts but also progression speed or higher order terms.

While in this work we have focused on Alzheimer's variants such as tAD and PCA, DKT can also be applied to other progressive neurodegenerative diseases of non-Alzheimer's type such as tauopathies (e.g. Frontotemporal dementia), synucleinopathies (e.g. Parkinson's disease), other neurodegenerative diseases such as Huntington's disease or Multiple Sclerosis, and even the normal ageing process. Cognitive tests can also be included in the disease-specific sub-models of DKT, or even allocated in the functional units of the regions that are responsible for those tasks, based on previous voxel-based morphometry studies. However, some care needs to be exercised when selecting the biomarkers and grouping them into functional units, as in some diseases the assumption of disease agnostic dynamics might not hold for some groups of molecular biomarkers. For example, some non-Alzheimer's tauopathies such as Frontotemporal dementia might show tau abnormalities but no corresponding amyloid abnormalities within the same region. In the case of Frontotemporal dementia, we recommend including higher-level biomarkers such as glucose metabolism from FDG, white matter degeneration from DTI or volume from structural MRI, but one should exclude amyloid markers. 

Our work has several limitations: 1) DKT assumes all subjects within a disease follow the same trajectory, without considering heterogeneity within the disease population, 2) the allocation of biomarkers into functional units has to be done using \emph{a-priori} human knowledge, 3) DKT currently works only on extracted brain features, discarding important information present in the brain morphometry, 4) for validation, the synthetic experiment we ran was limited to only one setting of the parameters and 5) the validation on patient data was also done only on a small set of 20 DTI scans, due to lack of multimodal data in PCA.

There are several potential avenues for further research: 1) to account for heterogeneity, DKT can also be easily extended to include subject-specific effects; 2) improved schemes for biomarker allocation to functional units can take connectivity into account, or derive it from the data automatically; 3) to account for brain morphometry and connectivity, DKT can be extended into a fully spatio-temporal model, by estimating continuous changes in volumetric brain images -- in this case, each voxel can have an associated dysfunctionality score that is derived from measurements of various modalities from that voxel; 4-5) DKT can be further validated on more complex synthetic experiments with variable parameter settings, and on patient data from ADNI, where the population could be \emph{a-priori} split into sub-groups with different progressions. On these subgroups, DKT can be used to transfer biomarker modalities that have been left out during training.

\section{Conclusion}
\label{sec:dktCon}

In this work I presented DKT, a novel method that can empower studies of rare dementias with limited biomarker data by leveraging data from larger datasets of related dementias. When applied to synthetic data with ground truth, I showed that DKT can robustly recover biomarker trajectories in two distinct diseases and also subject-specific time-shifts. I also applied DKT to multimodal imaging biomarkers from the TADPOLE Challenge dataset, where I showed that it can estimate plausible non-MRI biomarker trajectories for Posterior Cortical Atrophy in lack of such data for this disease. I validated the performance of DKT on a test set of 20 DTI scans from PCA and controls, and showed that DKT has similar or better performance compared to simpler models.

In the next chapter, I will present the TADPOLE  Challenge, which evaluates the performance of algorithms and features at predicting the future evolution of subjects at risk of AD. As opposed to the work performed in this chapter, the TADPOLE challenge aims to evaluate a much larger set of algorithms and features, comprising regression techniques, disease progression models, machine learning techniques and even manual predictions made by clinicians.

\chapter[TADPOLE Challenge: Prediction of Evolution in Alzheimer's Disease]{TADPOLE Challenge: Prediction of Longitudinal Evolution in Alzheimer's Disease}
\label{chapter:tadpole}

\section{Contributions}

In this chapter I present the design of \emph{The Alzheimer's Disease Progression Of Longitudinal Evolution} (TADPOLE) Challenge, which aims to predict the evolution of subjects at risk of Alzheimer's disease. The challenge was organised by the European Progression of Neurodegenerative (EuroPOND) consortium, in collaboration with the Alzheimer's disease Neuroimaging Initiative (ADNI). The key organisers of the challenge were, in alphabetical order: Daniel Alexander, Frederik Barkhof, Esther Bron, Nick Fox, Stefan Klein, Razvan Marinescu (myself), Neil Oxtoby and Alexandra Young. 

I contributed with suggestions to the challenge design, helped write the website, assembled the TADPOLE D2 longitudinal dataset and the data dictionary, and wrote benchmark prediction scripts. I also build the leaderboard system which performs live evaluation of the participants' submissions. I further helped promote the competition at several medical imaging conferences, and organised two mini-competitions at the PyConUK conference and at the CMIC summer school, 2018. 

Daniel Alexander proposed the main design of the challenge, secured funding, helped write the website, and wrote simple prediction scripts. Neil Oxtoby contributed to challenge design, helped me validate the D2 dataset, built the D3 cross-sectional dataset, helped write the website, organised webinars and promoted the competition. Alexandra Young contributed to challenge design, helped write the website, performed simulations to establish which target biomarkers are most suitable and promoted the competition. Esther Bron and Stefan Klein contributed to challenge design and helped write the website. Nick Fox and Frederik Barkhof provided valuable suggestions on the challenge design. Arthur Toga and Michael Weiner offered access to the ADNI database. 

\section{Publications}

\begin{itemize}
\item  R. V. Marinescu, N. P. Oxtoby, A. L. Young, E. E. Bron, A. W. Toga, M. W. Weiner, F. Barkhof, N. C. Fox, S. Klein, D. C. Alexander and the EuroPOND Consortium, TADPOLE Challenge: Prediction of Longitudinal Evolution in Alzheimer's Disease, arXiv, 2018 

I wrote this challenge design paper based on text and diagrams from the TADPOLE website. All collaborators contributed with feedback on the manuscript. The results of the challenge will be published in a separate paper in 2019, after enough data has been collected for the final evaluation.
\end{itemize}

\section{Introduction}
\label{intro}

As already mentioned in section \ref{chapter:bckDpm}, early diagnosis of dementia is important in order to enable the administration of treatments in early disease stages, before the onset of cognitive decline. While such early and accurate diagnosis of dementia can be challenging, this can be aided by quantitative biomarker measurements taken from magnetic resonance imaging (MRI), positron emission tomography (PET), and cerebro-spinal fluid (CSF) samples extracted from lumbar puncture. It has been hypothesised for AD \cite{jack2010hypothetical,jack2013update,aisen2010clinical,frisoni2010clinical} that all these biomarkers become abnormal at different intervals before symptom onset, suggesting that together they can be used for accurate prediction of onset and overall disease progression in individuals. In particular, some of the early biomarkers become abnormal decades before symptom onset, and can thus facilitate early diagnosis. 

Several approaches for predicting AD-related target variables (e.g. clinical diagnosis, cognitive/imaging biomarkers) have been proposed which leverage multimodal biomarker data available in AD. Traditional longitudinal approaches based on statistical regression model the relationship of the target variables with other known variables. Examples include regression of the target variables against clinical diagnosis \cite{scahill2002mapping}, cognitive test scores \cite{yang2011quantifying, sabuncu2011dynamics}, rate of cognitive decline \cite{doody2010predicting}, and retrospectively staging subjects by time to conversion between diagnoses \cite{guerrero2016instantiated}. Another approach involves supervised machine learning techniques such as support vector machines, random forests, and artificial neural networks, which use pattern recognition to learn the relationship between the values of a set of predictors (biomarkers) and their labels (diagnoses). These approaches have been used to discriminate AD patients from cognitively normal individuals \cite{kloppel2008automatic, zhang2011multimodal}, and for discriminating at-risk individuals who convert to AD in a certain time frame from those who do not \cite{young2013accurate, mattila2011disease}. The emerging approach of disease progression modelling aims to reconstruct biomarker trajectories or other disease signatures across the disease progression timeline, without relying on clinical diagnoses or estimates of time to symptom onset. Examples include models built on a set of scalar biomarkers to produce discrete \cite{fonteijn2012event, young2014data} or continuous \cite{jedynak2012computational, donohue2014estimating, villemagne2013amyloid} biomarker trajectories; richer but less comprehensive models that leverage structure in data such as MR images \cite{durrleman2013toward, lorenzi2015disentangling, bilgel2016multivariate}; and models of disease mechanisms \cite{seeley2009neurodegenerative, zhou2012predicting, raj2012network, iturria2016early}.

These models have shown promise for predicting AD biomarker progression when using existing test data, but few have been tested on truly unseen \emph{future} data. Moreover, different investigators test these models on different datasets (including subsets of a single dataset) and use different processing pipelines. Community challenges have proved effective, in the medical image analysis field and beyond, for providing unbiased comparative evaluations of algorithms and tools designed for a particular task. Previous challenges that focused on prediction of AD progression include the \emph{CADDementia challenge} \cite{bron2015standardized}, which aimed to predict clinical diagnosis from MRI scans. A similar challenge, the "\emph{International challenge for automated prediction of MCI from MRI data}" \cite{sarica2018machine} asked participants to predict diagnosis and conversion status from extracted MRI features of subjects from the ADNI study \cite{weiner2017recent}. Yet another challenge, The Alzheimer's Disease \emph{Big Data DREAM Challenge} \cite{allen2016crowdsourced}, asked participants to predict cognitive decline from genetic and MRI data. 

The Alzheimer's Disease Prediction Of Longitudinal Evolution (TADPOLE) Challenge aims to identify the data, features and approaches that are the most predictive of AD progression. In contrast to previous challenges, our motivation is to improve future clinical trials through identification of patients most likely to benefit from an effective treatment, i.e., those at early stages of disease who are likely to progress over the short-to-medium term (1-5 years). Identifying such subjects reliably helps cohort selection by focusing on groups that highlight positive treatment effects. The challenge thus focuses on forecasting three key features: clinical status, cognitive decline, and neurodegeneration (brain atrophy), over a five-year timescale. It uses \emph{rollover}\footnote{i.e. subjects who enrolled in the previous ADNI2 study and who will continue in the third phase.} subjects from the ADNI study for whom a history of measurements is available, and who are expected to continue in the study, providing future measurements for testing. Since the test data does not exist at the time of forecast submissions, the challenge provides a completely unbiased basis for performance comparison. TADPOLE goes beyond previous challenges by drawing on a vast set of multimodal measurements from ADNI which support prediction of AD progression.

\section{Competition Design}
\label{design}

\begin{figure}[h]
 \centering
 \includegraphics[width=0.7\textwidth]{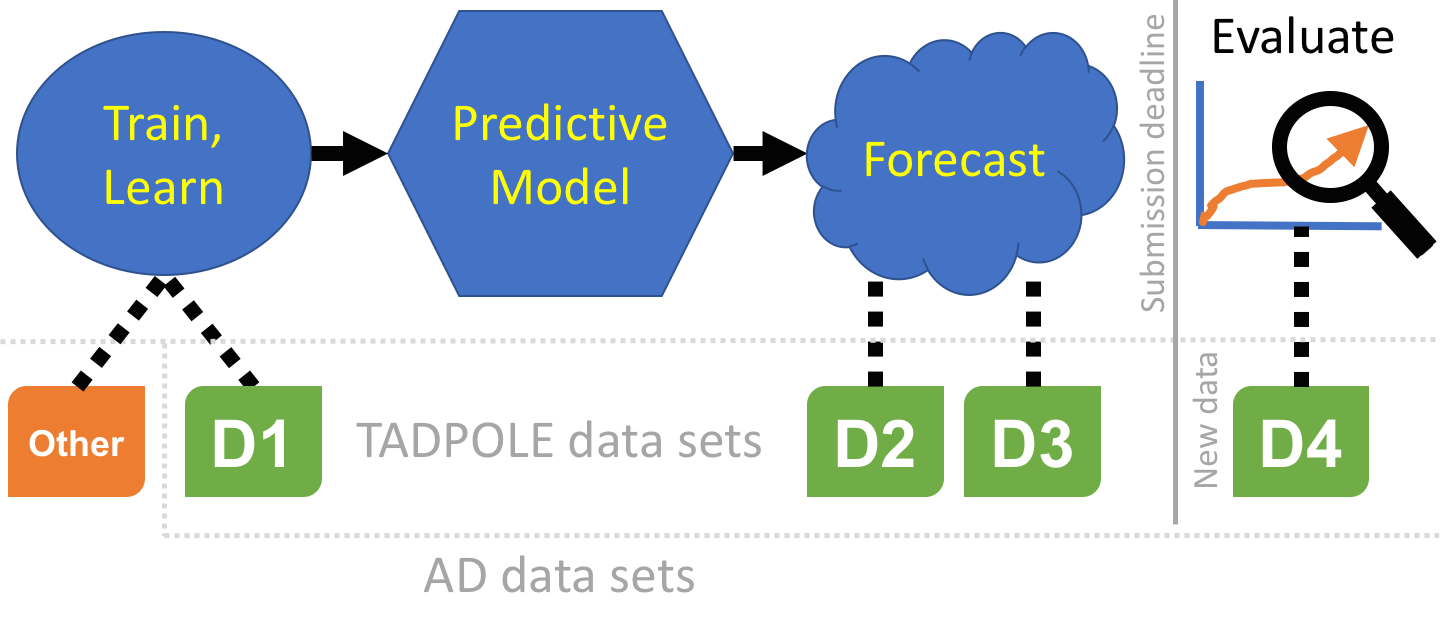}
 \caption[Diagram showing the TADPOLE Challenge design]{TADPOLE Challenge design. Participants are required to train a predictive model on a training dataset (D1 and/or others) and make forecasts for different datasets (D2, D3) by the submission deadline. Evaluation will be performed on a test dataset (D4) that is acquired after the submission deadline.}
 \label{fig:design}
\end{figure}

The aim of TADPOLE is to predict future outcome measurements of subjects at-risk of AD, enrolled in the ADNI study. A history of informative measurements from ADNI (imaging, psychology, demographics, genetics, etc.) from each individual is available to inform forecasts. TADPOLE participants are required to predict future measurements from these individuals and submit their predictions before a given submission deadline.  Evaluation of these forecasts occurs post-deadline, after the measurements have been acquired. A diagram of the TADPOLE flow is shown in Fig \ref{fig:design}. 

\section{Forecasts}

Since we do not know the exact time of future data acquisitions for any given individual, TADPOLE challenge participants are required to make, for every individual, month-by-month forecasts of three key biomarkers: (1) clinical diagnosis which can be either cognitively normal (CN), mild cognitive impairment (MCI) or probable Alzheimer's disease (AD); (2) ADAS-Cog13 (ADAS13) score; and (3) ventricle volume (divided by intra-cranial volume). Evaluation is performed using forecasts at the months that correspond to data acquisition. TADPOLE forecasts are required to be probabilistic and some evaluation metrics will account for forecast probabilities provided by participants. Methods or algorithms that do not produce probabilistic estimates can still be used, by setting binary probabilities (zero or one) and default confidence intervals.

Participants are required to submit forecasts in a standardised format (see Table \ref{tab:subFormat}). For clinical status, relative likelihoods of each option (CN, MCI, and AD) for each individual should be provided. These are normalised at evaluation time; negative likelihoods are set to zero. For ADAS13 and ventricle volume, participants need to provide a best-guess value as well as a 50\% confidence interval for each individual. This 50\% confidence interval (as opposed to the more standard 95\%) was chosen to provide a more symmetric and less noisy evaluation of over- and under-estimation of the confidence interval, because similar sample sizes of data fall inside and outside the interval. 

\newcommand{\wi}{1.4cm}
\setlength\tabcolsep{3pt} 

\begin{table}
\small
 \begin{tabular}{>{\centering\arraybackslash}m{0.8cm}  >{\centering\arraybackslash}m{1.2cm}  >{\centering\arraybackslash}m{1.3cm}  >{\centering\arraybackslash}m{1.0cm}  >{\centering\arraybackslash}m{1.0cm}  >{\centering\arraybackslash}m{1.0cm}  >{\centering\arraybackslash}m{0.9cm}  >{\centering\arraybackslash}m{1.2cm}  >{\centering\arraybackslash}m{1.2cm}  >{\centering\arraybackslash}m{1.3cm}  >{\centering\arraybackslash}m{1.3cm}  >{\centering\arraybackslash}m{1.3cm}}
  \textbf{RID} & \textbf{Month} & \textbf{Date} & \textbf{CN prob.} & \textbf{MCI prob.} & \textbf{AD prob.} & \textbf{ADAS} & \textbf{ADAS CI lower} & \textbf{ADAS CI upper} & \textbf{Vent.} & \textbf{Vent. CI lower} & \textbf{Vent. CI upper}\\
  \hline
  A & 1 & 2018-01 & 0 & 1 & 0 & 30 & 25 & 35 & 0.024 & 0.021 & 0.029\\
  B & 1 & 2018-01 & 3 & 2 & 0 & 25 & 21 & 26 & 0.023 & 0.021 & 0.025\\
  C & 1 & 2018-01 & 0.24 & 0.38 & 0.38 & 40 & 25 & 50 & 0.025 & 0.023 & 0.028\\
  
 \end{tabular}
  
 \caption{The format of the forecasts for three example subjects. Participants have to predict, for each subject, the probability of clinical diagnosis (CN/MCI/AD), the ADAS-Cog13 score and Ventricle volume, as well as the 50\% confidence range. RID - Roster ID is the unique identifier for ADNI subjects, ADAS - ADAS-Cog13, CI -  confidence range. Note that, even if the CN/MCI/AD probabilities don't sum to one, we will normalise them anyway.}
 \label{tab:subFormat}
\end{table}

\section{Data}

We provide participants with a standard ADNI-derived dataset (available via the Laboratory Of NeuroImaging: LONI) which they can use to train their algorithms, removing the need to pre-process the ADNI data themselves or merge different spreadsheets. However, participants are allowed to use a custom training set, by adding any other ADNI data or data from other studies. The software code used to generate the standard dataset is openly available in a GitHub repository\footnote{https://github.com/noxtoby/TADPOLE} and on the ADNI website, packaged with the standard dataset in the LONI ADNI database.

\subsection{ADNI Data}

Data used in the preparation of this article were obtained from the Alzheimer's Disease Neuroimaging Initiative (ADNI) database (\url{adni.loni.usc.edu}). The ADNI was launched in 2003 by the National Institute on Aging (NIA), the National Institute of Biomedical Imaging and Bioengineering (NIBIB), the Food and Drug Administration (FDA), private pharmaceutical companies and non-profit organisations, as a \$60 million, 5-year public-private partnership. The initial goal of ADNI was to recruit 800 subjects but ADNI has been followed by ADNI-GO and ADNI-2. To date these three protocols have recruited over 1500 adults, aged 55 to 90, to participate in the research, consisting of cognitively normal older individuals, people with early or late MCI, and people with early AD. The general ADNI inclusion criteria has been described in \cite{petersen2010alzheimer}. 

The data we used from ADNI consists of: (1) CSF markers of amyloid-beta and tau deposition; (2) various imaging modalities such as magnetic resonance imaging (MRI), positron emission tomography (PET) using several tracers: Fluorodeoxyglucose (FDG, hypometabolism), AV45 (amyloid), AV1451 (tau) as well as diffusion tensor imaging (DTI); (3) cognitive assessments acquired in the presence of a clinical expert; (4) genetic information such as Alipoprotein E4 (APOE4) status extracted from DNA samples; and (5) general demographic information. Extracted features from this data were merged together into a final spreadsheet and made available on the LONI ADNI website.

\subsection{Image Preprocessing}

The imaging data has been pre-processed with standard ADNI pipelines.  For MRI scans, this included correction for gradient non-linearity, B1 non-uniformity correction and peak sharpening\footnote{see MRI analysis on ADNI website: \url{http://adni.loni.usc.edu/methods/mri-analysis/
mri-pre-processing}}. Meaningful regional features such as volume and cortical thickness were extracted using the Freesurfer cross-sectional and longitudinal pipelines \cite{reuter2012within}. Each PET image (FDG, AV45, AV1451), which consists of a series of dynamic frames, had its frames co-registered, averaged across the dynamic range, standardised with respect to the orientation and voxel size, and smoothed to produce a uniform resolution of 8mm full-width/half-max (FWHM)\footnote{see PET analysis on ADNI website: \url{http://adni.loni.usc.edu/methods/pet-analysis/pre-processing}}. Standardised uptake value ratio (SUVR) measures for relevant regions-of-interest were extracted (see \cite{jagust2010alzheimer}) after registering the PET images to corresponding MR images using the SPM5 software \cite{ashburner2009computational}. DTI scans were corrected for head motion and eddy-current distortion, skull-stripped, EPI-corrected, and finally aligned to the T1 scans using the pipeline from \cite{nir2013effectiveness}. Diffusion tensor summary measures were estimated based on the Eve white-matter atlas by \cite{oishi2009atlas}. 

\section{TADPOLE Datasets}
\label{datasets}

The TADPOLE Challenge involves three kinds of data sets: (a) a \emph{training data set}, which is a collection of measurements with associated outcomes that can be used to fit models or train algorithms; (b) a \emph{prediction data set}, which contains only baseline measurements (possibly longitudinal), without associated outcomes --- this is the data that algorithms, models, or experts use as input to make their forecasts of later patient status and outcome; and (c) \emph{a test data set}, which contains the patient outcomes against which we will evaluate forecasts --- in TADPOLE, this data did not exist at the time of submitting forecasts.

In order to evaluate the effect of different methodological choices, we prepared three “standard” data sets for training and prediction: 
\begin{itemize}
 \item \textbf{D1}: The TADPOLE \underline{\smash{standard training set}} draws on longitudinal data from the entire ADNI history. The data set contains a set of measurements for every individual that has provided data to ADNI in at least two separate visits (different dates) across three phases of the study: ADNI1, ADNI GO, and ADNI2. 
 \item \textbf{D2}: The TADPOLE \underline{\smash{longitudinal prediction set}} contains as much available data as we could gather from the ADNI rollover individuals for whom challenge participants are asked to provide forecasts. D2 includes all available time-points for these individuals. 
 \item \textbf{D3}: The TADPOLE \underline{\smash{cross-sectional prediction set}} contains a single (most recent) time point and a limited set of variables from each rollover individual in D2. Although we expect worse forecasts from this data set than D2, D3 represents the information typically available when selecting a cohort for a clinical trial. 
\end{itemize}

\begin{figure}
 \centering
 \includegraphics[width=0.7\textwidth]{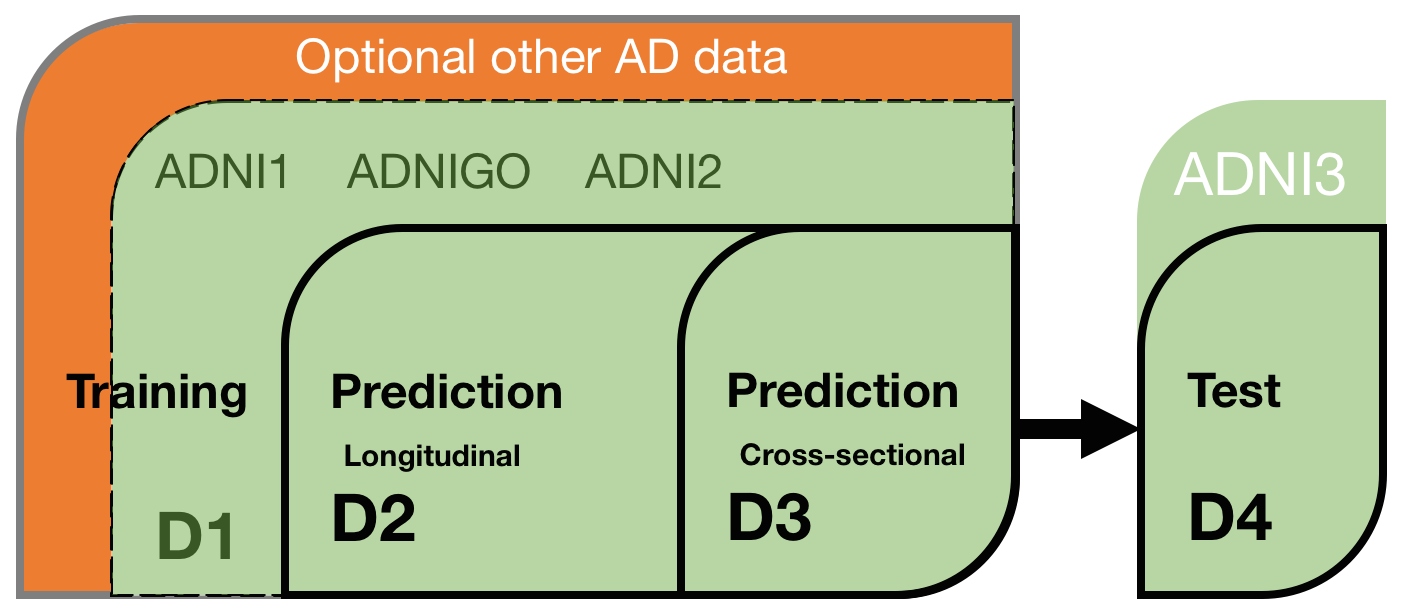}
 \caption[Venn diagram of the TADPOLE datasets derived from ADNI data.]{Venn diagram of the TADPOLE datasets derived from ADNI data, for training (D1), longitudinal prediction (D2), cross-sectional prediction (D3) and the test set (D4). D3 is a subset of D2, which in turn is a subset of D1. Other non-ADNI data can also be used for training.}
 \label{fig:venn_diagram}
\end{figure}

The forecasts will be evaluated on future data (D4 -- test set) from ADNI3 rollovers, acquired after the challenge submission deadline. In addition to the three standard datasets (D1, D2 and D3), challenge participants are allowed to use any other data sets that might serve as useful additional training data.  

Fig. \ref{fig:venn_diagram} shows a diagram highlighting the nested structure of datasets D1--D3. Table \ref{tab:biomk_data_available} shows the proportion of biomarker data available in each dataset. There are a considerable number of entries with missing data, especially for some biomarkers such as tau imaging (AV1451). We also estimated the expected number of subjects and available data for D4, using information from the ADNI3 procedures and using rollovers from previous ADNI studies (Table \ref{tab:biomk_data_available}, right-most column) -- See \ref{app:expectedD4} for more information on D4 estimates. Based on our estimates, we believe the size of D4 (around 330 subjects, 1 visit/subject) should be enough for a reliable evaluation of TADPOLE submissions.

\begin{table}
\centering
 \begin{tabular}{c | c | c c c c}
 \multicolumn{2}{c|}{\textbf{Subject statistics}} & D1 & D2 & D3 & D4 \\
 \hline
 \multicolumn{2}{c|}{Nr. of subjects } & 1667 & 896 & 896 & \emph{330}\\
 \multicolumn{2}{c|}{Visits per subject }&  $\mathbin{{7.6}{\pm}{3.8}}$  & $\mathbin{{8.5}{\pm}{4.2}}$ & $\mathbin{{1.0}{\pm}{0.0}}$ & $\mathit{\mathbin{{1.0}{\pm}{0.0}}}$\\
 & CN & 31 & 38 & 45 & \emph{39} \\
 Diagnosis* (\%) & MCI & 56 & 57 & 39 & \emph{49} \\
 & AD & 13 & 5 & 16 & \emph{12} \\
 \multicolumn{2}{l}{\textbf{Data availability**}}\\
 \hline
 \multicolumn{2}{c|}{Cognitive tests (\%) } & 70 & 68 & 84 & \emph{62} \\
 \multicolumn{2}{c|}{MRI (\%) } & 62 & 56 & 75 & \emph{69} \\
 \multicolumn{2}{c|}{FDG-PET (\%) } & 16 & 20 & 0 & \emph{20} \\
 \multicolumn{2}{c|}{AV45-PET (\%) } & 16 & 22 & 0 & \emph{19} \\
 \multicolumn{2}{c|}{AV1451-PET (\%) } & 0.7 & 1.1 & 0 & \emph{19} \\
 \multicolumn{2}{c|}{DTI (\%) } & 6 & 8 & 0 & \emph{15} \\
 \multicolumn{2}{c|}{CSF (\%) } & 18 & 19 & 0 & \emph{14} \\
 \end{tabular}
  
 \caption[Subject statistics and available data in the TADPOLE datasets D1, D2, D3 and D4.]{Subject statistics and available data in the TADPOLE datasets D1, D2 and D3. There is a considerable amount of missing data in some biomarkers such as AV1451. Numbers for D4 are estimated based on ADNI3 procedures (see ADNI3 procedures manual) and rollovers from previous ADNI studies. (*) Diagnosis at baseline visit. (**) Percentage of all visits (across all subjects) that have measurements for desired biomarker.}
 \label{tab:biomk_data_available}
\end{table}

\section{Submissions}
\label{submissions}

There are two kinds of submissions that challenge participants can make. A simple entry requires a minimal forecast and a description of methods; it makes participants eligible for the prizes but not co-authorship on the scientific paper documenting the results. A simple entry can use any training data or prediction sets and forecast at least one of the target outcome variables (clinical status, ADAS13 score, or ventricle volume). A full entry entitles participants for consideration as a co-author on the publication documenting the results. Such a full entry requires a complete forecast for all three outcome variables on all subjects from the D2 prediction set, along with a description of the methods. Each individual participant is limited to a maximum of three submissions. This restriction has been introduced to avoid the risk of participants “tuning” their method on the test set by submitting multiple predictions for a range of algorithm settings. Although not required for a full entry, participants are strongly encouraged to submit predictions also for D3. 

Prizes are awarded to the best submissions regardless of the choice of training sets (D1/custom) and prediction sets (D2/D3). However, the additional submissions support the key scientific aims of the challenge by allowing us to separate the influence of the choice of training data, post-processing pipelines, and modelling techniques or prediction algorithms. The target variables used for evaluation, in particular ventricle volume, will use the same post-processing pipeline as the standard data sets D1-D3.

Beyond the standard training dataset (D1), participants can include additional forecasts from "custom" (i.e. constructed by the participant) training data or custom post-processing of the raw data from subjects in the standard training set. The same applies to the prediction sets D2 and D3, which can be customised by the participants if desired, e.g. a prediction set with different features from the same individuals as in D2 and D3. Table \ref{tab:submissions} shows the twelve possible combinations of subject sets, processing and prediction sets, from which a full-entry submission must contain at least one of the first four (ID 1--4).

\begin{table}[h]
\centering
 \begin{tabular}{c | c | c | c}
\textbf{ID} & \multicolumn{2}{c|}{\textbf{Training set}} & \textbf{Prediction set}\\
& Subject set & Post-processing & \\
\hline
1 & D1 & standard & D2\\
2 & D1 & custom & D2\\
3 & custom & standard & D2\\
4 & custom & custom & D2\\
5 & D1 & standard & D3\\
6 & D1 & custom & D3\\
7 & custom & standard & D3\\
8 & custom & custom & D3\\
9 & D1 & standard & custom\\
10 & D1 & custom & custom\\
11 & custom & standard & custom\\
12 & custom & custom & custom\\
  
\end{tabular}
\caption[Types of TADPOLE submissions that can be made by participants.]{Types of submissions that can be made by participants, for different types of training sets, prediction sets and post-processing pipelines.}
\label{tab:submissions}
\end{table}

\section{Forecast Evaluation}
\subsection{Clinical Status Prediction}

For evaluation of clinical status predictions, we will use similar metrics to those that proved effective in the CADDementia challenge \cite{bron2015standardized}: (i) the multiclass area under the receiver operating curve (mAUC); and (ii) the overall balanced classification accuracy (BCA). The mAUC is independent of the group sizes and gives an overall measure of classification ability that accounts for relative likelihoods assigned to each class. The simpler BCA is also independent of group sizes, but does not exploit the probabilistic nature of the forecasts. 

\subsubsection{Multiclass Area Under the Receiver Operating Characteristic (ROC) Curve}

The multiclass Area Under the ROC Curve (mAUC) is a simple generalisation of the area under the ROC curve applicable to problems with more than two classes \cite{hand2001simple}. The AUC $\hat{A}(c_i|c_j)$ for classification of a class $c_i$ against another class $c_j$, is:
\begin{equation}
\hat{A}(c_i|c_j)=\frac{S_i-n_i(n_i+1)/2}{n_i n_j}
\end{equation}
where $n_i$ and $n_j$ are the number of points belonging to classes $i$ and $j$, respectively; while $S_i$ is the sum of the ranks of the class $i$ test points after ranking all the class $i$ and $j$ data points in increasing likelihood of belonging to class $i$. We further define the average AUC for classes $i$ and $j$ as $\hat{A}(c_i,c_j)= 0.5(\hat{A}(c_i|c_j)+\hat{A}(c_j|c_i))$. The overall mAUC is then obtained by averaging $\hat{A}(c_i,c_j)$ over all pairs of classes:
\begin{equation}
 mAUC = \frac{2}{L(L-1)}\sum_{i=2}^L\sum_{j=1}^{i}\hat{A}(c_i,c_j)
\end{equation}
where $L$ is the number of classes. The class probabilities that go into the calculation of $S_i$ in the first equation are $p_{CN}$, $p_{MCI}$ and $p_{AD}$, which are derived from the likelihoods of each ADNI subject being assigned to each diagnostic class, by normalising to have unity sum.

\subsubsection{Balanced Classification Accuracy}

The Balanced Classification Accuracy (see \cite{brodersen2010balanced}) is an extension of the classification accuracy measure that accounts for the imbalance in the numbers of datapoints belonging to each class. However, the measure is not probabilistic, so in TADPOLE the data points need to be assigned a hard classification to the class (CN, MCI, or AD) with the highest likelihood. The balanced accuracy for class $i$ is then:
\begin{equation}
 BCA_i = \frac{1}{2}\left[\frac{TP}{TP+FN}+\frac{TN}{TN+FP}\right]
\end{equation}
where TP, FP, TN, FN represent the number of true positives, false positives, true negatives and false negatives for classification as class $i$. In this case, true positives are data points with true label $i$ and correctly classified as such, while the false negatives are the data points with true label $i$ and incorrectly classified to a different class $j \ne i$. True negatives and false positives are defined similarly. The overall BCA is given by the mean of all the balanced accuracies for every class. 

\subsection{Continuous Feature Predictions}

For ADAS13 and ventricle volume, we will use three metrics: mean absolute error (MAE), weighted error score (WES) and coverage probability accuracy (CPA). The MAE focuses purely on accuracy of the best-guess prediction ignoring the confidence interval, whereas the WES incorporates confidence estimates into the error score. The CPA provides an assessment of the accuracy of the confidence estimates, irrespective of the best-guess prediction accuracy.

\subsubsection{Mean Absolute Error}

The mean absolute error (MAE) is:
\begin{equation}
 MAE = \frac{1}{N}\sum_{i=1}^{N}\left|{\tilde{M}_i-M_i}\right|
\end{equation}
where $N$ is the number of data points (forecasts) evaluated, $M_i$ is the actual biomarker value in individual $i$ in future data, and $\tilde{M}_i$ is the participant's best prediction for $M_i$.

\subsubsection{Weighted Error Score}

The weighted error score is defined as:
\begin{equation}
 WES=\frac{\sum_{i=1}^{N}\tilde{C}_i\left|\tilde{M}_i-M_i\right|}{\sum_{i=1}^{N}\tilde{C}_i}
\end{equation}
where $\tilde{C}_i$ is the participant's relative confidence in their $\tilde{M}_i$ estimate. We estimate $\tilde{C}_i$ as the inverse of the width of the 50\% confidence interval of their biomarker estimate:
\begin{equation}
\tilde{C}_i=\left(C_+-C_-\right)^{-1}
\end{equation}
where $[C-, C+]$ is the confidence interval provided by the participant.

\subsubsection{Coverage Probability Accuracy}

The coverage probability accuracy is:
\begin{equation}
CPA = |ACP - NCP| 
\end{equation}
where $NCP$ is the nominal coverage probability, the target for the confidence intervals, and $ACP$ is the actual coverage probability, defined as the proportion of measurements that fall within the corresponding confidence interval. In TADPOLE, we set $NCP$ to be 0.5, which means that ideally only 50\% of the measurements would fall inside the confidence interval. The CPA can take values between 0 and 1, and lower scores are better.

\section{Prizes}
We are extremely grateful to Alzheimer's Research UK, The Alzheimer's Society, and The Alzheimer's Association for sponsoring a prize fund of \pounds 30,000. At the time of first submission, we proposed six separate prizes, as outlined in Table \ref{tab:prizes}, but reserve the right to reallocate the prize money depending on the numbers of participants eligible for each prize. The first four are general categories (open to all challenge participants) and constitute one prize for the best forecast of each feature as well as one for overall best performance. The last two prizes are for two different student categories.

\begin{table}[h]
\centering
 \begin{tabular}{>{\centering\arraybackslash}m{1.5cm}  c  >{\centering\arraybackslash}m{3cm}  >{\centering\arraybackslash}m{3cm}}
\textbf{Prize amount} & \textbf{Outcome measure} & \textbf{Performance Metric} & \textbf{Eligibility} \\
\hline
\pounds 5,000 & Clinical status & mAUC & all \\
\pounds 5,000 & ADAS13 & MAE & all\\
\pounds 5,000 & Ventricle volume & MAE & all\\
\pounds 5,000 & Overall best & Lowest sum of ranks* & all\\
\pounds 5,000 & Clinical status & mAUC & University teams\\
\pounds 5,000 & Clinical status & mAUC & High-school teams\\
\end{tabular}
\caption[TADPOLE prize allocation scheme using funds from AD charities]{Prize allocation scheme using funds from Alzheimer's Research UK, The Alzheimer's Society and The Alzheimer's Association. There are 6 prizes awarded to different outcome measures, the last two of which are eligible only for university and high-school teams. (*) The overall best team will be the team that obtains the lowest sum of ranks in the clinical status, ADAS13 and ventricle volume categories. }
\label{tab:prizes}
\end{table}

\section{Discussion}

We have outlined the design of the TADPOLE Challenge, which aims to identify algorithms and features that can best predict the evolution of Alzheimer's disease. Challenge participants use historical data from ADNI in order to predict three key outcomes: clinical diagnosis, ADAS-Cog13 and ventricle volume. Determining which features and algorithms best predict AD evolution can aid refinement of cohorts and endpoint assessment for clinical trials, and can provide accurate prognostic information in clinical settings. 

The TADPOLE Challenge was designed to be transparent and accessible. To this end, all of our scripts are available in an open repository\footnote{TADPOLE repository: https://github.com/noxtoby/TADPOLE}. We also created a public forum\footnote{TADPOLE forum:  https://groups.google.com/forum/\#!forum/tadpolechallenge} where we answer participant questions. Finally, in order to enable participants to share algorithm performance results throughout the competition, we created a leaderboard system\footnote{Leaderboard: https://tadpole.grand-challenge.org/leaderboard/} that evaluates submissions on an existing test dataset and publishes the results live on our website.  

Going forward, we hope that by November 2018 sufficient data will be available from ADNI3 rollovers for a first meaningful evaluation of the forecasts. We plan to publish the results on the website in January 2019, and then submit a publication of the results soon after. However, we reserve the right to delay evaluation until sufficient data is available. At that time, we will also evaluate the impact and interest of the first phase of TADPOLE within the community, to guide decisions on whether to organise further submission and evaluation phases.

The fact that the D4 test set could have different properties from the training set is something that can affect the performance of certain algorithms. For example, some algorithms could perform better on different forecast time windows (short-term vs long-term) or on subjects with different properties (e.g. those with more follow-up training data vs those with less data). At the evaluation stage, we thus take into consideration doing the evaluation on different splits of the test set, in order to understand what kind of predictions algorithms perform best at. 

\section{Conclusion}

In this section I presented the TADPOLE Challenge, which aims to identify algorithms and features that best predict the evolution of subjects at risk of Alzheimer's disease. The outcomes of the challenge will be made available early in 2019, after sufficient data has been acquired. In the next chapter, I will present future work on the TADPOLE Challenge, as well as the other chapters of the thesis.

\chapter{Conclusions}
\label{chapter:conclusions}

\definecolor{darkred}{rgb}{0.7,0,0}
\newcommand{\hl}[1]{\textcolor{black}{#1}}

In this thesis I presented my work on disease progression model applications to typical Alzheimer's disease and Posterior Cortical Atrophy, as well as novel methodological developments. In this chapter I will present a summary of the thesis (section \ref{sec:conSum}), along with future research directions, both for applications to other neurodegenerative diseases (section \ref{sec:conNeu}) as well as further methodological developments (section \ref{sec:conMet}). 

\section{Summary}
\label{sec:conSum}

In chapter \ref{chapter:bck}, I first gave an overview of Alzheimer's disease (section \ref{sec:bckAd}) by describing its symptoms, disease causes and mechanisms, various risk factors involved, how it is currently diagnosed and the key biomarkers available to quantitatively measure Alzheimer's disease pathology. Afterwards, in section \ref{sec:bckProgAd} I described the progression of AD biomarkers and the Braak staging scheme. Finally, in section \ref{sec:bckPca} I performed a literature review on PCA, and described its symptoms, disease causes, diagnosis, management, neuroimaging and heterogeneity. Throughout the section, I compared and contrasted the differences between PCA and typical AD.

In chapter \ref{chapter:bckDpm}, I presented the state of the art in disease progression modelling. I started with the hypothetical model by Jack et al. \cite{jack2010hypothetical} (section \ref{sec:bckDpmHyp}), then presented early models of progression based on symptomatic groups (section \ref{sec:bckSym}), then moved to continuous models which regress against one biomarker (section \ref{sec:bckDpmReg}) and survival analysis models that compute time until an event such as clinical conversion occurs (section \ref{sec:bckDpmSur}). I then presented state of the art methods that combine multiple biomarker measurements and generally compute latent time shifts and other hidden variables. I categorised them into models based on scalar biomarker measurements (section \ref{sec:bckSca}), spatiotemporal models (section \ref{sec:bckSpa}) which model changes both in brain structure and over time, as well as mechanistic models (section \ref{sec:bckMec}) which can be used to infer underlying disease mechanisms. Finally, I presented a summary of key machine learning methods that have been frequently used in medical imaging, especially for diagnosis and prognosis (section \ref{sec:bckMac}). 

In chapter \ref{chapter:pca}, I presented a longitudinal comparison of Posterior Cortical Atrophy with typical Alzheimer's disease, analysing the progression of atrophy from MRI. I first presented the demographics (section \ref{sec:pcaParticipants}) of the cohort from the Dementia Research Centre, UK that I analysed, using data obtained by my collaborators. I then described the methodology I applied, which involved adaptations of the event-based model and the differential equation model to this specific dataset (section \ref{sec:pcaStaMet}). I showed that there were differences in the progression of brain volumes in PCA as opposed to  typical AD, where phenotype-specific areas were affected early in the disease process (section \ref{sec:pcaResPcaAd}). Moreover, I also showed that there were differences in atrophy progression in three cognitively-defined PCA subtypes, highlighting the amount of heterogeneity within PCA (section \ref{sec:pcaResPcaSub}). Finally, in section \ref{sec:pcaDis} I discussed the findings of our study, the strengths and limitations of our methods, and suggested directions for future research. 

In chapter \ref{chapter:perf}, I presented methodological advances in two disease progression models, the event-based model and the differential equation model. In section \ref{sec:perfEvalMethods}, I presented novel performance metrics that I designed, which enable us to compare the performance of these novel methods against the standard implementations. In section \ref{sec:perfRes}, I showed that novel EBM methods perform better than the standard EBM, while the novel DEM methods performs equally well to the standard method on those datasets. This also suggested that the novel metrics that we proposed are sensitive to these small changes in the EBM and DEM methodologies. 

In chapter \ref{chapter:dive} I presented Data-Driven Inference of Vertexwise Evolution (DIVE), a novel spatiotemporal disease progression model of brain pathology in neurodegenerative disorders. In section \ref{sec:diveInt} I first reviewed existing literature and motivated the need for such a model, due to the presence of dispersed atrophy patterns in AD caused by disruption in underlying brain connectomes \cite{seeley2009neurodegenerative}. I then presented the mathematical formulation of DIVE in section \ref{sec:diveMet}. I performed simulations to show that DIVE can reliably estimate cluster assignments, trajectory parameters and subject time-shifts in the presence of ground truth (section \ref{sec:diveSimulations}). Afterwards, I tested DIVE on four different datasets with distinct diseases (typical AD and PCA) and modalities (MRI and PET), and showed that it can recover meaningful patterns of pathology, which agree with previous findings in the literature, but offer us more spatial resolution, along with estimates of biomarker dynamics and subject-specific time shifts.  Finally, in section \ref{sec:diveEval} I validated DIVE by showing that the estimated clusters and trajectories are robust under 10-fold cross-validation, and that it has favourable predictive performance compared to simpler models.  

In chapter \ref{chapter:dkt} I presented Disease Knowledge Transfer (DKT), a novel model that robustly learns patterns of progression from several types of dementia combined. This allows the inference of biomarker signatures in rare, atypical types of dementia, which is otherwise difficult due to the lack of multimodal, longitudinal data. In section \ref{sec:dktMet}, I presented the DKT framework, which I designed to be flexible, allowing one to plug-in any disease progression model within each disease-agnostic and disease-specific unit. Using simulations, I then showed in section \ref{sec:dktResSyn} that DKT can accurately estimate biomarker trajectories in two distinct diseases, and even when there is a lack of data for one of the diseases, through correlations with other known markers. When applied to patient data (section \ref{sec:dktResTadDrc}), I showed that DKT can estimate plausible biomarker trajectories, and showed that is has favourable performance compared to standard models. Compared to previous deep transfer learning approaches, DKT is also interpretable and can predict the future evolution of subjects at risk of neurodegenerative diseases.

In chapter \ref{chapter:tadpole}, I presented the design of the TADPOLE Challenge, which aims to identify algorithms and features that can best predict the progression of subjects at risk of AD. The challenge was organised jointly by myself and my collaborators, and we had 33 international teams who made more than 90 submissions. For the challenge, I helped write the website, assembled the main training dataset, built a live leaderboard system that allowed instant evaluation of the predictions, and promoted the competition at various conferences. I also wrote the paper describing the design of the challenge \cite{marinescu2018tadpole}.

\section{Future Research Directions}
\label{sec:conFut}

There are several future research directions that can be pursued after this work. In section \ref{sec:conNeu}, I will present further applications of the methods I developed to neurodegenerative diseases, while in section \ref{sec:conMet} I will provide suggestions of improvements to the methods developed, along with ideas for new methods.

\subsection{Applications to Neurodegenerative Diseases}
\label{sec:conNeu}

The application of the models we developed to different neurodegenerative diseases is important for several key reasons. First of all, they allow us to understand underlying mechanisms underpinning phenotypic heterogeneity within PCA and the other diseases, which can provide more informed drug targets. Secondly, they enable better stratification and selection of endpoints for clinical trials. Third, they can be used to inform health policy, by predicting the future evolution of subjects who are at risk of developing such diseases. 

\subsubsection{Posterior Cortical Atrophy}
\label{sec:conTyp}

There are several further questions that need to be answered regarding the progression of Posterior Cortical Atrophy versus typical Alzheimer's Disease. To continue the work presented in chapter \ref{chapter:pca}, one can answer the following questions\footnote{The last three questions have been suggested by Sebastian Crutch}:
\begin{itemize}
 \item Differences in sub-populations: Are there differences in the estimated biomarker ordering of abnormality and trajectories for various sub-populations, such as APOE $\epsilon4$ positive vs negative or amyloid positive vs negative?
 \item Imaging predicting cognition: If we split the PCA population based on the discrepancy between occipital-hippocampal values at baseline, does that predict distinct patterns of cognitive impairment? One can hypothesise that relatively lower occipital volumes for basic visual-PCA predict early visual deficits, with memory deficits later on. On the other hand, relatively lower hippocampal volumes would predict early multi-domain cognitive deficits, with visual deficits later on.
 \item Relationship between posterior and anterior patterns of atrophy: Does greater inferior posterior atrophy predict greater inferior anterior atrophy, and vice-versa? Moreover, based on the cognitively-defined subgroups, is atrophy in dorsolateral prefrontal lobe different in the three cognitive subgroups, in the following manner: (highest) space $>$ object $>$ vision (lowest)? Similarly, is inferior prefrontal atrophy different between the three subgroups in the following manner: (highest) object $>$ space $>$ vision (lowest)?
 \item Asymmetry analysis: Are the PCA patterns of atrophy asymmetric? Previous analyses suggested relatively greater atrophy in the right superior parietal lobe, but this may not be the case for all patients, and cognitive tests suggest at least a minority have left-predominant atrophy.
\end{itemize}

 The above questions can explored not only using the EBM and DEM, but also using DIVE and DKT models. Moreover, another research direction would be to apply the models on biomarkers other than MRI brain volumes, such as cortical thickness from MRI, PET biomarkers (amyloid, tau, FDG), as well as DTI biomarkers (FA, MD, AD). The multimodal biomarker trajectories estimated in PCA with the EBM, DEM and DIVE models can also be compared with the ones inferred by DKT. 

\subsubsection{Typical Alzheimer's disease}

Several analyses can also be done to further understand typical Alzheimer's disease. \hl{For example, using DIVE one could test if the reason why disconnected vertices cluster together is due to underlying structural or functional connections. Such a hypothesis could be tested by computing the modularity (or other index of connectivity) of DIVE clusters on the weighted graph of white-matter connections between different vertices, where the weights are given by the number of tracts connecting the two vertices. The modularity coefficient could then be compared against a well-defined null hypothesis. This would help understand to what extent disruption of underlying connectomes affects neurodegeneration \cite{seeley2009neurodegenerative}. }

\subsubsection{Familial Alzheimer's disease}

The models and techniques we have developed here can also be applied to familial AD, using cohorts such as the Dominantly Inherited Alzheimer's Network (DIAN). So far, the EBM and DEM models have been applied to study familial AD \cite{oxtoby2018}, but more complex models are yet to be tested. 

Some adaptation of our the models should be done when modelling familial AD. As opposed to sporadic AD, in familial AD we have a relatively reliable estimate of the subjects' disease onset based on familial age of onset. Therefore, models such as DIVE and DKT should be adapted by setting a stronger prior distribution on the subjects' time-shift, centred on their parental age of onset, and having a standard deviation of around +/- 5 years, like the approach of \cite{oxtoby2018}. On the other hand, familial AD cohorts can also be used for model validation, by comparing the subjects' estimated time-shifts against the parental age of onset, this time using an uninformative prior on the subjects' time-shift.

\subsubsection{Other Alzheimer's variants}

The EBM, DEM, DIVE and DKT methodologies can be further applied to other types of Alzheimer's variants, such as the focal temporal lobe dysfunction, pure-amnestic AD with episodic memory impairment \cite{butters1996focal} or language variant AD \cite{green1990progressive, greene1996alzheimer, galton2000atypical}.

\subsubsection{Frontotemporal dementia}

Our models can also be applied to study the progression of Fronto-temporal dementia. Fronto-temporal dementia (FTD) is a clinically and pathologically heterogeneous group of non-Alzheimer's dementias that affect frontal and temporal lobes \cite{warren2013frontotemporal}. There are three main clinical syndromes: behavioural-variant FTD characterised by behavioural changes, primary progressive aphasia characterised by impaired speech, and semantic dementia, characterised by impaired semantic memory \cite{warren2013frontotemporal}. FTD also has a strong genetic component, due to mutations in the microtubule associated protein tau (MAPT), progranulin (GRN) and C9ORF72 genes. So far, an extension of the event-based model, which estimates multiple progression patterns in sub-populations, has been applied to FTD \cite{young2018uncovering}. Applying spatiotemporal models such as DIVE or multi-disease models such as DKT would help understand the heterogeneity and progression of FTD, find early biomarkers and allow better stratification in FTD clinical trials. Moreover, the heterogeneity present in FTD, combined with genetic information, can be used to further validate the DKT model by checking how robustly it can transfer biomarker trajectories between different FTD genetic groups. 

\subsubsection{Multiple Sclerosis}

Another disease where we can apply our models is Multiple Sclerosis (MS), which is a chronic autoimmune, inflammatory neurological disease that attacks myelinated axons and causes neurodegeneration \cite{goldenberg2012multiple}. MS can be of several types: (a) relapsing-remitting MS, marked by alterating periods of relapses (i.e. exacerbations of symptoms) and remission, (b) primary-progressive MS characterised by gradual worsening of symptoms, (c) secondary progressive MS where progressive MS develops in relapsing-remitting patients and (d) progressive-relapsing MS characterised by progressive disease with intermittent flare-ups of worsening symptoms \cite{goldenberg2012multiple}. When applying our models to MS data, some care needs to be taken as all our models assume monotonic biomarker trajectories. If there are biomarkers that are non-monotonic, the models can be extended to use non-parametric trajectories that enable non-monotonicity. However, this requires stronger priors on the subject-specific time-shifts and on the noise variable, otherwise the model will not be identifiable. In terms of the  disease progression models applied on MS, so far the event-based model has been applied by Eshaghi et al. \cite{eshaghi2018progression}. However, other data-driven disease progression models are yet to be tested on MS.

\subsubsection{Parkinson's Disease}

Our models can also be applied to study the progression of Parkinson's disease (PD). PD is a neurodegenerative disease characterised by atrophy in the substantia nigra, dopamine deficiency and aggregates of $\alpha$-synuclein \cite{poewe2017parkinson}. While the disease is diagnosed clinically based on bradykinesia and other motor features, PD also causes multi-domain cognitive decline \cite{caballol2007cognitive, poewe2017parkinson}.  It would be particularly useful to apply our models to estimate the progression of PD, help distinguish between PD and other types of degenerative parkinsonism, and also to identify early markers in prodromal disease stages, allowing novel disease-modifying therapies to be started as early as possible \cite{poewe2017parkinson}.

\subsubsection{Huntington's Disease}

Our models can also be applied to study the progression of Huntington's disease (HD). HD is a rare neurodegenerative disease characterised by jerky, involuntary movements, behavioural and psychiatric disturbances \cite{roos2010huntington}. The disease is caused by an elongated CAG repeat (36 repeats or more) on the short arm of chromosome 4p16.3 in the Huntingtin gene, with longer repeats causing earlier onset of disease \cite{roos2010huntington}. While different types of MRI \cite{georgiou2008magnetic} and PET \cite{feigin2001metabolic} imaging modalities have been central in identifying structural and functional abnormalities, there is still a need to identify quantitative biomarkers for early disease detection and for mapping its evolution\cite{georgiou2008magnetic}. In terms of data-driven disease progression models, only the event-based model has been applied to HD \cite{fonteijn2012event, wijeratne2018image}.

\subsection{Applications to Clinical Trials}
\label{sec:conCli}

The disease progression models that we have developed can be used to aid clinical trials in several ways. One area of application is for selecting the right subjects for enrolling in the clinical trial. For example, based on a few initial measurements such as cognitive tests and an MRI scan, our models can predict which subjects will develop dementia, along with the exact type of dementia, within a certain time window. \hl{This can help select a homogeneous group of patients for a clinical trial, which are otherwise estimated to develop the same type of dementia and at the same age/follow-up time. }

Another key area of application is evaluation of the effects of putative drugs. The subject-specific time shifts, estimated based on multimodal imaging measures, could provide more robust measures of disease stage compared to single imaging or cognitive markers. Finally, our models could also be applied as a safety endpoint in clinical trials, where they could detect very early changes that might be due to adverse effects of a drug, before the appearance of symptoms. Such early detection of adverse effects could be used to suggest the interruption of the clinical trial \cite{cash2014imaging}. 

\subsection{Methodological Developments}
\label{sec:conMet}

Further research can also focus on improvements in the models that I have developed, along with development of new models. Such methodological improvements are very important, as they enable understanding more complex disease mechanisms such as associations with genetics, and will enable more accurate predictions resulting in improved stratification for clinical trials. In the following sections I will present several key directions for further work, and will suggest concrete steps towards them.

\subsubsection{Towards Personalised Predictions}
\label{sec:conMetPer}
One key direction of these models is to enable them to perform personalised predictions, which will further enable personalised treatments to be delivered. To enable this in models such as DIVE or DKT, one can estimate subject-specific trajectories by adding random effects to the population trajectory. This will enable more accurate predictions and account for the heterogeneity in the modelled diseases. However, more longitudinal data is required for such personalised predictions, and model identifiability needs to be ensured.

Another extension to our models that can aid personalised predictions is to model distinct progressions for different sub-populations, in a data-driven way as done by the SuStaIn model \cite{young2018uncovering}. \hl{More precisely, one can assume that the population is made of unknown subgroups with different progressions, and each subject will have an associated latent variable denoting the subgroup it belongs to. This can still be optimised with the Expectation-Maximisation framework \cite{bishop2007pattern}.}

While estimating discrete subgroups with different progressions works well for some diseases such as FTD due to mutations in a few key genes, this might not work that well for diseases such as PCA or Huntington's, where it is believed that there is a continuum of phenotypic variability. In this case, our models should be extended to estimate a  continuous latent dimension of heterogeneity. This can be further extended to more than one latent dimension, to account for mixed pathologies, where each dimension could correspond to a different underlying pathology. While some studies have shown that a large proportion of healthy individuals and patients who receive a clinical diagnosis of AD actually have underlying mixed pathologies \cite{kovacs2013non, james2016tdp}, this analysis requires both in-vivo longitudinal data along with post-mortem pathological confirmation. This has recently become available in ADNI, which now has autopsy data for 56 AD and 52 age-matched controls \cite{trojanowski2010update}. 

\subsubsection{Spatio-temporal Modelling}
\label{sec:conMetSpa}

\hl{The spatio-temporal DIVE model we proposed can be further extended to cluster points on the brain based on a multi-modal signature. This could be done by extending the likelihood model from a single univariate Gaussian distribution to a multivariate Gaussian distribution with a small covariance matrix.  Parameter estimation can still be done using the Expectation-Maximisation framework. Such multimodal clusters could give further insights into the mechanisms underpinning Alzheimer's disease and other neurodegenerative diseases.}

The DKT model that we proposed can also be extended to estimate spatio-temporal changes in the brain. Such a spatiotemporal DKT model would be able to synthesize, based on e.g. a structural MRI scan, other types of scans such as PET or DTI, in patients with rare dementias, where there is a lack of such multimodal data. \hl{This could be done using deep learning methods, where the neural network could have, for each brain region independently, a shared 3D disease agnostic unit encompassing multimodal pathology across all diseases modelled. These shared units will then be used to estimate disease-specific dynamics by redirecting the training data along different pipelines. The disease specific parts could be implemented in an unsupervised manner (e.g. with autoencoder) or in a supervised manner, to predict e.g. cognitive tests. }

\subsubsection{Modelling Disease Mechanisms}
\label{sec:conMetMec}

One potential direction of research towards understanding underlying disease mechanisms is to model the dynamics of pathogenic proteins. The work of Raj et al. \cite{raj2012network, raj2015network} and Georgiadis et al. \cite{georgiadis2018computational} can be used as a starting point. Several concrete steps would be to extend the network diffusion model \cite{raj2015network} to estimate latent subject-specific time-shifts as in the work of Donohue et al. \cite{donohue2014estimating}. The model by \cite{georgiadis2018computational}, while simulating far more complex dynamics, needs to be validated using in-vitro studies, as well as using amyloid and tau PET imaging. 

One limitation of the diffusion models developed so far \cite{raj2015network, georgiadis2018computational} is that they assume a static structural connectome throughout the disease process. To this end, these models should be extended to account for changes in the connectome structure over the disease time-course, such as breakdown of key links or nodes, based on different kinds of selective vulnerabilities, e.g. those suggested by Zhou et al. \cite{zhou2012predicting}.

\subsubsection{Incorporating genetic data}
\label{sec:conMetGen}

Another key direction of further research is to connect our models with genetic data. In ADNI, genetic data is available to perform genome-wide association studies (GWAS). In particular, GWAS can help identify novel loci and genes involved in AD by finding associations with more robust and quantitative endophenotypes derived from imaging and other types of biomarkers. For example, very recent work by Sclesi et al. \cite{scelsi2018genetic} has used the disease progression model by Donohue et al. \cite{donohue2014estimating} to estimate a quantitative multimodal endophenotype from MRI and PET images, which identified a novel locus. Such associations were not significant for simpler hippocampal volume or cortical amyloid markers on their own \cite{scelsi2018genetic}. Extending such work by adding other types of biomarker data available in ADNI can identify further loci. Moreover, associations can also be found between genes and various regions in the brain, and even with pathology identified at voxelwise level from DIVE, using an approach similar to \cite{stein2010voxelwise}.

\subsubsection{Incorporating data from other datasets}
\label{sec:conMetOth}

Some of our models can be further extended to incorporate data from other dementia or normal ageing datasets. This can be done in a manner similar to DKT, but other transfer-learning approaches can also be used to this end. Some datasets that can be used include observational studies for sporadic AD (e.g. AIBL \cite{ellis2009australian}) familial AD (e.g. DIAN \cite{morris2012developing}), Multiple Sclerosis, Parkinson's disease (PPMI \cite{marek2011parkinson}), Huntington's disease (TRACK HD \cite{tabrizi2009biological}). For normal ageing, datasets such as the Rotterdam study \cite{ikram2017rotterdam} can also be used. 

Our models can also be further extended to use novel biomarker data from wearables or internet of things (IoT) devices. Data from smart watches or body sensors \cite{hsu2014gait}, eye-tracking devices \cite{hutton1984eye} or speech recorders \cite{hoffmann2010temporal} can be sensitive to dementia, and thus can be used to identify early signs and track its progression. 

\subsubsection{Estimating better features}
\label{sec:conMetFea}

An interesting direction of further research is to extract better features for scalar disease progression models such as EBM, DEM or DKT. This can be done by incorporating feature learning as part of the model itself, as in the case of DIVE. However, other methods based on dimensionality reduction using Principal Component Analysis or t-distributed Stochastic Neighbour Embedding (t-SNE) \cite{hinton2003stochastic} can also be used to extract more robust features by projecting the high-dimensional data into a lower-dimensional space. \hl{We hypothesize that  models with extracted features could perform better than vanilla models because it is harder for them to overfit the data.} Finally, deep learning approaches can also be used to learn complex non-local features from images, while also modelling the progression of the disease.

\subsection{Model Evaluation}
\label{sec:conEva}

Further work can also be done on model evaluation, by extending the TADPOLE Challenge. \hl{While submissions mostly used extracted features from imaging data, more complex non-local features can be extracted and used by spatiotemporal models or deep learning methods. The TADPOLE Challenge can also be extended to attempt to predict other types of target variables such as PET or DTI markers.} Yet another direction is to organise a competition similar to TADPOLE for AD related dementias, such as Posterior Cortical Atrophy, Frontotemporal Dementia, Huntington's disease, Parkinson's disease and Multiple Sclerosis.

Another different research direction in model evaluation is to evaluate models based on simulated brain images, with more reliable ground truth compared to patient data. Simulators based on biophysical principles such as \cite{khanal2016biophysical} can be used to this end, which generate realistic brain MRI images for a given spatial pattern of atrophy. Models can be evaluated in several tasks: prediction of future biomarkers or spatial structure at both population-level and subject-level, differential diagnosis, as well as disease staging. 

\appendix


\chapter[Longitudinal Neuroanatomical Progression of PCA]{Longitudinal Neuroanatomical Progression of Posterior Cortical Atrophy}
\label{sec:adni_extra_appendix}


\newcommand*{\scaleLabelImg}{0.7}
\begin{figure}
  \centering
  \includegraphics*[scale=\scaleLabelImg]{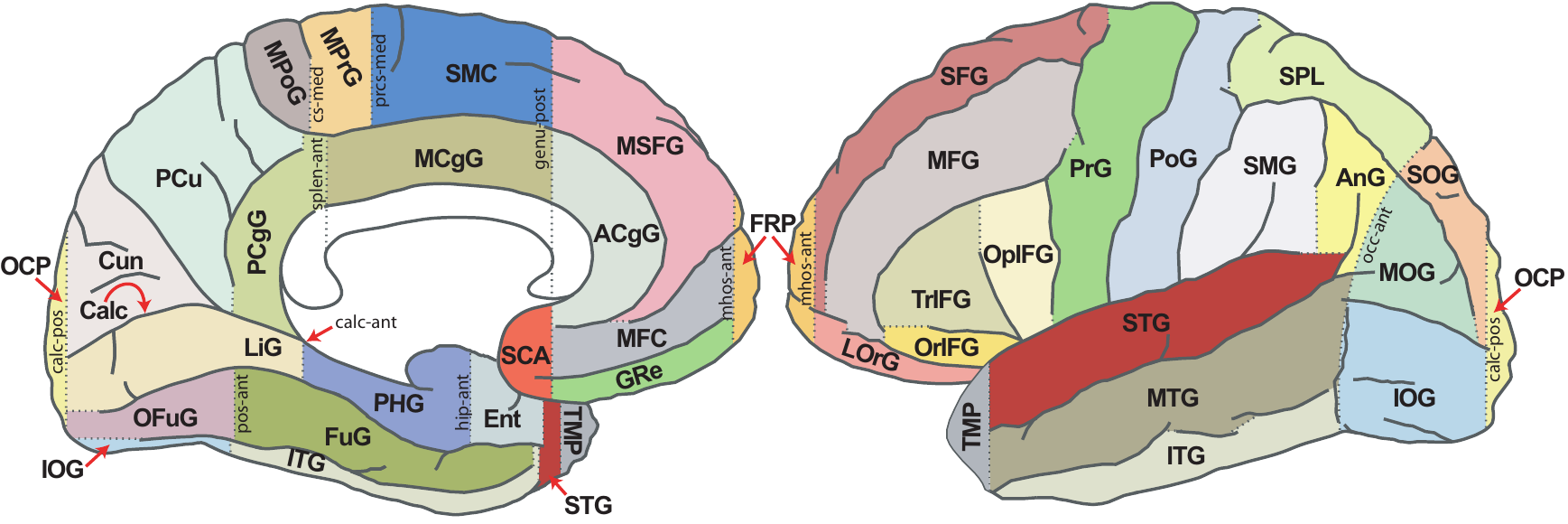}
  \caption[Labels of the different areas analysed in the EBM progression snapshots]{Labels of the different areas analysed in the EBM progression snapshots from chapter \ref{chapter:pca}. }
  \label{fig:ebmSnapLabels}
\end{figure}

Description of the labels shown in figure \ref{fig:ebmSnapLabels}:
\begin{itemize}
\item Frontal Lobe (FL)
        \begin{itemize}
        \item Lateral Surface
            \begin{itemize}
            \item Frontal Pole (FRP)
            \item Superior Frontal Gyrus (SFG)
            \item Middle Frontal Gyrus (MFG)
            \item Opercular part of the Inferior Frontal Gyrus (OpIFG)
            \item Orbital part of the Inferior Frontal Gyrus (OrIFG)
            \item Triangular part of the Inferior Frontal Gyrus (TrIFG)
            \item Precentral Gyrus (PrG)    
            \end{itemize}
        \item Medial Surface
            \begin{itemize}
            \item Superior Frontal Gyrus, medial segment (MSFG)
            \item Supplementary Motor Cortex (SMC)
            \item Medial Frontal Cortex (MFC)
            \item Gyrus Rectus (GRe) 
            \item Subcallosal Area (SCA)
            \item Precentral Gyrus (MPrG)
            \end{itemize}
        \item Inferior Surface
            \begin{itemize}
            \item Anterior Orbital Gyrus (AOrG)
            \item Medial Orbital Gyrus (MOrG)
            \item Lateral Orbital Gyrus (LOrG)
            \item Posterior Orbital Gyrus (POrG)
            \end{itemize}
        \item Opercular Region 
            \begin{itemize}
            \item Frontal Operculum (FO)
            \item Central Operculum (CO)
            \item Parietal Operculum (PO)
            \end{itemize}
        \item Insular Region
            \begin{itemize}
            \item Anterior Insula (AIns)
            \item Posterior Insula (PIns)
            \end{itemize}
        \end{itemize}
        
    \item Temporal Lobe (TL)
        \begin{itemize}
        \item Lateral Surface
            \begin{itemize}
            \item Temporal Pole (TMP)
            \item Superior Temporal Gyrus (STG)
            \item Middle Temporal Gyrus (MTG)
            \item Inferior Temporal Gyrus (ITG)
            \end{itemize}
        \item Supratemporal Surface
            \begin{itemize}
            \item Planum Polare (PP)
            \item Transverse Temporal Gyrus (TTG)
            \item Planum Temporal (PT)
             \end{itemize}
        \item Inferior Surface
            \begin{itemize}
            \item Fusiform Gyrus (FuG)
            \end{itemize}
        \end{itemize}
    \item Parietal lobe (PL)
        \begin{itemize}
        \item Lateral Surface
            \begin{itemize}
            \item Postcentral Gyrus (PoG)
            \item Supramarginal Gyrus (SMG)
            \item Superior Parietal Lobule (SPL)
            \item Angular Gyrus (AnG)
            \end{itemize}
        \item Medial Surface
            \begin{itemize}
            \item Postcentral Gyrus, medial segment (MPoG)
            \item Precuneus (PCu)
            \end{itemize}
        \end{itemize}
    
    \item Occipital Lobe (OL)
        \begin{itemize}
        \item Lateral Surface
            \begin{itemize}
            \item Superior Occipital Gyrus (SOG)
            \item Inferior Occipital Gyrus (IOG)
            \item Middle Occipital Gyrus (MOG)
            \item Occipital Pole (OCP)
            \end{itemize}
        \item Inferior Surface
            \begin{itemize}
            \item Occipital Fusiform Gyrus (OFuG)
            \end{itemize}
        \item Medial Surface
            \begin{itemize}
            \item Cuneus (Cun)
            \item Calcarine Cortex (Calc)
            \item Lingual Gyrus (LiG)
            \end{itemize}
        \end{itemize}
    
    \item Limbic Cortex (LC)
        \begin{itemize}
        \item Cingulate Cortex
            \begin{itemize}
            \item  Anterior cingulate gyrus (ACgG)
            \item Middle cingulate gyrus (MCgG)
            \item Posterior cingulate gyrus (PCgG)
            \end{itemize}
        \item Medial Temporal Cortex
            \begin{itemize}
            \item Parahippocampal Gyrus (PHG)
            \item Entorhinal Area (Ent)
            \end{itemize}
        \end{itemize}
    
\end{itemize}

\begin{figure}
\centering
  \begin{subfigure}{\textwidth}
  \centering
 \textbf{\large{\mbox{Posterior Cortical Atrophy}}}
 
 \includegraphics[width=0.7\textwidth,trim=100 30 0 50,clip]{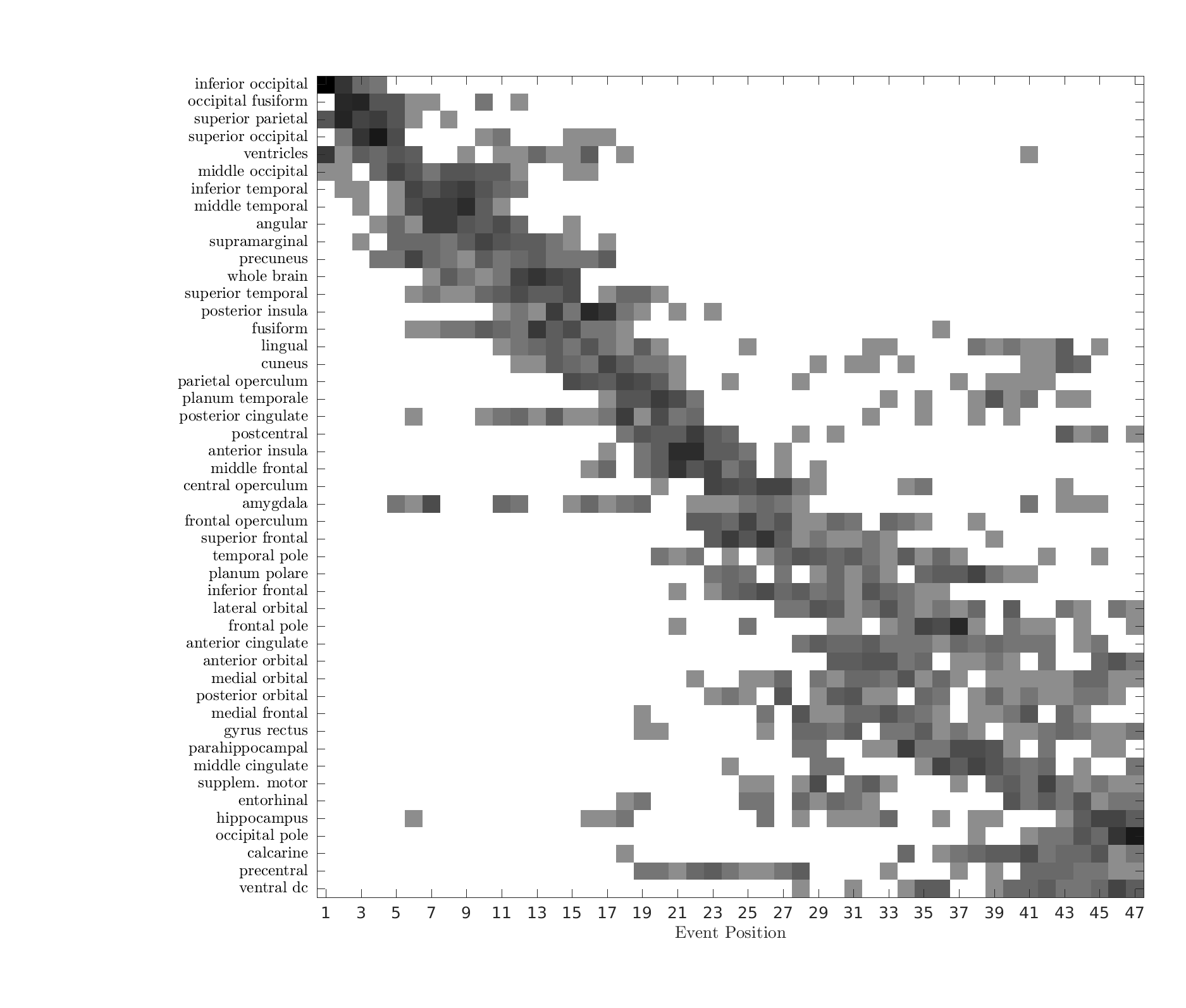}
 \end{subfigure}
 \vspace{1em}
 
 \begin{subfigure}{\textwidth}
  \centering
  \textbf{\large{\mbox{Typical Alzheimer's Disease}}}
  
 \includegraphics[width=0.7\textwidth,trim=100 30 0 50,clip]{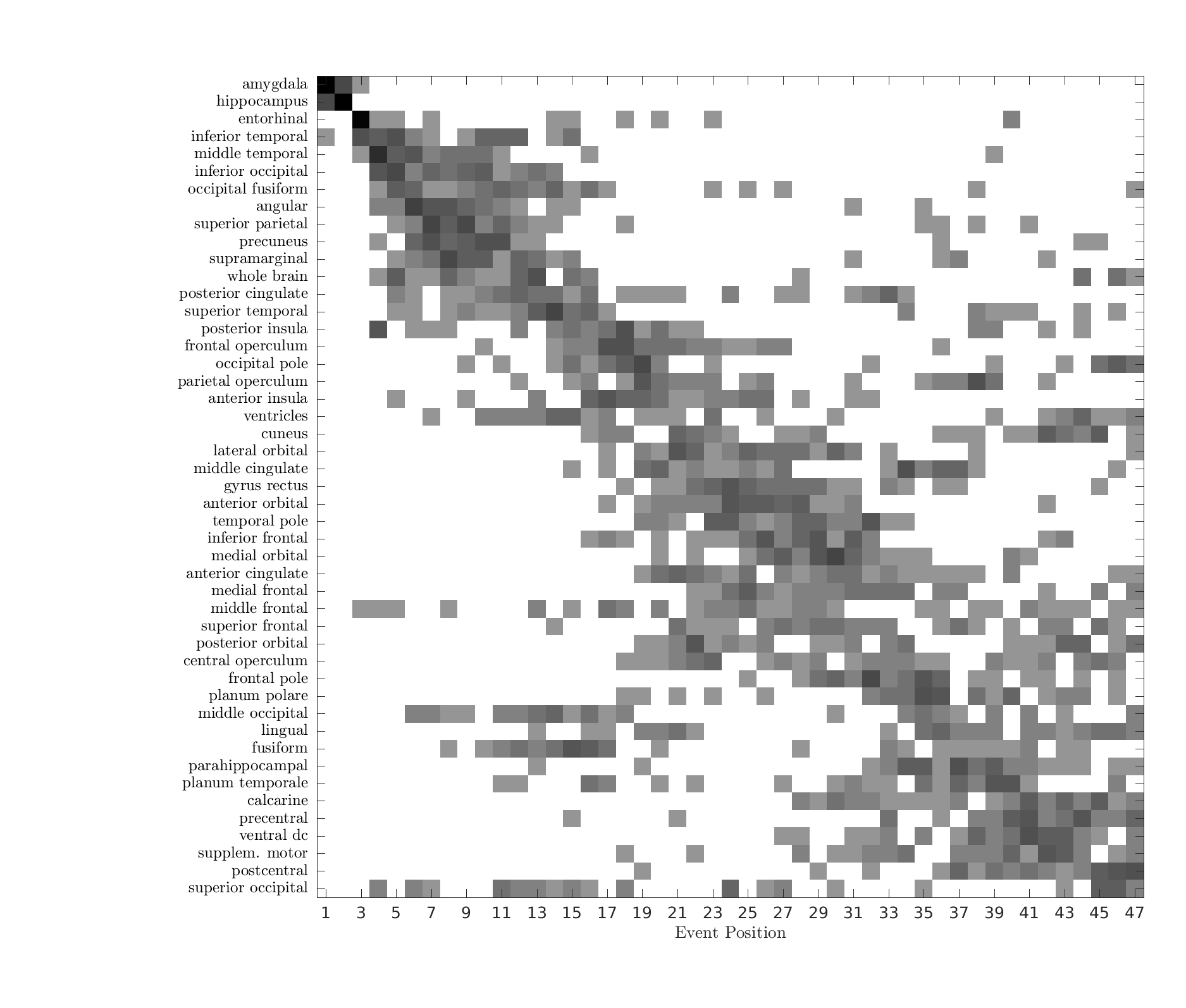}
 \end{subfigure}
 \caption[EBM bootstrap samples of the atrophy sequence for PCA and tAD]{Bootstrap samples of the atrophy sequence as estimated by the event-based model, for the PCA and typical AD cohorts. The maximum likelihood sequences were estimated using the EBM from 100 bootstrap datasets, with replacement, stratified by diagnosis.}
 \label{fig:bootPosVarAllPcaAd}
\end{figure}

\begin{figure}
\centering
  \begin{subfigure}{\textwidth}
  \centering
 \textbf{\large{\mbox{Posterior Cortical Atrophy}}}
 
 \includegraphics[width=0.7\textwidth,trim=50 30 0 30,clip]{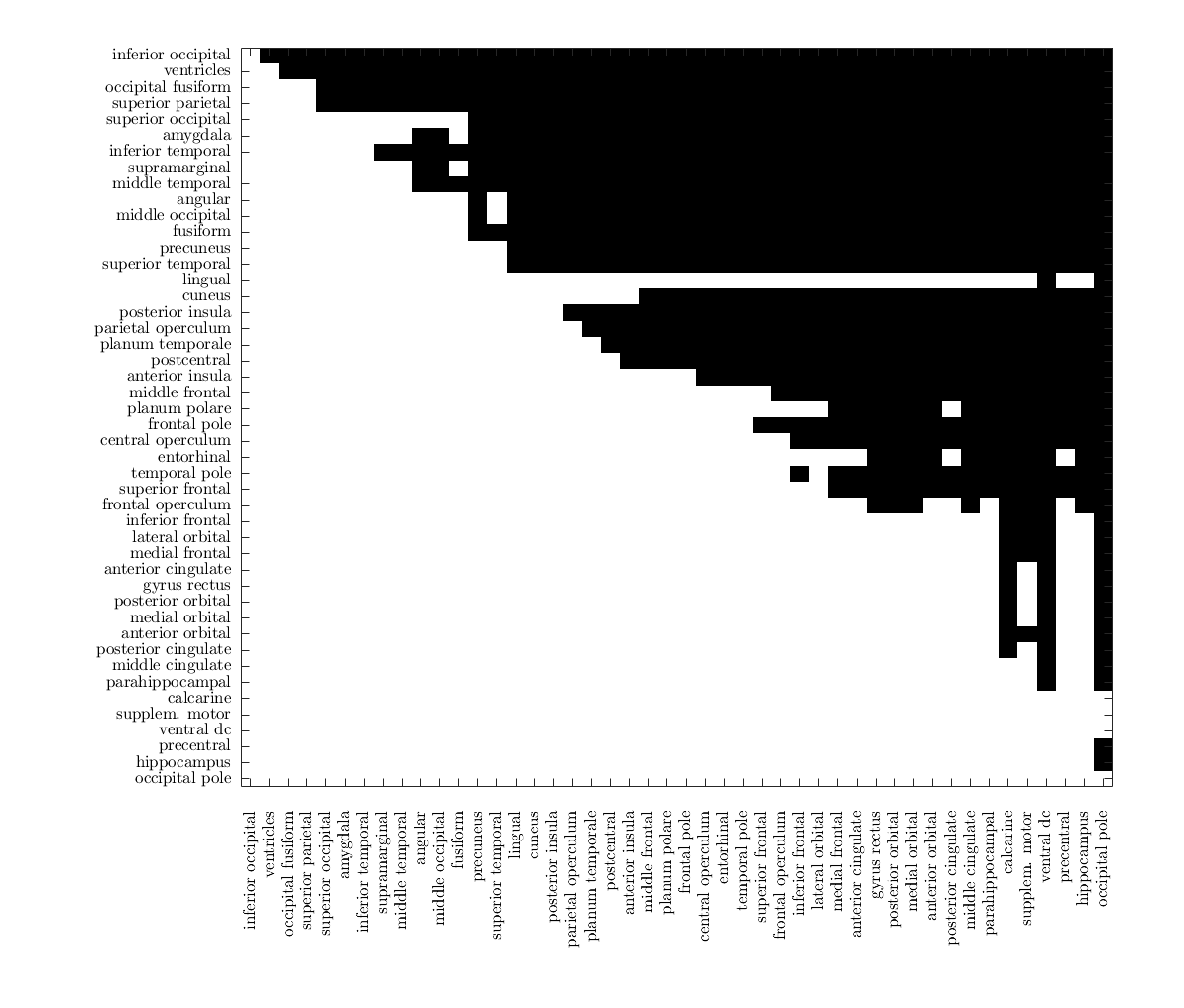}
 \end{subfigure}
 \vspace{1em}
 
 \begin{subfigure}{\textwidth}
  \centering
  \textbf{\large{\mbox{Typical Alzheimer's Disease}}}
  
 \includegraphics[width=0.7\textwidth,trim=50 30 0 30,clip]{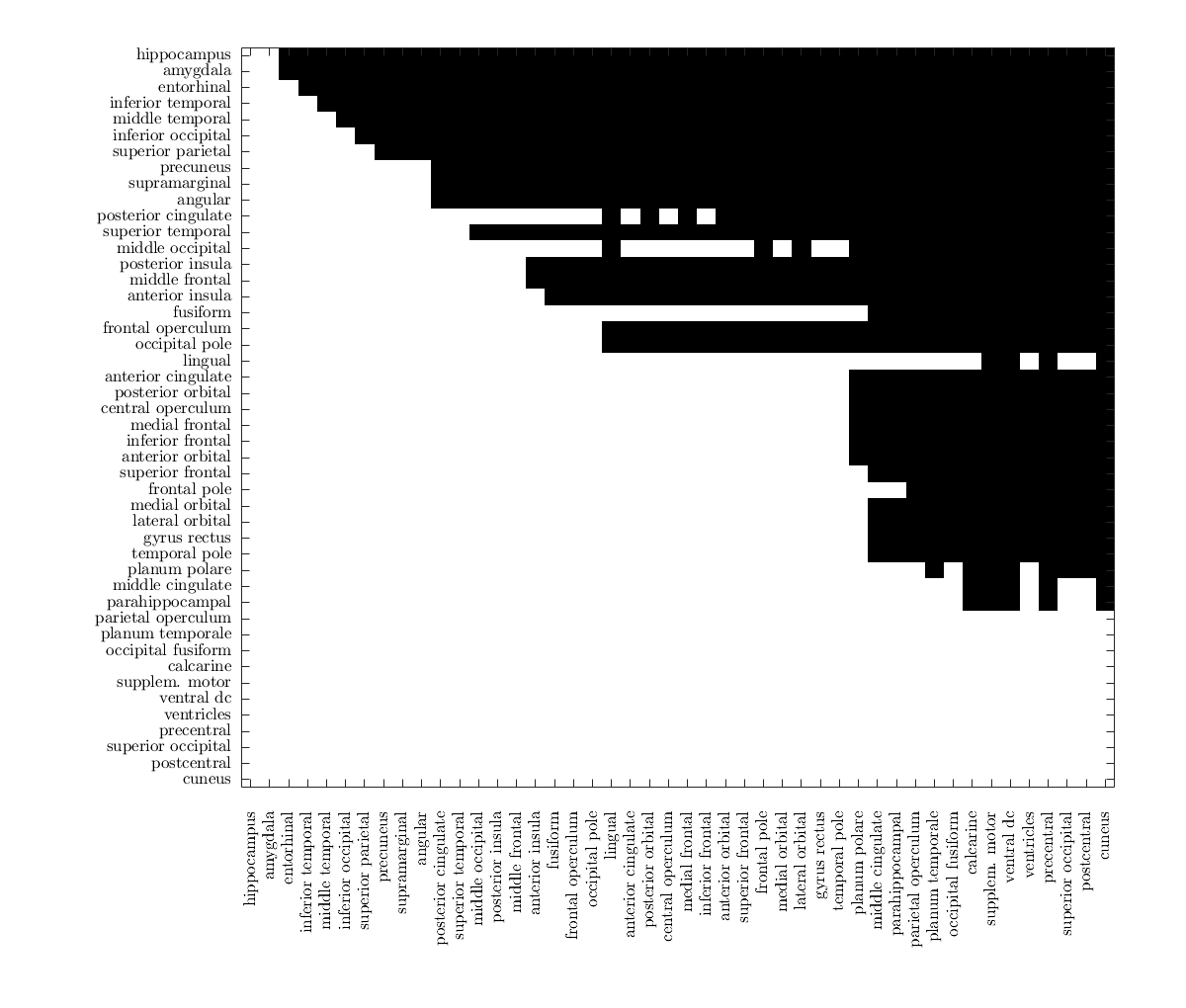}
 \end{subfigure}
 \caption[Hypothesis testing of ordering of events within PCA and tAD]{Hypothesis testing of ordering of events within PCA (top) and typical AD (bottom). We sampled 10,000 sequences from the EBM posterior using MCMC sampling and only kept every 1/100 in order to remove correlation between samples.  We applied the non-parametric paired Wilcoxon signed rank test for every pair of biomarkers (x,y). The null hypothesis is defined as H0: event A (Y-axis) becomes abnormal at the same time as event B (X-axis), while the alternative hypothesis H1: event A (Y-axis) become abnormal before event B (X-axis). The black squares show the pair of biomarkers where the null hypothesis was rejected at alpha=0.05/(N*(N-1)/2), thus surviving Bonferroni correction.}
  \label{fig:statTestPcaAd}
\end{figure}

\begin{figure}
\centering
  \begin{subfigure}[t]{0.48\textwidth}
  \centering
 \textbf{\large{\mbox{Basic visual impairment group}}}
    \includegraphics[width=1.2\textwidth,trim=50 30 0 50,clip]{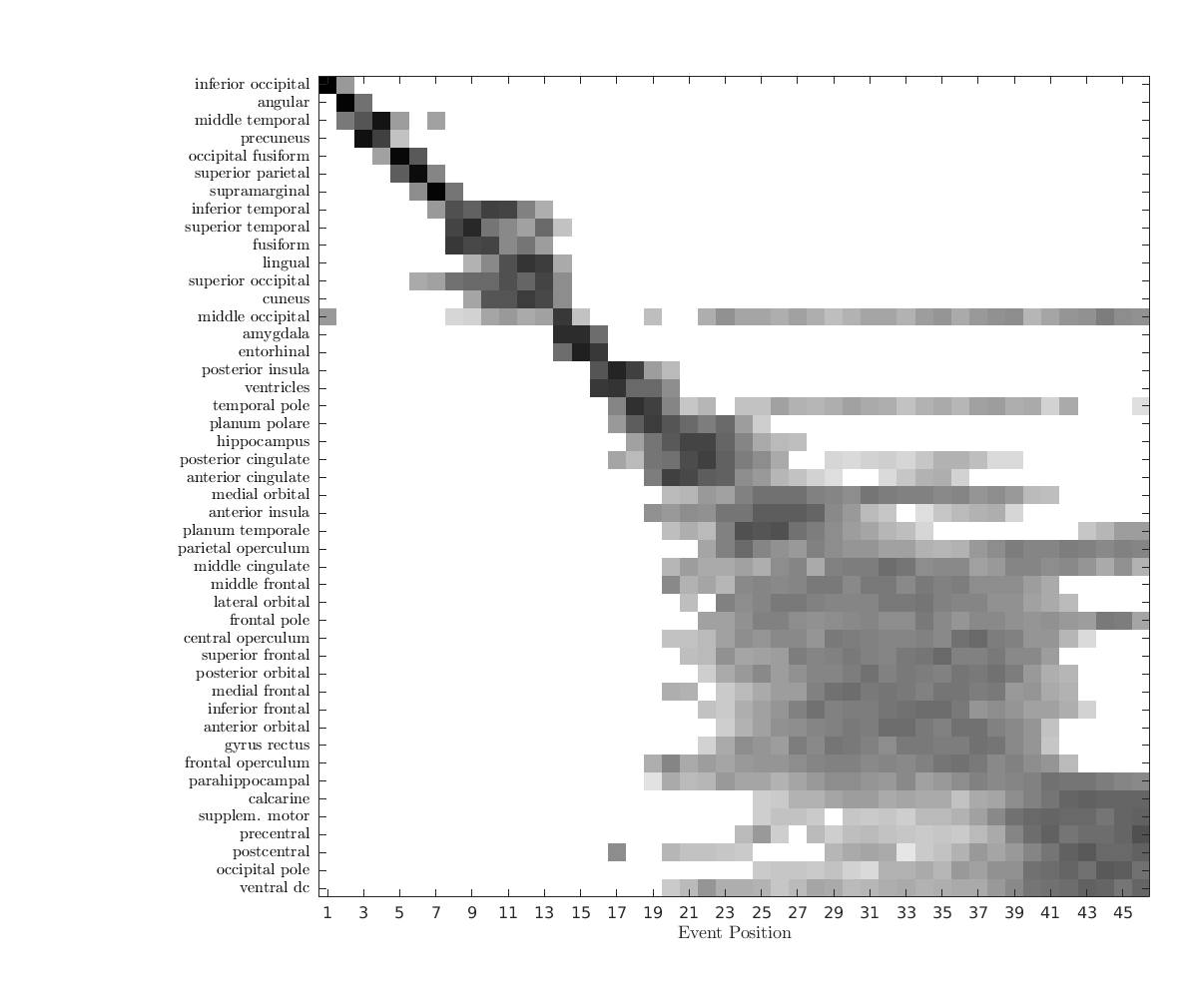}
 \end{subfigure}
 
  \begin{subfigure}[t]{0.48\textwidth}
  \centering
 \textbf{\large{\mbox{Space perception impairment group}}}
 \includegraphics[width=1.2\textwidth,trim=50 30 0 50,clip]{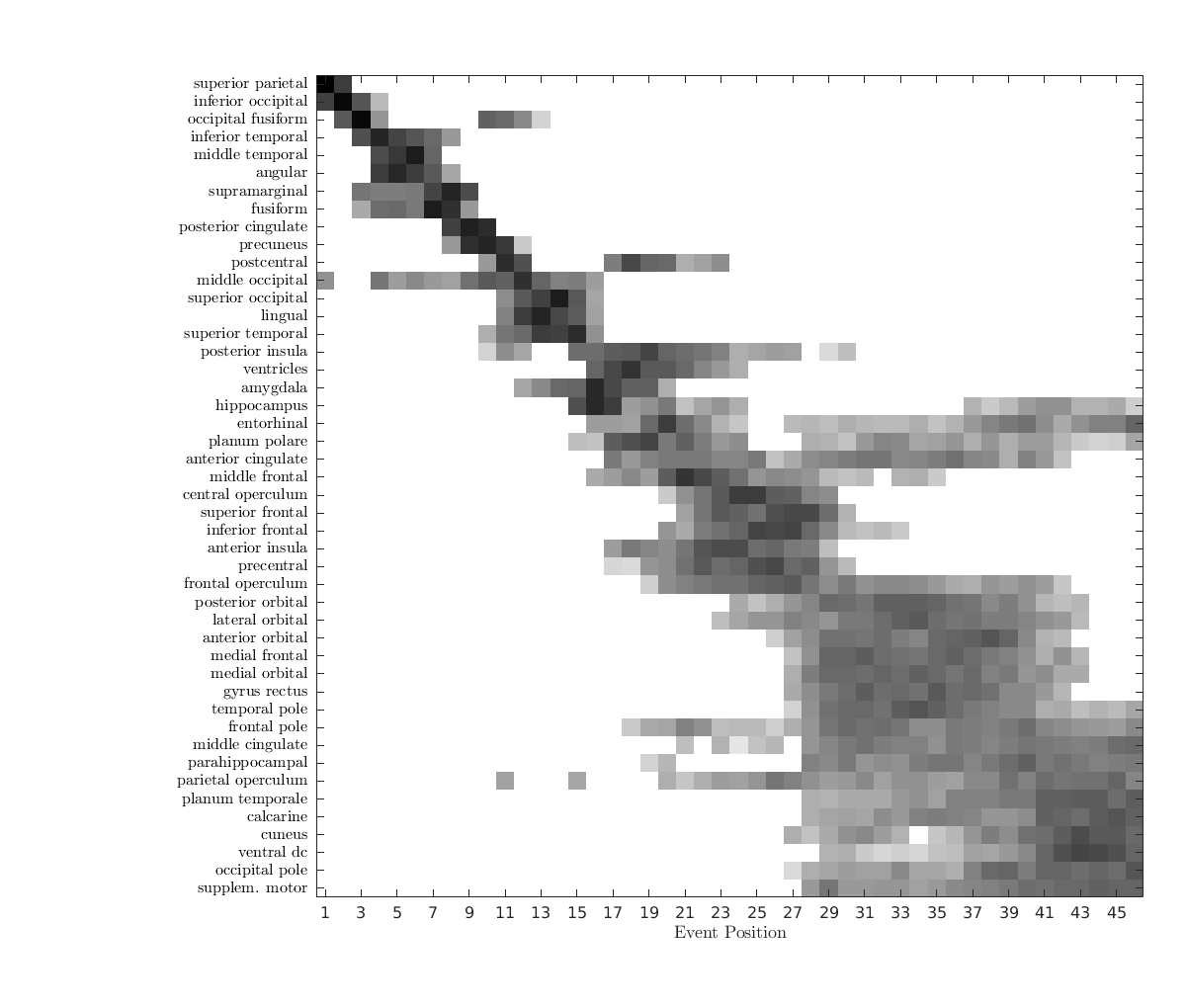}
 \end{subfigure}

\begin{subfigure}[t]{0.48\textwidth}
\centering
   \textbf{\large{\mbox{Object perception impairment group}}}
 \includegraphics[width=1.2\textwidth,trim=50 30 0 50,clip,valign=t]{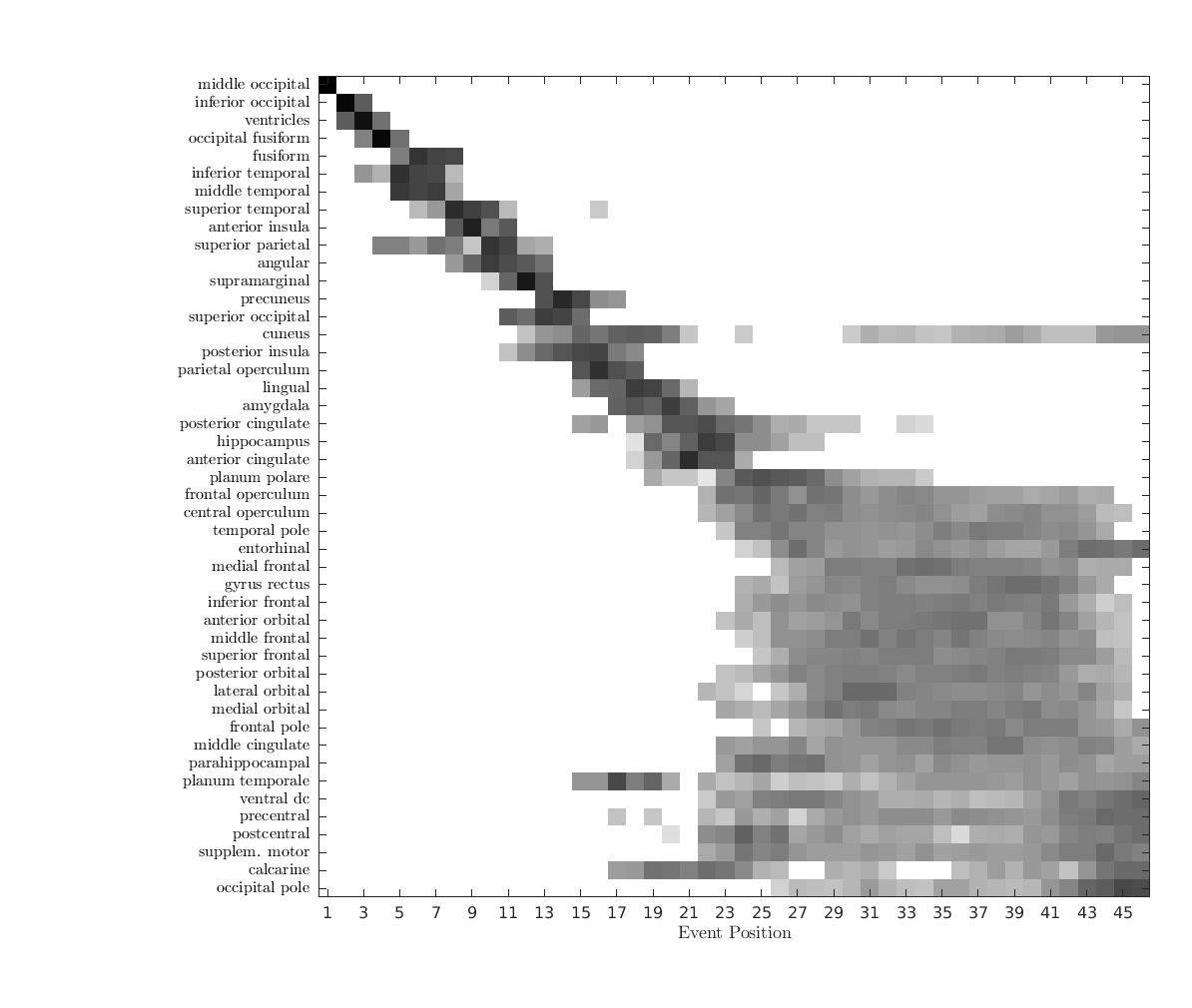}
 \end{subfigure}
 \caption[Positional variance diagram estimated by the event-based model, for three PCA sugroups]{Positional variance diagram estimated by the event-based model, for the three PCA sugroups: Basic visual impairment group, Space perception impairment and Object perception impairment.}
   \label{fig:posVarianceMatrixEarSpaPer}
\end{figure}

\begin{figure}
\centering
  \begin{subfigure}[t]{0.48\textwidth}
  \centering
 \textbf{\large{\mbox{Basic visual impairment group}}}
    \includegraphics[width=1.2\textwidth,trim=50 30 0 50,clip]{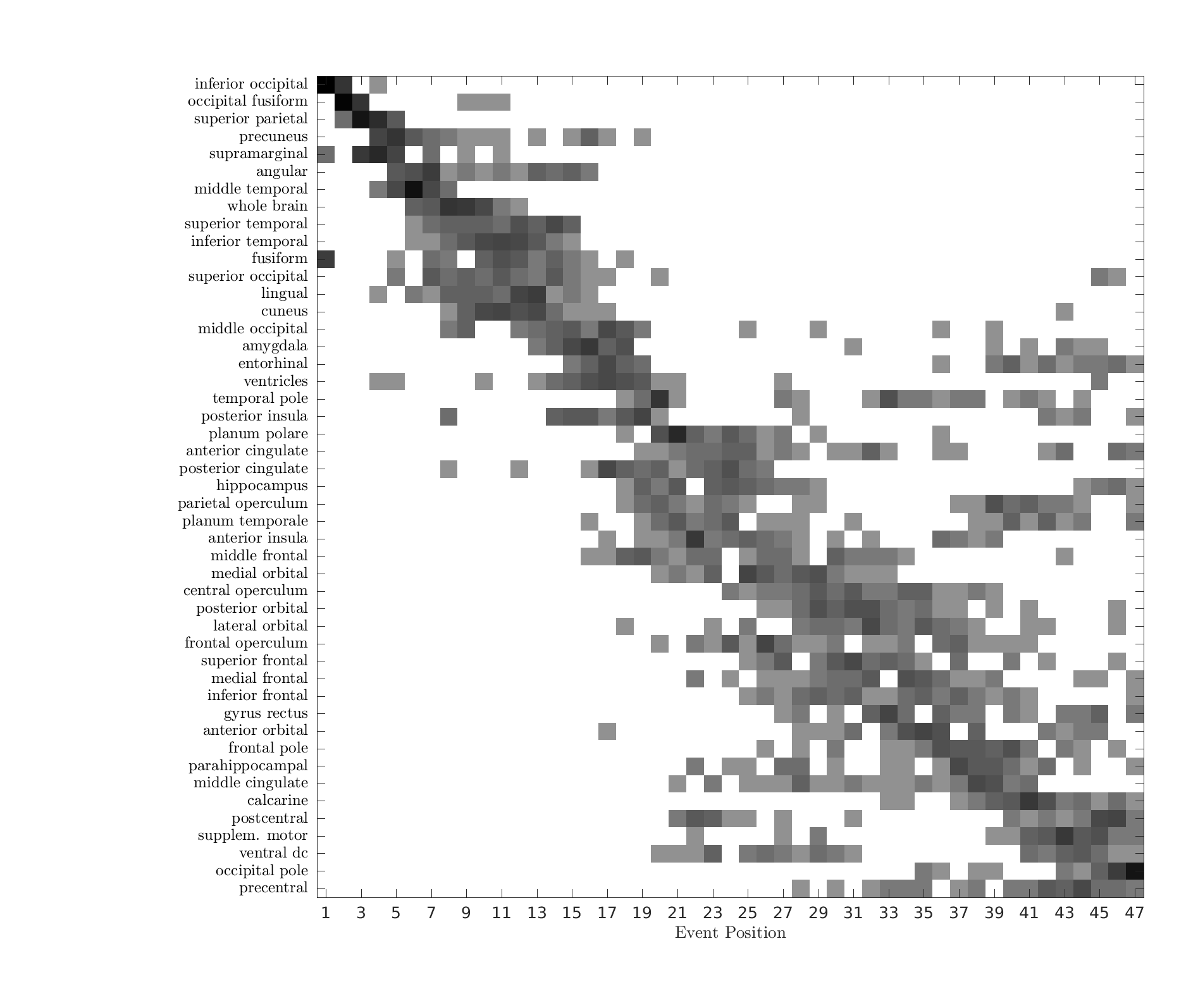}
 \end{subfigure}
 
  \begin{subfigure}[t]{0.48\textwidth}
  \centering
 \textbf{\large{\mbox{Space perception impairment group}}}
 \includegraphics[width=1.2\textwidth,trim=50 30 0 50,clip]{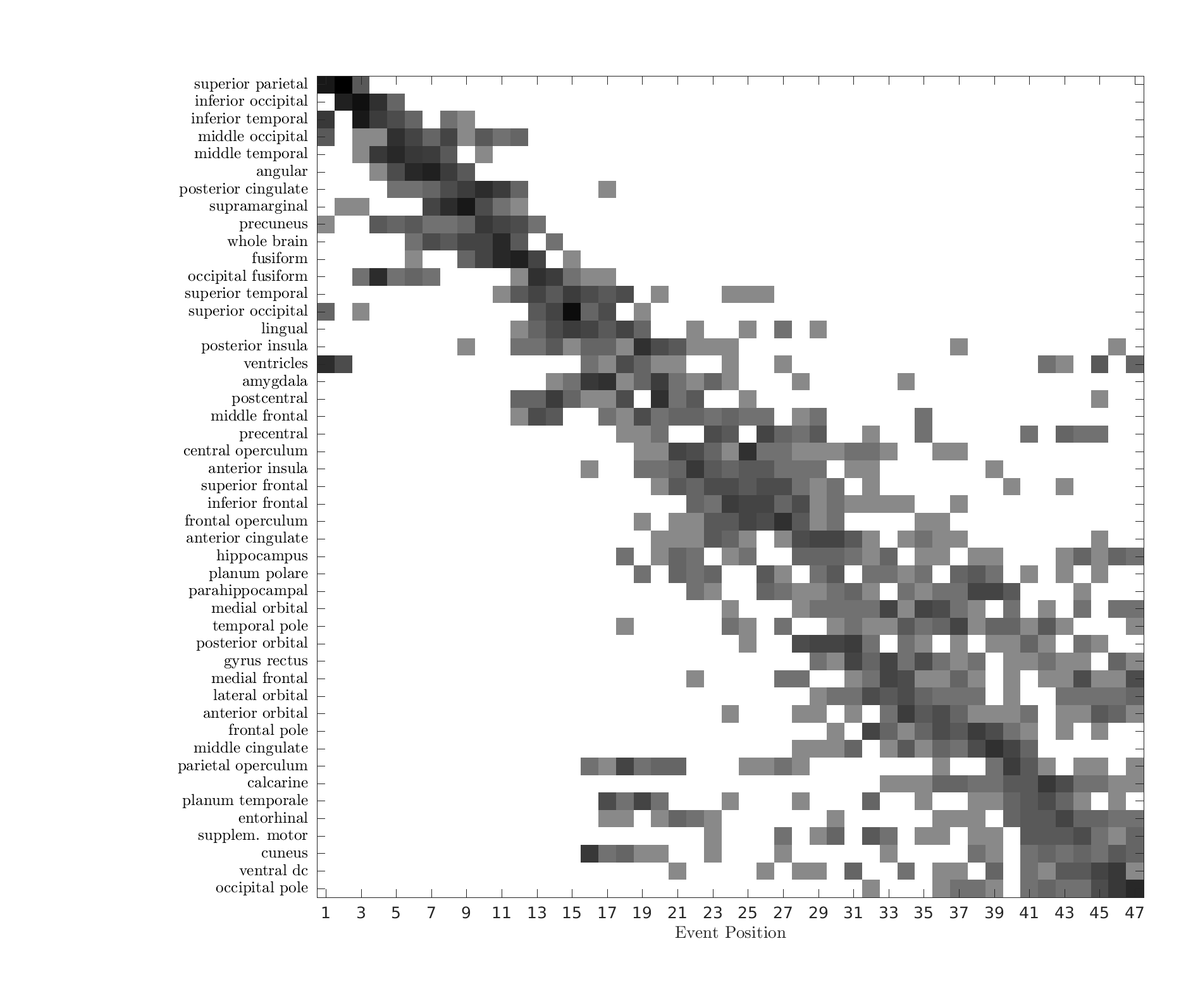}
 \end{subfigure}

\begin{subfigure}[t]{0.48\textwidth}
\centering
   \textbf{\large{\mbox{Object perception impairment group}}}
 \includegraphics[width=1.2\textwidth,trim=50 30 0 50,clip,valign=t]{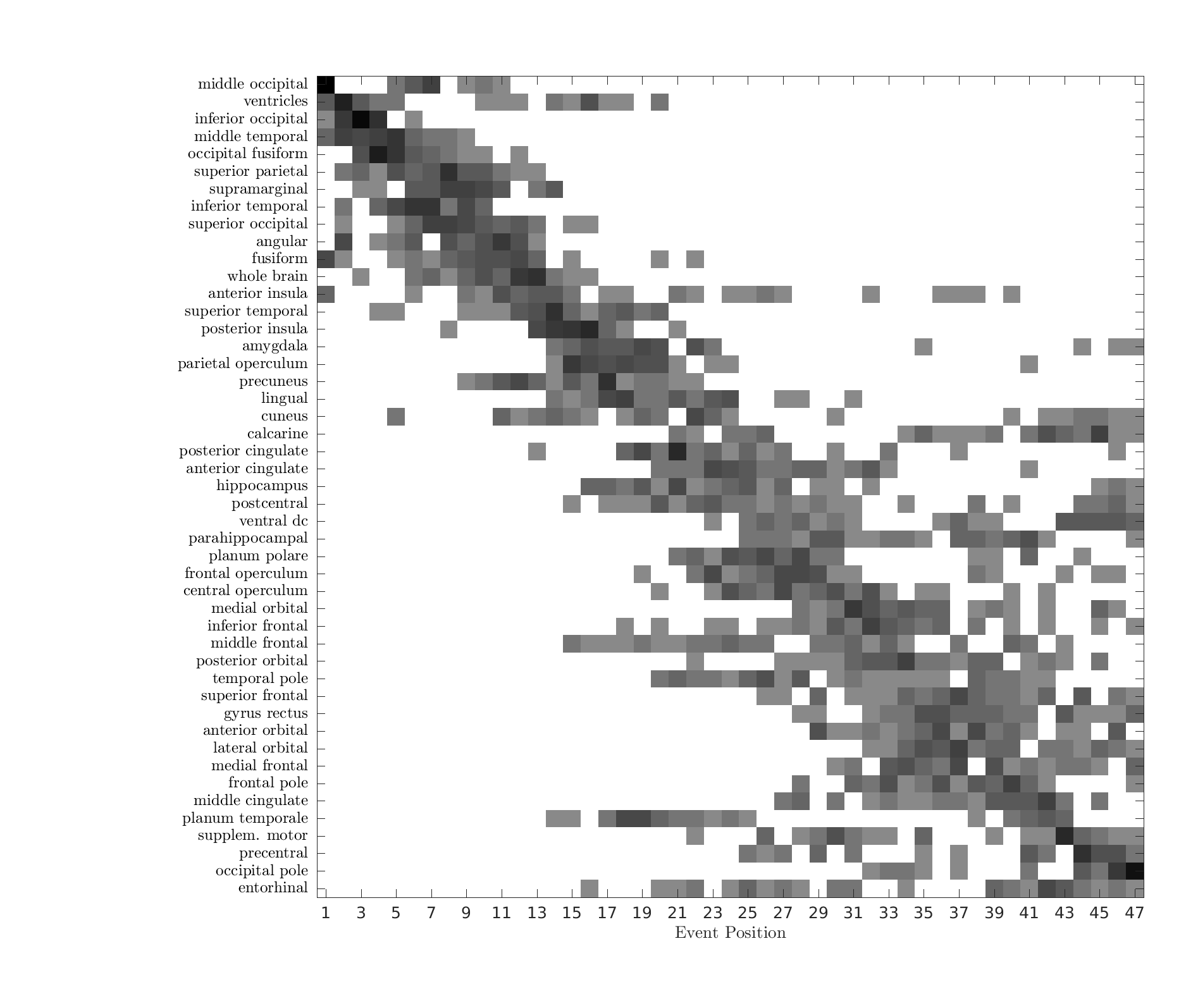}
 \end{subfigure}
 \caption[EBM bootstrap samples of the atrophy sequence, for three PCA subgroups]{Bootstrap samples of the atrophy sequence as estimated by the event-based model, for the three PCA sugroups: Basic visual impairment group, Space perception impairment and Object perception impairment.}
 \label{fig:bootPosVarAllEarSpaPer}
\end{figure}

\begin{figure}
\centering
  \begin{subfigure}[t]{0.48\textwidth}
  \centering
 \textbf{\large{\mbox{Basic visual impairment group}}}
    \includegraphics[width=1.2\textwidth,trim=50 30 0 30,clip]{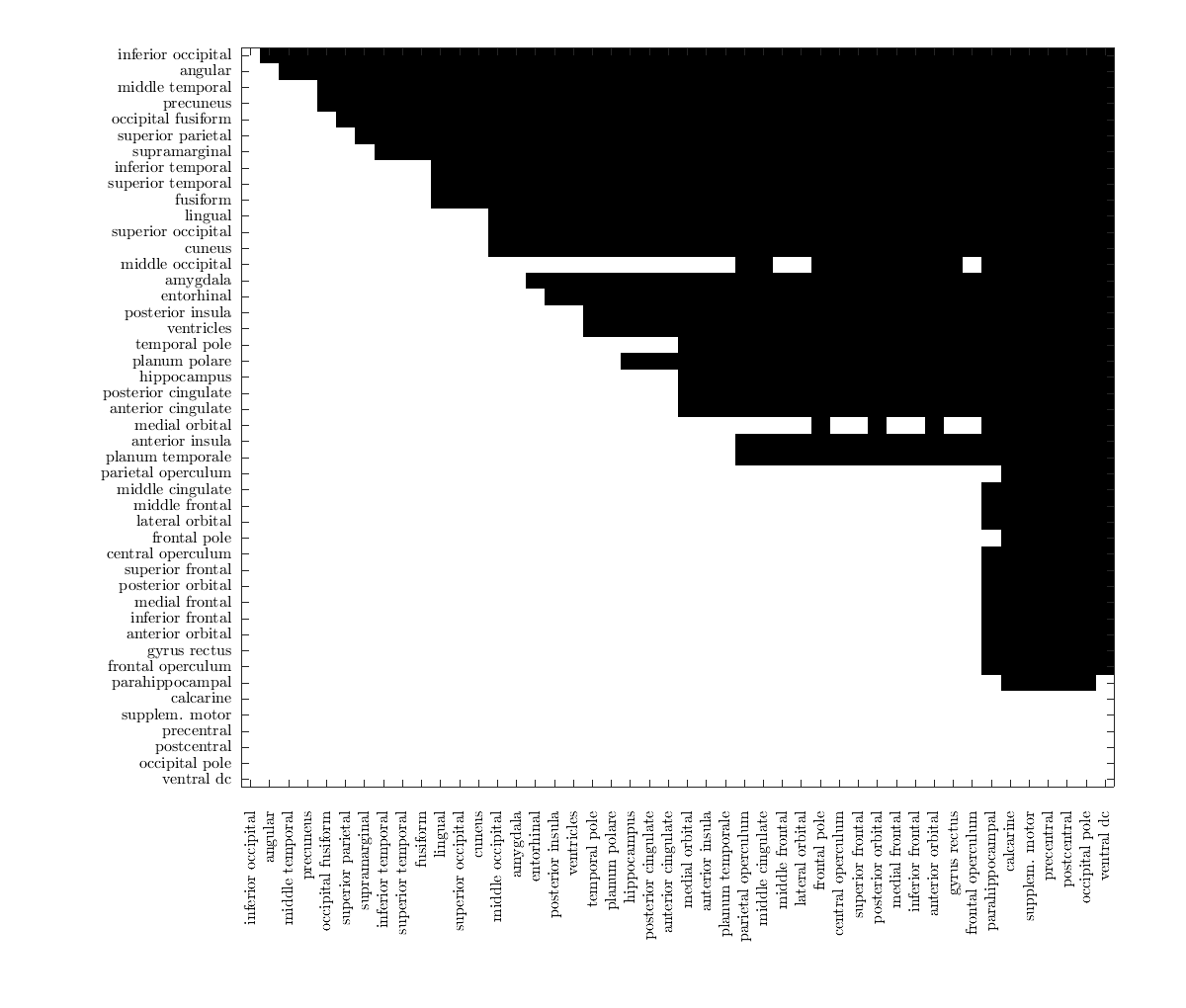}
 \end{subfigure}
 
  \begin{subfigure}[t]{0.48\textwidth}
  \centering
 \textbf{\large{\mbox{Space perception impairment group}}}
 \includegraphics[width=1.2\textwidth,trim=50 30 0 30,clip]{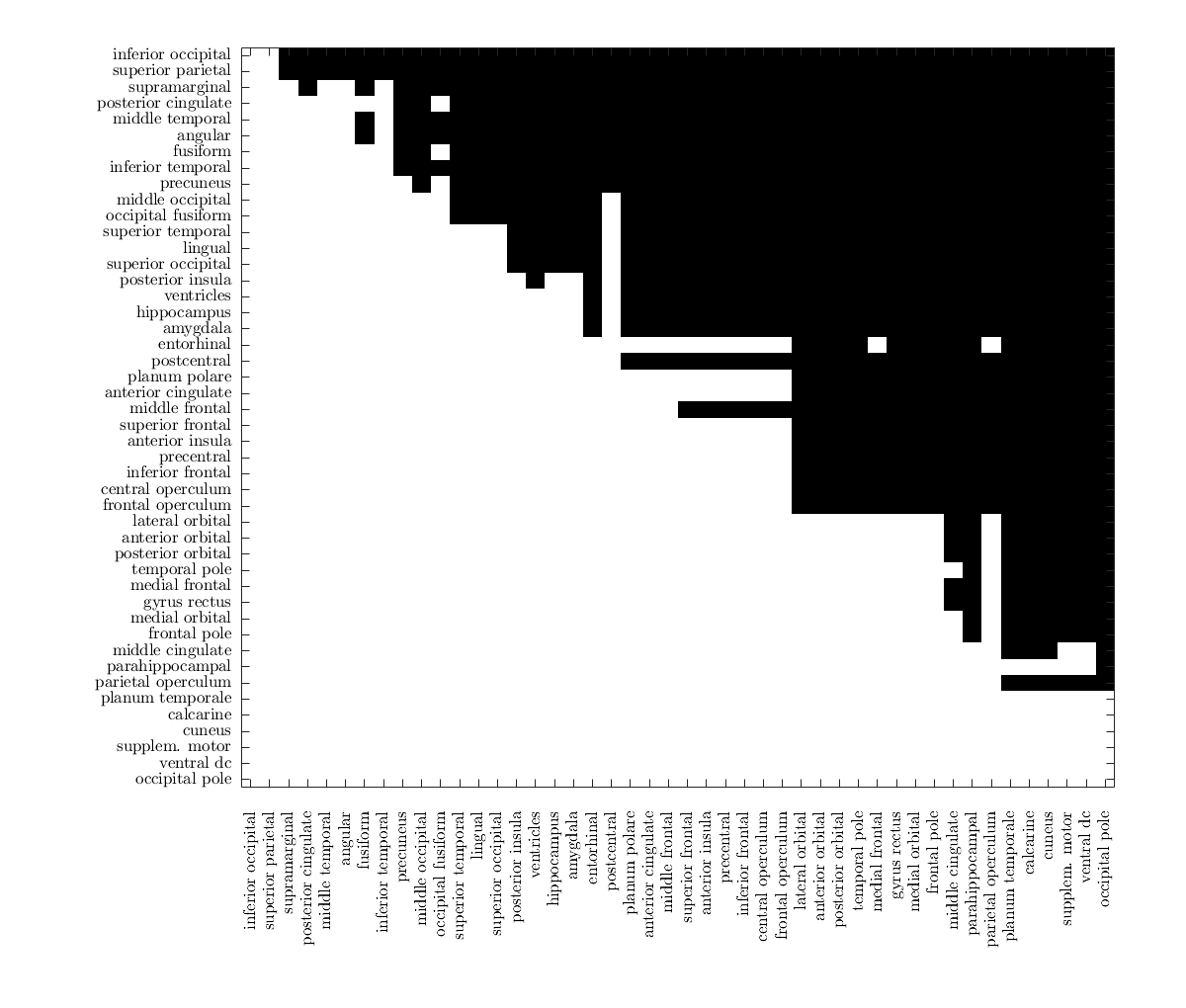}
 \end{subfigure}

\begin{subfigure}[t]{0.48\textwidth}
\centering
   \textbf{\large{\mbox{Object perception impairment group}}}
 \includegraphics[width=1.2\textwidth,trim=50 30 0 30,clip,valign=t]{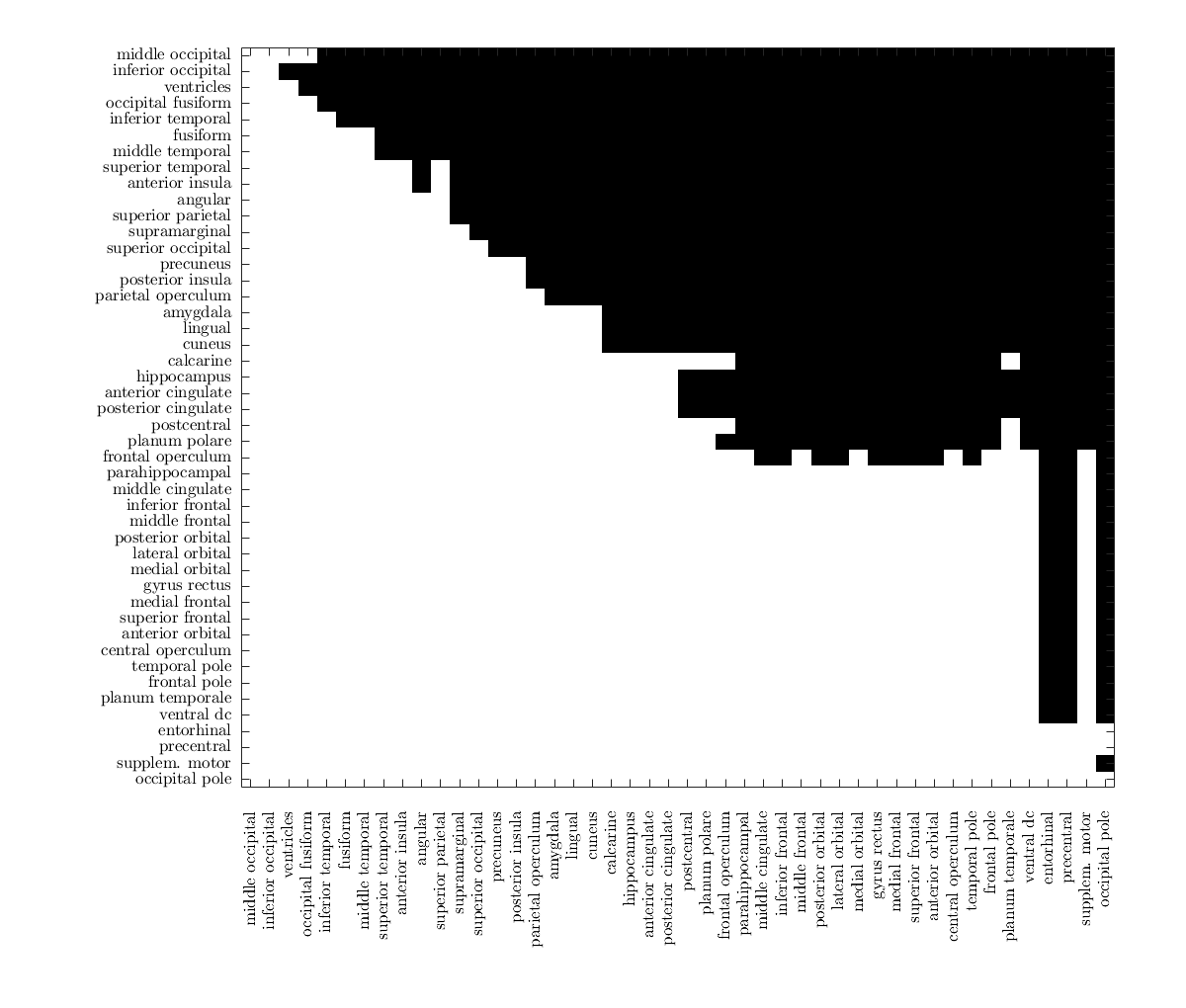}
 \end{subfigure}
 \caption[Hypothesis testing of the ordering of events within the three PCA subgroups.]{Hypothesis testing of the ordering of events within the three PCA subgroups. Hypothesis tests were designed as in \ref{fig:statTestPcaAd}}
\label{fig:statTestEarSpaPer}
\end{figure}

\FloatBarrier


\begin{table}
\centering
\begin{tabular}{c |c c c c c c c c }
Region & Whole Br. & Ventricles & Hippo. & Entorh. & Occipital & Temporal & Frontal & Parietal\\
\hline 
Whole Brain & - & - & - & - & - & - & - & -\\
Ventricles & 1.74e-04* & - & - & - & - & - & - & -\\
Hippo. & 1.20e-02 & 4.95e-02 & - & - & - & - & - & -\\
Entorhinal & 1.61e-12* & 1.27e-06* & 5.29e-10* & - & - & - & - & -\\
Occipital & 7.93e-03 & 4.16e-06* & 1.20e-04* & 9.44e-12* & - & - & - & -\\
Temporal & 2.66e-01 & 1.17e-02 & 3.12e-01 & 5.90e-10* & 1.81e-03 & - & - & -\\
Frontal & 9.58e-01 & 1.57e-04* & 1.07e-02 & 1.52e-12* & 8.68e-03 & 2.49e-01 & - & -\\
Parietal & 3.45e-04* & 1.31e-08* & 2.52e-07* & 3.17e-15* & 8.84e-01 & 7.91e-05* & 4.08e-04* & -\\
\end{tabular} 
\caption[Statistical testing for significant differences in volumes of different brain regions of PCA subjects at -10 years before$t_0$]{Statistical testing for significant differences in volumes of different brain regions of PCA subjects at -10 years before reference $t_0$. Shown here are p-values from two-tailed t-tests. (*) Statistically significant differences at significance level = 1.78e-3, Bonferroni corrected for all 28 comparisons. } 
\label{tab:demStatTestVolsPcaMinus10}
\end{table}


\begin{table}
\centering
\begin{tabular}{c |c c c c c c c c }
Region & Whole Br. & Ventricles & Hippo. & Entorh. & Occipital & Temporal & Frontal & Parietal\\
\hline 
Whole Brain & - & - & - & - & - & - & - & -\\
Ventricles & 1.52e-16* & - & - & - & - & - & - & -\\
Hippo. & 6.03e-13* & 8.95e-06* & - & - & - & - & - & -\\
Entorhinal & 4.78e-14* & 5.66e-01 & 9.60e-04* & - & - & - & - & -\\
Occipital & 1.32e-06* & 3.17e-17* & 1.45e-14* & 5.25e-16* & - & - & - & -\\
Temporal & 3.57e-01 & 1.75e-16* & 5.22e-13* & 2.90e-14* & 1.66e-05* & - & - & -\\
Frontal & 7.31e-12* & 1.67e-04* & 7.72e-01 & 4.38e-03 & 1.62e-14* & 3.50e-12* & - & -\\
Parietal & 1.53e-07* & 1.41e-21* & 3.30e-19* & 2.68e-19* & 2.20e-01 & 8.33e-06* & 3.39e-18* & -\\

\end{tabular} 
\caption[Statistical testing for significant differences in volumes of different brain regions of PCA subjects at $t_0$]{Statistical testing for significant differences in volumes of different brain regions of PCA subjects at $t_0$. See Supp. Table \ref{tab:demStatTestVolsPcaMinus10} for information on statistical testing.} 
\label{tab:demStatTestVolsPca0}
\end{table}


\begin{table}
\centering
\begin{tabular}{c |c c c c c c c c }
Region & Whole Br. & Ventricles & Hippo. & Entorh. & Occipital & Temporal & Frontal & Parietal\\
\hline 
Whole Brain & - & - & - & - & - & - & - & -\\
Ventricles & 5.97e-01 & - & - & - & - & - & - & -\\
Hippo. & 7.63e-13* & 4.14e-13* & - & - & - & - & - & -\\
Entorhinal & 5.88e-11* & 2.34e-11* & 1.23e-03* & - & - & - & - & -\\
Occipital & 4.04e-02 & 1.44e-01 & 3.00e-17* & 1.06e-15* & - & - & - & -\\
Temporal & 2.83e-03 & 1.22e-02 & 1.51e-15* & 2.66e-14* & 1.54e-01 & - & - & -\\
Frontal & 8.90e-15* & 5.73e-15* & 6.77e-02 & 7.35e-07* & 2.19e-19* & 2.99e-17* & - & -\\
Parietal & 7.38e-02 & 2.07e-01 & 1.25e-14* & 4.00e-13* & 9.44e-01 & 1.73e-01 & 1.91e-16* & -\\

\end{tabular} 
\caption[Statistical testing for significant differences in volumes of different brain regions of PCA subjects at 10 years after $t_0$.]{Statistical testing for significant differences in volumes of different brain regions of PCA subjects at 10 years after $t_0$. See Supp. Table \ref{tab:demStatTestVolsPcaMinus10} for information on statistical testing.} 
\label{tab:demStatTestVolsPcaPlus10}
\end{table}


\begin{table}
\centering
\begin{tabular}{c |c c c c c c c c }
Region & Whole Br. & Ventricles & Hippo. & Entorh. & Occipital & Temporal & Frontal & Parietal\\
\hline 
Whole Brain & - & - & - & - & - & - & - & -\\
Ventricles & 9.26e-03 & - & - & - & - & - & - & -\\
Hippo. & 2.04e-10* & 2.88e-14* & - & - & - & - & - & -\\
Entorhinal & 2.21e-02 & 3.40e-01 & 6.82e-09* & - & - & - & - & -\\
Occipital & 3.38e-03 & 2.01e-06* & 3.84e-04* & 9.98e-04* & - & - & - & -\\
Temporal & 3.93e-01 & 1.04e-01 & 4.72e-11* & 7.64e-02 & 7.51e-04* & - & - & -\\
Frontal & 8.30e-01 & 1.04e-02 & 4.79e-09* & 2.41e-02 & 1.15e-02 & 3.26e-01 & - & -\\
Parietal & 4.94e-03 & 2.13e-06* & 3.75e-05* & 4.63e-04* & 7.35e-01 & 8.64e-04* & 1.57e-02 & -\\

\end{tabular} 
\caption[Statistical testing for significant differences in volumes of different brain regions of tAD subjects at -10 years before $t_0$.]{Statistical testing for significant differences in volumes of different brain regions of tAD subjects at -10 years before $t_0$. See Supp. Table \ref{tab:demStatTestVolsPcaMinus10} for information on statistical testing.} 
\label{tab:demStatTestVolsAdMinus10}
\end{table}


\begin{table}
\centering
\begin{tabular}{c |c c c c c c c c }
Region & Whole Br. & Ventricles & Hippo. & Entorh. & Occipital & Temporal & Frontal & Parietal\\
\hline 
Whole Brain & - & - & - & - & - & - & - & -\\
Ventricles & 3.50e-11* & - & - & - & - & - & - & -\\
Hippo. & 4.12e-19* & 3.61e-25* & - & - & - & - & - & -\\
Entorhinal & 7.83e-02 & 2.64e-10* & 2.93e-13* & - & - & - & - & -\\
Occipital & 7.65e-02 & 2.51e-08* & 7.95e-10* & 7.94e-01 & - & - & - & -\\
Temporal & 6.29e-03 & 4.63e-13* & 1.84e-14* & 5.12e-01 & 8.07e-01 & - & - & -\\
Frontal & 2.01e-04* & 2.65e-04* & 4.16e-20* & 2.84e-05* & 1.98e-04* & 3.31e-07* & - & -\\
Parietal & 3.56e-03 & 6.00e-11* & 1.94e-10* & 2.45e-01 & 4.77e-01 & 4.81e-01 & 2.12e-06* & -\\
\end{tabular} 
\caption[Statistical testing for significant differences in volumes of different brain regions of tAD subjects at $t_0$.]{Statistical testing for significant differences in volumes of different brain regions of tAD subjects at $t_0$. See Supp. Table \ref{tab:demStatTestVolsPcaMinus10} for information on statistical testing.} 
\label{tab:demStatTestVolsAd0}
\end{table}


\begin{table}
\centering
\begin{tabular}{c |c c c c c c c c }
Region & Whole Br. & Ventricles & Hippo. & Entorh. & Occipital & Temporal & Frontal & Parietal\\
\hline 
Whole Brain & - & - & - & - & - & - & - & -\\
Ventricles & 2.92e-02 & - & - & - & - & - & - & -\\
Hippo. & 2.83e-01 & 1.67e-03* & - & - & - & - & - & -\\
Entorhinal & 8.13e-03 & 6.50e-01 & 2.63e-04* & - & - & - & - & -\\
Occipital & 8.40e-02 & 9.92e-01 & 1.13e-02 & 7.28e-01 & - & - & - & -\\
Temporal & 2.76e-12* & 1.46e-14* & 5.41e-11* & 6.35e-15* & 2.54e-10* & - & - & -\\
Frontal & 1.24e-09* & 1.91e-06* & 3.87e-11* & 4.81e-06* & 1.27e-04* & 2.43e-19* & - & -\\
Parietal & 7.92e-01 & 7.53e-02 & 1.98e-01 & 2.51e-02 & 1.57e-01 & 5.36e-11* & 3.13e-08* & -\\
\end{tabular} 
\caption[Statistical testing for significant differences in volumes of different brain regions of tAD subjects at 10 years after $t_0$.]{Statistical testing for significant differences in volumes of different brain regions of tAD subjects at 10 years after $t_0$. See Supp. Table \ref{tab:demStatTestVolsPcaMinus10} for information on statistical testing.} 
\label{tab:demStatTestVolsPlus10}
\end{table}

\begin{table}
\centering
\begin{tabular}{c |c c c }
Region & T1 = -10 years &  T2 = 0 years &  T3 = 10 years\\
\hline
Whole Brain & 2.52e-02 & 4.45e-05* & 4.19e-12*\\
Ventricles & 3.74e-05* & 3.06e-05* & 2.18e-13*\\
Hippocampus & 6.15e-15* & 1.13e-23* & 5.83e-04*\\
Entorhinal & 5.28e-08* & 7.71e-11* & 2.68e-03\\
Occipital & 5.72e-01 & 1.13e-05* & 2.44e-12*\\
Temporal & 2.48e-02 & 1.21e-02 & 4.02e-11*\\
Frontal & 2.91e-02 & 6.26e-03 & 2.12e-01\\
Parietal & 4.22e-01 & 2.95e-07* & 8.33e-12*\\

\end{tabular}
\caption[Statistical testing for significant differences in volumes of different brain regions between PCA and tAD at -10, 0 and 10 years from $t_0$.]{Statistical testing for significant differences in volumes of different brain regions between PCA and tAD at -10, 0 and 10 years from $t_0$. Shown here are p-values from two-tailed t-tests. (*) Statistically significant differences at significance level = 2.08e-3, Bonferroni corrected for all 28 comparisons.}
\label{tab:demStatTestPcaAd}
\end{table}

\begin{figure}
 \includegraphics[width=\textwidth,trim=50 50 50 0,clip]{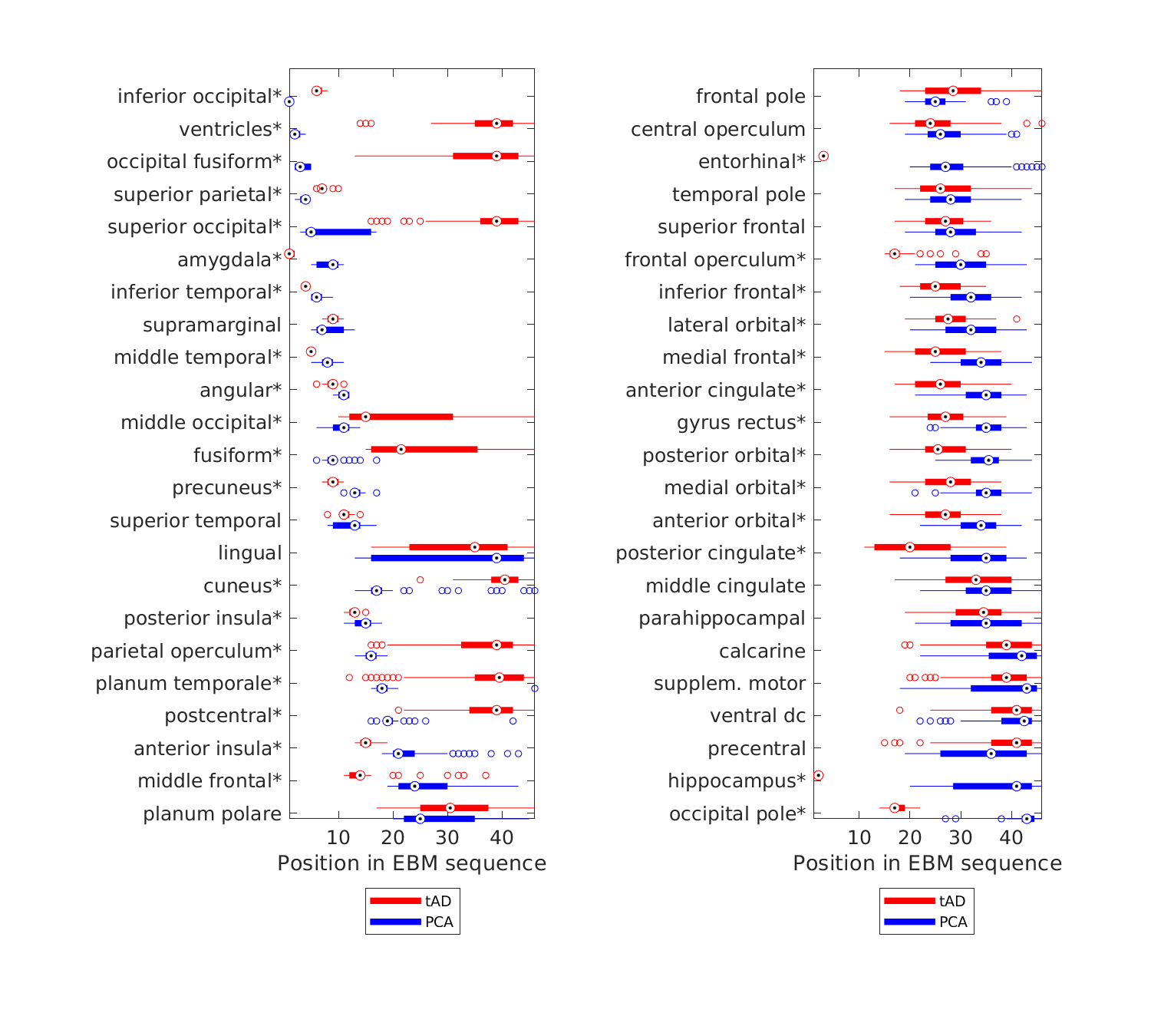}
 \caption[Testing for statistically significant differences in positions of each biomarker in the EBM abnormality sequences, for both PCA and typical AD.]{Testing for statistically significant differences in positions of each biomarker in the EBM abnormality sequences, for both PCA and typical AD. (*) Statistically significant differences in position of a biomarker in the EBM sequences for PCA and tAD at 99\% confidence, Bonferroni corrected for multiple comparisons (significance level = 5e-5). A non-parametric Mann-Whitney U test has been applied because of non-gaussianity of the data, which represents discrete ranks in a sequence. Most biomarkers show significant differences -- it is likely that there are differences in atrophy progression between PCA and tAD.}
 \label{fig:ebmStatTestPcaAd}
\end{figure}

\begin{figure}
\begin{subfigure}{\textwidth}
\includegraphics[width=1\textwidth,trim=0 20 0 0,clip]{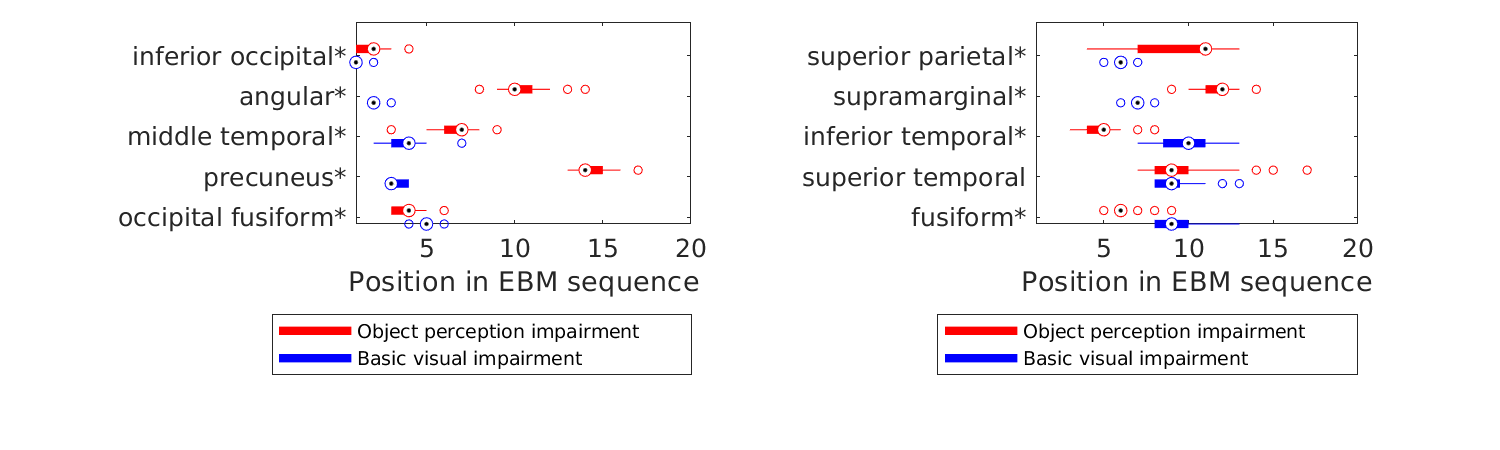}
\caption{Vision vs Object subgroups}
\end{subfigure}

\begin{subfigure}{\textwidth}
\includegraphics[width=1\textwidth,trim=0 20 0 0,clip]{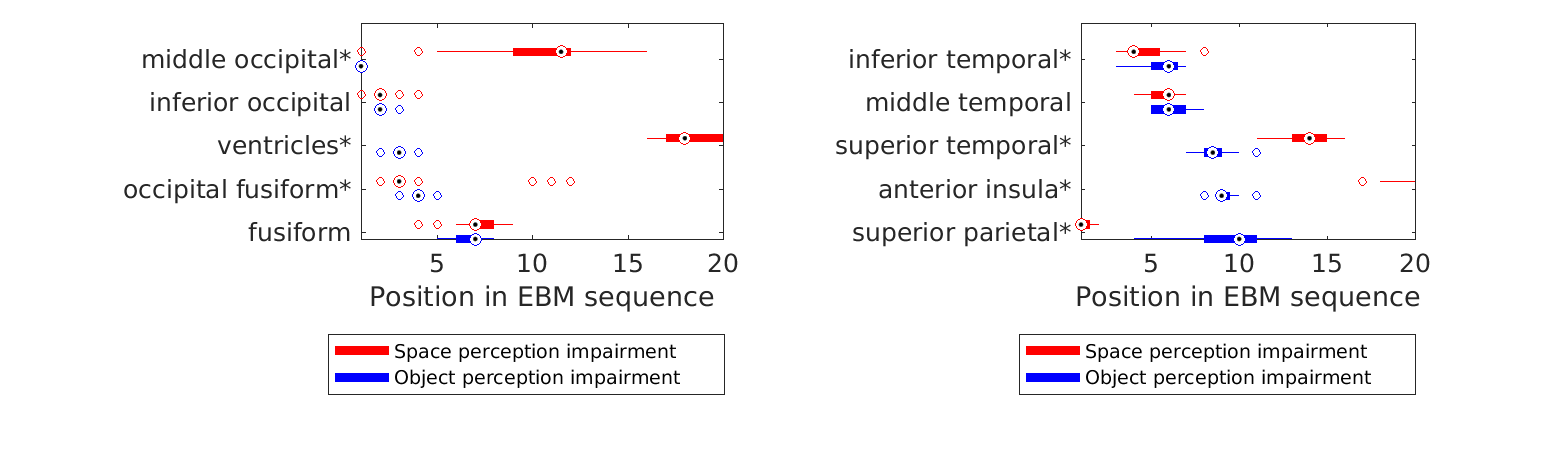}
\caption{Object vs Space subgroups}
\end{subfigure}

\begin{subfigure}{\textwidth}
\includegraphics[width=1\textwidth,trim=0 20 0 0,clip]{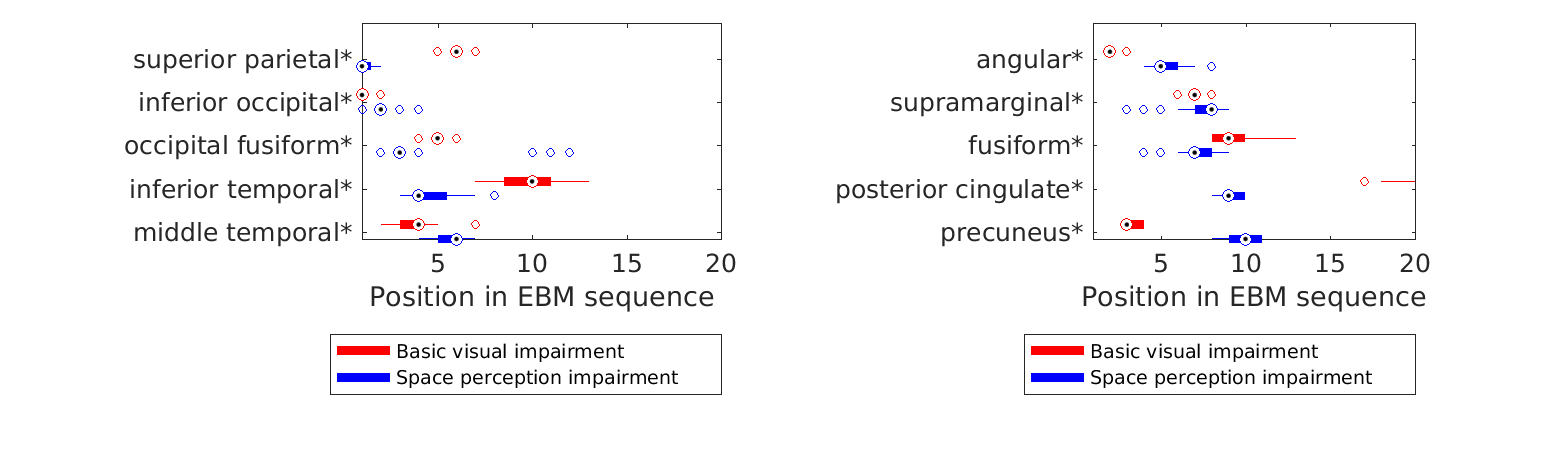}
\caption{Space vs Vision subgroups}
\end{subfigure}

\caption[Testing for statistically significant differences in biomarker positions in the EBM sequences of PCA subgroups.]{Testing for statistically significant differences in biomarker positions in the EBM sequences of PCA subgroups for (a) Object vs Visual (b) Space vs Object and (c) Visual vs Space subgroups.  Only the first 10 biomarkers from the EBM sequence of one disease (A -- visual B -- object C -- space) are shown. The images also show only the first 20 positions on the x-axis to aid visualisation. (*) Statistically significant differences in biomarker positions between pairs of PCA subgroups at 99\% confidence, Bonferroni corrected for multiple comparisons (significance level = 5e-5). A non-parametric Mann-Whitney U test has again been applied because of data non-gaussianity. All biomarkers show significant differences -- it is likely that there are differences in the progression of atrophy between PCA subgroups.}
\label{fig:ebmStatTestPcaSubgroups}
\end{figure}

\chapter[DIVE: A Spatiotemporal Progression Model of Brain Pathology]{DIVE: A Spatiotemporal Progression Model of Brain Pathology in Neurodegenerative Disorders}
\label{sec:diveAppendix}

\def\ci{\perp\!\!\!\perp}

\clearpage

\section{Simulations - Error in Estimated Trajectories and DPS}

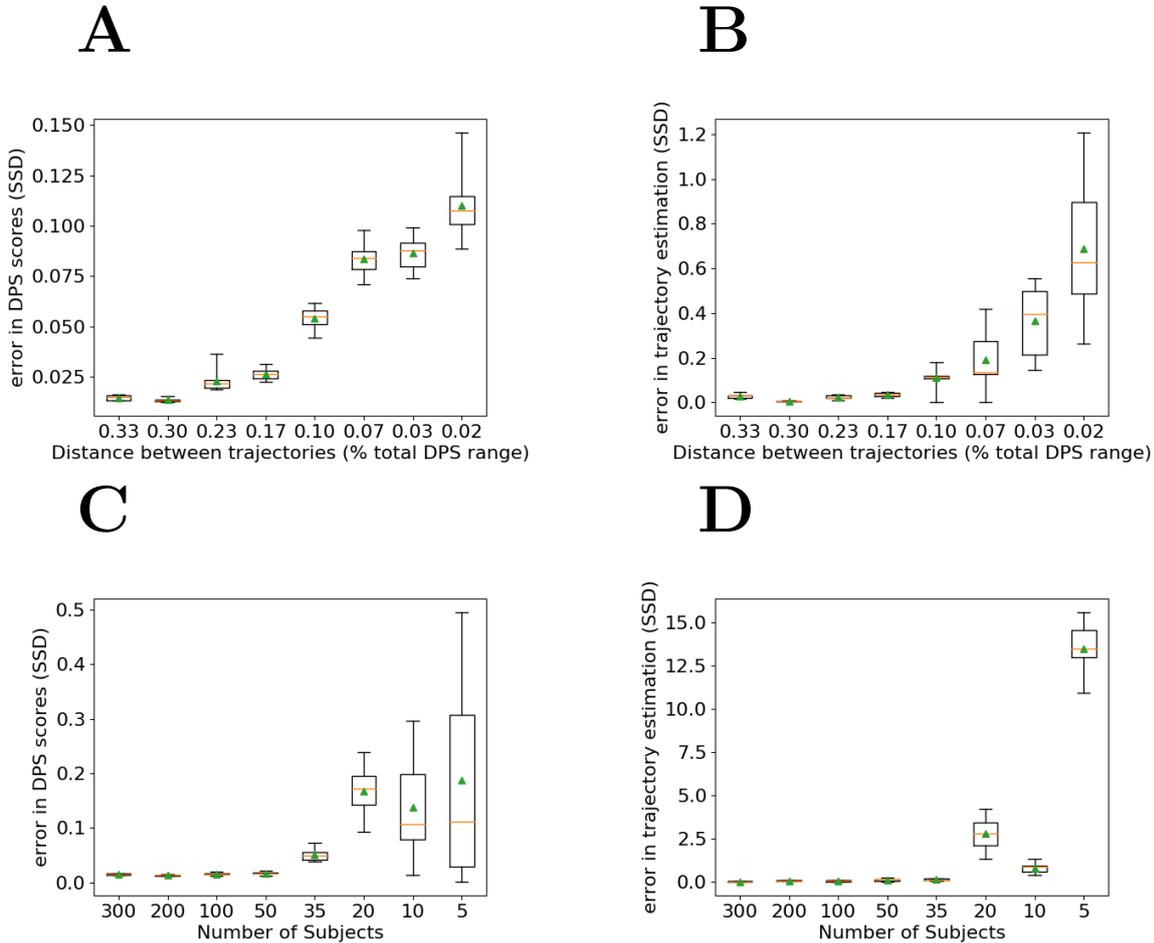
\begin{figure}[H]
\begin{picture}(230,180)
\put(0,0){\includegraphics[width=0.45\textwidth]{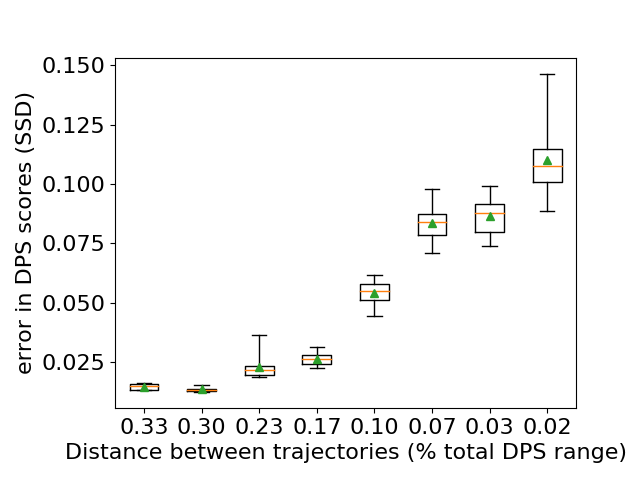}}
\put(30,160){\textbf{\huge{A}}}
\end{picture} 
\begin{picture}(230,180)
\put(0,0){\includegraphics[width=0.45\textwidth]{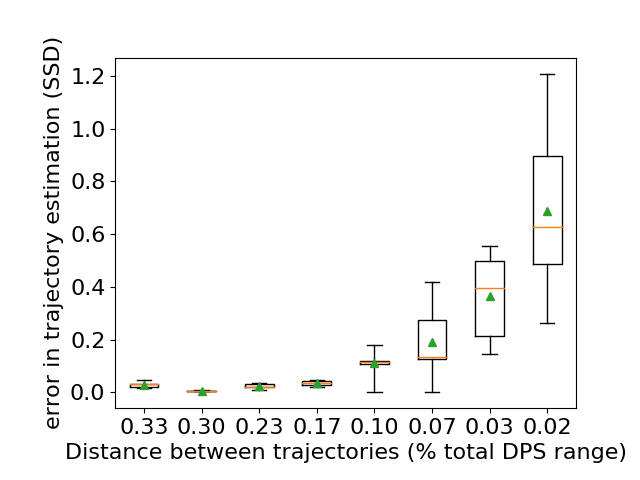}}
\put(30,160){\textbf{\huge{B}}}
\end{picture} 

\begin{picture}(230,180)
\put(0,0){\includegraphics[width=0.45\textwidth]{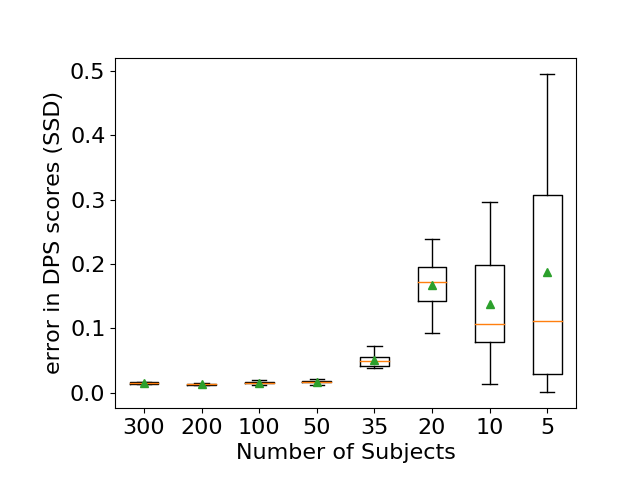}}
\put(30,160){\textbf{\huge{C}}}
\end{picture} 
\begin{picture}(230,180)
\put(0,0){\includegraphics[width=0.45\textwidth]{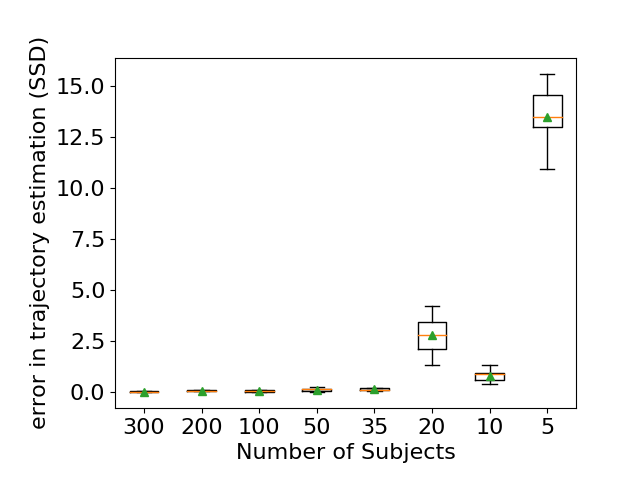}}
\put(30,160){\textbf{\huge{D}}}
\end{picture} 
\caption[DIVE: Error in DPS scores and trajectory estimation in simulations]{Error in DPS scores (A) and trajectory estimation (B) for Scenario 2 in simulation experiments. (C-D) The same error scores for Scenario 3. We notice that as the problem becomes more difficult, the errors in the DIVE estimated parameters increase. Errors were measured as sum of squared differences (SSD) between the true parameters and estimated parameters. For the trajectories, the SSD was calculated only based on the sigmoid centres, due to different scaling of the other sigmoidal parameters.}
\label{diveTrajError}
\end{figure}

\section{Comparison Between DIVE and Other Models}
\label{sec:diveCompAppendix}
 
\subsection{Motivation}

We were also interested to compare the performance of DIVE with other disease progression models. In particular, we were interested to test whether:
\begin{itemize}
 \item Modelling dynamic clusters on the brain surface improves subject staging and biomarker prediction
 \item Modelling subject-specific stages with a linear transformation (the $\alpha_i$ and $\beta_i$ terms) improves biomarker prediction
\end{itemize}

\subsection{Experiment Design}

We compared the performance of our model to two simplified models:
\begin{itemize}
 \item ROI-based model: groups vertices according to an a-priori defined ROI atlas. This model is equivalent to the model by Jedynak et al., Neuroimage, 2012 and is a special case of our model, where the latent variables $z_{lk}$ are fixed instead of being marginalised as in equation 6.
 \item No-staging model: This is a model that doesn't perform any time-shift of patients along the disease progression timeline. It fixes $\alpha_i=1$, $\beta_i=0$ for every subject, which means that the disease progression score of every subject is age.
\end{itemize}

We performed this comparison using 10-fold cross-validation. For each subject in the test set, we computed their DPS score and correlated all the DPS values with the same four cognitive tests used previously. We also tested how well the models can predict the future vertex-wise measurements as follows: for every subject i in the test set, we used their first two scans to estimate $\alpha_i=1$, $\beta_i=0$ and then used the rest of the scans to compute the prediction error. For one vertex location on the cortical surface, the prediction error was computed as the root mean squared error (RMSE) between its predicted measure and the actual measure. This was then averaged across all subjects and visits.

\subsection{Results}

Table \ref{tab1} shows the results of the model comparison, on ADNI MRI dataset. Each row represents one model tested, while each column represents a different performance measure: correlations with four different cognitive tests and accuracy in the prediction of future vertexwise measurements. In each entry, we give the mean and standard deviation of the correlation coefficients or RMSE across the 10 cross-validation folds.  

\begin{table}[H]
\centering
\begin{footnotesize}
 \begin{tabular}{c | c c c c | c}
  Model & CDRSOB ($\rho$) & ADAS13 ($\rho$) & MMSE ($\rho$) & RAVLT ($\rho$) & Prediction (RMSE)\\
  \hline 
DIVE & 0.37 +/- 0.09 & 0.37 +/- 0.10 & 0.36 +/- 0.11 & 0.32 +/- 0.12 & 1.021 +/- 0.008 \\
ROI-based model & 0.36 +/- 0.10 & 0.35 +/- 0.11 & 0.34 +/- 0.13 & 0.30 +/- 0.13 & 1.019 +/- 0.010 \\
No-staging model & *0.09 +/- 0.06 & *0.03 +/- 0.09 & *0.05 +/- 0.06 & *0.02 +/- 0.06 & *1.062 +/- 0.024 \\

 \end{tabular}
 \end{footnotesize}
 \caption[Comparison of DIVE with two more simplistic models on the ADNI MRI dataset.]{Comparison of DIVE with two more simplistic models on the ADNI MRI dataset. For each of the three models, we show the correlation of the disease progression scores (DPS) with respect to several cognitive tests: CDRSOB, ADAS13, MMSE and RAVLT. The correlation numbers represent the mean correlation across the 10 cross-validation folds. }
 \label{tab1}
\end{table}

\section{Derivation of the Generalised EM Algorithm}
\label{sec:diveEmDerivAppendix}

We seek to calculate $M^{(u)} = \argmax_{M} \mathbb{E}_{Z|V,M^{(u-1)}} \left[log\ p(V,Z|M)\right] + log\ p(M)$  where $M^{(u)} = (\alpha^{(u)}, \beta^{(u)}, \theta^{(u)}, \sigma^{(u)}, \lambda^{(u)})$ are the set of model parameters at iteration $u$ of the EM algorithm. Moreover, $p(M^{(u)})$ is a prior on these parameters that is chosen by the user. Expanding the expected value, we get:

\begin{equation}
\label{eq:EM1}
M^{(u)} = \argmax_{M} \sum_{z1,\dots, z_L}^K p(Z = (z_1, ..., z_L) | V, \Mu^{(u-1)}) \left[log\ p(V,Z|M)\right] + log\ p(M) 
\end{equation}

The E-step involves computing $p\left(Z = (z_1, ..., z_L)| V, \Mu^{(u-1)}\right)$, while the M-step comprises of solving the above equation.

\subsection{E-step}

In this step we need to estimate $p(Z | V, \Mu^{(u-1)})$. For notational simplificy we will drop the $(u-1)$ superscript from $\Mu$

\begin{equation}
 p(Z | V, \Mu) = \frac{1}{C}  p(V, Z | \Mu) =  \prod_l^L \left[ \prod_{(i,j) \in I} N(V_l^{ij} | f(\alpha_i t_{ij} + \beta_i | \theta_{Z_l}), \sigma_{Z_l})  \prod_{l_2 \in N_l} \Psi (Z_{l}, Z_{l_2}) \right]
\end{equation}
where $N_l$ is the set of neighbours of vertex $l$. However, this doesn't directly factorise over the vertices $l$ due to the MRF terms $\Psi (Z_{l}, Z_{l_2})$. It is however necessary to find a form that factorises over the vertices, otherwise we won't be able to represent in memory the joint distribution over all $Z$ variables. If we make the approximation $p(Z | V, \Mu) \approx \prod_l^L p(V_l|Z_l, \Mu)$ then we loose out all the MRF terms and the model won't account for spatial correlation. We instead do a first-degree approximation by conditioning on the values of $Z_{N_l}^{(u-1)}$, the labels of nearby vertices from the previous iteration. The approximation is thus:

\begin{equation}
\label{eq:app_e_approx}
 p(Z | V, \Mu) \approx \prod_l^L \mathbb{E}_{Z_{N_l}^{(u-1)}|V_l, M} \left[ p(Z_l|V_l, \Mu, Z_{N_l}^{(u-1)}) \right]
\end{equation}

This form allows us to factorise over all the vertices to get $p(Z_l | V_l, \Mu)$:

\begin{equation}
 p(Z_l | V_l, \Mu) \approx \frac{1}{C} \sum_{Z_{N_l}^{(u-1)}} p(V_l|Z_l, \Mu) p(Z_l|Z_{N_l}^{(u-1)}) p(Z_{N_l}^{(u-1)}|V_l, M) 
\end{equation}

where $C$ is a normalistion constant that can be dropped. We can now  further factorise $p(Z_l|Z_{N_l}^{(u-1)}) \approx  \prod_{m \in \{1, ..., N_l\}} p(Z_l | \Mu,  Z_{N_l(m)}^{(u-1)} = z_{N_l(m)})$ and apply a similar factorisation to the prior $p(Z_{N_l}^{(u-1)}|V_l, M) $, resulting in:

\begin{multline}
 p(Z_l | V_l, \Mu) \approx \frac{1}{C} p(V_l|Z_l, \Mu) \sum_{z_{N_l(1)}, .., z_{N_l(|N_l|)}}\ \ \  \prod_{m \in \{1, ..., N_l\}} p(Z_l | Z_{N_l(m)}^{(u-1)} = z_{N_l(m)}) \\ p(Z_{N_l(m)}^{(u-1)} = z_{N_l(m)}|V_l, M)
\end{multline}

Factorising the summation over $z_{N_l}$'s we get:

\begin{equation}
 p(Z_l | V_l, \Mu) = p(V_l| Z_l, \Mu)   \prod_{l_2 \in N_l} \sum_{z_{l_2}} p(Z_l | Z_{l_2}^{(u-1)} = z_{l_2}) p(Z_{l_2}^{(u-1)} = z_{l_2}|V_l, M)
\end{equation}

Replacing $z_{l_2}$ with $k_2$ we get:

\begin{equation}
 p(Z_l | V_l, \Mu) = p(V_l| Z_l, \Mu)   \prod_{l_2 \in N_l} \sum_{k_2} p(Z_l | Z_{l_2}^{(u-1)} = k_2) p(Z_{l_2}^{(u-1)} = k_2|V_l, M)
\end{equation}

We shall also denote $z_{lk} = p(Z_l | V_l, \Mu)$. Further simplifications result in:

\begin{equation}
 z_{lk}^{(u)} \propto  \left[ \prod_{i,j \in I} N(V_l^{ij} | f(\alpha_i t_{ij} + \beta_i | \theta_k), \sigma_k) \right] \left[ \prod_{l_2 \in N_l} \sum_{k_2 = 1}^K z_{l_2k_2}^{(u-1)}\ \Psi (Z_{l} = k, Z_{l_2} = k_2) \right]
\end{equation}

\begin{multline}
 log\ z_{lk}^{(u)} \propto  \left[ \sum_{i,j \in I} -\frac{log\ (2 \pi \sigma_k^2)}{2} - \frac{1}{2\sigma_k^2}(V_l^{ij} - f(\alpha_i t_{ij} + \beta_i | \theta_k))^2 \right] + \\ + \left[\sum_{l_2 \in N_l} log \sum_{k_2 = 1}^K z_{l_2k_2}^{(u-1)} (\delta_{k_2 k}\ exp(\lambda) + (1 - \delta_{k_2 k})\ exp(-\lambda^2)) \right]
\end{multline}

We further define the data-fit term $D_{lk}$ as follows:
\begin{equation}
\label{eq:app_e-step_Dlk}
D_{lk} = -\frac{log\ (2 \pi \sigma_k^2) |I|}{2} - \sum_{i,j \in I}  \frac{1}{2\sigma_k^2}(V_l^{ij} - f(\alpha_i t_{ij} + \beta_i | \theta_k))^2 
\end{equation}

This results in:

\small
\begin{equation}
 log\ z_{lk}^{(u)} \propto  D_{lk} + \left[\sum_{l_2 \in N_l} log \sum_{k_2 = 1}^K z_{l_2k_2}^{(u-1)} (\delta_{k_2 k}\ (exp(\lambda) - exp(-\lambda^2)) + exp(-\lambda^2)) \right]
\end{equation}
\normalsize

Finally, we simplify the sum over $k_2$ to get the update equation for $z_{lk}$:

\begin{equation}
\label{eq:app_e-step}
\begin{split}
 log\ z_{lk}^{(u)} \propto D_{lk}+ \left[ \sum_{l_2 \in N_l} log\ \left[ exp(-\lambda^2) + z_{l_2k}^{(u-1)} (exp(\lambda) - exp(-\lambda^2)) \right] \right]
\end{split}
\end{equation}

In practice, we cannot naively compute the exponential term $z_{lk} = exp(log(z_{lk}))$ due to precision loss. However, we go around this by recomputing the exponentiation and normalisation of $z_{lk}$ simultaneously. Denoting $x(k) = log\ z_{lk}$, for  $k \in [1 \dots K]$, we get: 

\begin{equation}
 z_{lk}  = \frac{e^{x(k)}}{e^{x(1)}+e^{x(2)} + \dots + e^{x(K)}} = \frac{1}{e^{x(1)-x(k)} +e^{x(2)-x(k)} + \dots + e^{x(K)-x(k)}}
\end{equation}

\subsection{M-step}

The M-step itself does not have a closed-form analytical solution. We choose to solve it by successive refinements of the cluster trajectory parameters and the subject time shifts. 

\subsection{Optimising Trajectory Parameters}

\textbf{Trajectory shape - $\theta$}\\

Taking equation \ref{eq:EM1} and fixing the subject time-shifts $\alpha$, $\beta$ and measurement noise $\sigma$, we can find its maximum with respect to $\theta$ only. More precisely, we want:

\begin{equation}
 \theta = \argmax_{\theta} \sum_{z1,\dots, z_L}^K p(Z = (z_1, ..., z_L) | V, \Mu^{(u-1)}) \left[ log\ p(V,Z|M) \right] + log\ p(\theta)
\end{equation}

We observe that for each cluster the individual $\theta_k$'s are conditionally independent, i.e. $\theta_k \ci \theta_m | \{Z, \alpha, \beta, \sigma$\} $\forall k,m$. We also assume that the prior factorizes for each $\theta_k$: $\log p(\theta) = \prod_k^K \log p(\theta_k)$. This allows us to optimise each $\theta_k$ independently:

\begin{equation}
 \theta_k = \argmax_{\theta_k} \sum_{z1,\dots, z_L}^K p(Z = (z_1, ..., z_L) | V, \Mu^{(u-1)}) \left[ log\ p(V,Z|M) \right] + log\ p(\theta_k)
\end{equation}

Replacing the full data log-likelihood, we get:

\begin{multline}
 \theta_k = \argmax_{\theta_k} \sum_{z1,\dots, z_L}^K p(Z = (z_1, ..., z_L) | V, \Mu^{(u-1)})\ \\ log \left[ \prod_{l=1}^L \prod_{(i,j) \in I} N(V_l^{ij} | f(\alpha_i t_{ij} + \beta_i | \theta_{z_l}), \sigma_{z_l}) \right]  + log\ p(\theta_k)
\end{multline}

Note that we didn't include the MRF clique terms, since they are not a function of $\theta_k$. We propagate the logarithm inside the products:

\begin{multline}
 \theta_k = \argmax_{\theta_k} \sum_{z1,\dots, z_L}^K p(Z = (z_1, ..., z_L) | V, \Mu^{(u-1)}) \sum_{l=1}^L \sum_{(i,j) \in I} log\ N(V_l^{ij} | f(\alpha_i t_{ij} + \beta_i | \theta_{z_l}), \sigma_{z_l}) + \\ + log\ p(\theta_k)
\end{multline}

\begin{sloppypar}
We next assume that $Z_l$, the hidden cluster assignment for vertex $l$, is conditionally independent of the other vertex assignments $Z_m$, $\forall m \neq l$ (See E-step approximation from Eq. \ref{eq:app_e_approx}). This independence assumption induces the following factorization: ${p(Z = (z_1, ..., z_L) | V, \Mu^{(u-1)}) = \prod_l^L p(Z_l = z_l | V, \Mu^{(u-1)})}$. Propagating this product inside the sum over the vertices, we get:
\end{sloppypar}

\begin{equation}
\label{eq:SupTheta5}
 \theta_k = \argmax_{\theta_k} \sum_{l=1}^L \sum_{z_l = 1}^K p(Z_l = z_l | V, \Mu^{(u-1)}) \sum_{(i,j) \in I} log\ N(V_l^{ij} | f(\alpha_i t_{ij} + \beta_i | \theta_{z_l}), \sigma_{z_l}) + log\ p(\theta_k)
\end{equation}

The terms which don't contain $\theta_k$ dissapear:

\begin{equation}
 \theta_k = \argmax_{\theta_k} \sum_{l=1}^L p(Z_l = k | V, \Mu^{(u-1)}) \sum_{(i,j) \in I} log\ N(V_l^{ij} | f(\alpha_i t_{ij} + \beta_i | \theta_k), \sigma_k) + log\ p(\theta_k)
\end{equation}

We further expand the Gaussian noise model:

\begin{multline}
 \theta_k = \argmax_{\theta_k} \sum_{l=1}^L p(Z_l = k | V, \Mu^{(u-1)}) \\ \sum_{(i,j) \in I} \left[ log\ (2 \pi \sigma_k)^{-1/2} - \frac{1}{2\sigma_k^2}(V_l^{ij} - f(\alpha_i t_{ij} + \beta_i | \theta_k))^2 \right] + log\ p(\theta_k)
\end{multline}

Constants dissapear due to the $\argmax$ and we get the final update equation for $\theta_k$:

\begin{equation}
 \theta_k = \argmax_{\theta_k} \sum_{l=1}^L p(Z_l = k | V, \Mu^{(u-1)}) \sum_{(i,j) \in I} \left[ - \frac{1}{2\sigma_k^2}(V_l^{ij} - f(\alpha_i t_{ij} + \beta_i | \theta_k))^2 \right] + log\ p(\theta_k)
\end{equation}

\textbf{Measurement noise - $\sigma$}\\

We first assume a uniform prior on the $\sigma$ parameters to simplify derivations. Using a similar approach as with $\theta$, after propagating the product inside the logarithm and removing the terms which don't contain $\sigma_k$, we get:

\begin{equation}
 \sigma_k = \argmax_{\sigma_k} \sum_{l=1}^L p(Z_l = k | V, \Mu^{(u-1)}) \sum_{(i,j) \in I} log\ N(V_l^{ij} | f(\alpha_i t_{ij} + \beta_i | \theta_k), \sigma_k)
\end{equation}

Note that, just as for $\theta$ above, the MRF clique terms were not included because they are not a function of $\sigma_k$. Expanding the noise model we get:

\begin{equation}
 \sigma_k = \argmax_{\sigma_k} \sum_{l=1}^L p(Z_l = k | V, \Mu^{(u-1)}) \sum_{(i,j) \in I} \left[ log\ (2 \pi \sigma_k^2)^{-1/2} - \frac{1}{2\sigma_k^2}(V_l^{ij} - f(\alpha_i t_{ij} + \beta_i | \theta_k))^2 \right]
\end{equation}

The maximum of a function $l(\sigma_k)$ can be computed by taking the derivative of the function $l$ and setting it to zero. This is under the assumption that $l$ is differentiable, which it is but we won't prove it here. This gives:

\begin{equation}
 \frac{\delta l(\sigma_k|.)}{\delta \sigma_k} =  \sum_{l=1}^L p(Z_l = k | V, \Mu^{(u-1)}) \sum_{(i,j) \in I} \frac{\delta}{\delta \sigma_k} \left[ log\ (2 \pi \sigma_k^2)^{-1/2} - \frac{1}{2\sigma_k^2}(V_l^{ij} - f(\alpha_i t_{ij} + \beta_i | \theta_k))^2 \right]
\end{equation}

Propagating the differential operator further inside the sums we get:

\begin{equation}
 \frac{\delta l(\sigma_k|.)}{\delta \sigma_k} =  \sum_{l=1}^L p(Z_l = k | V, \Mu^{(u-1)}) \sum_{(i,j) \in I} \left[ -\frac{\delta}{\delta \sigma_k} \frac{log\ \sigma_k^2}{2} - \frac{\delta}{\delta \sigma_k} \frac{1}{2\sigma_k^2}(V_l^{ij} - f(\alpha_i t_{ij} + \beta_i | \theta_k))^2 \right]
\end{equation}

We next perform several small manipulations to reach a more suitable form of the derivative and then set it to be equal to zero:

\begin{equation}
 \frac{\delta l(\sigma_k|.)}{\delta \sigma_k} =  \sum_{l=1}^L p(Z_l = k | V, \Mu^{(u-1)}) \sum_{(i,j) \in I} \left[ - \frac{1}{\sigma_k} - \frac{-2}{2\sigma_k^3}(V_l^{ij} - f(\alpha_i t_{ij} + \beta_i | \theta_k))^2 \right]
\end{equation}

\begin{equation}
 \frac{\delta l(\sigma_k|.)}{\delta \sigma_k} =  \sum_{l=1}^L p(Z_l = k | V, \Mu^{(u-1)}) \sum_{(i,j) \in I} \left[ - \frac{\sigma_k^2}{\sigma_k^3} + \frac{1}{\sigma_k^3}(V_l^{ij} - f(\alpha_i t_{ij} + \beta_i | \theta_k))^2 \right]
\end{equation}

\begin{equation}
 \frac{\delta l(\sigma_k|.)}{\delta \sigma_k} =  \sum_{l=1}^L p(Z_l = k | V, \Mu^{(u-1)}) \sum_{(i,j) \in I} \left[ - \sigma_k^2 + (V_l^{ij} - f(\alpha_i t_{ij} + \beta_i | \theta_k))^2 \right] = 0
\end{equation}


Finally, we solve for $\sigma_k$ and get its update equation: 

\begin{equation}
\label{eq:SupThetaFinal}
 \sigma_k^2 = \frac{1}{|I|} \sum_{l=1}^L p(Z_l = k | V, \Mu^{(u-1)}) \sum_{(i,j) \in I} (V_l^{ij} - f(\alpha_i t_{ij} + \beta_i | \theta_k))^2
\end{equation}

%

\subsection{Estimating Subject Time Shifts - $\alpha$, $\beta$}

For estimating $\alpha$, $\beta$, we adopt a similar strategy as in the case of $\theta$, up to Eq. \ref{eq:SupTheta5}. This gives us the following problem:

\begin{multline}
 \alpha_i, \beta_i = \argmax_{\alpha_i, \beta_i} \sum_{l=1}^L \sum_{k = 1}^K p(Z_l = k | V, \Mu^{(u-1)}) \sum_{(i',j) \in I} log\ N(V_l^{i'j} | f(\alpha_{i'} t_{i'j} + \beta_{i'} | \theta_{k}), \sigma_{k}) + \\ + log\ p(\alpha_i, \beta_i)
\end{multline}

The terms $\alpha_{i'}, \beta_{i'}$ for other subjects $i' \neq i$ dissappear:

\begin{equation}
 \alpha_i, \beta_i = \argmax_{\alpha_i, \beta_i} \sum_{l=1}^L \sum_{k = 1}^K p(Z_l = k | V, \Mu^{(u-1)}) \sum_{j \in I_i} log\ N(V_l^{ij} | f(\alpha_{i} t_{ij} + \beta_{i} | \theta_{k}), \sigma_{k}) + log\ p(\alpha_i, \beta_i)
\end{equation}

Expanding the Gaussian noise model we get:

\begin{multline}
 \alpha_i, \beta_i = \argmax_{\alpha_i, \beta_i} \sum_{l=1}^L \sum_{k = 1}^K p(Z_l = k | V, \Mu^{(u-1)}) \\ \sum_{j \in I_i} \left[ log\ (2 \pi \sigma_k^2)^{-1/2} - \frac{1}{2\sigma_k^2}(V_l^{ij} - f(\alpha_i t_{ij} + \beta_i | \theta_k))^2 \right] + \ log\ p(\alpha_i, \beta_i)
\end{multline}

After removing constant terms we end up with the final update equation for $\alpha_i$, $\beta_i$:

\begin{multline}
 \alpha_i, \beta_i = \argmin_{\alpha_i, \beta_i}  \left[ \sum_{l=1}^L \sum_{k=1}^K p(Z_l = k | V, \Mu^{(u-1)}) \frac{1}{2\sigma_k^2} \sum_{j \in I_i} (V_l^{ij} - f(\alpha_i t_{ij} + \beta_i | \theta_k))^2\right] - \\ - log\ p(\alpha_i, \beta_i)
\end{multline}

\subsection{Estimating MRF Clique Term - $\lambda$}

We optimise $\lambda$ using the following formula:

\begin{equation}
\lambda^{(u)} = \argmax_{\lambda} E_{p(Z|V, M^{(u-1)}, \lambda, Z^{(u-1)})}[log\ p(V,Z|M^{(u-1)})]
\end{equation}

Note that $p(Z|V, M^{(u-1)}, \lambda, Z^{(u-1)})$ is a function of $\lambda$, so for each lambda we estimate $z_{lk}$ through approximate inference. We do this because otherwise the optimisation of $\lambda$ will only take into account the clique terms and completely exclude the data terms. We further simplify the objective function for lambda as follows:

\begin{multline}
\lambda^{(u)} = \argmax_{\lambda} \sum_{z1,\dots, z_L}^K p(Z = (z_1, ..., z_L) | V, \Mu^{(u-1)}, \lambda, Z^{(u-1)})\ \\ log \left[ \prod_{l=1}^L \prod_{(i,j) \in I} N(V_l^{ij} | f(\alpha_i t_{ij} + \beta_i | \theta_{z_l}), \sigma_{z_l}) \prod_{l = 1}^{L}\prod_{l_2 \in N_l} \Psi (z_{l}, z_{l_2}) \right]
\end{multline}

We take the logarithm:

\begin{multline}
\lambda^{(u)} = \argmax_{\lambda} \sum_{z1,\dots, z_L}^K p(Z = (z_1, ..., z_L) | V, \Mu^{(u-1)}, \lambda, Z^{(u-1)})\ \\ \left[ \sum_{l=1}^L \sum_{(i,j) \in I} log\ N(V_l^{ij} | ..) + \sum_{l = 1}^{L}\sum_{l_2 \in N_l} log\ \Psi (z_{l}, z_{l_2}) \right]
\end{multline}

Let us denote $z_{lk} = p(Z_l = k | V, \Mu^{(u-1)}, \lambda, Z^{(u-1)})$. Assuming independence between the latent variables $Z_l$ we get:

\begin{multline}
\lambda = \argmax_{\lambda} \sum_{l=1}^L \sum_{k=1}^K z_{lk}\ \left[ \sum_{(i,j) \in I} log\ N(V_l^{ij} | ..) \right] + \\ + \sum_{l = 1}^{L} \sum_{k=1}^K z_{lk} \sum_{l_2 \in N_l} \sum_{k_2 = 1}^K z_{l_2k}\ log\ \Psi (Z_{l} = k, Z_{l_2} = k_2)  
\end{multline}

However, we now want to make $z_{lk}$ a function of $\lambda$ as previously mentioned, so $z_{lk}=\zeta_{lk}(\lambda)$, for some function $\zeta_{lk}$. More precisely, using the E-step update from Eq. \ref{eq:app_e-step} we define for each vertex $l$ and cluster $k$ a function $\zeta_{lk}(\lambda)$ as follows:

\begin{equation}
\zeta_{lk}(\lambda) = exp \left( D_{lk} +   \sum_{l_2 \in N_l} log\ \left[ exp(-\lambda^2) + z_{l_2k}^{(u-1)} (exp(\lambda) - exp(-\lambda^2)) \right] \right)
\end{equation}

where $D_{lk}$ is as defined in Eq \ref{eq:app_e-step_Dlk}. We then replace $z_{lk}$ with $\zeta_{lk}(\lambda)$ and introduce the chosen MRF clique model to get:

\begin{equation}
\lambda^{(u)} = \argmax_{\lambda}\ \sum_{l=1}^L \sum_{k=1}^K \zeta_{lk}(\lambda) D_{lk}  + \sum_{l = 1}^{L} \sum_{k}^K \sum_{l_2 \in N_l} \sum_{k_2 = 1}^K \zeta_{lk}(\lambda) \zeta_{l_2k}(\lambda)\ \left[ \delta_{kk_2} \lambda + (1-\delta_{kk_2}) (-\lambda^2)\right]
\end{equation}

We separate the cliques that have matching clusters to the ones that don't:

\begin{equation}
\lambda^{(u)} = \argmax_{\lambda}\ \sum_{l=1}^L \sum_{k=1}^K \zeta_{lk}(\lambda) D_{lk} + \sum_{l = 1}^{L} \sum_{l_2 \in N_l} \sum_{k}^K  \left[ \zeta_{lk}(\lambda) \zeta_{l_2k}(\lambda)\ \lambda + \sum_{k2 \neq k} \zeta_{lk}(\lambda) \zeta_{l_2k}(\lambda) (-\lambda^2)\right]
\end{equation}

We also factorise the clique terms:

\begin{multline}
\lambda^{(u)} = \argmax_{\lambda}\ \sum_{l=1}^L \sum_{k=1}^K \zeta_{lk}(\lambda) D_{lk} + \lambda \sum_{l = 1}^{L} \sum_{l_2 \in N_l} \sum_{k}^K   \zeta_{lk}(\lambda) \zeta_{l_2k}(\lambda)\ + \\ + (-\lambda^2) \sum_{l = 1}^{L} \sum_{l_2 \in N_l} \sum_{k}^K \zeta_{lk}(\lambda) (1- \zeta_{l_2k}(\lambda))
\end{multline}

Finally, we simplify to get the objective function for $\lambda$.

\begin{equation}
 \lambda^{(u)} = \argmax_{\lambda}\ \sum_{l=1}^L \sum_{k=1}^K \zeta_{lk}(\lambda) \left[  D_{lk} \  + \lambda \sum_{l_2 \in N_l}  \zeta_{l_2 k}(\lambda)\  -\lambda^2 \sum_{l_2 \in N_l} (1- \zeta_{l_2 k}(\lambda))  \right]
\end{equation}

For implementation speed-up, data-fit terms $D_{lk}$ can be pre-computed.

\section{Fast DIVE Implementation - Proof of Equivalence}
\label{sec:appDivFas}

Fitting DIVE can be computationally prohibitive, especially given that the number of vertices/voxels can be very high, e.g. more than 160,000 in our datasets. We derived a fast implementtion of DIVE, which is based on the idea that for each subject we compute a weighted mean of the vertices within a particular cluster, and then compare that mean with the corresponding trajectory value. This is in contrast with comparing the value at each vertex with the corresponding trajectory of its cluster. In the next few subsections, we will present the mathematical formulation of the fast implementation for parameters [$\theta$, $\alpha$, $\beta$]. Parameter $\sigma$ already has a closed-form update, while parameter $\lambda$ has a more complex update procedure for which this fast implementation doesn't work. For each parameter, we will also provide proofs of equivalence.

\subsection{Trajectory Parameters - $\theta$}

\subsection{Fast Implementation}

The fast implementation for $\theta$ implies that, instead of optimising Eq. \ref{eq:SupThetaFinal} we optimise the following problem:
\begin{equation}
\label{eq:supThetaFast1}
 \theta_k = \argmin_{\theta_k} \sum_{(i,j) \in I} (<V^{ij}>_{\hat{Z}_k} - f(\alpha_i t_{ij} + \beta_i | \theta_k))^2
\end{equation}

where $<V^{ij}>_{\hat{Z}_k}$ is the mean value of the vertices belonging to cluster $k$. Mathematically, we define $\hat{Z}_k = [z_{1k}\gamma_k,\ z_{2k}\gamma_k,\ \dots,\ z_{Lk}\gamma_k ]$ where  $\gamma_k = (\sum_{l=1}^L z_{lk})^{-1}$ is the normalisation constant. Moreover, we have that ${<V^{ij}>_{\hat{Z}_k} = \sum_{l=1}^L z_{lk} \gamma_k V^{ij}}$. We take the derivative of the likelihood function $l_{fast}$ of the fast implementation (Eq. \ref{eq:supThetaFast1}) with respect to $\theta_k$ and perform several simplifications:
\begin{equation}
\frac{\delta l_{fast}(\theta_k|.)}{\delta \theta_k} = \frac{\delta}{\delta \theta_k} \sum_{(i,j) \in I} \left( \sum_{l=1}^L z_{lk} \gamma_k V^{ij} - f(\alpha_i t_{ij} + \beta_i | \theta_k) \right)^2
\end{equation}

\begin{equation}
\frac{\delta l_{fast}(\theta_k|.)}{\delta \theta_k} = \sum_{(i,j) \in I} 2 \left( \left( \sum_{l=1}^L \gamma_k z_{lk} V^{ij} \right) - f(\alpha_i t_{ij} + \beta_i | \theta_k) \right) \frac{-\delta f(.)}{\delta \theta_k}
\end{equation}

using the fact that $\sum_{l=1}^L \gamma_k z_{lk} = 1$ we get:

\begin{equation}
\frac{\delta l_{fast}(\theta_k|.)}{\delta \theta_k} = \sum_{(i,j) \in I} 2 \left( \sum_{l=1}^L \gamma_k z_{lk} \left( V^{ij} - f(\alpha_i t_{ij} + \beta_i | \theta_k) \right) \right) \frac{-\delta f(.)}{\delta \theta_k}
\end{equation}

\begin{equation}
\frac{\delta l_{fast}(\theta_k|.)}{\delta \theta_k} = 2 \gamma_k \sum_{(i,j) \in I} \frac{-\delta f(.)}{\delta \theta_k} \left( \sum_{l=1}^L z_{lk} \left( V^{ij} - f(\alpha_i t_{ij} + \beta_i | \theta_k) \right) \right)
\end{equation}

By setting the derivative to zero, the optimal $\theta$ is thus a solution of the following equation:

\begin{equation}
\label{eq:supThetaFast2}
\sum_{(i,j) \in I} \frac{-\delta f(.)}{\delta \theta_k} \left( \sum_{l=1}^L z_{lk} \left( V^{ij} - f(\alpha_i t_{ij} + \beta_i | \theta_k) \right) \right) = 0
\end{equation}

\subsection{Slow Implementation}

We will prove that if theta is a solution of the slow implementation, it is also a solution of Eq. \ref{eq:supThetaFast2}, which will prove that the fast implementation is equivalent. The slow implementation is finding $\theta$ from the following equation:

\begin{equation}
 \theta_k = \argmin_{\theta_k} \sum_{l=1}^L z_{lk} \sum_{(i,j) \in I} (V_l^{ij} - f(\alpha_i t_{ij} + \beta_i | \theta_k))^2
\end{equation}

Taking the derivative of the function above ($l_{slow}$) with respect to $\theta_k$ we get:
\begin{equation}
\frac{\delta l_{slow}(\theta_k|.)}{\delta \theta_k} = \sum_{l=1}^L z_{lk} \sum_{(i,j) \in I} 2 (V_l^{ij} - f(\alpha_i t_{ij} + \beta_i | \theta_k)) \left(-\frac{\delta f(.)}{\delta \theta_k} \right) = 0
\end{equation}

After swapping terms around and using distributivity we get:

\begin{equation}
  \sum_{(i,j) \in I} \left(-\frac{\delta f(.)}{\delta \theta_k} \right) \sum_{l=1}^L z_{lk} (V_l^{ij} - f(\alpha_i t_{ij} + \beta_i | \theta_k))  = 0
\end{equation}

This is the same optimisation problem as in Eq. \ref{eq:supThetaFast2}, which proves that the two formulations are equivalent with respect to $\theta$.

\subsection{Noise Parameter - $\sigma$}

The noise parameter $\sigma$ can actually be computed in a closed-form solution for the original slow model implementation, so there is no benefit in implementing the fast update for $\sigma$. Moreover, the $\sigma$ in the fast implementation computed the standard deviation in the \emph{mean value} of the vertices within a certain cluster, and not the deviation withing the \emph{actual value} of the vertices.

\subsection{Subjects-specific Time Shifts - $\alpha$, $\beta$}

\subsection{Fast Implementation}

The equivalent fast formulation for the subject-specific time shifts is similar to the one for the trajectory parameters. It should be noted however that we need to weight the sums corresponding to each cluster by $\gamma_{k}^{-1}$. This gives the following equation for the fast formulation:

\begin{equation}
 \alpha_i, \beta_i = \argmin_{\alpha_i, \beta_i}  \sum_{k=1}^K \gamma_k^{-1} \frac{1}{2\sigma_k^2} \sum_{j \in I_i} (<V_l^{ij}>_{\hat{Z}_k} - f(\alpha_i t_{ij} + \beta_i | \theta_k))^2 = 0
\end{equation}

In order to prove that this is equivalent to the slow version, we need to take the derivative of the likelihood function ($l_{fast}$) from the above equation with respect to $\alpha_i$, $\beta_i$ and set it to zero:

\begin{equation}
 \frac{\delta l_{fast}(\alpha_i, \beta_i|.)}{\delta \alpha_i, \beta_i} =  \frac{\delta}{\delta \alpha_i, \beta_i} \sum_{k=1}^K \gamma_k^{-1} \frac{1}{2\sigma_k^2} \sum_{j \in I_i} (<V_l^{ij}>_{\hat{Z}_k} - f(\alpha_i t_{ij} + \beta_i | \theta_k))^2 = 0
\end{equation}

We expand the average across the vertices and slide the derivative operator inside the sums:

\begin{equation}
 \sum_{k=1}^K \gamma_k^{-1} \frac{1}{2\sigma_k^2} \sum_{j \in I_i} 2 \left( \sum_{l=1}^L \gamma_k z_{lk} V_l^{ij} - f(\alpha_i t_{ij} + \beta_i | \theta_k) \right) \frac{-\delta f(.)}{\delta \alpha_i, \beta_i}
\end{equation}

Since $ \sum_{l=1}^L \gamma_k z_{lk} = 1$ we get:

\begin{equation}
 2 \sum_{k=1}^K \gamma_k^{-1} \frac{1}{2\sigma_k^2} \sum_{j \in I_i} \frac{-\delta f(.)}{\delta \alpha_i, \beta_i}  \left( \sum_{l=1}^L \gamma_k z_{lk} (V_l^{ij} - f(\alpha_i t_{ij} + \beta_i | \theta_k)) \right) 
\end{equation}

Removing the factor 2 and sliding $\gamma_k$:

\begin{equation}
 \sum_{k=1}^K \gamma_k^{-1} \gamma_k \frac{1}{2\sigma_k^2} \sum_{j \in I_i} \frac{-\delta f(.)}{\delta \alpha_i, \beta_i}  \left( \sum_{l=1}^L  z_{lk} (V_l^{ij} - f(\alpha_i t_{ij} + \beta_i | \theta_k)) \right) 
\end{equation}

Further sliding $\sum_{l=1}^L z_{lk}$ to the left we get the final optimisation problem:
\begin{equation}
\label{eq:supAlphaFast2}
 \sum_{k=1}^K  \frac{1}{2\sigma_k^2} \sum_{l=1}^L z_{lk} \sum_{j \in I_i} \frac{- \delta f(.)}{\delta \alpha_i, \beta_i} (V_l^{ij} - f(\alpha_i t_{ij} + \beta_i | \theta_k))
\end{equation}

\subsection{Slow Implementation}

In a similar way to the trajectory parameters, we want to prove that solving the problem from Eq. \ref{eq:supAlphaFast2} (fast implementation) is the same as solving the original slow implementation problem, which is defined as:

\begin{equation}
 \alpha_i, \beta_i = \argmin_{\alpha_i, \beta_i}   \sum_{l=1}^L \sum_{k=1}^K z_{lk} \frac{1}{2\sigma_k^2} \sum_{j \in I_i} (V_l^{ij} - f(\alpha_i t_{ij} + \beta_i | \theta_k))^2
\end{equation}

Taking the derivative of the function above with respect to $\alpha_i, \beta_i$, we get:

\begin{equation}
 \frac{\delta l_{slow}(\alpha_i, \beta_i|.)}{\delta \alpha_i, \beta_i} = \sum_{k=1}^K \frac{1}{2\sigma_k^2} \sum_{l=1}^L z_{lk} \sum_{j \in I_i} \frac{- \delta f(.)}{\delta \alpha_i, \beta_i} (V_l^{ij} - f(\alpha_i t_{ij} + \beta_i | \theta_k))
\end{equation}

This is the same problem as the fast implementation one from Eq. \ref{eq:supAlphaFast2}, thus the fast model is equivalent to the slow model with respect to $\alpha$, $\beta$.

\chapter{Disease Knowledge Transfer across Neurodegenerative Diseases}



\begin{figure}
\includegraphics[width=\textwidth]{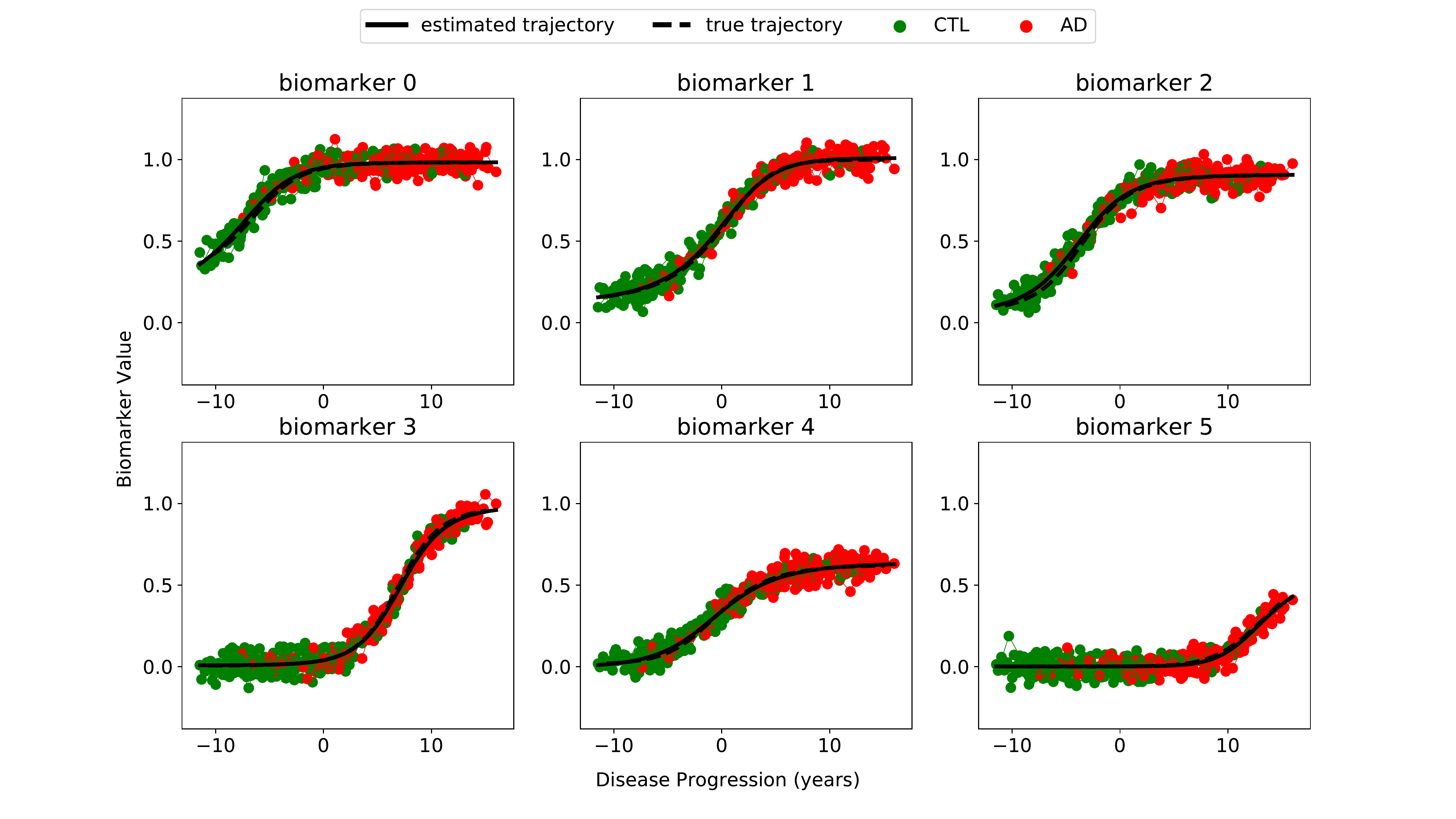}
 \caption[Estimated biomarker trajectories for the "synthetic AD" disease, plotted alongside true trajectories.]{Estimated biomarker trajectories for the "synthetic AD" disease, plotted alongside true trajectories. Estimation of the trajectories in biomarkers 0,1,4 and 5 has been done without any data from the "synthetic PCA" disease, only based on the disease-agnostic correlations with biomarkers 2 and 3.}
 \label{fig:dktSynthTrajADSpace}
\end{figure}

\chapter[Novel Extensions to the EBM and DEM]{Novel Extensions to the Event-based Model and Differential Equation Model}

\section{EBM Fitting using Expectation-Maximisation}
\label{sec:appEbmEm}

Let us assume that $p(x|E)$ and $p(x|\neg E)$ follow normal distributions, where $p(x|E_k) \sim N(\mu^a_k, \sigma^a_k)$ and $p(x|\neg E_k) \sim N(\mu^n_k, \sigma^n_k)$. Our vector of parameters is then given by $\theta = [[\mu_k^n, \sigma_k^n, \mu_k^a, \sigma_k^a]^{k=1..N}, S ]$ where $S$ is the event ordering. Moreover, we further define $Z = [Z_1, Z_2, \dots, Z_P]$ a vector of (latent) discrete random variables representing the stage of each subject which can take values $0 \dots N$, where $N$ is the number of biomarkers. The dataset is denoted by $X$ where $x_{ij}$ represents the data from subject $i$ for biomarker $j$ while $X_i = [x_{i1}, \dots, x_{iN}]$ is a vector of biomarker data for subject $i$. 

\subsection{M-step}

In the M-step we aim to find the arguments $\theta^*$ that maximise the expected log-likelihood of the complete data $\theta^* = \argmax_{\theta} Q(\theta | \theta^{old})$. 

\beq
Q(\theta | \theta^{old}) = \mathbb{E}_{Z|X,\theta^{old}}[log p(X,Z|\theta)]
\eeq

Assuming a uniform prior on $Z$, i.e. $\log p(Z = z) = C$ and expanding $Z$ we get: 
\beq
Q(\theta | \theta^{old}) = C + \sum_{z_1 = 0}^N \dots \sum_{z_P = 0}^N p(Z_1 = z_1, \dots, Z_P = z_P|X, \theta^{old}) log\ p(X|Z_1 = z_1, \dots, Z_P = z_P, \theta)
\eeq

Under the EBM model, the data from each subject is conditionally independent given the parameters, .i.e $X_i \independent X_j|\theta$ and $X_i \independent Z_j|\theta$ for $i \neq j$. A similar independence also holds for the latent variables $Z_i$ given the parameters. We therefore get:
\beq
Q(\theta | \theta^{old}) = C + \sum_{z_1, \dots, z_P} \prod_{i=1}^P  p(Z_i = z_i|X_i, \theta^{old})\ log\left[ \prod_{i=1}^{P} p(X_i|Z_i = z_i, \theta) \right]
\eeq

where $P$ is the number of patients. We further factorise the latent variables $Z$ to obtain:

\beq
Q(\theta | \theta^{old}) = C + \sum_{i=1}^P \sum_{z_i} p(Z_i = z_i|X_i, \theta^{old})\ log\left[ \prod_{i=1}^{P} p(X_i|Z_i = z_i, \theta) \right]
\eeq

After moving the log inside the products and removing the constant $C$ we get:
\beq
Q(\theta | \theta^{old}) = \sum_{i=1}^P \sum_{z_i} p(Z_i = z_i|X_i, \theta^{old}) \left[ \sum_{j=1}^{z_i} log\ p(x_{ij}|E_{S(j)}) + \sum_{j=z_i + 1}^N log\ p(x_{ij}| \neg E_{S(j)}) \right]
\eeq

Replacing $p(x|E)$ and $p(x|\neg E)$ with the pdf of a Gaussian distribution we get:
\begin{multline} 
\label{eq:eStep}
Q(\theta | \theta^{old}) = \sum_{i=1}^P \sum_{z_i}  p(Z_i = z_i|X_i, \theta^{old})\ \sum_{i=1}^{P} \\ \left[ \sum_{j=1}^{z_i} log\ N(x_{ij}|\mu_{S(j)}^a, \sigma_{S(j)}^a) + \sum_{j=z_i + 1}^N log\ N(x_{ij}|\mu_{S(j)}^n, \sigma_{S(j)}^n) \right] \\
\end{multline}

The function $Q(\theta | \theta^{old})$ is differentiable with respect to all parameters apart from $S$ (which is discrete). We can thus find $\theta^*$ by solving $\nabla_{\theta}Q(\theta|\theta_{old}) = 0$. We show the derivation for parameter $\mu_k^n$, which is the solution of  $\frac{d}{d\mu_k^n}Q(\theta | \theta^{old}) = 0 $. Using the result from equation \ref{eq:eStep} and moving the derivation operator inside the sums we get:
\begin{multline}
  \frac{d}{d\mu_k^n}Q(\theta | \theta^{old}) = \sum_{i=1}^P \sum_{z_i}p(Z_i = z_i|X_i, \theta^{old})\ \\ \left[ \sum_{j=1}^{z_i}  \frac{d}{d\mu_k^n}log\ N(x_{ij}|\mu_{S(j)}^a, \sigma_{S(j)}^a) + \sum_{j=z_i + 1}^N \frac{d}{d\mu_k^n}log\ N(x_{ij}|\mu_{S(j)}^n, \sigma_{S(j)}^n) \right] = 0
\end{multline}

The derivative term cancels all likelihood terms apart from the one where $S(j) = k$:

\beq\sum_{i=1}^P \sum_{z_i}p(Z_i = z_i|X_i, \theta^{old})\  \left[ \sum_{j=z_i + 1}^N \mathbb{I}[S(j) = k] \frac{d}{d\mu_k^n}log\ N(x_{ij}|\mu_k^n, \sigma_k^n) \right] = 0
\eeq
which can be rewritten as:
\beq \sum_{i=1}^P \sum_{z_i}p(Z_i = z_i|X_i, \theta^{old})\  \left[ \frac{d}{d\mu_k^n}log\ N(x_{ik}|\mu_k^n, \sigma_k^n) \sum_{j=z_i + 1}^N \mathbb{I}[j = S^{-1}(k)] \right] = 0
\eeq
\beq \sum_{i=1}^P \sum_{z_i} p(Z_i = z_i|X_i, \theta^{old})\  \left[ \frac{d}{d\mu_k^n}log\ N(x_{ik}|\mu_k^n, \sigma_k^n) \mathbb{I}[S^{-1}(k) > z_i] \right] = 0
\eeq
Further rearranging the sum terms we get:

\beq \sum_{i=1}^{P} \frac{d}{d\mu_k^n}log\ N(x_{ik}|\mu_k^n, \sigma_k^n) \sum_{z_i = 0}^N \mathbb{I}[S^{-1}(k) > z_i]\ p(Z_i = z_i | X, \theta^{old})\ = 0
\eeq

\beq \sum_{i=1}^{P} \frac{d}{d\mu_k^n}log\ N(x_{ik}|\mu_k^n, \sigma_k^n) \ p(S^{-1}(k) > Z_i | X, \theta^{old})\ = 0
\eeq
Inserting the formula for the Gaussian pdf we get:
\beq \sum_{i=1}^{P} \frac{d}{d\mu_k^n}\frac{(x_{ik} - \mu_k^n)^2}{2(\sigma_k^n)^2} \ p(S^{-1}(k) > Z_i | X, \theta^{old})\ = 0
\eeq
which results in the update rule for $\mu_k^n$, the mean of $p(x|\neg E_k)$
\beq \mu_k^n = \sum_{i=1}^P x_{ik} w_i^n\eeq\\
where
\beq 
w_i^n = \frac{p(S^{-1}(k) > Z_i | X, \theta^{old})}{\sum_{i=1}^P \ p(S^{-1}(k) > Z_i | X, \theta^{old})}
\eeq
and
\beq
p(S^{-1}(k) > Z_i | X, \theta^{old}) = \sum_{l=S^{-1}(k)+1}^{K} p(Z_i = l | X, \theta^{old})
\eeq
Using a similar approach we get the update rules for $\sigma_k^n$, $\mu_k^a$, $\sigma_k^a$:
\beq \sigma_k^n = \sqrt{\sum_{i=1}^P w_i^n (x_{ik} - \mu_k^n)^2} \eeq \\
\beq \mu_k^a = \sum_{i=1}^P x_{ik} w_i^a \eeq\\
\beq \sigma_k^a = \sqrt{\sum_{i=1}^P w_i^a (x_{ik} - \mu_k^a)^2} \eeq \\
where
\beq w_i^a = \frac{p(S^{-1}(k) \leq Z_i | X, \theta^{old})}{\sum_{i=1}^P \ p(S^{-1}(k) \leq Z_i | X, \theta^{old})}\eeq

Solving for $S$ in the M-step is intractable, so we use MCMC sampling where at each step of the sampling process we propose a new sequence $S^{new}$, find the optimal distribution parameters for each biomarker given $S^{new}$ using the EM update rules and then evaluate the likelihood $Q(\theta | \theta^{old})$. The sequence and parameters that maximise the likelihood are chosen and the EM proceeds to a new iteration. Although this approach might not guarantee that we truly find the optimal parameters, it still results in an increase of $Q(\theta | \theta^{old})$. This approach, called generalised EM, still guarantees that the method will still converge to a local maxima \cite{bishop2007pattern}. For parameter initialisation, we use the mean and standard deviation of the control and patient populations. 

\subsection{E-step}

In the E-step, we simply estimate the latent disease stages $Z_i$ for every subject $i$. The probability $p(Z_i = l|X, \theta^{old})$ that subject $i$ is at stage $l$ in the abnormality sequence, conditioned on the previous parameters $\theta^{old}$, has a closed-form solution given by:

\begin{equation}
p(Z_i = l|X, \theta^{old}) = \frac{\prod_{j=1}^{l} N(x_{i,s(j)}|\mu_{S(j)}^a, \sigma_{S(j)}^a)\prod_{j=l + 1}^N log\ N(x_{i,s(j)}|\mu_{S(j)}^n, \sigma_{S(j)}^n)}{\sum_{m=0}^K \prod_{j=1}^{m} N(x_{i,s(j)}|\mu_{S(j)}^a, \sigma_{S(j)}^a)\prod_{j=m + 1}^N log\ N(x_{i,s(j)}|\mu_{S(j)}^n, \sigma_{S(j)}^n) }
\end{equation}

\chapter[TADPOLE Challenge: Prediction of Longitudinal Evolution in AD]{TADPOLE Challenge: Prediction of Longitudinal Evolution in Alzheimer's Disease}

\section{Expected Number of Subjects and Available Data for D4}
\label{app:expectedD4}

We estimated the number of subjects and available data in D4 (Table \ref{tab:biomk_data_available}, last column) using information from the ADNI procedures manual and previous ADNI rollovers. For estimating the total number of subjects (first row) expected in D4, we computed the dropout rate (0.36) based on ADNI1 rollovers to ADNI2, then multiplied it by the total number of subjects in D2 (896). For estimating the proportions of each diagnostic category (third row), we used the proportion of diagnostic rates in D2 and multiplied them with conversion rates within 1 year from ADNI1/GO/2 (see website FAQ). For estimating the average number of visits per subject (mean $\pm$ std.) in D4 (second row), we used the proportions for each diagnostic group and considered one visit per subject (ADNI procedures). We set the standard deviation to be zero, although in practice this won't be the case. 

For estimating the available biomarker data (lower half of table), we used a 1-year time-frame from start of ADNI2 (July 2012 -- July 2013) and computed the proportion of available data in that time frame. For AV1451, we used the same estimate as for AV45, due to the fact that the scan was introduced later on in ADNI2, and we expect more subjects to undergo AV1451 scans in ADNI3. A Python script that computes all the data from Table \ref{tab:biomk_data_available} is given in the TADPOLE repository: \url{https://github.com/noxtoby/TADPOLE/blob/master/statistics/tadpoleStats.py}.

\cleardoublepage

\nocite{*} 
\bibliographystyle{unsrt}
\bibliography{report}


\end{document}